%% file: HBO5main.tex
\documentclass[11pt,a4paper]{article}
\pdfoutput=1
\usepackage{jheppub}
\usepackage{amsthm,amsbsy,amsfonts,mathrsfs,enumerate,float,wrapfig,amsmath}
\usepackage[utf8]{inputenc}
\DeclareUnicodeCharacter{2212}{-}
\usepackage[section]{placeins}
\usepackage{color,colortbl,multirow,longtable,caption,standalone} 
\usepackage{subcaption}
\definecolor{Grayy}{gray}{0.9} 
\usepackage{array}
\newcolumntype{L}[1]{>{\raggedright\let\newline\\\arraybackslash\hspace{0pt}}m{#1}}
\newcolumntype{C}[1]{>{\centering\let\newline\\\arraybackslash\hspace{0pt}}m{#1}}
\newcolumntype{R}[1]{>{\raggedleft\let\newline\\\arraybackslash\hspace{0pt}}m{#1}}

\DeclareMathOperator{\car}{char}
\DeclareMathOperator{\Pfc}{Pfc}

\newcommand{\be}{\begin{equation}}
\newcommand{\ee}{\end{equation}}
\newcommand{\bea}{\begin{eqnarray}}
\newcommand{\eea}{\end{eqnarray}}
\newcommand{\ba}{\begin{align}}
\newcommand{\ea}{\end{align}}

\newcommand{\bF}{\mathbf{F}}

\newcommand{\HS}{\text{HS}}
\newcommand{\Tr}{\text{Tr}}
\newcommand{\SI}{\text{SI}}

\usepackage{tikz}
\usetikzlibrary{decorations.pathmorphing}
\usetikzlibrary{positioning}
\usetikzlibrary{decorations.pathreplacing}
\usetikzlibrary{arrows}
\usetikzlibrary{shapes, shapes.geometric, shapes.symbols, shapes.arrows, shapes.multipart, shapes.callouts, shapes.misc}
\tikzset{7brane/.style={circle, draw=black, fill=black,ultra thick,inner sep=1.5 pt, minimum size=1 pt,}, c/.default={4pt}}

\tikzset{big7brane/.style={circle, draw=black, fill=black,ultra thick,inner sep=2.5 pt, minimum size=1 pt,}, c/.default={4pt}}
\tikzset{u/.style={circle, draw=black, fill=white,inner sep=2 pt, minimum size=2 pt,},f/.style={square, draw=black, fill=white,ultra thick,inner sep=4 pt, minimum size=2 pt,}}

\tikzset{so/.style={circle, draw=black, fill=red,inner sep=2 pt, minimum size=2 pt,},f/.style={square, draw=black, fill=white,ultra thick,inner sep=4 pt, minimum size=2 pt,}}

\tikzset{sp/.style={circle, draw=black, fill=blue,inner sep=2 pt, minimum size=2 pt,},f/.style={square, draw=black, fill=white,ultra thick,inner sep=4 pt, minimum size=2 pt,}}

\tikzset{uf/.style={rectangle, draw=black, fill=white,inner sep=2.5 pt, minimum size=4 pt,}}

\tikzset{spf/.style={rectangle, draw=black, fill=blue,inner sep=2.5 pt, minimum size=4 pt,}}

\tikzset{sof/.style={rectangle, draw=black, fill=red,inner sep=2.5 pt, minimum size=4 pt,}}
\tikzset{snake it/.style={decorate, decoration=snake}}
\title{Five-brane webs, Higgs branches and unitary/orthosymplectic magnetic quivers}
\author[a]{Mohammad Akhond,}
\author[b]{Federico Carta,}
\author[c]{Siddharth Dwivedi,}
\author[d]{Hirotaka Hayashi,}
\author[e]{Sung-Soo Kim,}
\author[f]{and Futoshi Yagi}
\affiliation[a]{Department of Physics, Swansea University, \\
 Singleton Park, Swansea, SA2 8PP, U.K.}
\affiliation[b]{Deutches Electronen-Synchrotron, DESY,\\
 Notkestra\ss e 85, 22607 Hamburg, Germany
}
\affiliation[c]{Center for Theoretical Physics, College of Physical Science and Technology, Sichuan University, \\Chengdu, 610064, China}
\affiliation[d]{Department of Physics, School of Science, Tokai University,\\ 4-1-1 Kitakaname, Hiratsuka-shi, Kanagawa 259-1292, Japan}
\affiliation[e]{School of Physics, University of Electronic Science and Technology of China, \\
No.4, Section 2, North Jianshe Road, Chengdu, Sichuan 610054, China}
\affiliation[f]{School of Mathematics, Southwest Jiaotong University,\\ 
West zone, High-tech district, Chengdu, Sichuan 611756, China}
\emailAdd{akhondmohammad@gmail.com}
\emailAdd{federico.carta@desy.de}
\emailAdd{sdwivedi@scu.edu.cn}
\emailAdd{h.hayashi@tokai.ac.jp}
\emailAdd{sungsoo.kim@uestc.edu.cn}
\emailAdd{futoshi\_yagi@swjtu.edu.cn}

\abstract{We study the Higgs branch of 5d superconformal theories engineered from brane webs with orientifold five-planes. We propose a generalization of the rules to derive magnetic quivers from brane webs pioneered in \cite{Bourget:2020gzi}, by analyzing theories that can be described with a brane web with and without O5 planes. Our proposed magnetic quivers include novel features, such as hypermultiplets transforming in the fundamental-fundamental representation of two gauge nodes, antisymmetric matter, and $\mathbb{Z}_2$ gauge nodes. We test our results by computing the Coulomb and Higgs branch Hilbert series of the magnetic quivers obtained from the two distinct constructions and find agreement in all cases. 

}
\begin{document}
\preprint{DESY 20-128}
\maketitle
\input{sec1intro.tex}
\input{sec2examples}

\input{sec3conclusions}
\acknowledgments
We thank Julius Grimminger, Kimyeong Lee, and Dominik Miketa for useful discussions. We would like to especially thank Amihay Hanany for suggesting half-integer sum of orthosymplectic magnetic quivers and also useful discussions. SSK thanks the hospitality of KIAS, POSTECH, and Sichuan University, where part of work is done, and also APCTP for hosting the Focus program “Strings, Branes and Gauge Theories 2020.” MA is grateful for the hospitality of Sichuan University during the earlier stages of this project. 
MA is supported by an STFC grant ST/S505778/1.
The work of HH is supported in part by JSPS KAKENHI Grant Number JP18K13543. 
FY is supported by the NSFC grant No. 11950410490, by Fundamental Research Funds for the Central Universities A0920502051904-48, by Start-up research grant A1920502051907-2-046, 
in part by NSFC grant No. 11501470 and No. 11671328, and by Recruiting Foreign Experts Program No. T2018050 granted by SAFEA. FC is supported by the ERC Consolidator Grant STRINGFLATION under the HORIZON 2020 grant agreement no. 647995.
\appendix 
\input{appendixA.tex}
\input{charge2examples.tex}

\input{appendixD.tex}
\listoffigures
\listoftables
\clearpage
\bibliographystyle{JHEP}
\bibliography{ref}
\end{document}

%% file: sec1intro.tex
\section{Introduction}\label{sec:intro}
It has been known for some time that there are interacting UV fixed points of the renormalization group (RG) in five dimensions~\cite{Seiberg:1996bd,Intriligator:1997pq,Morrison:1996xf}. 
Many of these 
superconformal theories (SCFTs) admit a relevant deformation whose low energy dynamics is captured effectively by an $\mathcal{N}=1$ gauge theory, despite the fact that Yang-Mills (YM) interactions are power-counting non-renormalizable in five dimensions. A generic feature of such an RG flow is that the global symmetries of the fixed point theory are enhanced with respect to the manifest global symmetries of the gauge theory description, which has been confirmed by various observables such as superconformal indices, Nekrasov partition functions and topological vertex~\cite{Kim:2012gu, Bashkirov:2012re, Iqbal:2012xm, Rodriguez-Gomez:2013dpa, Taki:2013vka, Bergman:2013ala, Bergman:2013aca,Bergman:2014kza,Hwang:2014uwa,Zafrir:2014ywa, Bao:2013pwa,Kim:2015jba,Hayashi:2013qwa,Mitev:2014jza,Kim:2014nqa,Hayashi:2015xla,Hayashi:2015fsa,Hayashi:2017btw,Hayashi:2018lyv,Hayashi:2019jvx}. It was also argued from  
the presence of instanton operators \cite{Lambert:2014jna, Rodriguez-Gomez:2015xwa, Tachikawa:2015mha,Zafrir:2015uaa,Yonekura:2015ksa}, defined as defect operators, that have the charges associated with the topological current $J_I=\frac{1}{8\pi^2}\Tr*(F\wedge F).$ This current can mix with the flavour symmetries in the UV to form a larger symmetry group. In the case of a quiver gauge theory, there are as many topological currents as the number of gauge nodes. Their algebra may be promoted to a non-abelian one, often without mixing with flavour symmetries. 

The 5-dimensional (5d) $\mathcal{N}=1$ gauge theories admit an embedding into type IIB string theory which is realized as 5-brane webs 
\cite{Aharony:1997ju,Aharony:1997bh}. The 5-brane webs have been a powerful tool to study 5d SCFTs, as they not only provides an effective description of the SCFT at low energy, but also reveal rich non-perturbative aspects of the SCFTs such as global symmetry enhancement \cite{DeWolfe:1999hj, Gaberdiel:1997ud, Gaberdiel:1998mv} and various dualities including S-duality as well as novel UV-dualities \cite{Bergman:2013aca, Zafrir:2014ywa, Bergman:2014kza, Gaiotto:2015una, Hayashi:2015fsa, Zafrir:2015rga, Hayashi:2015zka, Hayashi:2015vhy, Jefferson:2018irk, Hayashi:2018lyv, Hayashi:2019jvx}.

Of particular significance to this story is the Higgs branch of the moduli space of the 5d theory. The Higgs branch is both sensitive to the symmetry enhancement and computable at all values of the YM coupling. It also undergoes other dramatic effects along the flow, such as the appearance of new flat directions at the UV fixed point. 

 A program to study Higgs branches of 5d theories by relating them to Coulomb branches of 3d $\mathcal{N}=4$ quiver gauge theories, henceforth referred to as \emph{magnetic quivers}, was initiated in \cite{Cremonesi:2015lsa, Ferlito:2017xdq}, following earlier work \cite{Benini:2009gi, Benini:2010uu} observing similar connections. For related work on magnetic quivers, also see \cite{Eckhard:2020jyr}. 5-brane webs also play an important role in constructing the 3d quiver gauge theories associated to the Higgs branch. 
 A set of rules were established in \cite{Cabrera:2018jxt, Cabrera:2019izd} to derive the magnetic quivers directly from the 5-brane web. In particular, the stable intersection number from the substructure of the 5-brane web at the Higgs branch captures the multiplicity of edges connecting nodes of the 3d quiver~\cite{Cabrera:2018jxt}. This was later extended to brane webs with O5-planes in \cite{Bourget:2020gzi}.

 In this paper we continue along this line of logic. We generalize the construction of magnetic quivers from O5-planes by adding new entries to the list of rules established in \cite{Bourget:2020gzi}. We examine theories that can be constructed both using ordinary brane webs, as well as brane webs with O5-planes. We verify the equivalence of the Coulomb branch of the magnetic quivers obtained from the two distinct constructions by a Hilbert series computation. We view the agreement in these computations as a non-trivial test of our conjectured rules. We organize our study according to the asymptotics of the brane configuration. It will be convenient to distinguish configurations by the asymptotic charges of the O5-plane. Within a given set of asymptotic O5-plane charges, we further divide theories according to $(p,q)$ charges of the asymptotic 5-branes. We use naming conventions for the various cases inspired by \cite{Bergman:2018hin}. Our new rules translate to appearance of new qualitative features in the magnetic quivers. This includes exotic bifundamental matter and matter in the 2nd rank tensor representations. Our rules are obtained by examining several 5d theories which can be constructed both using a brane web with an O5-plane as well as brane web without the orientifold. We achieve this by considering 5d Orthosymplectic (OSp) quivers with an S-dual description as $D_3=A_3$ type Dynkin quiver. Upon identifying deformation parameters of the ordinary web description with those of the orientifold web one can produce many daughter theories by deforming the two sides in an equivalent way. After the deformation the unitary webs may or may not admit a simple gauge theory description, though this is not important for our purposes. We can then derive magnetic quivers for the unitary web constructions following \cite{Cabrera:2018jxt}, which serve as a consistency check of our conjectured rules for the OSp magnetic quivers obtained from orientifold webs. 

Although the original motivations for this work are as above, our study also hints towards implications for the magnetic quivers, viewed as 3d $\mathcal{N}=4$ gauge theories. In order to verify our results we performed Hilbert series computations for both the Coulomb and Higgs branches of these theories. In all cases we found an agreement between the two computations. Together with the fact that the 5d origin of these theories is identical, one is tempted to conjecture that the two theories are dual as 3d $\mathcal{N}=4$ theories. However our analysis here is too simple to determine exactly in which sense the two theories are dual to each other. 
 
 The organization of the paper is as follows. We divide the content by asymptotic behavior of orientifold planes. In section \ref{sec:O5-O5-} we start from examples which come from 5-brane web diagrams with asymptotic O5$^-$-planes on both ends and obtain magnetic quivers from the configurations. In the course of obtaining the magnetic quivers we observe new rules. We will also compute the Hilbert series of Coulomb branches of these magnetic quivers and compare them with those which arise from ordinary web diagrams. Section \ref{sec:O5+O5+} considers cases where the configurations have O5$^+$-planes on both ends, and  section \ref{sec:O5-O5+} considers examples with an O5$^-$-plane on one end and an O5$^+$-plane on the other end. In section \ref{sec:O5tilde+}, we consider some cases which involve an $\widetilde{\text{O5}}^+$-plane in the diagrams. 
Finally we summarise our conclusions together with a set of open problems that we find are worth further investigation in section \ref{sec:conclusions}. Appendix \ref{app1} summarizes the method for computing the Hilbert series of Coulomb branches and Higgs branches. Appendix \ref{sec:app2} gives some details of the Coulomb branch Hilbert series in the main sections. In appendix \ref{sec:charge2ex} we give more support for the rule about the number of charge $2$ hypermultiplets given in section \ref{sec:O5-O5-}. Appendix \ref{appYN} summarizes more examples from brane configurations with O5$^-$-planes on both ends.

\noindent \paragraph{Notation.} To avoid the cluttering of the quiver diagrams, we will use a color coding to represent the unitary and orthosymplectic nodes as given below: 
\begin{equation}
\begin{array}{|c|c|c|c|} \hline
\rowcolor{Grayy}
 \text{Node type} & \text{U($n$)} & \text{SO($m$)} & \text{USp($2k$)} \\ \hline
\text{Gauge} & \begin{array}{c}\begin{tikzpicture}
\node[label=right:{$n$}][u](u){};
\end{tikzpicture}\end{array} & \begin{array}{c}\begin{tikzpicture}
\node[label=right:{$m$}][so](so){};
\end{tikzpicture}\end{array} & \begin{array}{c}\begin{tikzpicture}
\node[label=right:{$2k$}][sp](sp){};
\end{tikzpicture}\end{array} \\
\text{Flavor}  & \begin{array}{c}\begin{tikzpicture} \node[label=right:{$n$}][uf](u){}; \end{tikzpicture}\end{array} & \begin{array}{c}\begin{tikzpicture} \node[label=right:{$m$}][sof](so){}; \end{tikzpicture}\end{array} & \begin{array}{c}\begin{tikzpicture} \node[label=right:{$2k$}][spf](sp){}; \end{tikzpicture}\end{array} \\ \hline
\end{array} ~.
\label{colornodes}
\end{equation}
In the above, the circular nodes denote the gauge group while the square nodes represent a global (rather than gauge) symmetry group.  In this work, we will have three kinds of links connecting the nodes: solid line, dashed line and wavy line. These links transform under the representations of the nodes it connects with the following dictionary. 
\begin{equation}
\begin{array}{|c|l|} \hline
\rowcolor{Grayy}
 \text{Link type} & \text{Interpretation} \\ \hline
\begin{tikzpicture}\node[u](u1){};
\node[u](u1')[right of=u1]{};\draw(u1)--(u1'); \end{tikzpicture} & \text{hypermultiplet transforming in the bifundamental representation} \\

\begin{tikzpicture}\node[u](u1){};
\node[so](u1')[right of=u1]{};\draw(u1)--(u1'); \end{tikzpicture} & \text{hypermultiplet transforming in the bifundamental representation} \\

\begin{tikzpicture}\node[u](u1){};
\node[sp](u1')[right of=u1]{};\draw(u1)--(u1'); \end{tikzpicture} & \text{hypermultiplet transforming in the bifundamental representation} \\

\begin{tikzpicture}\node[so](u1){};
\node[sp](u1')[right of=u1]{};\draw(u1)--(u1'); \end{tikzpicture} & \text{half-hypermultiplet transforming in the bifundamental representation} \\
\begin{tikzpicture}\node[u](u1){};
\node[u](u1')[right of=u1]{};\draw[dashed](u1)--(u1'); \end{tikzpicture}  & \text{hypermultiplet in the fundamental-fundamental representation} \\
\begin{tikzpicture}\node[u](u1){};
\node[uf](u1')[right of=u1]{}; \path [draw,snake it](u1)--(u1');\end{tikzpicture} & \text{charge 2 hypermultiplet} \\ \hline
\end{array}
\label{typesoflinks}
\end{equation}
In order to avoid confusion, we will denote 5d (electric) quivers as $\cdot \cdot \cdot-G-G_j-\cdot\cdot\cdot$ and use square braces $[F]$ to denote flavor nodes.

%% file: sec2examples.tex
\section{\texorpdfstring{Magnetic quivers from O5$^-$ - O5$^-$}{TEXT}}
\label{sec:O5-O5-}
We first consider examples whose brane configurations are accompanied with two asymptotic O5$^-$-planes. 
\subsection{\texorpdfstring{The $\#_{M,N}$ theory}{TEXT}}
\begin{figure}[!htb]
    \centering
      \begin{scriptsize}
     \begin{tikzpicture}[scale=0.4]
     
    \draw[thick,dashed](1,0)--(29,0); 
    \node at (2.5,-1) {O5$^-$};
    \node at (27.5,-1) {O5$^-$};
    \draw[thick](3,1)--(5,1);
    \draw[thick](3,1.5)--(4.5,1.5);
    \draw[thick](3,2)--(27,2);
    \draw[thick](3,4)--(27,4);
    \draw[thick](4.5,5)--(4.5,1.5);
    \draw[thick](4.5,1.5)--(5,1);
    \draw[thick](5,1)--(7,0);
    \draw [thick,decorate,decoration={brace,amplitude=6pt},xshift=0pt,yshift=0pt]
(2,0.65) -- (2,4.5)node [black,midway,xshift=-25pt] {
$M$ D5};
    \draw[thick](8,0)--(10,1);
    \draw[thick](10,1)--(12,1);
    \draw[thick](10,1)--(10.5,1.5);
    \draw[thick](10.5,1.5)--(11.5,1.5);
    \draw[thick](10.5,1.5)--(10.5,2);
    \draw[thick](10.5,2)--(10.5,5);
    \draw[thick](11.5,5)--(11.5,1.5);
    \draw[thick](11.5,1.5)--(12,1);
    \draw[thick](12,1)--(14,0);
    \draw[thick](15,0)--(17,1);
    \draw[thick](17,1)--(19,1);
    \draw[thick](17,1)--(17.5,1.5);
    \draw[thick](17.5,1.5)--(18.5,1.5);
    \draw[thick](17.5,1.5)--(17.5,2);
    \draw[thick](17.5,2)--(17.5,5);
    \draw[thick](18.5,5)--(18.5,1.5);
    \draw[thick](18.5,1.5)--(19,1);
    \draw[thick](19,1)--(21,0);
    \node at (22,1.5) {$\dotsb$};
    \draw [thick,decorate,decoration={brace,amplitude=6pt},xshift=0pt,yshift=0pt]
(4,6) -- (26,6)node [black,midway,xshift=0pt,yshift=20pt] {
$2N$ NS5};
        
    \draw[thick](23,0)--(25,1);
    \draw[thick](25,1)--(27,1);
    \draw[thick](25,1)--(25.5,1.5);
    \draw[thick](25.5,1.5)--(27,1.5);
    \draw[thick](25.5,1.5)--(25.5,2);
    \draw[thick](25.5,2)--(25.5,5);
    \node at (22,2.5){$\cdot$};
    \node at (22,3){$\cdot$};
    \node at (22,3.5){$\cdot$};
    \end{tikzpicture}
    \end{scriptsize}
    \caption[5-brane web for the $\#_{M,N}$ theory.]{5-brane web for the $\#_{M,N}$ theory.}
    \label{fig:+NM}
\end{figure}
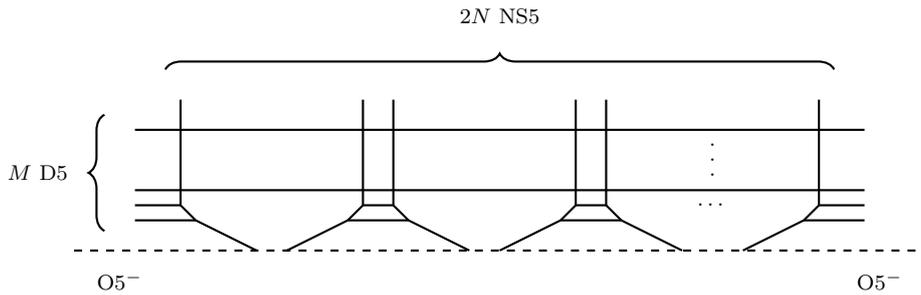
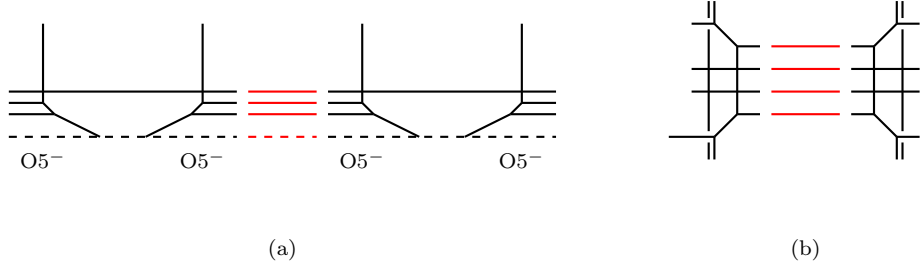
\begin{figure}[!htb]
    \centering
         \begin{scriptsize}
     \begin{tikzpicture}[scale=0.3]
     
    \draw[thick,dashed](3,0)--(13,0); 
    \node at (4.5,-1) {O5$^-$};
    \node at (11.5,-1) {O5$^-$};
    \draw[thick](3,1)--(5,1);
    \draw[thick](3,1.5)--(4.5,1.5);
    \draw[thick](3,2)--(13,2);
    \draw[thick](4.5,5)--(4.5,1.5);
    \draw[thick](4.5,1.5)--(5,1);
    \draw[thick](5,1)--(7,0);
    \draw[thick](9,0)--(11,1);
    \draw[thick](11,1)--(13,1);
    \draw[thick](11,1)--(11.5,1.5);
    \draw[thick](11.5,1.5)--(13,1.5);
    \draw[thick](11.5,1.5)--(11.5,2);
    \draw[thick](11.5,2)--(11.5,5);
    \draw[thick,red,dashed](13.5,0)--(16.5,0);
    \draw[thick,red](13.5,2)--(16.5,2);
    \draw[thick,red](13.5,1.5)--(16.5,1.5);
    \draw[thick,red](13.5,1)--(16.5,1);
    
    \draw[thick,dashed](17,0)--(27,0); 
    \node at (18.5,-1) {O5$^-$};
    \node at (25.5,-1) {O5$^-$};
    \draw[thick](17,1)--(19,1);
    \draw[thick](17,1.5)--(18.5,1.5);
    \draw[thick](17,2)--(27,2);
    \draw[thick](18.5,5)--(18.5,1.5);
    \draw[thick](18.5,1.5)--(19,1);
    \draw[thick](19,1)--(21,0);
    \draw[thick](23,0)--(25,1);
    \draw[thick](25,1)--(27,1);
    \draw[thick](25,1)--(25.5,1.5);
    \draw[thick](25.5,1.5)--(27,1.5);
    \draw[thick](25.5,1.5)--(25.5,2);
    \draw[thick](25.5,2)--(25.5,5);

    \node at (15,-5){(a)};
    \end{tikzpicture}
    \end{scriptsize}
    \hspace{1cm}
    \begin{scriptsize}
   \begin{tikzpicture}[scale=0.3]
    \draw[thick](2,1)--(3,1);
    \draw[thick](.75,.25)--(.75,4.75);
    \draw[thick](3,2)--(0,2);
    \draw[thick](3,3)--(0,3);
    \draw[thick](2,4)--(3,4);
    \draw[thick](2,1)--(2,4);
    \draw[thick](2,1)--(1,0);
    \draw[thick](2,4)--(1,5);
    \draw[thick](1,0)--(1,-1);
    \draw[thick](.75,-0.25)--(.75,-1);
    \draw[thick](1,0)--(-1,0);
    \draw[thick](1,5)--(0,5);
    \draw[thick](1,5)--(1,6);
    \draw[thick](.75,5.25)--(.75,6);
    \draw[thick,red](3.5,1)--(6.5,1);
        \draw[thick,red](3.5,2)--(6.5,2);
            \draw[thick,red](3.5,3)--(6.5,3);
                \draw[thick,red](3.5,4)--(6.5,4);
   \draw[thick](7,1)--(8,1);
   \draw[thick](8,1)--(9,0);
   \draw[thick](8,1)--(8,4);
   \draw[thick](8,4)--(9,5);
    \draw[thick](8,4)--(7,4);
    \draw[thick](7,2)--(10,2);
    \draw[thick](7,3)--(10,3);
    \draw[thick](9.25,.25)--(9.25,4.75);
    \draw[thick](9,5)--(9,6);
    \draw[thick](9.25,5.25)--(9.25,6);
    \draw[thick](9,5)--(10,5);
    \draw[thick](9,0)--(9,-1);
    \draw[thick](9.25,-0.25)--(9.25,-1);
    \draw[thick](9,0)--(10,0);
    
    \node at (5,-5) {(b)};
    \end{tikzpicture}
    \end{scriptsize}
    \caption[Obtaining web diagrams without O5-planes via gauging.]{(a) Gluing together 2 copies of $\text{USp}(2)+6\bF$ by gauging a common $\mathfrak{so}(6)$ subalgebra of their global symmetry. (b) Gluing together 2 copies of $\text{SU}(2)+6\bF$ by gauging a common $\mathfrak{su}(4)$ subalgebra of their global symmetry.}
    \label{fig:gauging su4=so6}
\end{figure}
\begin{figure}[!htb]
    \centering
    \begin{scriptsize}
    \begin{tikzpicture}[scale=.8]
    \draw[thick](0,0)--(8.5,0);
    \node[7brane]at(0,0){};
    \node[7brane]at(1,0){};
    \node[7brane]at(2,0){};
    \node[7brane]at(3,0){};
    \node[7brane]at(5.5,0){};
    \node[7brane]at(6.5,0){};
    \node[7brane]at(7.5,0){};
    \node[7brane]at(8.5,0){};
    \node at (.5,-.25){1};
    \node at (1.5,-.25){2};
    \node at (2.5,-.25){3};
    \node at (3.5,-.25){4};
    \node at (8,.25){1};
    \node at (7,.25){2};
    \node at (6,.25){3};
    \node at (5,.25){4};
    
    \draw[thick](4.25,-2)--(4.25,2);
    \node[7brane]at(4.25,-2){};
    \node[7brane]at(4.25,-1){};
    \node[7brane]at(4.25,1){};
    \node[7brane]at(4.25,2){};
    \node at (4.25,2.4){$.$};
    \node at (4.25,2.5){$.$};
    \node at (4.25,2.6){$.$};
    \node at (4.25,-2.4){$.$};
    \node at (4.25,-2.5){$.$};
    \node at (4.25,-2.6){$.$};
    \draw[thick](4.25,3)--(4.25,5);
    \draw[thick](4.25,-3)--(4.25,-5);
    \node[7brane]at(4.25,3){};
    \node[7brane]at(4.25,4){};
    \node[7brane]at(4.25,5){};
    \node[7brane]at(4.25,-3){};
    \node[7brane]at(4.25,-4){};
    \node[7brane]at(4.25,-5){};
    \node at(3.9,.7){$2N$};
    \node at(3.5,1.5){$2N-2$};
    \node at(3.9,3.5){$4$};
    \node at(3.9,4.5){$2$};
    \node at(4.7,-.7){$2N$};
    \node at(5,-1.5){$2N-2$};
    \node at(4.5,-3.5){$4$};
    \node at(4.5,-4.5){$2$};

    \end{tikzpicture}
    \end{scriptsize}
    \caption[Unitary web diagram of  $\#_{3,N}$ theory.]{A unitary web realization of the $\#_{3,N}$ theory. We depict here the web at the fixed point. Black dots represent 7-branes.}
    \label{fig:unitary web for +3M}
\end{figure}
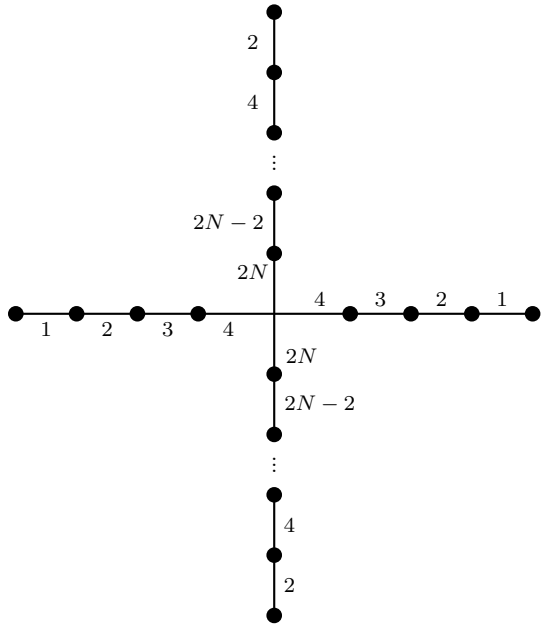

\paragraph{}
The first example we consider is the brane configuration obtained by intersecting $M$ D5 and $2N$ NS5 branes on top of an O5-plane, which in this section we take to be asymptotically an O5$^-$-plane. We call the theory on the web the $\#_{M,N}$ theory. The brane web for this theory is depicted in Figure \ref{fig:+NM}. The effective theory on the Coulomb branch is a linear orthosymplectic quiver
\be\label{electric +NM OSp}
\begin{array}{c}\begin{tikzpicture}
\node {$[M]-\text{USp}(2M-4)-\text{SO}(2M)-\text{USp}(2M-4)-\cdot\cdot\cdot-\text{SO}(2M)-\text{USp}(2M-4)-[M]$};
 \draw [thick,decorate,decoration={brace,amplitude=6pt},xshift=0pt,yshift=10pt]
(-5.75,0) -- (5.75,0)node [black,midway,xshift=0pt,yshift=20pt] {
$2N-1$};
\end{tikzpicture}
\end{array}
\ee 
The corresponding magnetic quiver was in fact already derived in \cite{Bourget:2020gzi}. We will not repeat the steps here and simply recall that it is given by
\be\label{OSp magnetic +NM}
\begin{array}{c}\begin{scriptsize}
\begin{tikzpicture}
\node[label=below:{2}][so](so2){};
\node[label=above:{2}][sp](sp2)[right of=so2]{};
\node[label=below:{$2M-2$}][so](so2n-2)[right of=sp2]{};
\node[label=above:{$2M-2$}][sp](sp2n-2)[right of=so2n-2]{};
\node[label=below:{$2M$}][so](so2n)[right of=sp2n-2]{};
\node[label=above:{$2M-2$}][sp](sp2n-2')[right of=so2n]{};
\node[label=below:{$2M-2$}][so](so2n-2')[right of=sp2n-2']{};
\node[label=above:{2}][sp](sp2')[right of=so2n-2']{};
\node[label=below:{2}][so](so2')[right of=sp2']{};
\node[label=left:{$2N$}][sp](sp2m)[above of=so2n]{};
\node[label=left:{$2N-1$}][u](u2m-1)[above of=sp2m]{};
\node[label=left:{2}][u](u2)[above of=u2m-1]{};
\node[label=left:{1}][u](u1)[above of=u2]{};
\draw(so2)--(sp2);
\draw[dotted](sp2)--(so2n-2);
\draw(so2n-2)--(sp2n-2);
\draw(sp2n-2)--(so2n);
\draw(so2n)--(sp2n-2');
\draw(sp2n-2')--(so2n-2');
\draw[dotted](so2n-2')--(sp2');
\draw(sp2')--(so2');
\draw(u1)--(u2);
\draw[dotted](u2)--(u2m-1);
\draw(u2m-1)--(sp2m);
\draw(sp2m)--(so2n);
\end{tikzpicture}
\end{scriptsize}
\end{array}
\ee
The 5d theory admits an S-dual description, as a $D$-type Dynkin quiver of special-unitary nodes~\cite{Zafrir:2016jpu,Hayashi:2015vhy}. In the special case when $M=3$, the S-dual theory on the Coulomb branch is
\be\label{electric +3M unitary}
\begin{array}{c}\begin{tikzpicture}
\node (horizontal){$\text{SU}(N)-\text{SU}(2N)-\text{SU}(N)$};
\node (flavour)[above of=horizontal]{$[2N]$};
\draw(flavour)--(horizontal);
\end{tikzpicture}
\end{array}
\ee
which can also be engineered via an ordinary web diagram, without an O5-plane. One way to see this is to consider gluing together $N$ copies of $\text{USp}(2)+6\bF$, by successive gauging $\mathfrak{so}(6)$ subalgebra of the flavour symmetry. This should be equivalent to gluing together $N$ copies of $\text{SU}(2)+6\bF$ by gauging $\mathfrak{su}(4)$ subalgebra of the global symmetry. Then we perform S-duality and the diagram yields the theory in \eqref{electric +3M unitary}. See Figure \ref{fig:gauging su4=so6} for the $N=2$ example. We will make use of this construction to obtain web diagrams without O5-planes. Furthermore there are various ways to realize $\text{USp}(2)$ gauge theory with six flavors depending on how we attach flavors to the diagram. Depending on situations we will use useful diagrams of $\text{SU}(2)+6\bF$ for the $\mathfrak{su}(4)$ gauging. 
The unitary web diagram for the $\#_{3,N}$ theory is shown in the Figure \ref{fig:unitary web for +3M}. 

Given this diagram, we can immediately obtain the magnetic quiver using the rules in \cite{Cabrera:2018jxt}. We are thus led to claim the equivalence of the Coulomb branch of \eqref{OSp magnetic +NM}, for $M=3$ with the following unitary magnetic quiver.

\be\label{+3,M unitary magnetic}
   \begin{array}{c}\begin{scriptsize}
       \begin{tikzpicture}
       \node[label=below:{2}][u](2) at (0,0){};
       \node[label=below:{4}][u](4)[right of=2]{};
       \node[label=below:{$2N-2$}][u](2k-2)[right of=4]{};
       \node[label=below:{$2N$}][u](2k)[right of=2k-2]{};
       \node[label=below:{$2N-2$}][u](2k-2')[right of=2k]{};
       \node[label=below:{4}][u](4')[right of=2k-2']{};
       \node[label=below:{2}][u](2')[right of=4']{};
       \node[label=above:{4}][u](44)[above of=2k]{};
       \node[label=above:{3}][u](333)[left of=44]{};
       \node[label=above:{2}][u](222)[left of=333]{};
       \node[label=above:{1}][u](111)[left of=222]{};
       \node[label=above:{3}][u](333')[right of=44]{};
       \node[label=above:{2}][u](222')[right of=333']{};
       \node[label=above:{1}][u](111')[right of=222']{};
       \draw(2)--(4);
       \draw[dotted](4)--(2k-2);
       \draw(2k-2)--(2k);
       \draw(2')--(4');
       \draw[dotted](4')--(2k-2');
       \draw(2k-2')--(2k);
       \draw(2k)--(44);
       \draw(44)--(333);
       \draw(333)--(222);
       \draw(222)--(111);
       \draw(44)--(333');
       \draw(333')--(222');
       \draw(222')--(111');
       \end{tikzpicture}
    \end{scriptsize}\end{array}
\ee
Both \eqref{+3,M unitary magnetic} and the $M=3$ case of \eqref{OSp magnetic +NM} hint at an enhanced $\text{SU}(2N)\times \text{SU}(4)^2$ flavour symmetry, which can be read off from the balanced nodes \cite{Gaiotto:2008ak}\footnote{We recall the balance condition for $\text{U}(r)$, $\text{USp}(2r)$ and $\text{SO}(m)$ is $n_f=2r$, $n_f=2r+1$ and $n_f=m-1$ respectively where $n_f$ is the number of effective flavors. From \cite{Gaiotto:2008ak}, chain of $p$ balanced alternating orthosymplectic nodes give rise to an $\text{SO}(p+1)$ isometry on the CB, which is further enhanced to $\text{SO}(p+2)$ if there is an $\text{SO}(2)$ node at the end of the chain. A set of $p$ balanced unitary nodes which form an ADE Dynkin diagram give rise to a symmetry of the corresponding type.}. The Coulomb and Higgs branch dimension of both quivers are also in agreement. A further non-trivial check of our discussion is the agreement of the Hilbert series, which we have explicitly computed for low values of $N$. For $N=1$, the unitary magnetic quiver in (\ref{+3,M unitary magnetic}) is well known with the Coulomb branch having $E_7$ as the enhanced global symmetry. We have tabulated the Coulomb branch Hilbert series for the unitary and the orthosymplectic magnetic quivers derived from the unitary and orientifold webs of $\#_{3,N}$ theory in Table  \ref{O5-O5-hashMNCoulombHS} for some small values of $N$. Note that the Hilbert series for $N=1$ is already known (\cite{zhongmaster:2018, Bourget:2020xdz}).
\begin{table}[!htb]
\centering
\begin{tabular}{|c|C{4.25cm}|C{4.25cm}|C{4.25cm}|} \hline
\rowcolor{Grayy}
   & Unitary magnetic quiver & \multicolumn{2}{c|}{Orthosymplectic magnetic quiver} \\ \cline{2-4}
	\rowcolor{Grayy}
  \multirow{-2}{*}{$\#_{3,N}$} & HS($t$)  & HS($t;\vec{m} \in \mathbb{Z}$) & HS($t;\vec{m} \in \mathbb{Z}+\tfrac{1}{2}$)  \\ \hline
		$\#_{3,1}$ & \footnotesize{$\begin{array}{l} \dfrac{(1+t)\,P_0(t)}{(1-t)^{34}} \\\\ = 1 + 133 t + 7371 t^2 \\+ 238602 t^3 + 5248750 t^4\\ + 85709988 t^5+\cdots  
	\end{array}$} & \footnotesize{$\begin{array}{l} \dfrac{P_1(t)}{(1-t)^{34}\, (1+t)^{17}} \\\\ = 1 + 69 t + 3723 t^2 + 119434 t^3\\ + 2625390 t^4 + 42857892 t^5+\cdots  
	\end{array}$} & \footnotesize{$\begin{array}{l} \dfrac{P_2(t)}{(1-t)^{34}\, (1+t)^{17}} \\\\ = 64 t + 3648 t^2 + 119168 t^3\\ + 2623360 t^4 + 42852096 t^5+\cdots  
	\end{array}$} \\ \hline
	$\#_{3,2}$ & \footnotesize{$1+45t+1277t^2+27399t^3+476864t^4+6979468t^5+87938113t^6+\cdots$} & \footnotesize{$1+45t+1085t^2+18951t^3+280320t^4+3739084t^5+45180033t^6+\cdots$} & \footnotesize{$192t^2+8448t^3+196544t^4+3240384t^5+42758080t^6+\cdots$}  \\ \hline
\end{tabular}
\caption[Coulomb branch HS for magnetic quivers of $\#_{3,N}$ theory.]{Coulomb branch Hilbert series of the unitary and orthosymplectic magnetic quivers for the $\#_{3,N}$ theory. The corresponding quivers are given in (\ref{+3,M unitary magnetic}) and \eqref{OSp magnetic +NM} respectively. For orthosymplectic quivers, we need to add the contributions of both integer and half integer fluxes. 
The explicit forms of the numerators $P_0(t), P_1(t), P_2(t)$ are provided in Appendix \ref{sec:app2}.}
\label{O5-O5-hashMNCoulombHS}
\end{table}
\subsection{\texorpdfstring{The $K_N$ family}{TEXT}}
\label{sec:KNO5-O5-}
\paragraph{}Decoupling flavors from, say, the rightmost gauge node in the $\#_{3,N}$ theory \eqref{electric +NM OSp}, we obtain a family of theories which we denote by $K_N^p$, where $p$ denotes the number of decoupled flavors. This family enjoys an IR quiver description as
\be\label{electric KNp OSp}
\begin{array}{c}\begin{tikzpicture}
\node {$[3]-\text{USp}(2)-\text{SO}(6)-\text{USp}(2)-\cdots-\text{SO}(6)-\text{USp}(2)-[3-p]$};
 \draw [thick,decorate,decoration={brace,amplitude=6pt},xshift=0pt,yshift=10pt]
(-4.75,0) -- (4.0,0)node [black,midway,xshift=0pt,yshift=20pt] {
$2N-1$};
\end{tikzpicture}
\end{array}
\ee 
Once again, it is possible to write down an ordinary web diagram for this theory, following a gluing procedure similar to Figure \ref{fig:gauging su4=so6}.\footnote{The cautious reader may be concerned about the non-uniqueness of this gauging procedure which is related to the Chern-Simons level of the gauging. One can remove the ambiguity by demanding that the OSp magnetic quiver agrees with the unitary quiver obtained from the unitary web after gauging. It is also possible to reproduce the same unitary web more rigorously by identifying the map between the deformation parameters in the orientifold and unitary web of the $\#_{3,N}$ theory.} We present the orientifold and the unitary web diagrams for the family $K_N^p$ for various number of decoupled flavors which can be found in the figures mentioned below.
\begin{equation*}
\begin{tabular}{|c|c|c|} \hline
\rowcolor{Grayy}
Theory & Orientifold web & Unitary web \\
$K_N^1$ & Figure \ref{fig:K3N1 O5 web} & Figure \ref{fig:KN1 ordinary web} \\
$K_N^2$ & Figure \ref{fig:K3N2 O5 web} & Figure \ref{fig:K3N2 unitary web} \\
$K_N^3$ & Figure \ref{fig:K3N3 O5 web} & Figure \ref{fig:KN3 unitary} \\ \hline
\end{tabular}
\end{equation*}
Here, we note that these are not the only possible subdivisions. We list some examples of subdivisions and their corresponding magnetic quivers. In this paper, our focus is on extracting the rules rather than an exhaustive analysis of the Higgs branch, so we consider some of the Higgs branches rather than exhausting all the branches.\footnote{We thank the authors of \cite{vanBeest:2020kou} for informing us that they were able to find some missing cones using their computer program.}

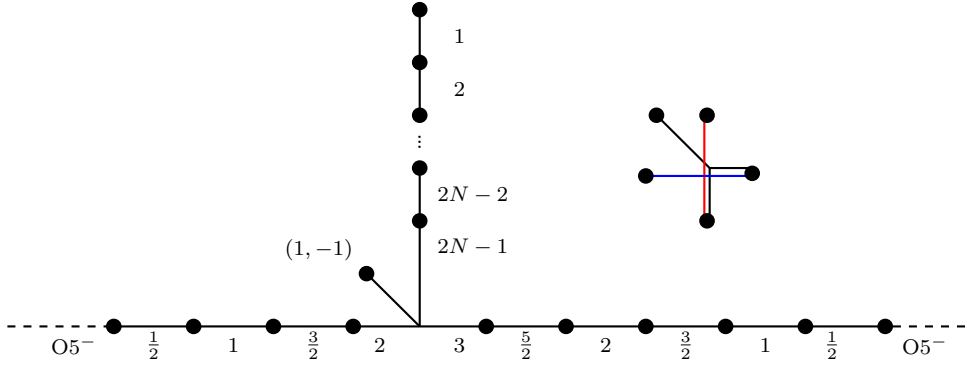
\begin{figure}[!htb]
    \centering
        \begin{scriptsize}
       \begin{tikzpicture}[scale=.7]
       \draw[thick](-6,0)--(8.5,0);
       \draw[thick,dashed](-8,0)--(-6,0);
       \draw[thick,dashed](10,0)--(8.5,0);
       \node[7brane] at (7,0){};
       \node[7brane] at (8.5,0){};
       \node[7brane] at (5.5,0){};
       \node[7brane] at (-6,0){};
       \node[7brane] at (4,0){};
       \node[7brane] at (2.5,0){};
       \node[7brane] at (1,0){};
       \node[7brane] at (-1.5,0){};
       \node[7brane] at (-3,0){};
       \node[7brane] at (-4.5,0){};
       \node[thick] at (-.25,3.4) {$.$};
       \node[thick] at (-.25,3.5) {$.$};
       \node[thick] at (-.25,3.6) {$.$};
       \node[7brane]at (-0.25,2){};
       \node[7brane]at (-.25,3){};
       \node[7brane]at (-.25,4){};
       \node[7brane]at (-.25,5){};
       \node[7brane]at (-.25,6){};
       \draw[thick](-.25,0)--(-.25,3);
       \draw[thick](-.25,4)--(-.25,6);
       \draw[thick](-.25,0)--(-1.25,1);
       \node at (0.5,5.5){1};
       \node at (0.5,4.5){2};
       \node at (.75,2.5){$2N-2$};
       \node at (.75,1.5){$2N-1$};
    
       \node[label=above left:{$(1,-1)$}][7brane]at(-1.25,1){};
       \node at (-6.75,-.35){O$5^-$};
       \node at (9.25,-.35){O$5^-$};
       \node at (-5.25,-.35){$\frac{1}{2}$};
       \node at (-3.75,-.35){1};
       \node at (-2.25,-.35){$\frac{3}{2}$};
       \node at (-1,-.35){2};
       \node at (.5,-.35) {3};
       \node at (1.75,-.35){$\frac{5}{2}$};
       \node at (3.25,-.35){$2$};
       \node at (4.75,-.35){$\frac{3}{2}$};
       \node at (6.25,-.35){$1$};
       \node at (7.5,-.35){$\frac{1}{2}$};
       
       \draw[thick](5.2,3)--(6,3);
       \draw[thick](5.2,3)--(5.2,2);
       \draw[thick](5.2,3)--(4.2,4);
       \draw[thick,red](5.1,2)--(5.1,4);
       \draw[thick,blue](4,2.85)--(6,2.85);
       \node[7brane]at(6,2.9){};
       \node[7brane]at(4.2,4){};
       \node[7brane]at(4,2.85){};
       \node[7brane]at(5.151,2){};
       \node[7brane]at(5.151,4){};
       \end{tikzpicture}
       \end{scriptsize}
    \caption[Orientifold web diagram of $\text{K}_{N}^{1}$ theory.]{An orientifold web for the $\text{K}_{N}^{1}$ theory and the maximal subdivision at the centre of the junction.}
    \label{fig:K3N1 O5 web}
\end{figure}

For reading off the magnetic quivers from the unitary web diagrams in Figures \ref{fig:KN1 ordinary web}, \ref{fig:K3N2 unitary web}, \ref{fig:KN3 unitary}, we can use the rules established in \cite{Cabrera:2018jxt}. For the magnetic quivers originated from the orientifold web diagrams in Figures \ref{fig:K3N1 O5 web}, \ref{fig:K3N2 O5 web}, \ref{fig:K3N3 O5 web}, a large part of the magnetic quivers can be obtained by the rules in \cite{Bourget:2020gzi}, but in fact, some part already requires an extension of the rule. In \cite{Bourget:2020gzi}, it has been argued that a subweb associated with a U(1) gauge node in a magnetic quiver which passes through the O5$^-$-plane may have charge $2$ hypermultiplets coupled to the U(1). Such a subweb appears in the $K_N^p \; (p=1,2,3)$ family at the center of the junction in the orientifold diagrams, and it is depicted as the subweb in black in each maximal subdivision in Figures \ref{fig:K3N1 O5 web}, \ref{fig:K3N2 O5 web}, \ref{fig:K3N3 O5 web}. 

The subweb configuration of the maximal subdivision in Figure \ref{fig:K3N1 O5 web} has already appeared in \cite{Bourget:2020gzi}, for example, for the magnetic quiver of the rank-1 $E_6$ theory. In this case, the number of the charge $2$ hypermultiplets attached to the U(1) node is zero. For the subweb configurations of the $K_N^2$ and $K_N^3$ theories, we find that the number of the charge $2$ hypermultiplets is zero and one respectively to match the Coulomb branch Hilbert series for their magnetic quiver theories with the Coulomb branch Hilbert series for the corresponding unitary magnetic quivers. Based on these examples as well as the other examples which we will see later, we observe that the number of the charge $2$ hypermultiplets may be counted by 
\begin{equation}\label{SI-charge-2}
\frac{\text{SI of subweb with its own mirror image}}{2} - \text{SI of subweb with O5$^-$},
\end{equation}
where SI represents the stable intersection number discussed in \cite{Cabrera:2018jxt}.\footnote{In this paper, we use ``SI'' to denote the generalized stable intersection number, which includes the contribution from the common 7-branes, for simplicity. We call usual stable intersection without the contribution from the common 7-branes as ``bare SI''.} 

Let us then illustrate how the rule \eqref{SI-charge-2} works for the subwebs of the $K_N^2$ and $K_N^3$ theories. From the maximal subdivision in Figure \ref{fig:K3N2 O5 web}, the subweb in black at the center of the junction yields a U(1) gauge node. The stable intersection number of the subweb with its own mirror image is given by
\begin{equation}\label{SIwithmirrorKN2}
\text{SI of subweb with its own mirror image} = 4 - 2 = 2.
\end{equation}
On the other hand, the stable intersection number of the subweb with O5 needs some care. The subweb configuration with the orientifold is depicted in \eqref{subweb_KN2}.
\begin{equation}\label{subweb_KN2}
\begin{array}{c}
\begin{scriptsize}
\begin{tikzpicture}
            \draw[thick](-1.45,0.05)--(-1.5,0.1);
            \draw[thick](-1.5,0.1)--(0,0.1);
            \draw[thick](-1.45,0.05)--(1.5,0.05);
            \draw[thick](-1.45,0.05)--(-1.45,-1);
            \draw[thick](-1.5,0.1)--(-3,0.85);
            \draw[thick,dashed](-3,0)--(6,0);
            \node[7brane] at (0,0){};
            \node[7brane] at (1.5,0){};
            \node[7brane] at (3,0){};
            \node at (-2.25,-.35){O$5^-$};
            \node at (0.75,-.35){$\widetilde{\text{O5}}^-$};
            \node at (2.25,-.35){O$5^-$};            
            \end{tikzpicture}\end{scriptsize}
\end{array}
\end{equation}
Note here that the RR charge of O5$^-$-plane is $-1$ and that of $\widetilde{\text{O5}}^-$-plane is $-\frac{1}{2}$ due to the half D5-brane. Then the stable intersection number of the subweb with the O5-planes becomes
\begin{equation}\label{SIwithO5KN2}
\text{SI of subweb with O5$^-$} = \left(1 - 1\right) + \left(\frac{1}{2}  - \frac{1}{2}\right) + 1 = 1
\end{equation}
The first bracket in \eqref{SIwithO5KN2} is the stable intersection number between the left O5$^-$-plane and the subweb in \eqref{subweb_KN2}, the second bracket in \eqref{SIwithO5KN2} is the stable intersection number between the left $\widetilde{\text{O5}}^-$-plane and the subweb in \eqref{subweb_KN2}, and the last $1$ is the stable intersection number between the right O5$^-$-plane and the subweb in \eqref{subweb_KN2}. Namely we consider the net contribution of the stable intersection numbers between the subweb and each piece of the orientifold. Putting together the result of \eqref{SIwithmirrorKN2} and \eqref{SIwithO5KN2}, the \eqref{SI-charge-2} becomes
\begin{equation}
\frac{2}{2} - 1 = 0,
\end{equation}
which is the right number of the charge $2$ hypermultiplet coupled to the $U(1)$ gauge node associated to the subweb in \eqref{subweb_KN2}. 

We can also do the same computation for the subweb in the maximal subdivision at the center of the junction in the $K_N^3$ theory depicted in Figure \ref{fig:K3N3 O5 web}. The stable intersection number of the subweb with its own mirror is given by 
\begin{equation}\label{SIwithmirrorKN3}
\text{SI of subweb with its own mirror image} = 6 - 3 = 3.
\end{equation}
For computing the stable intersection number of the subweb with O5$^-$, we consider the configuration around the subweb depicted in \eqref{subweb_KN3}. 
\begin{equation}\label{subweb_KN3}
\begin{array}{c}
\begin{scriptsize}
\begin{tikzpicture}
            \draw[thick](3,-0.05)--(-1.3,-0.05); 
            \draw[thick](-1.4,0.05)--(-1.3,-0.05);
            \draw[thick](-1.4,0.05)--(-1.5,0.1);
            \draw[thick](-1.5,0.1)--(0,0.1);
            \draw[thick](-1.4,0.05)--(1.5,0.05);
            \draw[thick](-1.3,-0.05)--(-1.3,-1);
            \draw[thick](-1.5,0.1)--(-3,0.6);
            \draw[thick,dashed](-3,0)--(6,0);
            \node[7brane] at (0,0){};
            \node[7brane] at (1.5,0){};
            \node[7brane] at (3,0){};
            \node at (-2.25,-.35){O$5^-$};
            \node at (0.75,-.35){$\widetilde{\text{O5}}^-$};
            \node at (2.25,-.35){O$5^-$};  
            \node at (3.75,-.35){$\widetilde{\text{O5}}^-$};            
            \end{tikzpicture}\end{scriptsize}
\end{array}
\end{equation}
Then the stable intersection number of the subweb with the O5$^-$-planes becomes
\begin{equation}\label{SIwithO5KN3}
\text{SI of subweb with O5$^-$} = \left(1 - 1\right) + \left(\frac{1}{2}  - \frac{1}{2}\right) + \left(1 - 1\right) + \frac{1}{2} = \frac{1}{2}.
\end{equation}
Hence the number of the charge $2$ hypermultiplets counted by \eqref{SI-charge-2} is 
\begin{equation}
\frac{3}{2} - \frac{1}{2} = 1,
\end{equation}
which is the correct number for the charge $2$ hypermultiplets coupled to the U(1) gauge node associated to the subweb in \eqref{subweb_KN3}. The rule \eqref{SI-charge-2} also works for the subweb in the maximal subdivision at the center of the junction in the $K_N^1$ theory depicted in Figure \ref{fig:K3N1 O5 web}. 

The other parts of the magnetic quivers can be obtained from the rules established in \cite{Cabrera:2018jxt,Bourget:2020gzi}. We summarize the unitary and the orthosymplectic magnetic quiver theories derived from the unitary and the orientifold web diagrams of $K_N^p$ ($p=1,2,3$) family in the Table \ref{QuiversKNpfamily}. It is possible to compute the Coulomb branch Hilbert series for these magnetic quivers for each family in Table \ref{QuiversKNpfamily}. We present some results in Table \ref{O5MMKNfamilyHS}, and we see that the Hilbert series of the unitary and orthosymplectic quivers agree with each other.
\begin{figure}[!htb]
    \centering
    \begin{scriptsize}
       \begin{tikzpicture}[scale=.85]
       \draw[thick](0,0)--(0,2);
       \node at (-.75,.3) {$2N-2$};
       \node at (-.75,-.5) {$2N-2$};
       \node at (-.25,-2.25) {$4$};
       \node at (-.25,-3.25) {$2$};
         \node at (.75,2.5) {$2N-2$};
       \node at (.25,5.25) {$2$};
       \node at (.25,4.25) {$4$};
       \draw[thick,blue](.1,1)--(.1,2);
       \draw[thick](.05,2)--(.05,3);
       \node[7brane] at (.05,3){};
       \node[thick] at (0.05,3.45) {$.$};
       \node[thick] at (0.05,3.25) {$.$};
       \node[thick] at (0.05,3.35) {$.$};
       \node[7brane] at (.05,3.7){};
       \draw[thick](.05,3.7)--(.05,5.7);
       \node[7brane] at (.05,4.7){};
       \node[7brane] at (.05,5.7){};
       \draw[thick,blue](.1,1)--(1.1,0);
       \draw[thick](1.1,0)--(2.1,-1);
       \node[7brane] at (2.1,-1){};
       \node at (1.6,-.75){1};
       \draw[thick,blue](.1,1)--(-1.1,1);
       \draw[thick,red]((-1.1,.9)--(1.1,.9);
       \node at (.6,.2){2};
       \node at (.75,1.1) {2};
       \node[7brane] at (0,0){};
       \node[7brane] at (-1.1,.95){};
       \node[7brane] at (1.1,.9){};
       \node[7brane] at (1.1,0){};
       \node[7brane] at (0.05,2){};
       \draw (-1,.95)--(-4,.95);
       \node[7brane]at(-2,.95){};
       \node[7brane]at(-3,.95){};
       \node[7brane]at(-4,.95){};
       \draw (1,.9)--(2,.9);
       \node[7brane] at (2,.9){};
       \node at (1.5,1.1){1};
       \node at (-3.5,1.15){1};
       \node at (-2.5,1.15){2};
       \node at (-1.5,1.15){3};
       \node at (-1.5,1.1){};
       \node[thick] at (0,-1.25) {$.$};
       \node[thick] at (0,-1.35) {$.$};
       \node[thick] at (0,-1.45) {$.$};
       \node[7brane] at (0,-1){};
       \draw[thick](0,0)--(0,-1);
       \node[7brane] at (0,-1.7){};
       \draw[thick] (0,-1.7)--(0,-3.7);
       \node[7brane] at (0,-2.7){};
       \node[7brane] at (0,-3.7){};
       \end{tikzpicture}
    \end{scriptsize}
    \caption[Unitary web diagram of $\text{K}_{N}^{1}$ theory.]{A 
unitary web for the $\text{K}_{N}^{1}$ theory. The maximal subdivision leading to the magnetic quiver is indicated by use of colours.}
    \label{fig:KN1 ordinary web}
\end{figure}
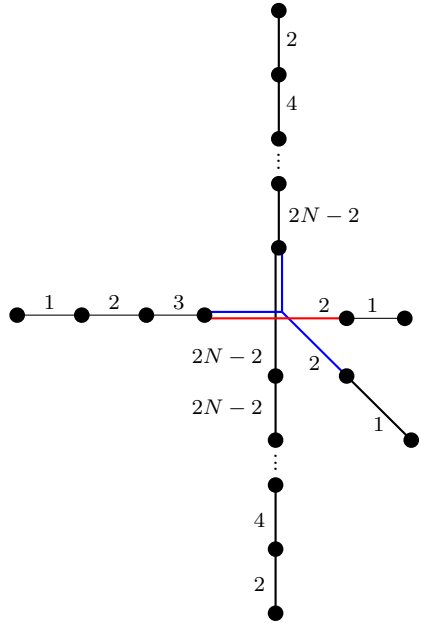
\begin{figure}[!htb]
    \centering
        \begin{scriptsize}
       \begin{tikzpicture}[scale=.7]
       \draw[thick](-5.5,0)--(8.5,0);
       \draw[thick,dashed](-7,0)--(-5.5,0);
       \draw[thick,dashed](10,0)--(8.5,0);
       \node[7brane] at (7,0){};
       \node[7brane] at (8.5,0){};
       \node[7brane] at (5.5,0){};
       \node[7brane] at (-5.5,0){};
       \node[7brane] at (4,0){};
       \node[7brane] at (2.5,0){};
       \node[7brane] at (1,0){};
       \node[7brane] at (-3,0){};
       \node[thick] at (-.5,3.4) {$.$};
       \node[thick] at (-.5,3.5) {$.$};
       \node[thick] at (-.5,3.6) {$.$};
       \node[7brane]at (-.5,2){};
       \node[7brane]at (-.5,3){};
       \node[7brane]at (-.5,4){};
       \node[7brane]at (-.5,5){};
       \node[7brane]at (-.5,6){};
       \draw[thick](-.5,0)--(-.5,3);
       \draw[thick](-.5,4)--(-.5,6);
       \draw[thick](-.5,0)--(-2.5,1);
       \node at (0,5.5){1};
       \node at (0,4.5){2};
       \node at (.25,2.5){$2N-2$};
       \node at (.25,1.5){$2N-1$};
    
       \node[label=above left:{$(2,-1)$}][7brane]at(-2.5,1){};
       \node at (-6.25,-.35){O$5^-$};
       \node at (9.25,-.35){O$5^-$};
       \node at (-4,-.35){$\frac{1}{2}$};
       \node at (-1.5,-.35){1};
       \node at (.5,-.35) {3};
       \node at (1.75,-.35){$\frac{5}{2}$};
       \node at (3.25,-.35){$2$};
       \node at (4.75,-.35){$\frac{3}{2}$};
       \node at (6.25,-.35){$1$};
       \node at (7.5,-.35){$\frac{1}{2}$};
       
       \draw[thick](4,3.5)--(5,3);
       \draw[thick](5,3)--(6,3);
       \draw[thick](5,3)--(5.3,2.7);
       \draw[thick](5.3,2.7)--(6.5,2.7);
       \draw[thick](5.3,2.7)--(5.3,2);
       \draw[thick,red](5.1,2)--(5.1,4);
       \draw[thick,blue](4,2.85)--(6,2.85);
       \node[7brane]at(6,2.9){};
       \node[7brane]at(4,3.5){};
       \node[7brane]at(6.5,2.7){};
       \node[7brane]at(4,2.85){};
       \node[7brane]at(5.2,2){};
       \node[7brane]at(5.1,4){};
       \end{tikzpicture}
       \end{scriptsize}
    \caption[Orientifold web diagram of $\text{K}_{N}^{2}$ theory.]{An orientifold web for the $\text{K}_{N}^{2}$ theory and the maximal subdivision at the centre of the junction.}
    \label{fig:K3N2 O5 web}
\end{figure}
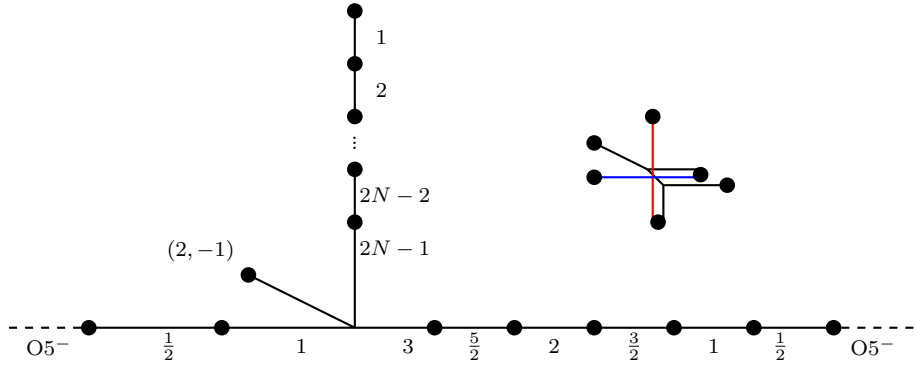
\begin{figure}[!htb]
    \centering
    \begin{scriptsize}
    \begin{tikzpicture}[scale=.7]
\draw[thick,](-2.25,10.25)--(+2.25,5.75);
\node[thick][7brane] at (1.5,6.5) {};
\node[thick][7brane] at (-1.5,9.5) {};
\node[thick][7brane] at (2.25,5.75) {};
\node[thick][7brane] at (-2.25,10.25) {};
\node at (-2.5,10.5) {.};
\node at (-2.6,10.6) {.};
\node at (-2.7,10.7) {.};
\node[thick][7brane] at (-2.95,10.95) {};
\node[thick][7brane] at (-3.7,11.7) {};
\node[thick][7brane] at (-4.45,12.45) {};
\draw[thick](-2.95,10.95)--(-4.45,12.45);
\node at (-3.8,12.25) {$2$};
\node at (-3,11.5) {$4$};
\node at (-1.3,10.2) {$2N-4$};
\node at (-2,9) {$2N-2$};

\node at (2.5,5.5) {.};
\node at (2.6,5.4) {.};
\node at (2.7,5.3) {.};
\node[thick][7brane] at (2.95,5.05) {};
\node[thick][7brane] at (3.7,4.3) {};
\node[thick][7brane] at (4.45,3.55) {};
\draw[thick](2.95,5.05)--(4.45,3.55);
\node at (3.8,3.75) {$2$};
\node at (3,4.5) {$4$};
\node at (1.3,5.8) {$2N-2$};
\node at (.8,6.8) {$2N$};

\draw[thick](-5,8)--(2,8);
\node at (1.5,7.75) {$1$};
\node at (1.25,8.9) {$1$};
\node at (0.2,9.5) {$1$};
\node at (-4.5,7.75) {$1$};
\node at (-3.5,7.75) {$2$};
\node at (-2.5,7.75) {$3$};
\node at (-1.5,7.75) {$4$};
\node[thick][7brane] at (-2,8) {};
\node[thick][7brane] at (-3,8) {};
\node[thick][7brane] at (-4,8) {};
\node[thick][7brane] at (-5,8) {};
\node[thick][7brane] at (2,8) {};

\draw[thick](0,8)--(1.5,9.5);
\node[7brane]at (1.5,9.5){};

\draw[thick](0,8)--(0,10);
\node[thick][7brane] at (0,10) {};
        \end{tikzpicture}\end{scriptsize}
        \hspace{1 cm}
        \begin{scriptsize}
        \begin{tikzpicture}
        \node at (0,-2){};
        \draw[thick,blue](0,0)--(-1,1);
        \draw[thick,blue](-1,1)--(-1,1.2);
        \draw[thick,blue](-1,1.2)--(0,2.2);
        \node[7brane]at(0,2.2){};
        \draw[thick,blue](-1,1.2)--(-2,1.2);
        \draw[thick,blue](-1,1)--(-2.5,1);
        \node[7brane]at(-2.5,1){};
        \node[7brane]at(-2,1.2){};
        \draw[thick,red](-2,1.3)--(.5,1.3);
        \node[7brane]at(.5,1.3){};
        \draw[thick,green](-2,1.1)--(-.75,1.1);
        \draw[thick,green](-.75,2.5)--(-.75,1.1);
        \node[7brane] at (-.75,2.5){};
        \draw[thick,green](.25,0.1)--(-.75,1.1);
        \draw[thick,black](0.2,0)--(-1.8,2);
        \node[big7brane] at (.15,0){};
        \node[7brane] at (-1.8,2){};
        \end{tikzpicture}
        \end{scriptsize}
    \caption[Unitary web diagram of $\text{K}_{N}^{2}$ theory.]{A unitary 
web description for the $\text{K}_{N}^{2}$ theory, together with the maximal subdivision for the Higgs branch at infinite coupling.}
    \label{fig:K3N2 unitary web}
\end{figure}
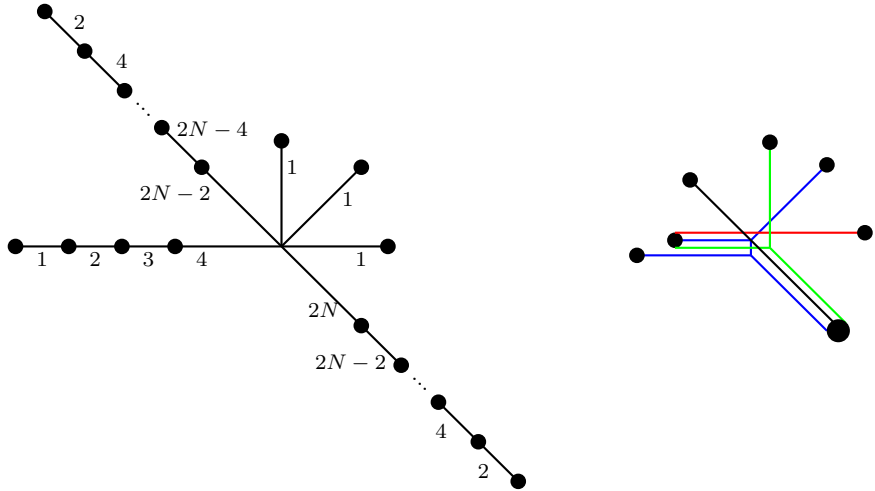
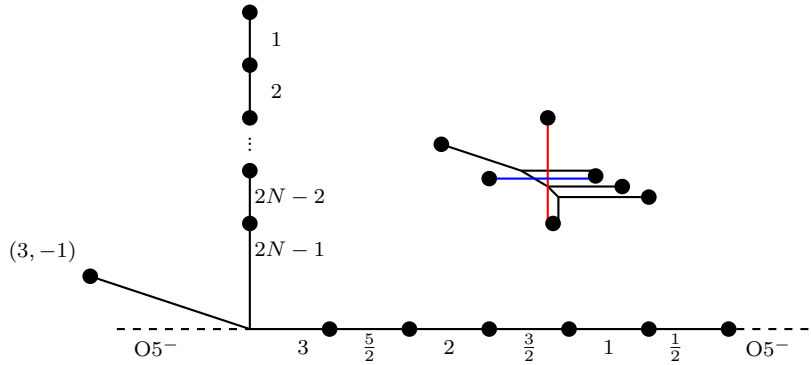
\begin{figure}[!htb]
    \centering
        \begin{scriptsize}
       \begin{tikzpicture}[scale=.7]
       \draw[thick](-0.5,0)--(8.5,0);
       \draw[thick,dashed](-3,0)--(-.5,0);
       \draw[thick,dashed](10,0)--(8.5,0);
       \node[7brane] at (7,0){};
       \node[7brane] at (8.5,0){};
       \node[7brane] at (5.5,0){};
       \node[7brane] at (4,0){};
       \node[7brane] at (2.5,0){};
       \node[7brane] at (1,0){};
       \node[thick] at (-.5,3.4) {$.$};
       \node[thick] at (-.5,3.5) {$.$};
       \node[thick] at (-.5,3.6) {$.$};
       \node[7brane]at (-.5,2){};
       \node[7brane]at (-.5,3){};
       \node[7brane]at (-.5,4){};
       \node[7brane]at (-.5,5){};
       \node[7brane]at (-.5,6){};
       \draw[thick](-.5,0)--(-.5,3);
       \draw[thick](-.5,4)--(-.5,6);
       \draw[thick](-.5,0)--(-3.5,1);
       \node at (0,5.5){1};
       \node at (0,4.5){2};
       \node at (.25,2.5){$2N-2$};
       \node at (.25,1.5){$2N-1$};
    
       \node[label=above left:{$(3,-1)$}][7brane]at(-3.5,1){};
       \node at (-2.25,-.35){O$5^-$};
       \node at (9.25,-.35){O$5^-$};
       \node at (.5,-.35) {3};
       \node at (1.75,-.35){$\frac{5}{2}$};
       \node at (3.25,-.35){$2$};
       \node at (4.75,-.35){$\frac{3}{2}$};
       \node at (6.25,-.35){$1$};
       \node at (7.5,-.35){$\frac{1}{2}$};

       \draw[thick](3.1,3.5)--(4.6,3);
       \draw[thick](4.6,3)--(6,3);
       \draw[thick](5.1,2.7)--(6.5,2.7);
       \draw[thick](5.3,2.5)--(7,2.5);
       \draw[thick](5.3,2.5)--(5.3,2);
       \draw[thick](5.3,2.5)--(5.1,2.7);
       \draw[thick](4.6,3)--(5.1,2.7);
       \draw[thick,red](5.1,2)--(5.1,4);
       \draw[thick,blue](4,2.85)--(6,2.85);
       \node[7brane]at(6,2.9){};
       \node[7brane]at(3.1,3.5){};
       \node[7brane]at(6.5,2.7){};
       \node[7brane]at(7,2.5){};
       \node[7brane]at(4,2.85){};
       \node[7brane]at(5.2,2){};
       \node[7brane]at(5.1,4){};
  \end{tikzpicture}
       \end{scriptsize}
    \caption[Orientifold web diagram of $\text{K}_{N}^{3}$ theory.]{An orientifold web for the $\text{K}_{N}^{3}$ theory and the maximal subdivision at the centre of the junction.}
    \label{fig:K3N3 O5 web}
\end{figure}
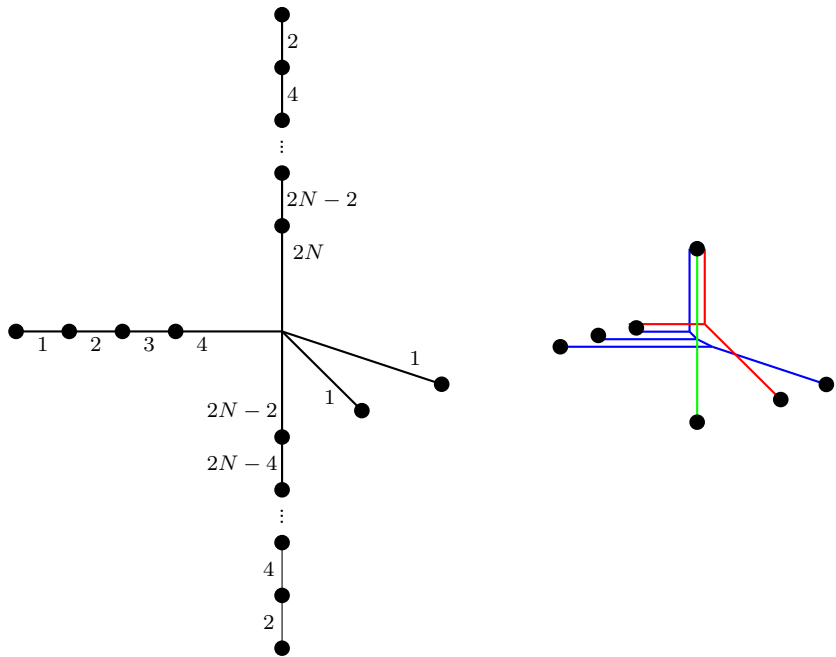
\begin{figure}[!htb]
    \centering
    \begin{scriptsize}
    \begin{tikzpicture}[scale=.7]
\draw[thick,](0,8)--(1.5,6.5);
\node[thick][7brane] at (1.5,6.5) {};
\draw[thick](0,8)--(3,7);
\node[7brane]at(3,7){};
\draw[thick](0,8)--(0,5);
\node[7brane]at(0,6){};
\node[7brane]at(0,5){};
\node at (0,4.6){$.$};
\node at (0,4.5){$.$};
\node at (0,4.4){$.$};
\draw(0,4)--(0,2);
\node[7brane]at(0,4){};
\node[7brane]at(0,3){};
\node[7brane]at(0,2){};
\node at (-0.25,2.5) {$2$};
\node at (-0.25,3.5) {$4$};
\node at (-0.75,5.5) {$2N-4$};
\node at (-0.75,6.5) {$2N-2$};
\draw[thick](-5,8)--(0,8);
\node at (2.5,7.5) {$1$};
\node at (.9,6.75) {$1$};
\node at (0.2,13.5) {$2$};
\node at (0.2,12.5) {$4$};
\node at (.75,10.5) {$2N-2$};
\node at (0.5,9.5) {$2N$};
\node at (-4.5,7.75) {$1$};
\node at (-3.5,7.75) {$2$};
\node at (-2.5,7.75) {$3$};
\node at (-1.5,7.75) {$4$};
\node[thick][7brane] at (-2,8) {};
\node[thick][7brane] at (-3,8) {};
\node[thick][7brane] at (-4,8) {};
\node[thick][7brane] at (-5,8) {};

\draw[thick](0,8)--(0,11);
\node[thick][7brane] at (0,10) {};
\node[thick][7brane] at (0,11) {};
\node at (0,11.4){$.$};
\node at (0,11.5){$.$};
\node at (0,11.6){$.$};
\draw[thick](0,12)--(0,14);
\node[7brane]at(0,12){};
\node[7brane]at(0,13){};
\node[7brane]at(0,14){};
        \end{tikzpicture}\end{scriptsize}
        \hspace{1 cm}
        \begin{scriptsize}
        \begin{tikzpicture}
        \node at (0,-4){};
  \draw[thick,blue](0,0)--(1.5,-.5);
  \draw[thick,blue](0,0)--(-2,0);
  \draw[thick,blue](0,0)--(-.2,.1);
  \draw[thick,blue](-.3,0.2)--(-.2,.1);
  \draw[thick,blue](-1.5,0.1)--(-.2,.1);
  \draw[thick,blue](-.3,0.2)--(-1,.2);
  \draw[thick,blue](-.3,0.2)--(-.3,1.3);
  \draw[thick,red](-1.1,.3)--(-.1,.3);
  \draw[thick,red](.9,-.7)--(-.1,.3);
  \draw[thick,red](-.1,1.3)--(-.1,.3);
  \draw[thick,green](-.2,1.3)--(-.2,-1);
  \node[7brane]at(-2,0){};
  \node[7brane]at(-1.5,0.15){};
  \node[7brane]at(-1,0.25){};
  \node[7brane]at(-.2,-1){};
  \node[7brane]at(-.2,1.3){};
  \node[7brane]at(1.5,-.5){};
  \node[7brane]at(.9,-.7){};
        \end{tikzpicture}
        \end{scriptsize}
    \caption[Unitary web diagram of $\text{K}_{N}^{3}$ theory.]{A unitary 
web description for the $\text{K}_{N}^{3}$ theory, together with the maximal subdivision for the Higgs branch at infinite coupling.}
    \label{fig:KN3 unitary}
\end{figure}
\begin{table}[htbp]
\centering 
\begin{tabular}{|c|C{7.1cm}|C{5.9cm}|} \hline
\rowcolor{Grayy}
$K_N^p$ & Unitary magnetic & Orthosymplectic magnetic \\ \hline
$K_N^1$ & \includestandalone[width=0.48\textwidth]{KN1U} & \includestandalone[width=0.41\textwidth]{KN1OSp} \\ \hline
$K_N^2$ & \includestandalone[width=0.48\textwidth]{KN2U} & \includestandalone[width=0.40\textwidth]{KN2OSp} \\ \hline
$K_N^3$ & \includestandalone[width=0.48\textwidth]{KN3U} & \includestandalone[width=0.40\textwidth]{KN3OSp} \\ \hline
\end{tabular} 
\caption[Magnetic quivers for $\text{K}_{N}^{p}$ family.]{Magnetic quivers for the $K_N^p$ family. The unitary quivers are derived from the unitary web diagrams of figures \ref{fig:KN1 ordinary web}, \ref{fig:K3N2 unitary web}, \ref{fig:KN3 unitary}. The orthosymplectic quivers on the other hand come from the orinetifold web diagrams of figures \ref{fig:K3N1 O5 web},  \ref{fig:K3N2 O5 web} and \ref{fig:K3N3 O5 web} respectively.}
\label{QuiversKNpfamily}
\end{table}
\begin{table}
\centering
\begin{tabular}{|c|C{4.25cm}|C{4.25cm}|C{4.25cm}|} \hline
\rowcolor{Grayy}
   & Unitary magnetic quiver & \multicolumn{2}{c|}{Orthosymplectic magnetic quiver} \\ \cline{2-4}
	\rowcolor{Grayy}
  \multirow{-2}{*}{$K_N^p$} & HS($t$)  & HS($t;\vec{m} \in \mathbb{Z}$) & HS($t;\vec{m} \in \mathbb{Z}+\tfrac{1}{2}$)  \\ \hline
	$K_1^1$ & \footnotesize{$1+78t+2430t^2+43758t^3+537966t^4+4969107t^5+\ldots$} & \footnotesize{$1+46t+1278t^2+22254t^3+270798t^4+2491731t^5+\ldots$} & \footnotesize{$32t+1152t^2+21504t^3+267168t^4+2477376t^5+\ldots$}  \\ \hline
	$K_2^1$ & \footnotesize{$1+30t+592t^2+8867t^3+106965t^4+1073577t^5+\ldots$}  & \footnotesize{$1+30t+496t^2+6083t^3+63477t^4+586537t^5+\ldots$} & \footnotesize{$96t^2+2784t^3+43488t^4+487040t^5+\ldots$}  \\ \hline
	$K_1^2$ & \footnotesize{$\begin{array}{l} \dfrac{P_3(t)}{(1-t)^{14}} \\\\ = 1+45t+770t^2+7644t^3\\+52920t^4+282744t^5+\ldots  
	\end{array}$} & \footnotesize{$\begin{array}{l} \dfrac{P_4(t)}{(1-t)^{14}\,(1+t)^7} \\\\ = 1+29t+434t^2+4060t^3\\+27384t^4+144312t^5+\ldots  
	\end{array}$} & \footnotesize{$\begin{array}{l} \dfrac{P_5(t)}{(1-t)^{14}\,(1+t)^7} \\\\ = 16t+336t^2+3584t^3\\+25536t^4+138432t^5+\ldots  
	\end{array}$} \\ \hline
	$K_2^2$ & \footnotesize{$1+25t+392t^2+4590t^3+42387t^4+320549t^5+\ldots$} & \footnotesize{$1+25t+344t^2+3438t^3+27843t^4+191957t^5+\ldots$} & \footnotesize{$48t^2+1152t^3+14544t^4+128592t^5+\ldots$}  \\ \hline
	$K_1^3$ & \footnotesize{$\begin{array}{l} \dfrac{1+16t+36t^2+16t^3+t^4}{(1-t)^{8}} \\\\ = 1+24t+200t^2+1000t^3\\+3675t^4+10976t^5+\ldots  
	\end{array}$} & \footnotesize{$\begin{array}{l} \dfrac{P_6(t)}{(1-t)^8\, (1+t)^4} \\\\ = 1+16t+120t^2+560t^3\\+1995t^4+5824t^5+\ldots  
	\end{array}$} & \footnotesize{$\begin{array}{l} \dfrac{P_7(t)}{(1-t)^8\, (1+t)^4} \\\\ = 8t+80t^2+440t^3\\+1680t^4+5152t^5+\ldots  
	\end{array}$} \\ \hline
	$K_2^3$ & \footnotesize{$1+24t+296t^2+2510t^3+16374t^4+87306t^5+\ldots$} & \footnotesize{$1+24t+272t^2+2078t^3+12294t^4+60450t^5+\ldots$}  & \footnotesize{$24t^2+432t^3+4080t^4+26856t^5+\ldots$} \\ \hline
\end{tabular}
\caption[Coulomb branch HS for magnetic quivers of $K_N^p$ family.]{Coulomb branch Hilbert series of the unitary and orthosymplectic magnetic quivers for the $K_N^p$ family listed in Table \ref{QuiversKNpfamily}. For orthosymplectic quivers, we need to add the contributions of both integer and half integer fluxes. The total Hilbert series then matches with that of the unitary quivers. The explicit forms of $P_3(t), P_4(t), P_5(t), P_6(t), P_7(t)$ are given in Appendix \ref{sec:app2}. }
\label{O5MMKNfamilyHS}
\end{table}
\FloatBarrier
\subsection{\texorpdfstring{The $\text{Y}_{N}$ family}{TEXT}}
We then consider a different type of decoupling from the $\#_{3,N}$ theory to arrive at different examples which show some new features. 
\subsubsection{The Y$_N^{1,1}$ theory}
\begin{figure}[!htb]
    \centering\begin{scriptsize}
    \begin{tikzpicture}[scale=.7]
    \draw[thick](-2,2)--(2,-2);
    \node[7brane] at (-1.25,1.25){};
    \node[7brane] at (1.25,-1.25){};
    \node[7brane] at (2,-2){};
    \node[7brane] at (-2,2){};    
    \node at (-1.5,1.8) {1};
    \node at (-.8,1.1) {2};
    \node at (1.7,-1.4) {1};
    \draw[thick](-2.5,0)--(2.5,0);
    \node[7brane] at (-1.5,0){};
    \node[7brane] at (1.5,0){};
    \node[7brane] at (2.5,0){};
    \node[7brane] at (-2.5,0){};
    \node at (2,0.25) {1};
    \node at (1,0.25) {2};
    \node at (-2,0.25) {1};
    \draw[thick](0,-2.25)--(0,2.25);
    \node[7brane] at (0,1.5){};
    \node[7brane] at (0,-1.5){};
    \node[7brane] at (0,-2.25){};
    \node[7brane] at (0,2.25){};    
    \draw[thick](0,3.25)--(0,5.25);
    \node[7brane] at (0,3.25){};
    \node[7brane] at (0,4.25){};
    \node[7brane] at (0,5.25){};    
    \draw[thick](0,-3.25)--(0,-5.25);
    \node[7brane] at (0,-3.25){};
    \node[7brane] at (0,-4.25){};
    \node[7brane] at (0,-5.25){};    
        \node[thick] at (0,2.65) {$.$};
\node[thick] at (0,2.75) {$.$};
\node[thick] at (0,2.85) {$.$};
    \node[thick] at (0,-2.65) {$.$};
\node[thick] at (0,-2.75) {$.$};
\node[thick] at (0,-2.85) {$.$};    
    \node at (0.25,4.75) {2};
    \node at (0.25,3.75) {4};
    \node at (0.75,1.85) {$2N-2$};
    \node at (0.35,1) {$2N$};    
    \node at (-0.25,-4.75) {2};
    \node at (-0.25,-3.75) {4};
    \node at (-0.75,-1.85) {$2N-2$};
        \end{tikzpicture}\end{scriptsize}
        \hspace{1cm}
    \begin{tikzpicture}[scale=.8]
    \draw[thick](0,1.5)--(0,-1.5);
    \node[7brane] at(0,1.5){};
    \node[7brane] at(0,-1.5){};    
    \draw[thick, red] (-1.25,1.25)--(1.25,-1.25);
    \node[7brane]at (-1.25,1.25){};
    \node[7brane] at(1.25,-1.25){};    
    \draw[thick,blue](-1.5,0)--(1.5,0);
    \node[7brane] at(1.5,0){};
    \node[7brane] at(-1.5,0){};    
    \node at (0,-2.5) {(I)};    
      \draw[thick,blue](.2,-6.5)--(.2,-3.5);    
    \draw[thick] (-1.5,-5)--(0,-5);
    \draw[thick](0,-5)--(0,-3.5);
    \draw[thick] (0,-5)--(1.25,-6.25);
    \node[7brane] at (1.25,-6.25){};
    \node[7brane] at (-1.5,-5){};
    \node[7brane] at (.1,-3.5){};    
    \draw[thick,red] (0,-6.5)--(0,-5);
    \draw[thick,red] (0,-5)--(1.5,-5);
    \draw[thick,red] (0,-5)--(-1.25,-3.75);
    \node[7brane] at (-1.25,-3.75){};
    \node[7brane] at (1.5,-5){};
    \node[7brane] at (.1,-6.5){};    
    \node at (5,-2.5) {(II)};
          \draw[thick,blue](5.2,-1.5)--(5.2,1.5);     
    \draw[thick] (3.5,0)--(5,0);
    \draw[thick](5,0)--(5,1.5);
    \draw[thick] (5,0)--(6.25,-1.25);
    \node[7brane] at (6.25,-1.25){};
    \node[7brane] at (5.1,1.5){};
    \draw[thick,pink] (3.5,0.2)--(6.5,0.2);
      \node[7brane] at (3.5,0.1){};
    \draw[thick,red] (5,-1.5)--(5,0);
    \draw[thick,red] (5,0)--(6.5,0);
    \draw[thick,red] (5,0)--(3.75,1.25);
    \draw[thick,green] (6.25,-1.05)--(3.75,1.45);
    \node[7brane] at (3.75,1.35){};
    \node[7brane] at (6.5,0.1){};
    \node[7brane] at (5.1,-1.5){};    
    \node at (0.1,-7.5) {(III)};    
\end{tikzpicture}
    \caption[Unitary web diagram of $\text{Y}_{N}^{1,1}$ theory.]{A unitary 
web for the $\text{Y}_{N}^{1,1}$ theory at the fixed point, along with the three possible distinct maximal subdivisions of the centre of the junction.}
    \label{fig:YN11 ordinary web}
\end{figure}
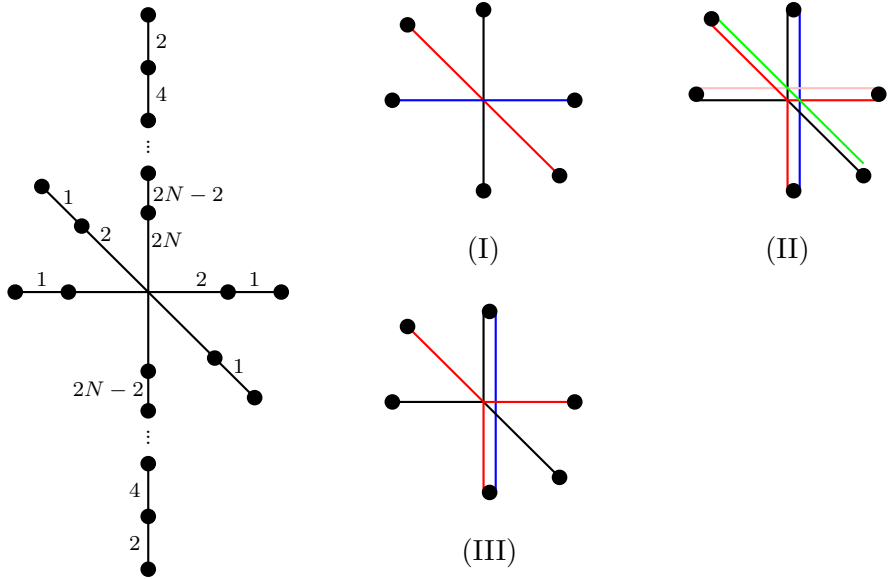

In section \ref{sec:KNO5-O5-}, we decouple flavors of the $\text{USp}(2)$ gauge node on one end. Here we decouple one flavor from the $\text{USp}(2)$ gauge nodes on the two ends and call the theory $\text{Y}_N^{1,1}$ theory. An IR description of the theory is 
\be\label{electric YN11 OSp}
\begin{array}{c}\begin{tikzpicture}
\node {$[2]-\text{USp}(2)-\text{SO}(6)-\text{USp}(2)-\cdots-\text{SO}(6)-\text{USp}(2)-[2]$};
 \draw [thick,decorate,decoration={brace,amplitude=6pt},xshift=0pt,yshift=10pt]
(-4.75,0) -- (4.0,0)node [black,midway,xshift=0pt,yshift=20pt] {
$2N+1$};
\end{tikzpicture}
\end{array} ~.
\ee 
An orientifold web diagram of the $\text{Y}_{N}^{1,1}$ theory is obtained by intersecting 2 D5, $2N$ NS5, one $(1,1)$ and one $(1,-1)$ 5-brane on top of an O5-plane, here taken to be asymptotically O5$^-$-plane (Figure \ref{fig:YN11 O5}). 
The theory also admits a description in terms of an ordinary web which we have shown in Figure \ref{fig:YN11 ordinary web}. The ordinary web description follows either by reading off the low energy gauge theory from an S-dual description 
or by following a gluing procedure similar to the one described in Figure \ref{fig:gauging su4=so6}. 
\begin{figure}[!htb]
    \centering\begin{scriptsize}
    \begin{tikzpicture}[scale=.7]
    \draw[thick](-8,0)--(8,0);
    \draw[thick,dashed](-10,0)--(-8,0);
    \draw[thick,dashed](10,0)--(8,0);
    \node[7brane] at (2,0){};
    \node[7brane] at (4,0){};
    \node[7brane] at (6,0){};
    \node[7brane] at (8,0){};
    \node[7brane] at (-2,0){};
    \node[7brane] at (-4,0){};
    \node[7brane] at (-6,0){};
    \node[7brane] at (-8,0){};
    \draw[thick](0,0)--(0,3);
    \node[7brane] at (0,2){};
    \node[7brane] at (0,3){};
    \node[thick] at (0,3.4) {$.$};
\node[thick] at (0,3.5) {$.$};
\node[thick] at (0,3.6) {$.$};
\node[7brane] at (0,4){};
\node[7brane] at (0,5){};
\node[7brane] at (0,6){};
\node at (.5,5.5) {1};
\node at (.5,4.5) {2};
\node at (.75,2.5) {$2N-1$};
\node at (.35,1.5) {$2N$};
\draw[thick](0,4)--(0,6);
\draw[thick](0,0)--(1.5,1.5);
\draw[thick](0,0)--(-1.5,1.5);
\node[7brane,label=above right:{$(1,1)$}] at (1.5,1.5){};
\node[7brane,label=above left:{$(1,-1)$}] at (-1.5,1.5){};

\node at (9,-.25) {$\text{O}5^-$};
\node at (7,0.35) {$\frac{1}{2}$};
\node at (5,0.35) {1};
\node at (3,0.35) {$\frac{3}{2}$};
\node at (1.5,0.35) {2};
\node at (-1.5,0.35) {2};
\node at (-9,-.25) {$\text{O}5^-$};
\node at (-7,0.35) {$\frac{1}{2}$};
\node at (-5,0.35) {1};
\node at (-3,0.35) {$\frac{3}{2}$};

    \end{tikzpicture}\end{scriptsize}
    \vspace{.5cm}
    \begin{tikzpicture}[scale=.6]
\draw[thick](-2,-7)--(2,-7);
\draw[thick,green](0,-5.25)--(0,-8.75);
\draw[thick,dashed,red](-1.5,-5.5)--(1.5,-8.5);
\draw[thick,red](1.5,-5.5)--(-1.5,-8.5);
\node[7brane]at (-1.5,-5.5){};
\node[7brane]at (1.5,-5.5){};
\node[7brane]at (1.5,-8.5){};
\node[7brane]at (-1.5,-8.5){};
\node[7brane]at (0,-8.75){};
\node[7brane]at (0,-5.25){};    
\node[7brane]at (2,-7){};
\node[7brane]at (-2,-7){};

\node at (0,-10){(I)};

\draw[thick,green](5.8,-8.5)--(5.8,-5.25);
\draw[thick](5.6,-7)--(5.6,-5.25);
\draw[thick](5.6,-7)--(7.1,-7);
\draw[thick](5.6,-7)--(4.1,-8.5);
\draw[thick,red](5.4,-7)--(5.4,-5.25);
\node[7brane] at (5.6,-5.25){};
\draw[thick,red](5.4,-7)--(3.9,-7);
\draw[thick,red](5.4,-7)--(6.9,-8.5);
\draw[thick,cyan](3.9,-6.8)--(7.1,-6.8);
\node[7brane] at (7.1,-6.9){};
\node[7brane] at(6.9,-8.5){};
\node[7brane] at(4.1,-8.5){};
\node[7brane] at(3.9,-6.9){};
\node[7brane] at(5.8,-8.5){};

\node at (5.5,-10){(II)};

\draw[thick](11,-8.75)--(11,-5.25);
\draw[thick,red](10.85,-7)--(10.85,-4.5);
\node[7brane] at (10.9,-4.5){};
\draw[thick,red](11.15,-7)--(11.15,-5.25);
\draw[thick,red](10.85,-7)--(11.15,-7);
\draw[thick,red](10.85,-7)--(9.35,-8.5);
\draw[thick,red](11.15,-7)--(12.65,-8.5);
\draw[thick,blue](9,-6.8)--(13,-6.8);
\node[7brane]at (11.1,-5.25){};
\node[7brane]at (11,-8.75){};
\node[7brane] at (9,-6.8){};
\node[7brane] at (13,-6.8){};
\node[7brane] at (12.65,-8.5){};
\node[7brane] at (9.35,-8.5){};
\node at (11,-10){(III)};
    \end{tikzpicture}
    \caption[Orientifold web diagram of $\text{Y}_{N}^{1,1}$ theory.]{An orientifold web for the $\text{Y}_{N}^{1,1}$ theory at the fixed point. We show the three possible maximal subdivisions of the centre of the junction at the bottom.}
    \label{fig:YN11 O5}
\end{figure}
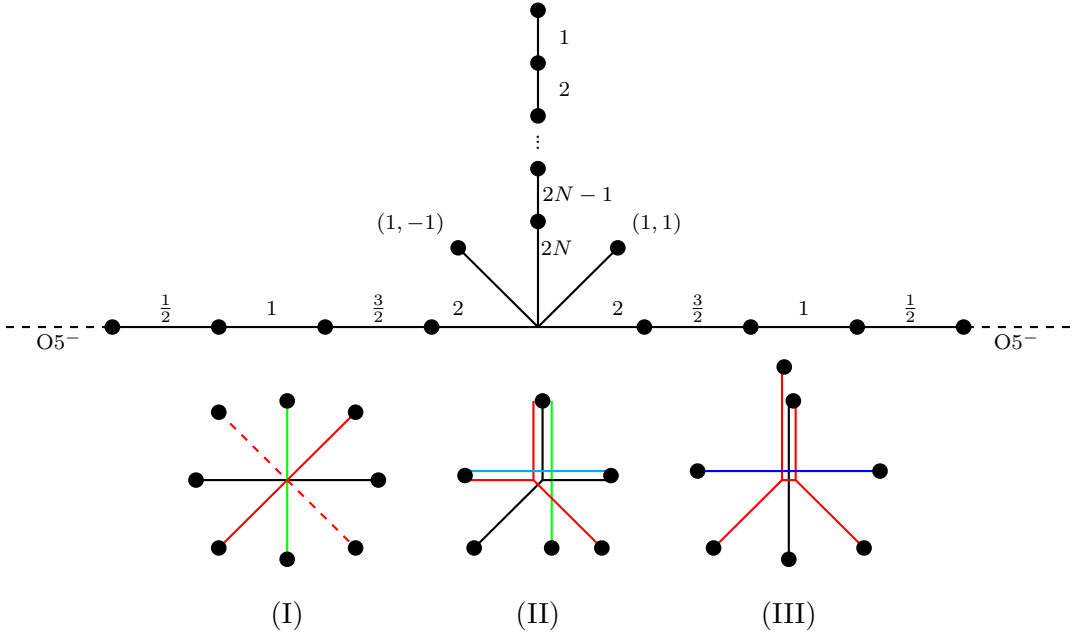

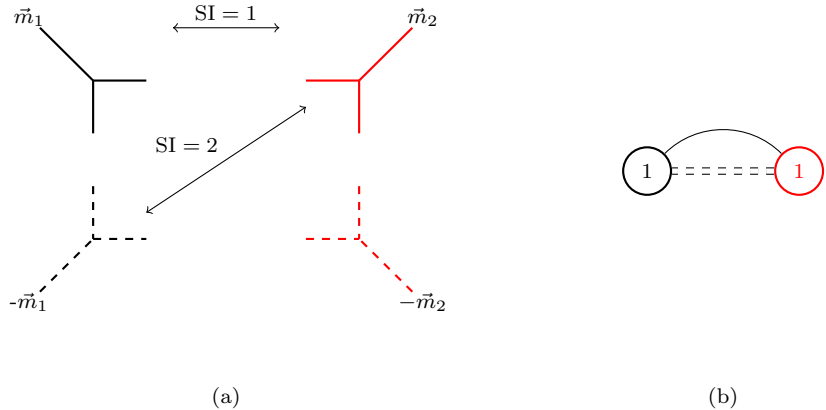
\begin{figure}[!htb]
    \centering
    \begin{scriptsize}
    \begin{tikzpicture}[scale=.7]
    \draw[thick,red](0,0)--(1,0);
    \draw[thick,red](1,0)--(2,1);
    \draw[thick,red](1,0)--(1,-1);
    \node at(2.2,1.2) {$\Vec{m}_2$};
    \draw[<->](-.5,1)-- node[above] {$\SI=1$} (-2.5,1);
    
    \draw[dashed,red,thick](1,-2)--(1,-3);
    \draw[dashed,red,thick](0,-3)--(1,-3);
    \draw[dashed,red,thick](2,-4)--(1,-3);
    \node at(2.2,-4.2) {$-\Vec{m}_2$};
    
    \draw[<->](0,-.5)-- node[above left] {$\SI=2$} (-3,-2.5);
    
    \draw[thick](-4,0)--(-3,0);
    \draw[thick](-4,0)--(-5,1);
    \draw[thick](-4,-1)--(-4,0);
    \node at(-5.2,1.2) {$\Vec{m}_1$};
    
    \draw[thick,dashed](-4,-2)--(-4,-3);
    \draw[thick,dashed](-4,-3)--(-3,-3);
    \draw[thick,dashed](-5,-4)--(-4,-3);
    \node at(-5.2,-4.2) {-$\Vec{m}_1$};
    \node at (-1.5,-6) {(a)};
    \end{tikzpicture}
    \hspace{2cm}
    \begin{tikzpicture}
    \node[circle,inner sep=4pt,draw,thick](black) at (0,2){1};
    \node[circle,inner sep=4pt,draw,red,thick](red) at (2,2){1};
    \draw[double distance=2pt,dashed](red)--(black);
    \draw (black)to[out=45,in=135](red);
    \node at (1,-1){(b)};
    \end{tikzpicture}
    \end{scriptsize}
    \caption[Origin of the exotic hypermultiplets.]{The origin of the exotic bi-fundamental hypermultiplets. (a) The maximal subdivision of the relevant web. The dashed lines correspond to the mirror images of the solid line subwebs. For ease of presentation we have not included the O5-plane in the picture. (b) The corresponding magnetic quiver with the exotic bi-fundamental hypermultiplet.}
    \label{fig:exotic bifundamental}
\end{figure}

Given the maximal subdivisions in Figure \ref{fig:YN11 O5} and Figure \ref{fig:YN11 ordinary web}, we can write down the corresponding orthosymplectic and unitary magnetic quivers, the results are collected in Table  \ref{tab:magnetic quivers YN11}. For the maximal subdivisions labeled as (I) and (III), the magnetic quivers are straightforward to derive. The subdivision (II) requires further clarification. Here we encounter another instance of a new feature appearing in the OSp magnetic quiver. The appearance of an exotic bi-fundamental, denoted by a dashed link in the orthosymplectic quiver in Table  \ref{tab:magnetic quivers YN11} corresponding to maximal subdivision (II). To explain the origin, as well as the meaning of this link, we refer to Figure \ref{fig:exotic bifundamental}. 
Here we show the subdivisions responsible for the two U(1) nodes from which this link emanates. Denoting the coordinates of the $x^{7,8,9}$ directions, which are the directions where 7-branes extend but 5-branes do not extend, of a given subweb by $\Vec{m}_i$, its mirror image must be at coordinate $-\Vec{m}_i$. The distance between the upper right and upper left subwebs in, say the $x^7$ direction, in Figure \ref{fig:exotic bifundamental} is therefore given by $|m_1^{(7)}-m_2^{(7)}|$. We claim that this gives rise to an ordinary bi-fundamental hypermultiplet transforming in the $(1,-1)$ representation of $\text{U}(1)\times \text{U}(1)$. An intuitive explanation of this is that a D3-brane extended between these two subwebs does not feel the presence of the orientifold and is oriented. In contrast, the distance between the upper right subweb and the lower-left subweb (mirror image to upper left) is given by $|m_1^{(7)}+m_2^{(7)}|$. This gives rise to an exotic hypermultiplet transforming as $(1,1)$ under $\text{U}(1)\times \text{U}(1)$. Since a D3-brane extending between these two subwebs must cross the orientifold, it is unoriented, which gives an intuitive explanation for the hypermultiplet's democratic nature. The number of each type of hypermultiplet follows, as is standard, by computing the  
stable intersection number. Altogether, this leads to the magnetic quiver shown in Figure \ref{fig:exotic bifundamental}. This explains the appearance of the dashed lines in the second row of Table  \ref{tab:magnetic quivers YN11}. To make the proposal more convincing, we compute the Coulomb branch Hilbert series for the OSp and unitary magnetic quivers. The results are collected in Table  \ref{YN11HS}, and they agree with each other.
\begin{table}[!htb]
\centering 
\begin{tabular}{|c|C{8.3cm}|C{5cm}|} \hline
\rowcolor{Grayy}
MS&Unitary magnetic&Orthosymplectic magnetic \\ \hline
 (I) & \includestandalone[width=0.45\textwidth]{YN11PhaseIU} & \includestandalone[width=0.3\textwidth]{YN11PhaseIOSp} \\ \hline
(II) & \includestandalone[width=0.55\textwidth]{YN11PhaseIIU} & \hspace*{-1.25cm}\includestandalone[width=0.5\textwidth]{YN11PhaseIIOSp} \\ \hline
(III) & \includestandalone[width=0.45\textwidth]{YN11PhaseIIIU} & \includestandalone[width=0.3\textwidth]{YN11PhaseIIIOSp} \\ \hline
\end{tabular} 
\caption[Magnetic quivers for $\text{Y}_{N}^{1,1}$ theory.]{The unitary and the orthosymplectic magnetic quivers derived from various maximal subdivisions (MS) corresponding to the unitary and the orientifold web diagrams in Figure \ref{fig:YN11 ordinary web} and Figure \ref{fig:YN11 O5} respectively for the Higgs branch of $\text{Y}_{N}^{1,1}$ theory.}
\label{tab:magnetic quivers YN11}
\end{table}
\begin{table}[!htb]
\centering
\begin{tabular}{|c|C{4.1cm}|C{4.1cm}|C{4.1cm}|} \hline
\rowcolor{Grayy}
   & Unitary magnetic quiver & \multicolumn{2}{c|}{Orthosymplectic magnetic quiver} \\ \cline{2-4}
	\rowcolor{Grayy}
  \multirow{-2}{*}{MS} & HS($t$)  & HS($t;\vec{m} \in \mathbb{Z}$) & HS($t;\vec{m} \in \mathbb{Z}+\tfrac{1}{2}$)  \\ \hline
	$(\text{I})_{N=1}$ & \footnotesize{$1+16t+185t^2+1585t^3+10919t^4+62648t^5+308937t^6+1338676t^7+5192925t^8+18300090t^9+59307538t^{10}+\ldots$}  & \footnotesize{$1+16t+153t^2+1105t^3+6759t^4+35992t^5+169449t^6+713140t^7+2714621t^8+9447450t^9+30359666t^{10}+\ldots$} & \footnotesize{$32t^2+480t^3+4160t^4+26656t^5+139488t^6+625536t^7+2478304t^8+8852640t^9+28947872t^{10}+\ldots$} \\ \hline
	$(\text{I})_{N=2}$ & \footnotesize{$1+28t+419t^2+4519t^3+39592t^4+298310t^5+\ldots$}  & \footnotesize{$1+28t+419t^2+4423t^3+37000t^4+261190t^5+\ldots$} & \footnotesize{$96t^3+2592t^4+37120t^5+\ldots$} \\ \hline
	$(\text{II})_{N=1}$ & \footnotesize{$1+18t+246t^2+2266t^3+15910t^4+89506t^5+422730t^6+1728642t^7+6272807t^8+20573244t^9+61888524t^{10}+\ldots$}  & \footnotesize{$1+18t+198t^2+1530t^3+9574t^4+50466t^5+229338t^6+914946t^7+3266279t^8+10596380t^9+31638956t^{10}+\ldots$} & \footnotesize{$48t^2+736t^3+6336t^4+39040t^5+193392t^6+813696t^7+3006528t^8+9976864t^9+30249568t^{10}+\ldots$} \\ \hline
	$(\text{II})_{N=2}$ & \footnotesize{$1+30t+476t^2+5465t^3+51395t^4+416458t^5+\ldots$}  & \footnotesize{$1+30t+476t^2+5305t^3+46915t^4+350474t^5+\ldots$} & \footnotesize{$160t^3+4480t^4+65984t^5+\ldots$} \\ \hline
	$(\text{III})_{N=1}$ & \footnotesize{$\begin{array}{l} \dfrac{P_8(t)}{(1-t)^{14}\,(1+t)^7} \\\\ = 1+13t+121t^2+797t^3\\+4240t^4+18760t^5+\ldots  
	\end{array}$}  & \footnotesize{$\begin{array}{l} \dfrac{P_9(t)}{(1-t)^{14}\,(1+t+t^2+t^3)^7} \\\\ = 1+13t+105t^2+605t^3\\+2864t^4+11640t^5+\ldots  \end{array}$} & \footnotesize{$\begin{array}{l} \dfrac{P_{10}(t)}{(1-t)^{14}\,(1+t+t^2+t^3)^7} \\\\ = 16t^2+192t^3+1376t^4\\+7120t^5+\ldots  
	\end{array}$} \\ \hline
	$(\text{III})_{N=2}$ & \footnotesize{$1+28t+419t^2+4452t^3+37756t^4+270816t^5+\ldots$}  & \footnotesize{$1+28t+419t^2+4388t^3+36028t^4+246496t^5+\ldots$} & \footnotesize{$64t^3+1728t^4+24320t^5+\ldots$} \\ \hline
\end{tabular}
\caption[Coulomb branch HS for magnetic quivers of $\text{Y}_{N}^{1,1}$ theory.]{Coulomb branch Hilbert series of the unitary and the orthosymplectic magnetic quivers for different maximal subdivisions of the $\text{Y}_{N}^{1,1}$ theory. The corresponding quivers are presented in Table \ref{tab:magnetic quivers YN11}. The explicit forms of $P_8(t), P_9(t), P_{10}(t)$ are given in Appendix \ref{sec:app2}.}
\label{YN11HS}
\end{table}
\FloatBarrier
\subsubsection{The Y$_N^{2\times1,2\times1}$ theory}
The examples we consider are brane configurations obtained by intersecting $2N$ NS5, 2 $(1,1)$, 2 $(1,-1)$ and one D5 brane on top of an O5-plane which is asymptotically an O5$^-$-plane. We call the theory on the web the $Y_{N}^{2\times1,2\times1}$ theory. At low energies the theory is described by the following quiver,
\be\label{electric Y2121 OSp}
\begin{array}{c}\begin{tikzpicture}
\node {$[1S+1C]-\text{SO}(6)-\text{USp}(2)-\text{SO}(6)-\cdots-\text{USp}(2)-\text{SO}(6)-[1S+1C]$};
 \draw [thick,decorate,decoration={brace,amplitude=6pt},xshift=0pt,yshift=10pt]
(-4.25,0) -- (4.0,0)node [black,midway,xshift=0pt,yshift=20pt] {
$2N+1$};
\end{tikzpicture}
\end{array} ~.
\ee 
An orientifold web diagram is given in Figure \ref{fig:YN2x1 orientifold web}.

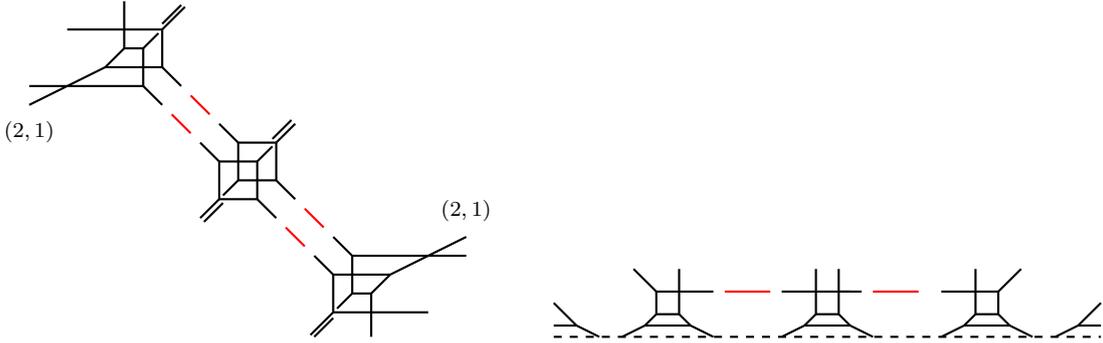
\begin{figure}[!htb]
    \centering
    \begin{scriptsize}
        \begin{tikzpicture}[scale=.5]
\draw[thick] (0,1)--(0,2);
\draw[thick] (0,2)--(-.5,2);
\draw[thick] (0.5,2.5)--(-2,2.5);
\draw[thick] (-3,1)--(0,1);
        
        \draw[thick] (-.5,3.25)--(-.5,2);
\draw[thick] (-.5,2)--(-1,1.5);
\draw[thick] (-1,1.5)--(-3,.5);
\node[label=below:{$(2,1)$}] at (-3,.5) {};
\draw[thick] (0,2)--(.4,2.4);
\draw[thick] (.5,2.5)--(1.1,3.1);
\draw[thick] (.5,2.65)--(1.,3.15);

\draw[thick] (0.5,1.5)--(0.5,2.5);
\draw[thick] (-1,1.5)--(.5,1.5);
\draw[thick] (1,1)--(.5,1.5);
\draw[thick] (.5,.5)--(0,1);

        \draw[thick,red](.75,.25)--(1.25,-.25);
       \draw[thick,red](1.25,.75)--(1.75,.25);
       
        \draw[thick](2,-2)--(3,-2);
        \draw[thick](2,-2)--(2,-1);
        \draw[thick](3,-1)--(3,-2);
        \draw[thick](3,-1)--(2,-1);
        
        \draw[thick](3,-2)--(3.5,-2.5);
        \draw[thick](2,-2)--(1.5,-2.5);
        \draw[thick](2.1,-2.1)--(1.6,-2.6);
        \draw[thick](2,-1)--(1.5,-.5);
        \draw[thick](3,-1)--(3.4,-.6);

        \draw[thick](2.5,-1.5)--(3.5,-1.5);
        \draw[thick](2.5,-1.5)--(2.5,-.5);
        \draw[thick](3.5,-.5)--(3.5,-1.5);
        \draw[thick](3.5,-.5)--(2.5,-.5);
        
        \draw[thick](3.5,-1.5)--(4,-2);
        \draw[thick](2.5,-1.5)--(2.1,-1.9);
        \draw[thick](2.5,-.5)--(2,0);
        \draw[thick](3.5,-.5)--(4,0);
        \draw[thick](3.4,-.4)--(3.9,0.1);
        
        \draw[thick,red](3.75,-2.75)--(4.25,-3.25);
       \draw[thick,red](4.25,-2.25)--(4.75,-2.75);

\draw[thick] (5.5,-3.5)--(8.5,-3.5);
\draw[thick] (5.5,-3.5)--(5.5,-4.5);
\draw[thick] (5.5,-3.5)--(5,-3);
\draw[thick] (5.5,-4.5)--(6,-4.5);
\draw[thick] (6,-4.5)--(6,-5.65);
\draw[thick] (6,-4.5)--(6.5,-4);
\draw[thick] (6.5,-4)--(8.5,-3);
\node[label=above:{$(2,1)$}] at (8.5,-3) {};

\draw[thick] (5,-4)--(6.5,-4);
\draw[thick] (5,-4)--(5,-5);
\draw[thick] (5,-4)--(4.5,-3.5);
\draw[thick] (5,-5)--(7.5,-5);

\draw[thick] (5,-5)--(4.4,-5.6);
\draw[thick] (5,-5.15)--(4.5,-5.65);
\draw[thick](5.5,-4.5)--(5.1,-4.9);
        \end{tikzpicture}
    \end{scriptsize}
    \hspace{.25cm}
    \begin{scriptsize}
        \begin{tikzpicture}[scale=.6]
        \draw[thick,dashed](-2,0)--(10,0);
        \draw[thick](-1,0)--(-1.5,.25);
        \draw[thick](-2,0.25)--(-1.5,.25);
        \draw[thick](-2,0.75)--(-1.5,.25);
        
        \draw[thick](-.5,0)--(0,.25);
        \draw[thick](1,0.25)--(0,.25);
        \draw[thick](.25,.5)--(0,.25);
        \draw[thick](.25,.5)--(.75,.5);
        \draw[thick](1,.25)--(.75,.5);
        \draw[thick](1,.25)--(1.5,0);
        \draw[thick](.75,1.5)--(.75,.5);
        \draw[thick](.25,.5)--(0.25,1);
        \draw[thick](-.25,1.5)--(0.25,1);
        \draw[thick](1.5,1)--(0.25,1);
        
        \draw[thick,red](1.75,1)--(2.75,1);
        
        \draw[thick](5,0)--(4.5,.25);
        \draw[thick](3.5,0.25)--(4.5,.25);
        \draw[thick](4.25,.5)--(4.5,.25);
        \draw[thick](4.25,.5)--(3.75,.5);
        \draw[thick](3.5,.25)--(3.75,.5);
        \draw[thick](3.5,.25)--(3,0);
        \draw[thick](3.75,1.5)--(3.75,.5);
        \draw[thick](4.25,.5)--(4.25,1);
        \draw[thick](4.25,1.5)--(4.25,1);
        \draw[thick](4.5,1)--(4.25,1);
        \draw[thick](3,1)--(4.75,1);
        
        \draw[thick,red](5,1)--(6,1);

        \draw[thick](9,0)--(9.5,.25);
        \draw[thick](10,0.25)--(9.5,.25);
        \draw[thick](10,0.75)--(9.5,.25);
        
        \draw[thick](8.5,0)--(8,.25);
        \draw[thick](7,0.25)--(8,.25);
        \draw[thick](7.75,.5)--(8,.25);
        \draw[thick](7.75,.5)--(7.25,.5);
        \draw[thick](7,.25)--(7.25,.5);
        \draw[thick](7,.25)--(6.5,0);
        \draw[thick](7.25,1.5)--(7.25,.5);
        \draw[thick](7.75,.5)--(7.75,1);
        \draw[thick](8.25,1.5)--(7.75,1);
        \draw[thick](6.5,1)--(7.75,1);
        
        \end{tikzpicture}
    \end{scriptsize}
    \caption{Constructing web diagrams for Y$_N^{2\times1,2\times1}$ theory by gauging SU(2)'s.}
    \label{fig:gauging su2 vs usp2}
\end{figure}

It may also be understood as gluing $N-1$ copies of $\text{SO}(6)$ with 2 vectors and 2 copies of $\text{SO}(6)$ with one vector, one spinor, and one conjugate spinor, by successive gauging of $\text{USp}(2)$ subgroups of the global symmetry. This latter viewpoint allows us to construct a unitary web for the same theory, by gluing $N-1$ copies of $\text{SU}(4)_0$ with two second rank antisymmetric hypermultiplets, and 2 copies of $\text{SU}(4)_0$ with 2 fundamentals and one 2nd rank antisymmetric hypermultiplet, via gauging common $\text{SU}(2)$ subgroups of the global symmetry. The construction is illustrated in Figure \ref{fig:gauging su2 vs usp2}.
Using this method, we obtain a unitary web diagram for the theory of \eqref{electric Y2121 OSp}, and it is depicted in Figure \ref{fig:YN2x1 unitary web}.

At infinite coupling, there are 10 maximal subdivisions of the unitary web, of which we only show 8 explicitly in Figure \ref{fig:YN2x1 unitary web}. Two further subdivisions are obtained, by 180-degree rotation of those labeled (V) and (VIII) in Figure \ref{fig:YN2x1 unitary web}. The unitary magnetic quivers follow straightforwardly and are listed in the second column of Table \ref{YN2121quivers}. In the following, we provide a guide to extract the magnetic quivers from the orientifold web.

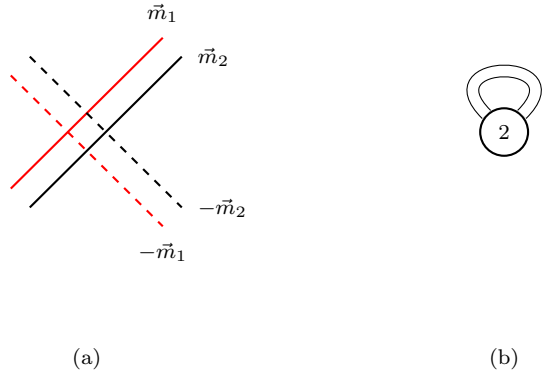
\begin{figure}[!htb]
    \centering
    \begin{scriptsize}
        \begin{tikzpicture}
        \draw[thick,red](-1,-.75)--(1,1.25);
        \node[label=above:{$\vec{m}_1$}]at(1,1.25){};
        \draw[thick](-.75,-1)--(1.25,1);
        \node[label=right:{$\vec{m}_2$}]at(1.25,1){};
        \draw[thick,dashed](-.75,1)--(1.25,-1);
        \draw[thick,red,dashed](-1,.75)--(1,-1.25);
        \node[label=below:{$-\vec{m}_1$}]at(1,-1.25){};
        \node[label=right:{$-\vec{m}_2$}]at(1.25,-1){};
        \node at (0,-3){(a)};
        \end{tikzpicture}
    \hspace{2cm}
    \begin{tikzpicture}
    \node[circle,inner sep=4pt,draw,thick](black) at (0,2){2};
     \draw[double distance=4pt] (black)to[out=135,in=45,loop,looseness=6](black);
     \node at (0,-1){(b)};
    \end{tikzpicture}
     \end{scriptsize}
    \caption[Origin of the antisymmetric matter.]{The origin of the antisymmetric matter appearing in some magnetic quivers.}
    \label{fig:antisymmetric hyper}
\end{figure}

Consider the maximal subdivision (I) corresponding to the orientifold web in Figure \ref{fig:YN2x1 orientifold web}. The corresponding orthosymplectic magnetic quiver appears in Table \ref{YN2121quivers}. It is a fairly tame object, except for the appearance of the 2 antisymmetric hypermultiplets attached to one of the U(2) nodes. Figure \ref{fig:antisymmetric hyper} shows the subweb responsible for the U(2) gauge node.
There are two $(1,1)$ 5-branes, whose position in the transverse $x^{7,8,9}$ directions are denoted by $\vec{m}_1$ and $\vec{m}_2$ respectively. Consider a D3-brane which is suspended between the $(1,1)$ 5-brane at position $m_1^{(7)}$, and the mirror of the $(1,1)$ 5-brane at position $m_2^{(7)}$, which is located at position $-m_2^{(7)}$ along the $x^7$ direction. Clearly, the distance between the two subwebs is $|m_1^{(7)}+m_2^{(7)}|$, which is the weight corresponding to the second rank antisymmetric representation of U(2).\footnote{In this specific case, the antisymmetric representation of U(2) is a singlet under the SU(2) factor and carries charge 2 under the U(1) factor of the gauge group. However, this should be distinguished from the rule on charge 2 hypermultiplets. For higher rank groups we expect a hypermultiplet transforming under the antisymmetric of U($N$).
} The fact that there are two such multiplets follows, as is standard, from the stable intersection of the $(1,1)$ and $(1,-1)$ 5-branes. One can repeat this exercise in the presence of $n$ $(1,1)$ 5-branes and their mirror images, and identify the weight system for the second rank antisymmetric representation of U($n$) in a similar manner. To make this proposal more concrete, we computed the Coulomb and Higgs branch Hilbert series of the full orthosymplectic magnetic quiver corresponding to maximal subdivision (I) in Table \ref{YN2121quivers}. The results are in agreement with the unitary magnetic quiver (see Table \ref{YN2121HS} for the matching of the Coulomb branch Hilbert series).

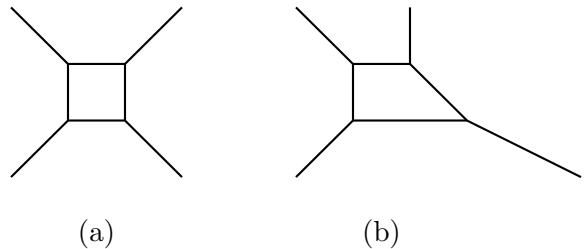
\begin{figure}[!htb]
\centering
\begin{tikzpicture}[scale=.75]
\draw[thick](0,0)--(1,0);
\draw[thick](0,0)--(0,-1);
\draw[thick](1,0)--(1,-1);
\draw[thick](0,-1)--(1,-1);
\draw[thick](0,0)--(-1,1);
\draw[thick](0,-1)--(-1,-2);
\draw[thick](1,-1)--(2,-2);
\draw[thick](1,0)--(2,1);
\node at (.5,-3){(a)};
\draw[thick](5,0)--(6,0);
\draw[thick](5,0)--(5,-1);
\draw[thick](6,0)--(7,-1);
\draw[thick](5,-1)--(7,-1);
\draw[thick](5,0)--(4,1);
\draw[thick](5,-1)--(4,-2);
\draw[thick](7,-1)--(9,-2);
\draw[thick](6,0)--(6,1);
\node at (5.5,-3){(b)};
\end{tikzpicture}
\caption[Usual 5-brane web diagrams for $E_1$ and $\tilde{E}_1$ SCFT.]{Usual 5-brane web diagram for (a) $E_1$ SCFT and (b) $\tilde{E}_1$ SCFT .}
\label{fig:E1-E1t}
\end{figure}

Next, consider the maximal subdivision (III) of the orientifold web. Here, we encounter the first example of orientifold web diagrams with identical shapes, that are actually inequivalent. In order to clarify this situation, we start from a much simpler example, known as the rank 1 $E_1$ SCFT and $\tilde{E}_1$ SCFT \cite{Morrison:1996xf}. They correspond to the 5d $\mathcal{N}=1$ pure SU(2) gauge theories with discrete theta angle 0 and $\pi$, respectively. When we describe them in terms of the ordinary 5-brane web diagram, there are clear differences, as depicted in Figure \ref{fig:E1-E1t}  \cite{Aharony:1997bh}. 
This difference is interpreted as two inequivalent ways of decomposing an $O7^{-}$-plane into two $(p,q)$ 7-branes \cite{Bergman:2015dpa}. 

The situation in the 5-brane web diagram with O5-plane for these two theories is much subtler. These two web diagrams cannot be distinguished in the weakly coupled phase. However, there are clear difference in another phase as depicted in Figure \ref{fig:E1O5} and \ref{fig:E1tO5} \cite{Hayashi:2017btw}.  This phase is the counterpart of the phase denoted in \cite{Aharony:1997bh} as ``past infinite coupling''. 
These claims are justified from the analysis of the decompactification limit of the Seiberg-Witten curve obtained from the M5-brane configuration corresponding to these 5-brane webs with O5-plane.
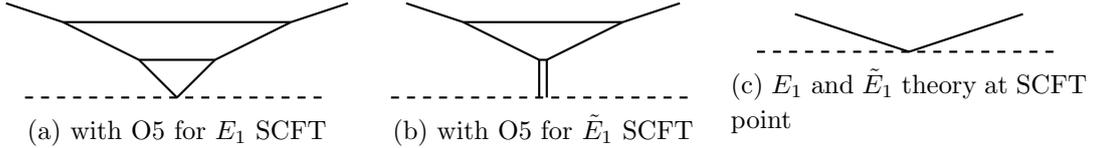
\begin{figure}[!htb]
\centering
   \begin{subfigure}{0.31\linewidth} \centering
     \begin{tikzpicture}[scale=.5]
\draw[thick,dashed](-4,0)--(4,0);
\draw[thick](0,0)--(1,1);
\draw[thick](0,0)--(-1,1);
\draw[thick](-1,1)--(1,1);
\draw[thick](3,2)--(1,1);
\draw[thick](-3,2)--(-1,1);
\draw[thick](-3,2)--(3,2);
\draw[thick](-3,2)--(-4.5,2.5);
\draw[thick](3,2)--(4.5,2.5);
\end{tikzpicture}
     \caption{with O5 for $E_1$ SCFT}\label{fig:E1O5}
   \end{subfigure}
   \begin{subfigure}{0.31\linewidth} \centering
     \begin{tikzpicture}[scale=.5]
\draw[thick,dashed](-4,0)--(4,0);
\draw[thick] (-.1,0)--(-.1,1);
\draw[thick] (.1,0)--(.1,1);
\draw[thick] (-.1,1)--(.1,1);
\draw[thick] (-2.1,2)--(-.1,1);
\draw[thick] (2.1,2)--(.1,1);
\draw[thick] (2.1,2)--(-2.1,2);
\draw[thick] (3.6,2.5)--(2.1,2);
\draw[thick] (-3.6,2.5)--(-2.1,2);
\end{tikzpicture}
     \caption{with O5 for $\tilde{E}_1$ SCFT}\label{fig:E1tO5}
   \end{subfigure}
	\begin{subfigure}{0.31\linewidth} \centering
     \begin{tikzpicture}[scale=.5]
\draw[thick,dashed](-4,0)--(4,0);
\draw[thick](0,0)--(3,1);
\draw[thick](0,0)--(-3,1);
\end{tikzpicture}
     \caption{$E_1$ and $\tilde{E}_1$ theory at SCFT point}\label{fig:E1E1tSCFT}
   \end{subfigure}
\caption{5-brane web diagrams with O5-plane for $E_1$ and $\tilde{E}_1$ SCFTs.} \label{fig:twofigs}
\end{figure}
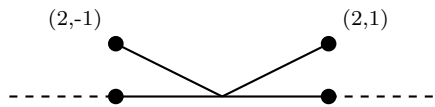
\begin{figure}
\centering
\begin{scriptsize}
\begin{tikzpicture}[scale=.7]
\draw[thick,dashed](0,0)--(2,0);
\draw[thick](4,0)--(6,1);
\draw[thick](4,0)--(2,1);
\node[7brane]at(2,0){};
\draw[thick](2,0)--(6,0);
\node[7brane]at(6,0){};
\node[7brane][label=above right:{(2,1)}]at(6,1){};
\node[7brane][label=above left:{(2,-1)}]at(2,1){};
\draw[thick,dashed](6,0)--(8,0);
\end{tikzpicture}
\end{scriptsize}
\caption{5-brane web diagram with O5-plane for $E_3$ SCFT.}
\label{fig:E3O5}
\end{figure}

Suppose that we start from these two different webs and go to SCFT point. Then, the difference disappears at the level of the 5-brane web diagram, as depicted in Figure \ref{fig:E1E1tSCFT}. However, they should still be distinguished, taking into account that they correspond to two inequivalent 5d SCFT. 
That is, we should distinguish the 5-brane web diagram in Figure \ref{fig:E1E1tSCFT} as a limit of Figure \ref{fig:E1O5}, from the 5-brane web diagram in Figure \ref{fig:E1E1tSCFT} as a limit of Figure \ref{fig:E1tO5}. 
Since we have already known that the former should give non-trivial Higgs branch while the latter should not give any continuous Higgs branch \cite{Cremonesi:2015lsa},  
we denote these two webs as ``decomposable'' web and ``not decomposable'' web, respectively.
This discussion can be generalized to the web diagram where a $(p,1)$-5 brane and its mirror image are intersecting on top of the O5$^-$-plane for any $p$. This may be either decomposable to give non-trivial Higgs branch, or not decomposable to give no continuous Higgs branch.

As discussed in \cite{Bourget:2020gzi}, we can see only one Higgs branch of the rank 1 $E_3$ SCFT from naive analysis with the 5-brane web with O5-plane. However, once we accept the claims above, we can reproduce the two different branches of rank 1 $E_3$ SCFT. Depending on whether the $(2,1)$ 5-brane in Figure \ref{fig:E3O5} is either decomposable or not decomposable, they lead to two different magnetic quivers:
\begin{equation}
\begin{array}{c}
\begin{scriptsize}
    \begin{tikzpicture}
    \node[label=below:{2}][so](so2){};
    \node[label=below:{1}][u](u1)[right of=so2]{};
    \node[label=below:{1}][uf](uf)[right of=u1]{};
    \draw(so2)--(u1);
    \path [draw,snake it](u1)--(uf);
\end{tikzpicture}
\end{scriptsize}
\end{array} \quad \text{or} \quad
    \begin{array}{c}
		\begin{scriptsize}
    \begin{tikzpicture}
    \node[label=below:{2}][so](so2'){};
    \node[label=below:{2}][spf](spf)[right of=so2']{};
    \draw(so2')--(spf);
    \end{tikzpicture}  
					\end{scriptsize}
    \end{array}
\end{equation}
Here, not decomposable web, which cannot be detached from the O5-plane, contribute as a flavor.
A different example of treating a subweb which cannot be detached from the O5-plane as a flavor is discussed in \cite{Bourget:2020gzi}. 

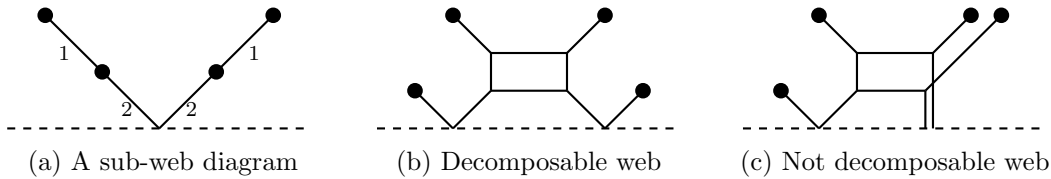
\begin{figure}[!htb]
\centering
   \begin{subfigure}{0.31\linewidth} \centering
     \begin{scriptsize}
    \begin{tikzpicture}[scale=.5]
    \draw[thick,dashed](-4,0)--(4,0);
    \draw[thick](0,0)--(3,3);
    \draw[thick](0,0)--(-3,3);
    \node[7brane]at (1.5,1.5){};
    \node[7brane]at (3,3){};
    \node[7brane]at (-1.5,1.5){};
    \node[7brane]at (-3,3){};
    \node at (2.5,2){1};
    \node at (.85,.5){2};
    \node at (-2.5,2){1};
    \node at (-.85,.5){2};
    \end{tikzpicture}
\end{scriptsize}
     \caption{A sub-web diagram}\label{fig:VSCFT}
   \end{subfigure}
   \begin{subfigure}{0.31\linewidth} \centering
     \begin{tikzpicture}[scale=.5]
\draw[thick,dashed](-4,0)--(4,0);
\draw[thick](-2,0)--(-3,1);
\draw[thick](2,0)--(3,1);
\draw[thick](2,0)--(1,1);
\draw[thick](-2,0)--(-1,1);
\draw[thick](1,1)--(-1,1);
\draw[thick](1,1)--(1,2);
\draw[thick](-1,1)--(-1,2);
\draw[thick](1,2)--(-1,2);
\draw[thick](2,3)--(1,2);
\draw[thick](-2,3)--(-1,2);
\node[7brane] at (2,3){};
\node[7brane] at (-2,3){};
\node[7brane] at (3,1){};
\node[7brane] at (-3,1){};
\end{tikzpicture}
     \caption{Decomposable web}\label{fig:VE1}
   \end{subfigure}
	\begin{subfigure}{0.31\linewidth} \centering
    \begin{tikzpicture}[scale=.5]
\draw[thick,dashed](-4,0)--(4,0);
\draw[thick](-2,0)--(-3,1);
\draw[thick](-2,0)--(-1,1);
\draw[thick](.8,1)--(-1,1);
\draw[thick](.8,1)--(2.8,3);
\draw[thick](.8,1)--(.8,0);
\draw[thick](1,0)--(1,2);
\draw[thick](-1,1)--(-1,2);
\draw[thick](1,2)--(-1,2);
\draw[thick](2,3)--(1,2);
\draw[thick](-2,3)--(-1,2);
\node[7brane] at (2,3){};
\node[7brane] at (-2,3){};
\node[7brane] at (-3,1){};
\node[7brane] at (2.8,3){};
\end{tikzpicture}
     \caption{Not decomposable web}\label{fig:VE1t}
   \end{subfigure}
\caption{Decomposable and not decomposable orientifold web diagrams in $Y^{2 \times 1, 2 \times 1}_N$.} \label{Subweb-YN2121}
\end{figure}

Analogous discussion is possible for still another type of 5-brane web.
We should distinguish the diagram in Figure \ref{fig:VSCFT} as a limit of the diagram in Figure \ref{fig:VE1}, which is decomposable, and the diagram in Figure \ref{fig:VSCFT} as a limit of the diagram in Figure \ref{fig:VE1t}, which is not decomposable. The former can be decomposed in a natural way to give $\text{U}(1) \times \text{U}(2) \times \text{U}(1)$ gauge group in the magnetic quiver while the latter cannot be detached from the O5-plane and thus should be treated as a flavor. 
This claim is justified by comparing them with the equivalent ordinary 5-brane web diagrams.


Now we go back to the $Y^{2 \times 1, 2 \times 1}_N$ theory. The orientifold web in Figure \ref{fig:YN2x1 orientifold web} includes the subweb in Figure \ref{fig:VSCFT}. The maximal subdivision (I) includes the decomposable web, and the maximal subdivision (III) includes a subweb that is not decomposable.



Among the eight maximal subdivisions, the maximal subdivision (V) in Figure \ref{fig:YN2x1 orientifold web} yields a magnetic quiver with a charge $2$ hypermultiplet. The maximal subdivision (V) contains the subweb in red in the maximal subdivision in Figure \ref{fig:YN2x1 orientifold web}. We compute the number of the charge $2$ hypermultiplets associated with the U(1) gauge node from the subweb by the rule in \eqref{SI-charge-2}. The stable intersection number of the subweb with its mirror is 
\begin{equation}\label{SIwithmirrorYN2121}
\text{SI of subweb with its own mirror image} = 10 - 2 = 8,
\end{equation}
and the stable intersection number of the subweb with O5 is 
\begin{equation}\label{SIwithO5YN2121}
\text{SI of subweb with O5} = 3.
\end{equation}
Then the number of the charge $2$ hypermultiplets from \eqref{SI-charge-2} is 
\begin{equation}
\frac{8}{2} - 3 = 1,
\end{equation}
for the U(1) node associated with the subweb in red of the maximal subdivision (V) in Figure \ref{fig:YN2x1 orientifold web}.

\begin{figure}[!htb]
    \centering
		\begin{scriptsize}
    \begin{tikzpicture}[scale=.7]
    \draw[thick](-8,0)--(8,0);
    \draw[thick,dashed](-10,0)--(-8,0);
    \draw[thick,dashed](10,0)--(8,0);
    \node[7brane] at (4,0){};
    \node[7brane] at (8,0){};
    \node[7brane] at (-4,0){};
    \node[7brane] at (-8,0){};
    \draw[thick](0,0)--(0,3);
    \node[7brane] at (0,2){};
    \node[7brane] at (0,3){};
    \node[thick] at (0,3.4) {$.$};
\node[thick] at (0,3.5) {$.$};
\node[thick] at (0,3.6) {$.$};
\node[7brane] at (0,4){};
\node[7brane] at (0,5){};
\node[7brane] at (0,6){};
\node at (.5,5.5) {1};
\node at (.5,4.5) {2};
\node at (.75,2.5) {$2N-1$};
\node at (.35,1.5) {$2N$};
\draw[thick](0,4)--(0,6);
\draw[thick](0,0)--(2.5,2.5);
\draw[thick](0,0)--(-2.5,2.5);
\node[7brane] at (1.5,1.5){};
\node[7brane] at (-1.5,1.5){};
\node[7brane,label=above right:{$(1,1)$}] at (2.5,2.5){};
\node[7brane,label=above left:{$(1,-1)$}] at (-2.5,2.5){};
\node at (9,-.25) {$\text{O}5^-$};
\node at (7,0.3) {$\frac{1}{2}$};
\node at (3,0.25) {1};
\node at (-9,-.25) {$\text{O}5^-$};
\node at (-7,0.3) {$\frac{1}{2}$};
\node at (-3,0.25) {1};
\node at (-2,1.7) {1};
\node at (2,1.75) {1};
\node at (-1,.7) {2};
\node at (1,.75) {2};
\end{tikzpicture}
\vspace{0.5cm}
\begin{tikzpicture}[scale=.6]
\draw[thick](-2,-5)--(2,-5);
\draw[thick,green](0,-7.25)--(0,-2.75);
\draw[thick,red,dashed](-1.75,-3.25)--(1.75,-6.75);
\draw[thick,red](1.75,-3.25)--(-1.75,-6.75);
\node[7brane]at (-1.75,-3.25){};
\node[7brane]at (1.75,-3.25){};
\node[7brane]at (0,-2.75){};    
\node[7brane]at (2,-5){};
\node[7brane]at (-2,-5){};
\node[7brane]at (1.75,-6.75){};
\node[7brane]at (-1.75,-6.75){};
\node[7brane]at (0,-7.25){};
\node at (0,-8) {(I)};

\draw[thick](-9,-5)--(-7.1,-5);
\draw[thick](-7.1,-7)--(-7.1,-5);
\draw[thick,green](-7,-7)--(-7,-3);
\draw[thick](-5.35,-3.25)--(-7.1,-5);
\draw[thick,red](-8.65,-3.25)--(-6.9,-5);
\draw[thick,red](-6.9,-7)--(-6.9,-5);
\draw[thick,red](-4.9,-5)--(-6.9,-5);
\draw[thick,cyan] (-8.75,-6.85)--(-5.25,-3.35);
\draw[thick,cyan,dashed] (-5.25,-6.85)--(-8.75,-3.35);
\node[7brane]at (-8.75,-3.25){};
\node[7brane]at (-5.25,-6.85){};
\node[7brane]at (-8.75,-6.85){};
\node[7brane]at (-5.25,-3.25){};
\node[7brane]at (-7,-3){};
\node[7brane]at (-7,-7){};    
\node[7brane]at (-5,-5){};
\node[7brane]at (-9,-5){};
\node at (-7,-8) {(II)};

\draw[thick](5,-5)--(9,-5);
\draw[thick,green](7,-7.25)--(7,-2.75);
\draw[thick,red](6,-5)--(5,-4);
\draw[thick,red](6,-5)--(5,-6);
\draw[thick,red](8,-5)--(9,-4);
\draw[thick,red](8,-5)--(9,-6);
\node[7brane]at (7,-2.75){};    
\node[7brane]at (9,-5){};
\node[7brane]at (5,-5){};
\node[7brane]at (5,-4){};
\node[7brane]at (5,-6){};
\node[7brane]at (9,-4){};
\node[7brane]at (9,-6){};
\node[7brane]at (7,-7.25){};
\node at (7,-8) {(III)};

\draw[thick](-9,-12.3)--(-4.9,-12.3);
\draw[thick,red](-7.1,-10.5)--(-7.1,-12);
\draw[thick,green](-7,-14.3)--(-7,-10.5);
\draw[thick,red](-5.15,-13.75)--(-6.9,-12);
\draw[thick,red](-8.85,-13.75)--(-7.1,-12);
\draw[thick,red](-7.1,-12)--(-6.9,-12);
\draw[thick,red](-6.9,-9.75)--(-6.9,-12);

\draw[thick,cyan](-8.7,-13.9)--(-5.35,-10.55);
\draw[thick,cyan,dashed](-5.3,-13.9)--(-8.65,-10.55);
\node[7brane]at (-8.65,-13.75){};
\node[7brane]at (-5.35,-13.75){};
\node[7brane]at (-5.25,-10.55){};
\node[7brane]at (-8.75,-10.55){};
\node[7brane]at (-7.1,-10.5){};
\node[7brane]at (-6.9,-9.75){};
\node[7brane]at (-7,-14.3){};
\node[7brane]at (-5,-12.3){};
\node[7brane]at (-9,-12.3){};
\node at (-7,-15) {(IV)};

\draw[thick](2,-12.3)--(-2,-12.3);
\draw[thick,red](-.1,-10.5)--(-.1,-12);
\draw[thick,green](0,-14.3)--(0,-10.5);
\draw[thick,red](1.85,-13.75)--(.1,-12);
\draw[thick,red](-1.85,-13.75)--(-.1,-12);
\draw[thick,red](-.1,-12)--(.1,-12);
\draw[thick,red](.1,-9.75)--(.1,-12);

\node[7brane]at (-1.75,-13.75){};
\node[7brane]at (1.75,-13.75){};
\node[7brane]at (-.1,-10.5){};
\node[7brane]at (.1,-9.75){};
\node[7brane]at (0,-14.3){};
\node[7brane]at (2,-12.3){};
\node[7brane]at (-2,-12.3){};
\node at (1.25,-13.25) {2};
\node at (0,-15) {(VII) $N\geq2$};
\draw[thick,cyan](5.2,-12.3)--(7.2,-12.3);
\draw[thick,cyan](7.2,-12.3)--(8.95,-10.45);
\draw[thick,cyan](7.2,-12.3)--(7.2,-14.3);
\draw[thick](8.8,-12.3)--(6.8,-12.3);
\draw[thick](6.8,-12.3)--(5.05,-10.55);
\draw[thick](6.8,-12.3)--(6.8,-14.3);
\draw[thick,red](6.9,-10.5)--(6.9,-12);
\draw[thick,green](7,-14.3)--(7,-10.5);
\draw[thick,red](8.85,-13.75)--(7.1,-12);
\draw[thick,red](5.15,-13.75)--(6.9,-12);
\draw[thick,red](6.9,-12)--(7.1,-12);
\draw[thick,red](7.1,-9.75)--(7.1,-12);

\node[7brane]at (5.25,-13.75){};
\node[7brane]at (8.75,-13.75){};
\node[7brane]at (6.9,-10.5){};
\node[7brane]at (7.1,-9.75){};

\node[7brane]at (-1.75,-13.75){};
\node[7brane]at (1.75,-13.75){};
\node[7brane]at (-.1,-10.5){};
\node[7brane]at (.1,-9.75){};

\node[7brane]at (5,-10.45){};
\node[7brane]at (9,-10.45){};
\node[big7brane]at (7,-14.3){};    
\node[7brane]at (9,-12.3){};
\node[7brane]at (5,-12.3){};
\node at (7,-15) {(VIII) $N\geq2$};

\draw[thick](-2,-19)--(2,-19);
\draw[thick,green](0,-21.25)--(0,-16.75);
\draw[thick,red](0.15,-19)--(0.15,-16.75);
\draw[thick,red](-0.15,-19)--(-0.15,-16.75);
\draw[thick,red](0,-19)--(1.75,-20.75);
\draw[thick,red](0,-19)--(1.75,-17.25);
\draw[thick,red](0,-19)--(-1.25,-20.25);
\draw[thick,red](0.15,-19.15)--(-1.6,-20.9);

\node[7brane]at (1.75,-17.25){};
\node[7brane]at (-1.6,-20.9){};
\node[7brane]at (0,-16.75){};    
\node[7brane]at (2,-19){};
\node[7brane]at (-2,-19){};
\node[7brane]at (1.75,-20.75){};
\node[7brane]at (-1.25,-20.25){};
\node[7brane]at (0,-21.25){};
\node at (0,-22) {(V)};
    \end{tikzpicture}\end{scriptsize}
    \caption[Orientifold web diagram for $Y_{N}^{2\times1,2\times1}$ theory.]{An orientifold web for the $Y_{N}^{2\times1,2\times1}$ theory along with the possible maximal subdivisions at the centre of the junction. The number 2 in figure (VII) indicates that all the red lines correspond to two 5-branes.}
    \label{fig:YN2x1 orientifold web}    
\end{figure}
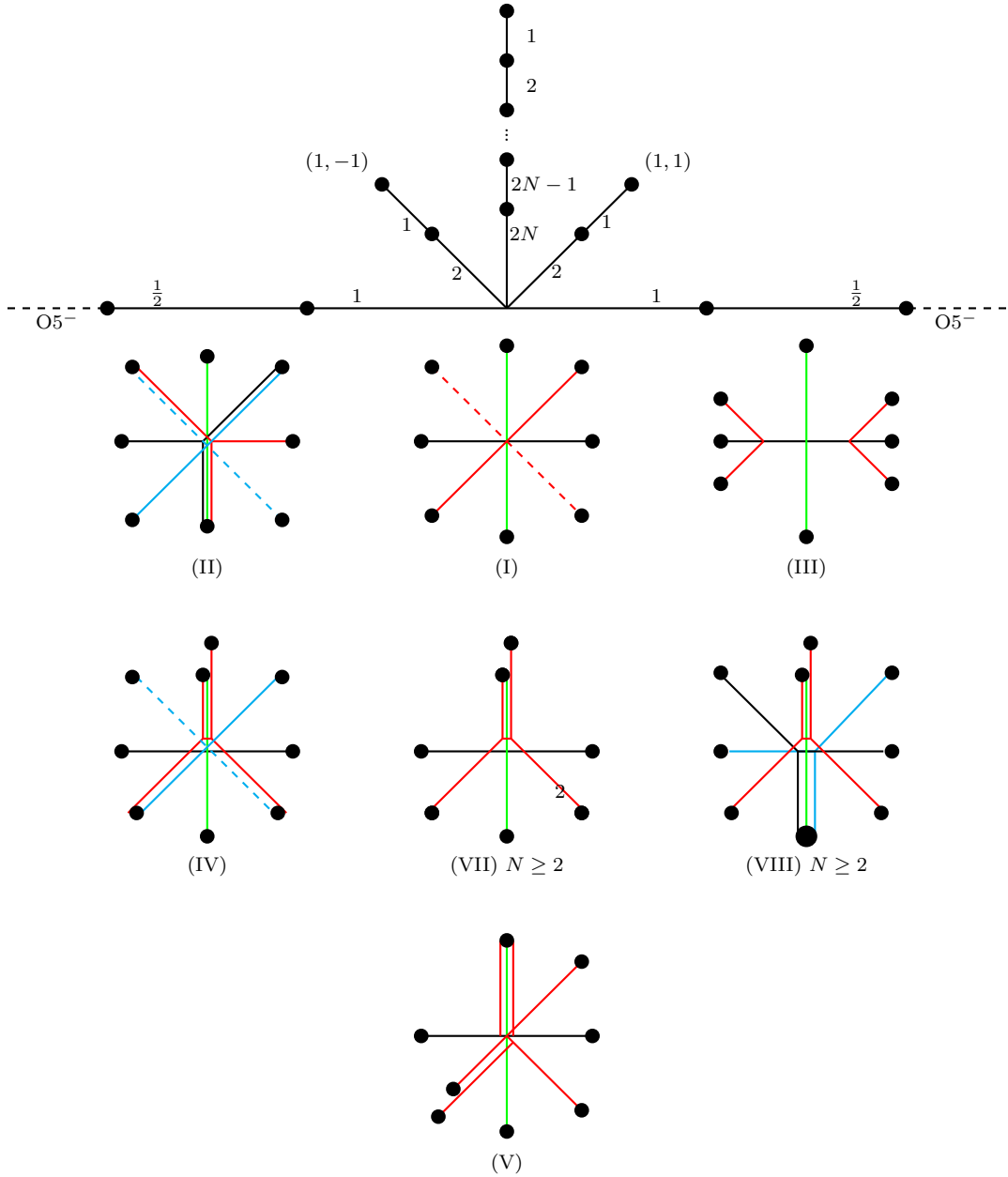
\begin{figure}[!htb]
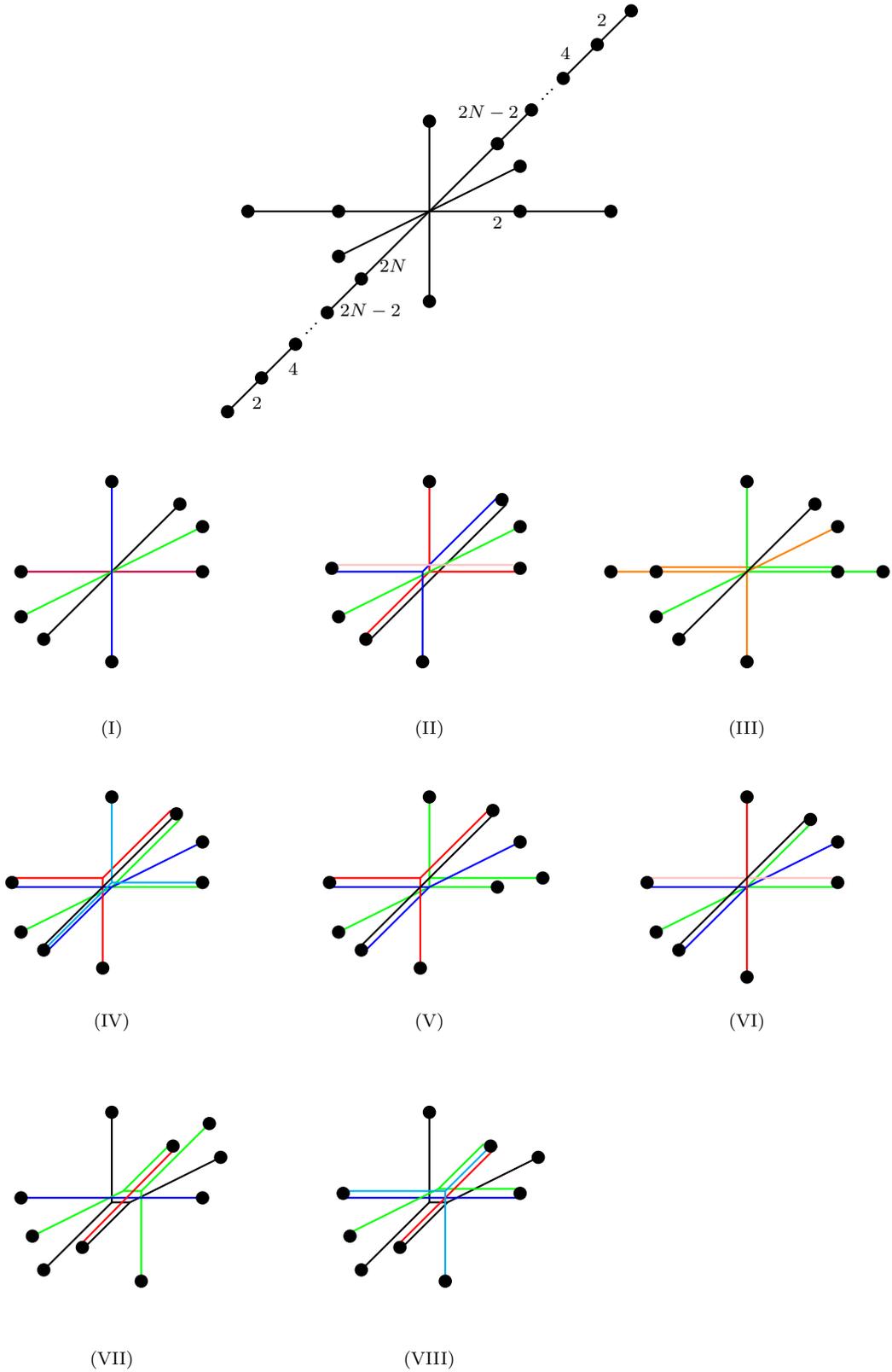

    \centering
    \includestandalone{YN2121UnitaryWeb}
    \caption[Unitary web diagram for $Y_{N}^{2\times1,2\times1}$ theory.]{A unitary web diagram for the $Y_{N}^{2\times1,2\times1}$ theory along with the possible maximal subdivisions at the centre of the junction.}
    \label{fig:YN2x1 unitary web}
\end{figure}

With the rules described above, we propose the magnetic quivers from the orientifold web for the eight maximal subdivisions and the result is summarized in Table \ref{YN2121quivers}.
\begin{table}
\centering
\vspace*{-2cm}\hspace*{-1.25cm}
\scalebox{0.95}{%
\begin{tabular}{|c|C{8.2cm}|C{7.3cm}|} \hline
\rowcolor{Grayy}
MS & Unitary & Orthosymplectic\\\hline
(I) &  \includestandalone{YN2121PhaseIU} & \includestandalone{YN2121PhaseIOSp} \\\hline
\shortstack{(II)\\ \& \\(VI)} & \includestandalone{YN2121PhaseIIU} & \includestandalone{YN2121PhaseIIOSp} \\\hline 
(III) & \includestandalone{YN2121PhaseIIIU} & \includestandalone{YN2121PhaseIIIOSp}  \\\hline 
(IV) & \includestandalone{YN2121PhaseIVU} & \includestandalone{YN2121PhaseIVOSp} \\\hline 
(V) & \includestandalone{YN2121PhaseVU} & \includestandalone{YN2121PhaseVOSp} \\\hline 
(VII) & \includestandalone{YN2121PhaseVIIU} & \includestandalone{YN2121PhaseVIIOSp} \\\hline 
(VIII) & \includestandalone{YN2121PhaseVIIIU} & \includestandalone{YN2121PhaseVIIIOSp} \\\hline 
\end{tabular}%
}
\caption[Magnetic quivers for $Y_{N}^{2\times1,2\times1}$ theory.]{Magnetic quivers for various maximal subdivisions (MS) of $Y_{N}^{2\times1,2\times1}$ theory.}
\label{YN2121quivers}
\end{table}
We have checked the matching of the Coulomb branch Hilbert series of the unitary and orthosymplectic magnetic quivers which can be seen from Table \ref{YN2121HS}.

\begin{table}
\centering
\begin{tabular}{|c|C{3.9cm}|C{3.9cm}|C{3.9cm}|} \hline
\rowcolor{Grayy}
   & Unitary & \multicolumn{2}{c|}{Orthosymplectic} \\ \cline{2-4}
	\rowcolor{Grayy}
  \multirow{-2}{*}{MS} & HS($t$)  & HS($t;\vec{m} \in \mathbb{Z}$) & HS($t;\vec{m} \in \mathbb{Z}+\tfrac{1}{2}$)  \\ \hline
	$(\text{I})_{N=0}$ & \footnotesize{$\begin{array}{l} \dfrac{P_{11}(t)}{(1-t)^{10}\,(1+t)^5} \\\\ = 1+16t+132t^2+735t^3\\+3134t^4+10974t^5+\ldots  
	\end{array}$}  & \footnotesize{$\begin{array}{l} \dfrac{P_{12}(t)}{(1-t)^{10}\,(1+t)^5\,(1+t^2)} \\\\ = 1+8t+72t^2+371t^3\\+1598t^4+5510t^5+\ldots  
	\end{array}$} & \footnotesize{$\begin{array}{l} \dfrac{P_{13}(t)}{(1-t)^{10}\,(1+t)^5\,(1+t^2)} \\\\ = 8t+60t^2+364t^3\\+1536t^4+5464t^5+\ldots  
	\end{array}$} \\ \hline
	$(\text{I})_{N=1}$ & \footnotesize{$1+11t+84t^2+485t^3+2346t^4+9738t^5+\cdots$}  & \footnotesize{$1+11t+68t^2+317t^3+1346t^4+5290t^5+\cdots$} & \footnotesize{$16t^2+168t^3+1000t^4+4448t^5+\cdots$} \\ \hline
	$(\text{I})_{N=2}$ & \footnotesize{$1+23t+290t^2+2653t^3+19602t^4+123630t^5+\ldots$}  & \footnotesize{$1+23t+290t^2+2605t^3+18522t^4+110470t^5+\ldots$} & \footnotesize{$48t^3+1080t^4+13160t^5+\ldots$} \\ \hline
	$\begin{array}{l} (\text{II})_{N=1} \\ \& \\ (\text{VI})_{N=1} \end{array}$ & \footnotesize{$1+12t+91t^2+8t^{5/2}+484t^3+104t^{7/2}+2032t^4+720t^{9/2}+7152t^5+\cdots$}  & \footnotesize{$1+12t+75t^2+336t^3+16t^{7/2}+1268t^4+208t^{9/2}+4220t^5+\cdots$} & \footnotesize{$16t^2+8t^{5/2}+148t^3+88t^{7/2}+764t^4+512t^{9/2}+2932t^5+\cdots$} \\ \hline
	$\begin{array}{l} (\text{II})_{N=2} \\ \& \\ (\text{VI})_{N=2} \end{array}$ & \footnotesize{$1+24t+313t^2+2943t^3+32t^{7/2}+22157t^4+768t^{9/2}+140921t^5+\cdots$}  & \footnotesize{$1+24t+313t^2+2895t^3+21089t^4+32t^{9/2}+128073t^5+\cdots$} & \footnotesize{$48t^3+32t^{7/2}+1068t^4+736t^{9/2}+12848t^5+\cdots$} \\ \hline
	$(\text{III})_{N=1}$ & \footnotesize{$\begin{array}{l} \dfrac{P_{14}(t)}{(1-t)^2\,(1-t^3)\,(1-t^4)^3} \\\\ = 1+4t+13t^2+33t^3\\+80t^4+165t^5+\cdots  
	\end{array}$}  & \footnotesize{$\begin{array}{l} \dfrac{P_{14}(t)}{(1-t)^2\,(1-t^3)\,(1-t^4)^3} \\\\ = 1+4t+13t^2+33t^3\\+80t^4+165t^5+\cdots  
	\end{array}$} & \footnotesize{not required} \\ \hline
	$(\text{III})_{N=2}$ & \footnotesize{$1+16t+151t^2+1039t^3+5750t^4+26954t^5+\cdots$}  & \footnotesize{$1+16t+151t^2+1039t^3+5750t^4+26954t^5+\cdots$} & \footnotesize{not required} \\ \hline
	$(\text{IV})_{N=1}$ & \footnotesize{$1+9t+43t^2+16t^{5/2}+157t^3+128t^{7/2}+488t^4+560t^{9/2}+1400t^5+\cdots$}  & \footnotesize{$1+9t+43t^2+157t^3+488t^4+1400t^5+\cdots$} & \footnotesize{$16t^{5/2}+128t^{7/2}+560t^{9/2}+\cdots$} \\ \hline
	$(\text{IV})_{N=2}$ & \footnotesize{$1+24t+313t^2+2860t^3+64t^{7/2}+20297t^4+1472t^{9/2}+118722t^5+\cdots$}  & \footnotesize{$1+24t+313t^2+2860t^3+20297t^4+118722t^5+\cdots$} & \footnotesize{$64t^{7/2}+1472t^{9/2}+\cdots$} \\ \hline
	$(\text{V})_{N=1}$ & \footnotesize{$\begin{array}{l} \dfrac{P_{15}(t)}{(1-t)^8\,(1-t^3)\,(1-t^5)^3} \\\\ = 1+8t+34t^2+8t^{5/2}\\+106t^3+56t^{7/2}+275t^4\\+216t^{9/2}+646t^5+\cdots  
	\end{array}$}  & \footnotesize{$\begin{array}{l} \dfrac{P_{16}(t)}{(1-t)^8\,(1-t^3)\,(1-t^{10})^3} \\\\ = 1+8t+34t^2+106t^3\\+275t^4+646t^5+\cdots  
	\end{array}$} & \footnotesize{$\begin{array}{l} \dfrac{8t^{5/2}P_{17}(t)}{(1-t)^8\,(1-t^3)\,(1-t^{10})^3} \\\\ = 8t^{5/2}+56t^{7/2}\\+216t^{9/2}+\cdots  
	\end{array}$} \\ \hline
	$(\text{V})_{N=2}$ & \footnotesize{$1+20t+224t^2+1803t^3+32t^{7/2}+11510t^4+608t^{9/2}+61468t^5+\cdots$}  & \footnotesize{$1+20t+224t^2+1803t^3+11510t^4+61468t^5+\cdots$} & \footnotesize{$32t^{7/2}+608t^{9/2}+\cdots$} \\ \hline
	$(\text{VII})_{N=2}$ & \footnotesize{$1 + 23 t + 259 t^2 + 1992 t^3 + 11927 t^4 + 59343 t^5+\cdots$}  & \footnotesize{$1 + 23 t + 259 t^2 + 1944 t^3 + 11075 t^4 + 51523 t^5+\cdots$} & \footnotesize{$48 t^3 + 852 t^4 + 7820 t^5+\cdots$} \\ \hline
	$(\text{VII})_{N=3}$ & \footnotesize{$1 + 43 t + 980 t^2 + 15615 t^3 + 194028 t^4+\cdots$}  & \footnotesize{$1 + 43 t + 980 t^2 + 15615 t^3 + 193868 t^4+\cdots$} & \footnotesize{$160 t^4+\cdots$} \\ \hline
	$(\text{VIII})_{N=2}$ & \footnotesize{$1+24t+297t^2+2560t^3+32t^{7/2}+17229t^4+736t^{9/2}+95960t^5+\cdots$}  & \footnotesize{$1+24t+297t^2+2512t^3+16299t^4+32t^{9/2}+86552t^5+\cdots$} & \footnotesize{$48t^3+32t^{7/2}+930t^4+704t^{9/2}+9408t^5+\cdots$} \\ \hline
\end{tabular}
\caption[Coulomb branch HS for magnetic quivers of $Y_{N}^{2\times1,2\times1}$ theory.]{Coulomb branch Hilbert series of the unitary and orthosymplectic magnetic quivers for different maximal subdivisions (MS) of $Y_{N}^{2\times1,2\times1}$ theory. The corresponding quivers are presented in Table \ref{YN2121quivers}. The explicit forms of $P_{11}(t), \cdots, P_{17}(t)$ are given in Appendix \ref{sec:app2}.}
\label{YN2121HS}
\end{table}

 There are a few more possible configurations in the $Y_N^{p,q}$ family. They do not give rise to any new rules, in addition to those already mentioned so far. They do however serve as working examples that demonstrate the validity of the rules proposed above. We refer the curious reader to Appendix \ref{appYN}.
 \FloatBarrier
\subsection{The \texorpdfstring{$H_N$}{TEXT} family}
The final example in section \ref{sec:O5-O5-} arises from 
decoupling flavors from the unitary electric quiver for the $\#_{3,N}$ theory \eqref{electric +3M unitary}. We denote by $H_N^{p,q}$, the theory obtained by decoupling flavors attached to the central node in \eqref{electric +3M unitary}. Here $p$ ($q$) are the number of flavors integrated out with positive (negative) mass such that $M=2N-p-q$. In other words, the $H_N^{p,q}$ theory is nothing but
\be\label{electric HN unitary}
\begin{array}{c}\begin{tikzpicture}
\node (horizontal){$\text{SU}(N)-\text{SU}(2N)_{\frac{p-q}{2}}-\text{SU}(N)$};
\node (flavour)[above of=horizontal]{$[M]$};
\draw(flavour)--(horizontal);
\end{tikzpicture}
\end{array} ~.
\ee
The unitary and orientifold web diagram for the $H_N^{p,q}$ theory is obtained from those of $+_{3,N}$ theory, i.e. Figure \ref{fig:unitary web for +3M} and Figure \ref{fig:+NM} respectively. Note that in the orientifold web of Figure \ref{fig:+NM}, the desirable mass deformation corresponds to the position of the external NS5 branes along the horizontal axis, a fact which is more transparent in the S-dual frame. Thus the orientifold web description of the $H_N^{p,q}$ theory is obtained from that of $+_{3, N}$ theory by decoupling, say, $p$ of the external NS5 branes to the left, and $q$ to the right. 

Let us focus on the case with $p = q = N$. The 5d theory \eqref{electric HN unitary} becomes 
\be\label{electric_HN_theory}
\begin{array}{c}\begin{tikzpicture}
\node (horizontal){$\text{SU}(N)-\text{SU}(2N)_{0}-\text{SU}(N).$};
\end{tikzpicture}
\end{array}
\ee
When $N=1$, the 5d theory is simply the SU(2) gauge theory with four flavors. 
The orientifold web diagram and the unitary web diagram at the infinitely strong coupling of the theory \eqref{electric_HN_theory} are depicted in Figure \ref{fig:HNU} and Figure \ref{fig:HNOSp} respectively. 
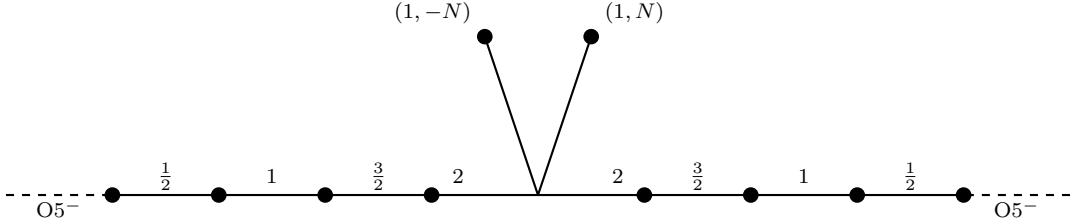
\begin{figure}[!ht]
    \centering\begin{scriptsize}
    \begin{tikzpicture}[scale=.7]
    \draw[thick](-8,0)--(8,0);
    \draw[thick,dashed](-10,0)--(-8,0);
    \draw[thick,dashed](10,0)--(8,0);
    \node[7brane] at (2,0){};
    \node[7brane] at (4,0){};
    \node[7brane] at (6,0){};
    \node[7brane] at (8,0){};
    \node[7brane] at (-2,0){};
    \node[7brane] at (-4,0){};
    \node[7brane] at (-6,0){};
    \node[7brane] at (-8,0){};
\draw[thick](0,0)--(1,3);
\draw[thick](0,0)--(-1,3);
\node[7brane,label=above right:{\vspace{-1cm}$(1,N)$}] at (1,3){};
\node[7brane,label=above left:{$(1,-N)$}] at (-1,3){};

\node at (9,-.25) {$\text{O}5^-$};
\node at (7,0.35) {$\frac{1}{2}$};
\node at (5,0.35) {1};
\node at (3,0.35) {$\frac{3}{2}$};
\node at (1.5,0.35) {2};
\node at (-1.5,0.35) {2};
\node at (-9,-.25) {$\text{O}5^-$};
\node at (-7,0.35) {$\frac{1}{2}$};
\node at (-5,0.35) {1};
\node at (-3,0.35) {$\frac{3}{2}$};
    \end{tikzpicture}\end{scriptsize}
\caption[Orientifold web diagram for $H_N^{N, N}$ theory.]{The orientifold web diagram for the $H_N^{N, N}$ theory at the infinitely strong coupling. }
\label{fig:HNU}
\end{figure}
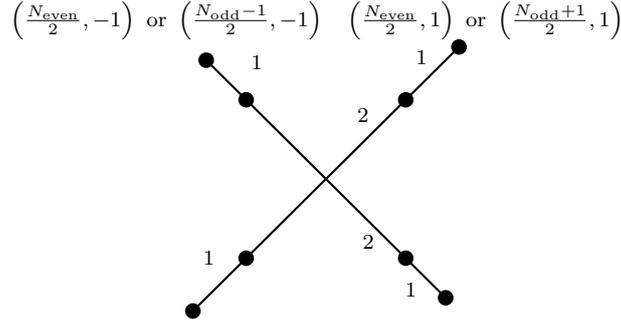
\begin{figure}[!ht]
\centering\begin{scriptsize}
    \begin{tikzpicture}[scale=.7]
\draw[thick,](-2.25,10.25)--(+2.25,5.75);
\node[thick][7brane] at (1.5,6.5) {};
\node[thick][7brane] at (-1.5,9.5) {};
\node[thick][7brane] at (2.25,5.75) {};
\node[thick][7brane] at (-2.25,10.25) {};
\node at (-1.3,10.2) {$1$};
\node at (-3,11) {$\left(\frac{N_{\text{even}}}{2}, -1\right) \text{ or } \left(\frac{N_{\text{odd}}-1}{2}, -1\right)$};

\node at (1.6,5.9) {$1$};
\node at (.8,6.8) {$2$};

\draw[thick](-2.5,5.5)--(2.5,10.5);
\node[7brane]at (-1.5,6.5){};
\node[7brane]at (1.5,9.5){};
\node[7brane]at (-2.5,5.5){};
\node[7brane]at (2.5,10.5){};
\node at (1.8,10.3) {$1$};
\node at (3, 11) {$\left(\frac{N_{\text{even}}}{2}, 1\right) \text{ or } \left(\frac{N_{\text{odd}} + 1}{2}, 1\right) $};
\node at (.7,9.2) {$2$};
\node at (-2.2,6.5) {$1$};
        \end{tikzpicture}\end{scriptsize}
\caption[Unitary web diagram for $H_N^{N, N}$ theory.]{The unitary web diagram for the $H_N^{N, N}$ theory at the infinitely strong coupling. }
\label{fig:HNOSp}
\end{figure}
An interesting point about the orientifold web diagram in Figure \ref{fig:HNU} is that the diagram has $(1, N)$ and $(1, -N)$ 5-branes where $N$ can be larger than $1$ and they intersect on the orientifold plane. This is a new feature which has not appeared in the past web diagrams. Hence this example is important for checking the rule \eqref{SI-charge-2} for the number of charge $2$ hypermultiplets attached to the U(1) gauge node originated from the $(1, N)$ and $(1, -N)$ 5-branes with $N \geq 2$\footnote{We will consider cases which involve $(p, q)$ and $(p, -q)$ 5-branes with both $p$ and $q$ larger than $1$ in Appendix \ref{sec:charge2ex}.}. From the web diagrams it is possible to infer the corresponding magnetic quiver theories and we argue that they are the ones given in Table \ref{tb:QuiversHNfamily}.
\begin{table}
\centering 
\begin{tabular}{|c|C{7.1cm}|C{5.9cm}|} \hline
\rowcolor{Grayy}
$H_N^{p, q}$ & Unitary magnetic & Orthosymplectic magnetic \\ \hline
$H_N^{N,N}$ & \includestandalone[width=0.48\textwidth]{HNU} & \hspace{-0.3cm}\includestandalone[width=0.41\textwidth]{HNOSp} \\ \hline
\end{tabular} 
\caption[Magnetic quivers for $H_N^{N, N}$ family.]{Magnetic quivers for the $H_N^{N, N}$ family. The index $N$ in each figure denotes the number of the hypermultiplets.}
\label{tb:QuiversHNfamily}
\end{table}
Note that the number of the charge $2$ hypermultiplets attached to the $U(1)$ gauge node in the orthosymplectic magnetic quiver 
is zero due to \eqref{SI-charge-2}. Indeed with this number for the charge $2$ hypermultiplets we find perfect agreement between the Coulomb branch Hilbert series of the unitary and orthosymplectic magnetic quiver theories. We summarize the Coulomb branch Hilbert series of the magnetic quivers in Table \ref{tb:QuiversHNfamily} for various $N$ in Table \ref{tb:HSHN}.
\begin{table}
\centering
\begin{tabular}{|c|C{4.1cm}|C{4.1cm}|C{4.1cm}|} \hline
\rowcolor{Grayy}
   & Unitary magnetic quiver & \multicolumn{2}{c|}{Orthosymplectic magnetic quiver} \\ \cline{2-4}
	\rowcolor{Grayy}
  \multirow{-2}{*}{$H_N^{N,N}$}
 & HS($t$)  & HS($t;\vec{m} \in \mathbb{Z}$) & HS($t;\vec{m} \in \mathbb{Z}+\tfrac{1}{2}$)  \\ \hline
	$H_2^{2,2}$ & \footnotesize{$1 + 13 t + 121 t^2 + 797 t^3 + 4240 t^4 + 18760 t^5 + \cdots$}  & \footnotesize{$1 + 13 t + 105 t^2 + 605 t^3 + 2864 t^4 + 11640 t^5 + \cdots$} & \footnotesize{$16 t^2 + 192 t^3 + 1376 t^4 + 7120 t^5+\cdots$} \\ \hline
	$H_3^{3,3}$ & \footnotesize{$1 + 13 t + 89 t^2 + 461 t^3 + 2007 t^4 + 7579 t^5+\cdots$}  & \footnotesize{$1 + 13 t + 89 t^2 + 445 t^3 + 1815 t^4 + 6347 t^5+\cdots$} & \footnotesize{$16 t^3 + 192 t^4 + 1232 t^5+\cdots$} \\ \hline
	$H_4^{4.4}$ & \footnotesize{$1 + 13 t + 89 t^2 + 429 t^3 + 1671 t^4 + 5659 t^5+\cdots$}  & \footnotesize{$1 + 13 t + 89 t^2 + 429 t^3 + 1655 t^4 + 5467 t^5+\cdots$} & \footnotesize{$16 t^4 + 192 t^5+\cdots$} \\ \hline
	$H_5^{5,5}$ & \footnotesize{$1 + 13 t + 89 t^2 + 429 t^3 + 1639 t^4 + 5323 t^5+\cdots$}  & \footnotesize{$1 + 13 t + 89 t^2 + 429 t^3 + 1639 t^4 + 5307 t^5+\cdots$} & \footnotesize{$16 t^5+\cdots$} \\ \hline
\end{tabular}
\caption[Coulomb branch HS for magnetic quivers of $H_N^{N, N}$ family.]{Coulomb branch Hilbert series of the unitary and orthosymplectic magnetic quivers for the $H_{N}^{N, N}$ family whose quivers are presented in Table \ref{tb:QuiversHNfamily}.}
\label{tb:HSHN}
\end{table}

\section{\texorpdfstring{Magnetic quivers from O5$^+$ - O5$^+$}{TEXT}}
\label{sec:O5+O5+}
Next we consider examples which arise from brane configurations with asymptotic O5$^+$-planes on both the ends. 
\subsection{The \texorpdfstring{$+_{1,N}$}{TEXT} theory}
\label{sec:+1NO5+O5+}
\begin{figure}[!ht]
    \centering
    \begin{scriptsize}
    \begin{tikzpicture}[scale=.7]
    \draw[thick](-4,0)--(4,0);
    \draw[thick,dashed](-6,0)--(-4,0);
    \draw[thick,dashed](6,0)--(4,0);
    \node[7brane] at (2,0){};
    \node[7brane] at (4,0){};
    \node[7brane] at (-2,0){};
    \node[7brane] at (-4,0){};
    \draw[thick](0,0)--(0,3);
    \node[7brane] at (0,2){};
    \node[7brane] at (0,3){};
    \node[thick] at (0,3.4) {$.$};
\node[thick] at (0,3.5) {$.$};
\node[thick] at (0,3.6) {$.$};
\node[7brane] at (0,4){};
\node[7brane] at (0,5){};
\node[7brane] at (0,6){};
\node at (.5,5.5) {1};
\node at (.5,4.5) {2};
\node at (.75,2.5) {$2N-1$};
\node at (.25,1.5) {$2N$};
\draw[thick](0,4)--(0,6);

\node at (5,-.25) {$\text{O}5^+$};
\node at (3,0.35) {$1$};
\node at (1.5,0.35) {1};
\node at (-1.5,0.35) {1};
\node at (-5,-.25) {$\text{O}5^+$};
\node at (-3,0.35) {$1$};

    \end{tikzpicture}\end{scriptsize}
    \caption[Orientifold web diagram for $+_{1,N}$ theory.]{
An orientifold web of the $+_{1,N}$ theory with asymptotically O5$^+$ orientifold planes at strong coupling.}
    \label{fig:+1M O5+-O5+}
    \end{figure}
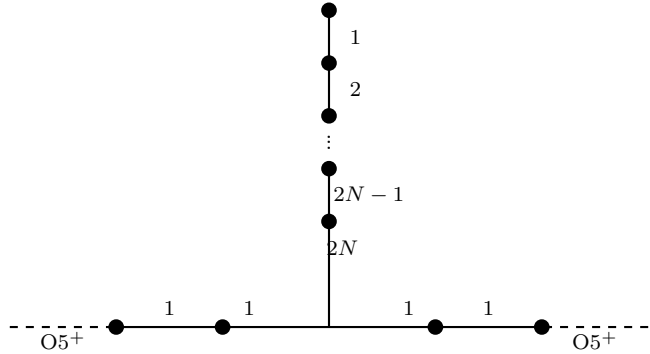

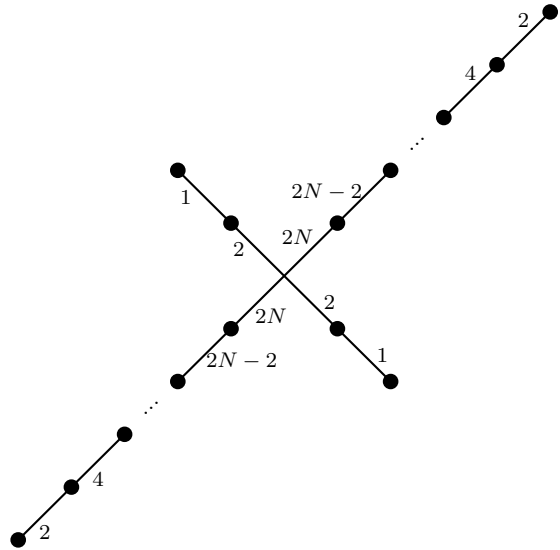
\begin{figure}
        \centering
        \begin{scriptsize}
        \begin{tikzpicture}[scale=.7]
        \draw[thick](-2,2)--(2,-2);
        \node[7brane]at(-2,2){};
        \node[7brane]at(-1,1){};
        \node[7brane]at(1,-1){};
        \node[7brane]at(2,-2){};
        \draw[thick](2,2)--(-2,-2);
        \node[7brane]at(-2,-2){};
        \node[7brane]at(2,2){};
        \node[7brane]at(-1,-1){};
        \node[7brane]at(1,1){};
        \node at (2.4,2.4){$.$};
        \node at (2.5,2.5){$.$};
        \node at (2.6,2.6){$.$};
        \node at (-2.4,-2.4){$.$};
        \node at (-2.5,-2.5){$.$};
        \node at (-2.6,-2.6){$.$};
        \draw[thick](3,3)--(5,5);
        \node[7brane]at(3,3){};
        \node[7brane]at(4,4){};
        \node[7brane]at(5,5){};
        \draw[thick](-3,-3)--(-5,-5);
        \node[7brane]at(-3,-3){};
        \node[7brane]at(-4,-4){};
        \node[7brane]at(-5,-5){};
        \node at(.85,-.5){2};
        \node at(1.85,-1.5){1};
        \node at(-.85,+.5){2};
        \node at(-1.85,1.5){1};
        \node at(.25,.75){$2N$};
        \node at(.8,1.6){$2N-2$};
        \node at(3.5,3.85){$4$};
        \node at(4.5,4.85){$2$};
        \node at(-.25,-.75){$2N$};
        \node at(-.8,-1.6){$2N-2$};
        \node at(-3.5,-3.85){$4$};
        \node at(-4.5,-4.85){$2$};
        
        \end{tikzpicture}
        \end{scriptsize}
        \caption[Unitary web diagram for $+_{1,N}$ theory.]{A unitary web for the $+_{1,N}$ theory with asymptotically O5$^+$ orientifold planes.}
        \label{fig:+1M unitary web}
    \end{figure}

Intersecting $2N$ NS5s, and a single D5 on top of on orientifold plane that is asymptotically an O5$^+$-plane, we arrive at the $+_{1,N}$ theory (Figure \ref{fig:+1M O5+-O5+}). It has an IR gauge theory description as 
\be\label{electric +1N O5+O5+ OSp}
\begin{array}{c}\begin{tikzpicture}
\node {$[1]-\text{SO}(6)-\text{USp}(2)-\text{SO}(6)-\cdots-\text{USp}(2)-\text{SO}(6)-[1]$};
 \draw [thick,decorate,decoration={brace,amplitude=6pt},xshift=0pt,yshift=10pt]
(-4.25,0) -- (4.0,0)node [black,midway,xshift=0pt,yshift=20pt] {
$2N-1$};
\end{tikzpicture}
\end{array} ~.
\ee
It can also be understood as gluing $N$ copies of SO(6) with two vector hypermultiplets by successive gauging of $\text{USp}(2)$ subgroups of the flavour symmetry. One can therefore engineer the same theory with an ordinary web diagram by gluing together $N$ copies of $\text{SU}(4)_0$ with 2 antisymmetric hypermultiplets, via successive gauging of SU(2) subgroups of the global symmetry (Figure \ref{fig:+1M unitary web}).

In this setup, the $2N$ NS5-branes intersecting with the O5$^+$-plane contribute to the magnetic quiver as a $\text{USp}(2N)$ gauge node.
Here, we claim that there is a new feature in this case, which did not appear for NS5-branes intersecting with O5$^-$-plane. 
In order for the orthosymplectic magnetic quiver to give consistent results with the corresponding unitary quiver, 
we find that we need to add three fundamental half-hypermultiplets on this $\text{USp}(2N)$ gauge node. 

We would like to interpret these three half-hypermultiplets as follows.
First, we observe that the RR charge of the O5$^+$-plane is identical to the sum of the RR charges of O5$^-$-plane and of four half D5-branes.
This motivates us to treat O5$^+$-plane as if it is the composite of them:
\begin{align}
( \text{O5}^+ \text{-plane} ) = ( \text{O5}^-  \text{-plane} ) + 4 \times (  \text{Half D5 branes}).
\label{eq:O5+-}
\end{align}
Here, we assume that the half D5 branes cannot be detached from the O5$^-$-plane. 
Basically, the charge 1 hypermultiplet can be reinterpreted as coming from 
the D3-branes suspended between the NS5-branes and these half D5-branes. 
However, we need a further explanation of why the number of half-hypermultiplets is three instead of four.
In this setup, there are two half D5-branes on top of the O5$^+$-plane, producing $\text{SO}(3)$ gauge group in the magnetic quiver.
We would like to interpret that one out of the four half D5-branes inside the O5$^+$-plane is used for constructing an $\text{SO}(3)$ gauge group.
In other words, the D3-brane suspended between the NS5-branes and this half D5-brane contributes as 
a part of the bi-fundamental hypermultiplets between the $\text{USp}(2N)$ node and the $\text{SO}(3)$ node.
This indicates that only the remaining three out of the four half D5-branes contribute as fundamental half-hypermultiplets. 

Under this assumption, we find the agreement between the Hilbert series of the orthosymplectic quiver in Figure \ref{O5++(+1N)OSpquiver}
and the Hilbert series of the unitary quiver in Figure \ref{O5++(+1N)unitaryquiver} both for the Coulomb branches and for the Higgs branches. In particular, the Higgs branch Hilbert series matching is crucial to settle the question about whether the $\text{O}(1)\simeq \mathbb{Z}_2$ nodes in the orthosymplectic quiver are flavor or gauge nodes. Such a match is only obtained if $\mathbb{Z}_2$ gaugings are assumed. The obtained result is:
\begin{equation}
\text{HS}_{H}(t^2)=1+3t^2+15t^4+36t^6+98t^8+\cdots .
\end{equation}
The results of the Coulomb branch Hilbert series is tabulated in Table \ref{O5++(+1N)}.

Encouraged by this agreement, we propose the following rule for the $\text{USp}(2N)$ gauge node
coming from the $2N$ NS5-branes intersecting with O5$^+$-plane.
If this $\text{USp}(2N)$ gauge node has a bi-fundamental hypermultiplet with the SO(odd) gauge node
coming from the D5-branes on the O5$^+$-plane, there are three fundamental half-hypermultiplets.
Otherwise, there are four fundamental half-hypermultiplets. 

		\begin{figure}[!htb]
\centering
   \begin{subfigure}{0.45\linewidth} \centering
     \begin{scriptsize}
    \begin{tikzpicture}
    \node[label=below:{2}][u](2){};
    \node[label=below:{4}][u](4)[right of=2]{};
    \node[label=below:{$2N-2$}][u](2m-2)[right of=4]{};
    \node[label=below:{$2N$}][u](2m)[right of=2m-2]{};
    \node[label=below:{$2N-2$}][u](2m-2')[right of=2m]{};
    \node[label=below:{$4$}][u](4')[right of=2m-2']{};
    \node[label=below:{$2$}][u](2')[right of=4']{};
    \node[label=above:{2}][u](u2)[above of=2m]{};
    \node[label=above:{1}][u](u1)[left of=u2]{};
    \node[label=above:{1}][u](u1')[right of=u2]{};
    \draw(2)--(4);
    \draw[dotted](2m-2)--(4);
    \draw(2m-2)--(2m);
    \draw(2')--(4');
    \draw[dotted](2m-2')--(4');
    \draw(2m-2')--(2m);
    \draw[double distance=2pt](2m)--(u2);
    \draw(u1)--(u2);
    \draw(u1')--(u2);
    \end{tikzpicture}
    \end{scriptsize}
     \caption{Unitary quiver}\label{O5++(+1N)unitaryquiver}
   \end{subfigure}
   \begin{subfigure}{0.45\linewidth} \centering
     \begin{scriptsize}
    \begin{tikzpicture}
    \node[label=left:{1}][u](1){};
    \node[label=left:{$2N-1$}][u](2m-1)[below of=1]{};
    \node[label=left:{$2N$}][sp](sp2m)[below of=2m-1]{};
    \node[label=right:{3}][sof](sof)[right of=sp2m]{};
    \node[label=below:{3}][so](so3)[below of=sp2m]{};
    \node[label=below:{2}][sp](sp2)[left of=so3]{};
    \node[label=below:{2}][sp](sp2')[right of=so3]{};
    \node[label=below:{1}][so](o1)[left of=sp2]{};
    \node[label=below:{1}][so](o1')[right of=sp2']{};
    \draw[dotted](1)--(2m-1);
    \draw(2m-1)--(sp2m);
    \draw(sp2m)--(sof);
    \draw(sp2m)--(so3);
    \draw(so3)--(sp2);
    \draw(so3)--(sp2');
    \draw(sp2)--(o1);
    \draw(sp2')--(o1');
    \end{tikzpicture}
    \end{scriptsize}
     \caption{Orthosymplectic quiver}\label{O5++(+1N)OSpquiver}
   \end{subfigure}
\caption[Magnetic quivers for $+_{1,N}$ theory.]{Magnetic quivers for the $+_{1,N}$ theory} \label{O5++(+1N)magneticquivers}
\end{figure}
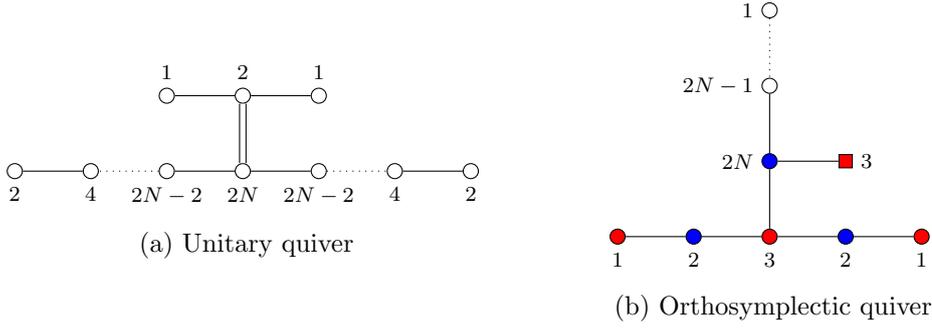
	\begin{table}
\centering
\begin{tabular}{|c|C{4.9cm}|C{4.9cm}|C{2.8cm}|} \hline
\rowcolor{Grayy}
   & Unitary magnetic quiver & \multicolumn{2}{c|}{Orthosymplectic magnetic quiver} \\ \cline{2-4}
	\rowcolor{Grayy}
  \multirow{-2}{*}{$+_{1,N}$}
 & HS($t$)  & HS($t;\vec{m} \in \mathbb{Z}$) & HS($t;\vec{m} \in \mathbb{Z}+\tfrac{1}{2}$)  \\ \hline
	$+_{1,1}$ & \footnotesize{$\begin{array}{l} \dfrac{P_{18}(t)}{(1-t)^{10}\,(1+t)^5} \\\\ = 1 + 13 t + 100 t^2 + 527 t^3 \\+ 2174 t^4 + 7425 t^5+\cdots  
	\end{array}$}  & \footnotesize{$\begin{array}{l} \dfrac{P_{18}(t)}{(1-t)^{10}\,(1+t)^5} \\\\ = 1 + 13 t + 100 t^2 + 527 t^3 \\+ 2174 t^4 + 7425 t^5+\cdots  
	\end{array}$} & \footnotesize{not required} \\ \hline
	$+_{1,2}$ & \footnotesize{$1 + 21 t + 249 t^2 + 2188 t^3 + 15657 t^4 + 95340 t^5+\cdots$}  & \footnotesize{$1 + 21 t + 249 t^2 + 2188 t^3 + 15657 t^4 + 95340 t^5+\cdots$} & \footnotesize{not required} \\ \hline
\end{tabular}
\caption[Coulomb branch HS for magnetic quivers of $+_{1,N}$ theory.]{Coulomb branch Hilbert series for the unitary and orthosymplectic magnetic quivers in Figure \ref{O5++(+1N)unitaryquiver} and Figure \ref{O5++(+1N)OSpquiver} for the $+_{1,N}$ theory. The explicit form of $P_{18}(t)$ is given in Appendix \ref{sec:app2}.}
\label{O5++(+1N)}
\end{table}
\subsection{The \texorpdfstring{$\hat{K}_{N}^1$}{TEXT} theory}
\begin{figure}[!htb]
    \centering
    \begin{scriptsize}
    \begin{tikzpicture}[scale=.7]
        \draw[thick](0,0)--(4,0);
        \draw[thick](0,0)--(-1.5,1.5);
    \draw[thick,dashed](-4,0)--(-0,0);
    \draw[thick,dashed](6,0)--(4,0);
    \node[7brane] at (2,0){};
    \node[7brane] at (4,0){};
    \node[label=above left:{$(1,-1)$}][7brane] at (-1.5,1.5){};
    \draw[thick](0,0)--(0,3);
    \node[7brane] at (0,2){};
    \node[7brane] at (0,3){};
    \node[thick] at (0,3.4) {$.$};
\node[thick] at (0,3.5) {$.$};
\node[thick] at (0,3.6) {$.$};
\node[7brane] at (0,4){};
\node[7brane] at (0,5){};
\node[7brane] at (0,6){};
\node at (.5,5.5) {1};
\node at (.5,4.5) {2};
\node at (.75,2.5) {$2N-2$};
\node at (.75,1.5) {$2N-1$};
\node at (-1.2,.8){1};
\draw[thick](0,4)--(0,6);

\node at (5,-.25) {$\text{O}5^+$};
\node at (3,0.35) {$1$};
\node at (1.5,0.35) {1};
\node at (-3,-.25) {$\text{O}5^+$};
    
    \end{tikzpicture}
    \end{scriptsize}
    \caption[Orientifold web diagram for $\hat{K}_{N}^1$ theory.]{An orientifold web for the $\hat{K}_{N}^1$ theory with asymptotically O5$^+$ planes}
    \label{fig:KN1 O5+O5+ orientifold web}
\end{figure}
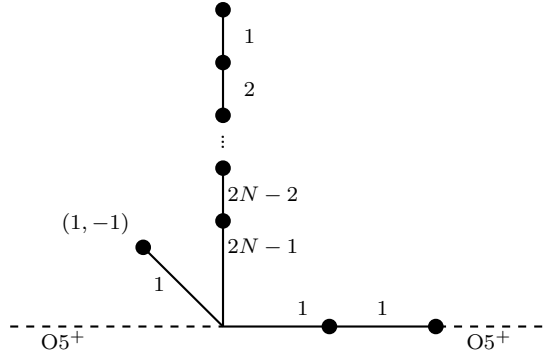

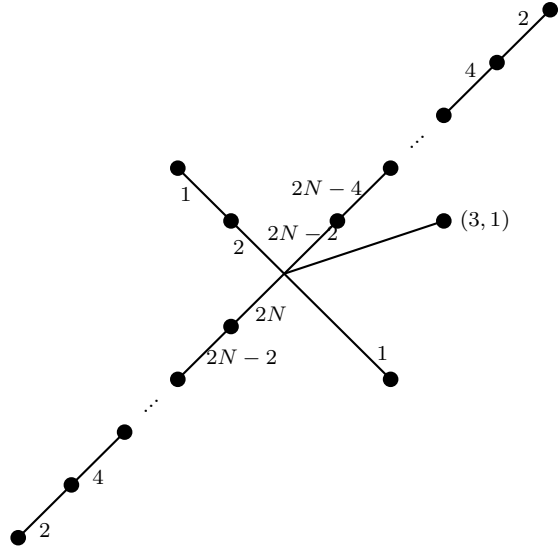
\begin{figure}[!htb]
        \centering
        \begin{scriptsize}
        \begin{tikzpicture}[scale=.7]
        \draw[thick](-2,2)--(2,-2);
        \draw[thick](0,0)--(3,1);
        \node[label=right:{$(3,1)$}][7brane]at(3,1){};
        \node[7brane]at(-2,2){};
        \node[7brane]at(-1,1){};
        \node[7brane]at(2,-2){};
        \draw[thick](2,2)--(-2,-2);
        \node[7brane]at(-2,-2){};
        \node[7brane]at(2,2){};
        \node[7brane]at(-1,-1){};
        \node[7brane]at(1,1){};
        \node at (2.4,2.4){$.$};
        \node at (2.5,2.5){$.$};
        \node at (2.6,2.6){$.$};
        \node at (-2.4,-2.4){$.$};
        \node at (-2.5,-2.5){$.$};
        \node at (-2.6,-2.6){$.$};
        \draw[thick](3,3)--(5,5);
        \node[7brane]at(3,3){};
        \node[7brane]at(4,4){};
        \node[7brane]at(5,5){};
        \draw[thick](-3,-3)--(-5,-5);
        \node[7brane]at(-3,-3){};
        \node[7brane]at(-4,-4){};
        \node[7brane]at(-5,-5){};
        \node at(1.85,-1.5){1};
        \node at(-.85,+.5){2};
        \node at(-1.85,1.5){1};
        \node at(.35,.75){$2N-2$};
        \node at(.8,1.6){$2N-4$};
        \node at(3.5,3.85){$4$};
        \node at(4.5,4.85){$2$};
        \node at(-.25,-.75){$2N$};
        \node at(-.8,-1.6){$2N-2$};
        \node at(-3.5,-3.85){$4$};
        \node at(-4.5,-4.85){$2$};
        
        \end{tikzpicture}
        \end{scriptsize}
        \caption[Unitary web diagram for $\hat{K}_{N}^1$ theory.]{A unitary web for the $\hat{K}_{N}^1$ theory with asymptotically O5$^+$ orientifold planes.}
        \label{fig:KN1 unitary web O5+-O5+}
    \end{figure}

Decoupling a single flavor from, say, the leftmost node in the $+_{1,N}$ theory \eqref{electric +1N O5+O5+ OSp}, one arrives at the $\hat{K}_N^1$ theory. It has an IR gauge theory description as 
\be\label{electric KN1 O5+O5+ OSp}
\begin{array}{c}\begin{tikzpicture}
\node {$\text{SO}(6)-\text{USp}(2)-\text{SO}(6)-\cdots-\text{USp}(2)-\text{SO}(6)-[1]$};
 \draw [thick,decorate,decoration={brace,amplitude=6pt},xshift=0pt,yshift=10pt]
(-4.25,0) -- (4.0,0)node [black,midway,xshift=0pt,yshift=20pt] {
$2N-1$};
\end{tikzpicture}
\end{array}
\ee
We depict the orientifold and unitary web in Figure \ref{fig:KN1 O5+O5+ orientifold web} and Figure \ref{fig:KN1 unitary web O5+-O5+} respectively.

Also, in this setup, we find $\text{USp}(2N-2)$ gauge node coming from the $2N-2$ NS5-branes intersecting with the O5$^+$-plane.
According to the proposal in the previous subsection, there are four fundamental half hypermultiplets for this node
because there is no SO(odd) gauge node coupled to this $\text{USp}(2N-2)$ gauge node. 

In addition, there is one subweb intersecting with the O5$^+$-plane, contributing as a U(1) gauge node of the magnetic quiver.
In general,
when a subweb is intersecting with the O5$^+$-plane, 
it would be reasonable to expect that 
there are contributions from the D3-branes suspended between the considered subweb and its mirror image.
Analogous to the case with O5$^-$-plane,
such contribution is the hypermultiplets with charge 2 coupled to the corresponding U(1) node.
The number of such charge 2 hypermultiplets would be given schematically by
\begin{align}
\frac{(\text{SI with its mirror image})}{2} - (\text{SI with O5}^+),
\label{charge2O5+}
\end{align}
as discussed around \eqref{SI-charge-2}.
In this specific setup, the number of the charge 2 hypermultiplets should be zero, in order to agree with the unitary magnetic quiver.

However, again, the situation for such U(1) node is different from the case with O5$^-$-plane.
Analogous to the case with NS5-branes intersecting with the O5$^+$-plane, 
the idea discussed around \eqref{eq:O5+-} implies that there is a contribution from the D3-branes 
suspended between this subweb and the half D5-branes included in the O5$^+$-plane. 
Such D3-branes correspond to hypermultiplets with charge 1 instead of charge 2 
because the distance between the considered subweb and the O5$^+$-plane 
is half the distance between the original subweb and its mirror image. 
In order for the orthosymplectic magnetic quiver to be consistent with the unitary quiver,
we need three hypermultiplets with charge $1$ coupled to the U(1) node.
This three is interpreted as 
\begin{align}\label{SI-charge-1}
 4 (\text{bare SI with half D5}) - (\text{Contribution from the common half D7 on the O5}^+).
\end{align}
In our case, bare SI with half D5-brane is one. The contribution from the half D7-brane is also one
because the considered subweb and the half D5 are both attached to the common half D7-brane from the same direction.\footnote{When we compute the contribution from the common half D7-brane on the O5$^+$-plane, we treat it as if there were only a single half D5-brane terminated at the common half D7-brane. 
We do not multiply 4 for this contribution contrary to the contribution from SI.}
We propose that the number of charge 1 hypermultiplets coupled to the U(1) node is given by \eqref{SI-charge-1} in general. 
The contribution from the common half D7-brane on the O5$^+$-plane is computed analogously to the case for the unitary quiver discussed in \cite{Cabrera:2018jxt}.

To support this proposed rule, we match the Higgs branch Hilbert series for the unitary and the orthosymplectic magnetic quiver, especially to settle the question of whether the $\mathbb{Z}_2$ node is gauge or flavor. We find that only choosing the $\mathbb{Z}_2$ node to be gauge we recover the correct match. We computed the Hilbert series for both $N=2$ and $N=3$, and the result is tabulated in Table \ref{O5++KN1HiggsHS}.


		\begin{figure}[!htb]
\centering
   \begin{subfigure}{0.49\linewidth} \centering
    \begin{scriptsize}
       \begin{tikzpicture}
       \node[label=below:{2}][u](2){};
       \node[label=below:{4}][u](4)[right of=2]{};
       \node[label=below:{$2N-4$}][u](2n-4)[right of=4]{};
       \node[label=below:{$2N-2$}][u](2n-2)[right of=2n-4]{};
       \node[label=below:{$2N-2$}][u](2n-2')[right of=2n-2]{};
       \node[label=below:{$2N-4$}][u](2n-4')[right of=2n-2']{};
       \node[label=below:{4}][u](4')[right of=2n-4']{};
       \node[label=below:{2}][u](2')[right of=4']{};
       \node[label=left:{1}][u](u1)[above of=2n-2]{};
       \node[label=right:{1}][u](u1')[above of=2n-2']{};
       \node[label=above:{1}][u](u11)[above right of=u1]{};
       \draw(2)--(4);
       \draw[dotted](4)--(2n-4);
       \draw(2n-4)--(2n-2);
       \draw(2n-2)--(2n-2');
       \draw(2')--(4');
       \draw[dotted](4')--(2n-4');
       \draw(2n-4')--(2n-2');
       \draw[double distance=2pt](2n-2)--(u1);
       \draw[double distance=2pt](2n-2')--(u1');
       \draw[double distance =3pt](u1)--(u1');
       \draw(u1)--(u1');
       \draw(u1)--(u11);
       \draw(u1')--(u11);
       \end{tikzpicture}
       \end{scriptsize}
     \caption{Unitary quiver}\label{O5++KN1unitaryquiver}
   \end{subfigure}
   \begin{subfigure}{0.49\linewidth} \centering
     \begin{scriptsize}
      \begin{tikzpicture}
      \node[label=below:{1}][u](1){};
      \node[label=below:{$2N-2$}][u](2n-2)[right of=1]{};
      \node[label=left:{$2N-2$}][sp](sp2n-2)[above of=2n-2]{};
      \node[label=above:{4}][sof](so4f)[above of=sp2n-2]{};
      \node[label=below:{1}][u](u1)[right of=2n-2]{};
      \node[label=below:{2}][sp](sp2)[right of=u1]{};
      \node[label=above:{3}][uf](uf)[above of=u1]{};
      \node[label=above:{1}][sof](sof)[above of=sp2]{};
      \node[label=below:{1}][so](o1)[right of=sp2]{};
      \draw[dotted](1)--(2n-2);
      \draw(2n-2)--(sp2n-2);
      \draw(sp2n-2)--(so4f);
      \draw(2n-2)--(u1);
      \draw(u1)--(uf);
      \draw(u1)--(sp2);
      \draw(sp2)--(sof);
      \draw(sp2)--(o1);
      \end{tikzpicture}
      \end{scriptsize}
     \caption{Orthosymplectic quiver}\label{O5++KN1OSpquiver}
   \end{subfigure}
\caption[Magnetic quivers for $\hat{K}_{N}^1$ theory.]{Magnetic quivers for the $\hat{K}_{N}^1$ theory} \label{O5++KN1magneticquivers}
\end{figure}
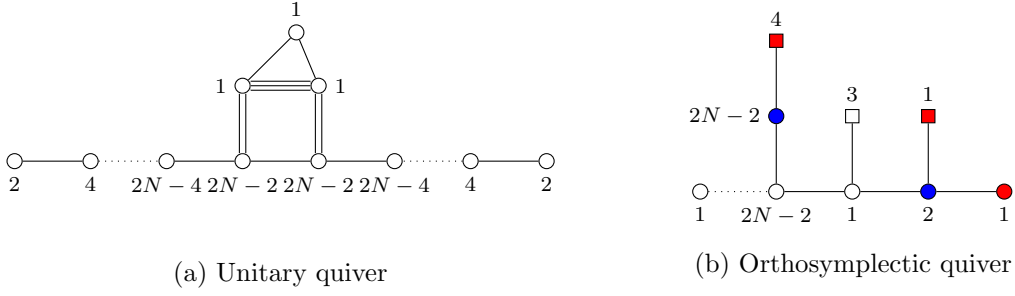
\begin{table}[!htb]
\centering
\begin{tabular}{|c|C{6.25cm}|C{6.25cm}|} \hline
\rowcolor{Grayy}
   & Unitary magnetic quiver & Orthosymplectic magnetic quiver \\ \cline{2-3}
	\rowcolor{Grayy}
  \multirow{-2}{*}{$\hat{K}_{N}^1$} & $\text{HS}_{\mathbb{H}}(t)$  & $\text{HS}_{\mathbb{H}}(t)$  \\ \hline
	$\hat{K}_2^1$ & \footnotesize{$1 + 9 t^2 + 6 t^3 + 36 t^4 + 36 t^5 + 112 t^6 + 120 t^7 + 285 t^8+\ldots$} & \footnotesize{$1 + 9 t^2 + 6 t^3 + 36 t^4 + 36 t^5 + 112 t^6 + 120 t^7 + 285 t^8+\ldots$}  \\ \hline
	$\hat{K}_3^1$ & \footnotesize{$1 + 16 t^2 + 6 t^3 + 150 t^4 + 86 t^5 + 981 t^6+\ldots$} & \footnotesize{$1 + 16 t^2 + 6 t^3 + 150 t^4 + 86 t^5 + 981 t^6+\ldots$}  \\ \hline
\end{tabular}
\caption[Higgs branch HS for magnetic quivers of $\hat{K}_{N}^1$ theory.]{Higgs branch Hilbert series of the unitary and orthosymplectic magnetic quivers presented in Figure \ref{O5++KN1magneticquivers}.}
\label{O5++KN1HiggsHS}
\end{table}

\subsection{The \texorpdfstring{$X_{N}^{1,1}$}{TEXT} theory}
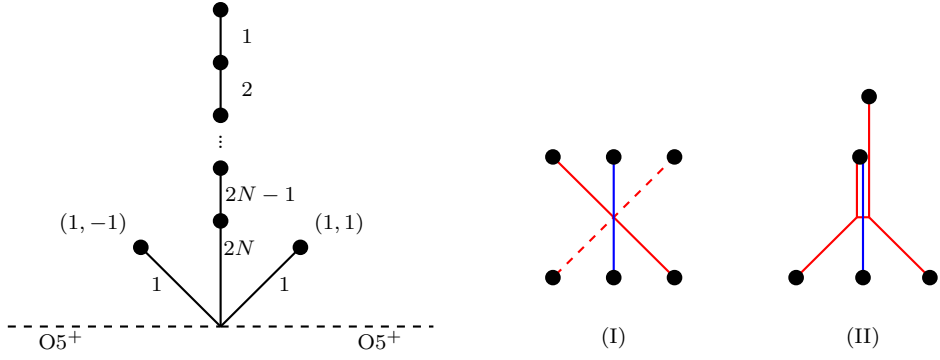
\begin{figure}
    \centering
    \begin{scriptsize}
    \begin{tikzpicture}[scale=.7]
        \draw[thick](0,0)--(1.5,1.5);
        \draw[thick](0,0)--(-1.5,1.5);
    \draw[thick,dashed](-4,0)--(-0,0);
    \draw[thick,dashed](0,0)--(4,0);
    \node[label=above left:{$(1,-1)$}][7brane] at (-1.5,1.5){};
        \node[label=above right:{$(1,1)$}][7brane] at (1.5,1.5){};
    \draw[thick](0,0)--(0,3);
    \node[7brane] at (0,2){};
    \node[7brane] at (0,3){};
    \node[thick] at (0,3.4) {$.$};
\node[thick] at (0,3.5) {$.$};
\node[thick] at (0,3.6) {$.$};
\node[7brane] at (0,4){};
\node[7brane] at (0,5){};
\node[7brane] at (0,6){};
\node at (.5,5.5) {1};
\node at (.5,4.5) {2};
\node at (.75,2.5) {$2N-1$};
\node at (.35,1.5) {$2N$};
\node at (-1.2,.8){1};
\node at (1.2,.8){1};
\draw[thick](0,4)--(0,6);

\node at (3,-.25) {$\text{O}5^+$};
\node at (-3,-.25) {$\text{O}5^+$};
    
    \end{tikzpicture}
    \end{scriptsize}
    \hspace{1cm}
    \begin{scriptsize}
    \begin{tikzpicture}[scale=.8]
    \draw[thick,red](-1,1)--(1,-1);
    \draw[thick,red,dashed](1,1)--(-1,-1);
    \draw[thick,blue](0,-1)--(0,1);
    \node[7brane]at(1,1){};
    \node[7brane]at(-1,-1){};
    \node[7brane]at(1,-1){};
    \node[7brane]at(-1,1){};
    \node[7brane]at(0,1){};
    \node[7brane]at(0,-1){};
    \node at(0,-2){(I)};
    
    \draw[thick,red](4,0)--(4.2,0);
    \draw[thick,red](4,0)--(3,-1);
    \draw[thick,red](5.2,-1)--(4.2,0);
    \draw[thick,red](4,0)--(4,1);
    \draw[thick,red](4.2,2)--(4.2,0);
    \draw[thick,blue](4.1,-1)--(4.1,1);
    \node[7brane]at(4.1,-1){};
    \node[7brane]at(4.05,1){};
    \node[7brane]at(4.2,2){};
    \node[7brane]at(3,-1){};
    \node[7brane]at(5.2,-1){};
     \node at(4.1,-2){(II)};
    \end{tikzpicture}
    \end{scriptsize}
    \caption[Orientifold web diagram for $X_{N}^{1,1}$ theory.]{An orientifold web for the $X_{N}^{1,1}$ theory with asymptotically O5$^+$ planes. We show the two possible maximal subdivisions on the right.}
    \label{fig:YM11 orientifold web}
\end{figure}

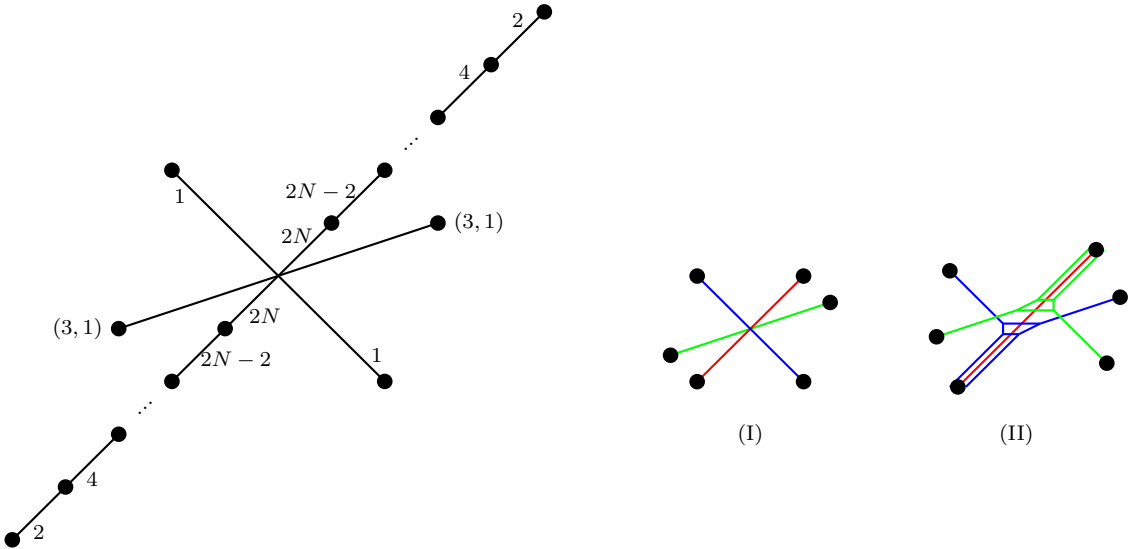
\begin{figure}[!htbp]
        \centering
        \begin{scriptsize}
        \begin{tikzpicture}[scale=.7]
        \draw[thick](-2,2)--(2,-2);
        \draw[thick](-3,-1)--(3,1);
        \node[label=right:{$(3,1)$}][7brane]at(3,1){};
        \node[label=left:{$(3,1)$}][7brane]at(-3,-1){};
        \node[7brane]at(-2,2){};
        \node[7brane]at(2,-2){};
        \draw[thick](2,2)--(-2,-2);
        \node[7brane]at(-2,-2){};
        \node[7brane]at(2,2){};
        \node[7brane]at(-1,-1){};
        \node[7brane]at(1,1){};
        \node at (2.4,2.4){$.$};
        \node at (2.5,2.5){$.$};
        \node at (2.6,2.6){$.$};
        \node at (-2.4,-2.4){$.$};
        \node at (-2.5,-2.5){$.$};
        \node at (-2.6,-2.6){$.$};
        \draw[thick](3,3)--(5,5);
        \node[7brane]at(3,3){};
        \node[7brane]at(4,4){};
        \node[7brane]at(5,5){};
        \draw[thick](-3,-3)--(-5,-5);
        \node[7brane]at(-3,-3){};
        \node[7brane]at(-4,-4){};
        \node[7brane]at(-5,-5){};
        \node at(1.85,-1.5){1};
        \node at(-1.85,1.5){1};
        \node at(.35,.75){$2N$};
        \node at(.8,1.6){$2N-2$};
        \node at(3.5,3.85){$4$};
        \node at(4.5,4.85){$2$};
        \node at(-.25,-.75){$2N$};
        \node at(-.8,-1.6){$2N-2$};
        \node at(-3.5,-3.85){$4$};
        \node at(-4.5,-4.85){$2$};
        \end{tikzpicture}
        \end{scriptsize}
        \hspace{1cm}
        \begin{scriptsize}
        \begin{tikzpicture}[scale=.7]
        \draw[thick,green](-1.5,-.5)--(1.5,.5);
        \draw[thick,red](-1,-1)--(1,1);
        \draw[thick,blue](1,-1)--(-1,1);
        \node[7brane]at(1,1){};
        \node[7brane]at(-1,-1){};
        \node[7brane]at(-1,1){};
        \node[7brane]at(1,-1){};
        \node[7brane]at(1.5,.5){};
        \node[7brane]at(-1.5,-.5){};
        \node at(0,-2){(I)};
        
        \draw[thick,red](3.9,-1.1)--(6.5,1.5);
        \draw[thick,blue](4.75,-.1)--(5.05,-.1);
        \draw[thick,blue](4.75,-.1)--(3.75,-1.1);
        \draw[thick,blue](4.05,-1.1)--(5.05,-.1);
        \draw[thick,blue](4.75,-.1)--(4.75,.1);
        \draw[thick,blue](3.75,1.1)--(4.75,.1);
        \draw[thick,blue](5.45,.1)--(5.05,-.1);
        \draw[thick,blue](5.45,.1)--(4.75,.1);
        \draw[thick,blue](5.45,.1)--(6.95,.6);
        
        \node[7brane]at(6.95,.6){};
   \node[7brane]at(3.75,1.1){};
      \draw[thick,green](5.4,.55)--(5.7,.55);
      \draw[thick,green](6.65,1.5)--(5.7,.55);
      \draw[thick,green](5.4,.55)--(6.4,1.55);
      \draw[thick,green](5.4,.55)--(5,.35);
      \draw[thick,green](5.7,.35)--(5.7,.55);
      \draw[thick,green](5.7,.35)--(5,.35);
      \draw[thick,green](5.7,.35)--(6.7,-.65);
      \draw[thick,green](3.5,-.15)--(5,.35);
      
      \node[7brane]at(6.7,-.65){};
      \node[7brane]at(3.5,-.15){};
        \node[7brane]at(3.9,-1.1){};
        \node[7brane]at(6.5,1.5){};
\node at(5,-2){(II)};
\node at(0,-4){};
        \end{tikzpicture}
        \end{scriptsize}
        \caption[Unitary web diagram for $X_{N}^{1,1}$ theory.]{A unitary web for the $X_{N}^{1,1}$ theory with asymptotically O5$^+$ orientifold planes. The two possible maximal subdivisions are shown on the right.}
        \label{fig:YM11 unitary web O5+-O5+}
    \end{figure}

We then consider the configuration obtained by intersecting $2N$ NS5s, one $(1,1)$, and one $(1,-1)$ on top of an O5$^+$-plane (Figure \ref{fig:YM11 orientifold web}). We call the theory on the web the $X_N^{1,1}$ theory. There is a corresponding IR gauge theory description as

\be\label{electric XN11 O5+O5+ OSp}
\begin{array}{c}\begin{tikzpicture}
\node {$\text{SO}(6)-\text{USp}(2)-\text{SO}(6)-\cdots-\text{USp}(2)-\text{SO}(6)$};
 \draw [thick,decorate,decoration={brace,amplitude=6pt},xshift=0pt,yshift=10pt]
(-4.25,0) -- (4.0,0)node [black,midway,xshift=0pt,yshift=20pt] {
$2N+1$};
\end{tikzpicture}
\end{array}
\ee
Alternatively, it may be understood as gluing $N-1$ copies of $\text{SO}(6)$ with 2 vectors and two copies of $\text{SO}(6)$ with one vector, via successive gauging of $\text{USp}(2)$ subgroups of the flavour symmetry. This allows us to construct a unitary web for this theory by gluing $N-1$ copies of $\text{SU}(4)_0$ with 2 antisymmetric hypermultiplets and two copies of $\text{SU}(4)_0$ with one antisymmetric hypermultiplet, via successive gauging of $\text{SU}(2)$ subgroups of the global symmetry (Figure \ref{fig:YM11 unitary web O5+-O5+}). 

    \begin{table}[!htb]
        \centering
      \begin{tabular}{|c|C{6.5cm}|C{4.5cm}|}
			\rowcolor{Grayy}
         \hline MS&Unitary & Orthosymplectic\\\hline
        (I) &\begin{scriptsize}
        \begin{tikzpicture}
        \node[label=below:{2}][u](2){};
        \node[label=below:{4}][u](4)[right of=2]{};
        \node[label=below:{$2N-2$}][u](2n-2)[right of=4]{};
        \node[label=below:{$2N$}][u](2n)[right of=2n-2]{};
        \node[label=below:{$2N-2$}][u](2n-2')[right of=2n]{};
        \node[label=below:{$4$}][u](4')[right of=2n-2']{};
        \node[label=below:{$2$}][u](2')[right of=4']{};
        \node[label=left:{$1$}][u](u1)[above left of=2n]{};
        \node[label=right:{$1$}][u](u1')[above right of=2n]{};
        \draw(2)--(4);
        \draw[dotted](4)--(2n-2);
        \draw(2n-2)--(2n);
        \draw(2')--(4');
        \draw[dotted](4')--(2n-2');
        \draw(2n-2')--(2n);
        \draw[double distance=2pt](u1)--(2n);
        \draw[double distance=2pt](u1')--(2n);
        \draw[double distance=4pt](u1)--(u1');
        \draw[double distance=1pt](u1)--(u1');
        \end{tikzpicture}
        \end{scriptsize} &\begin{scriptsize}
        \begin{tikzpicture}
        \node[label=below:{1}](1)[u]{};
        \node[label=below:{$2N-1$}][u](2n-1)[right of=1]{};
        \node[label=below:{$2N$}][sp](sp2n)[right of=2n-1]{};
        \node[label=below:{1}][u](so2)[right of=sp2n]{};
        \node[label=above:{4}][sof](sof)[above of=sp2n]{};
        \node[label=above:{4}][uf](spf)[above of=so2]{};
        \draw(1)[dotted]--(2n-1);
        \draw(2n-1)--(sp2n);
        \draw(sp2n)--(so2);
        \draw(sp2n)--(sof);
        \draw(so2)--(spf);
        \end{tikzpicture}
        \end{scriptsize}\\\hline
        (II)&\begin{scriptsize}
        \begin{tikzpicture}
        \node[label=below:{2}][u](2){};
        \node[label=below:{4}][u](4)[right of=2]{};
        \node[label=below:{$2N-2$}][u](2n-2)[right of=4]{};
        \node[label=below:{$2N-2$}][u](2n)[right of=2n-2]{};
        \node[label=below:{$2N-2$}][u](2n-2')[right of=2n]{};
        \node[label=below:{$4$}][u](4')[right of=2n-2']{};
        \node[label=below:{$2$}][u](2')[right of=4']{};
        \node[label=left:{$1$}][u](u1)[above of=2n-2]{};
        \node[label=right:{$1$}][u](u1')[above of=2n-2']{};
        \node[label=above:{8}](8)[right of=u1]{};
        \draw(2)--(4);
        \draw[dotted](4)--(2n-2);
        \draw(2n-2)--(2n);
        \draw(2')--(4');
        \draw[dotted](4')--(2n-2');
        \draw(2n-2')--(2n);
        \draw[double distance=2pt](u1)--(2n-2);
        \draw[double distance=2pt](u1')--(2n-2');
        \draw[double distance=4pt](u1)--(u1');
        \draw[double distance=1pt](u1)--(u1');
        \end{tikzpicture}
        \end{scriptsize}&\begin{scriptsize}
        \begin{tikzpicture}
        \node[label=below:{1}][u](1){};
        \node[label=below:{$2N-3$}][u](2n-3)[right of=1]{};
        \node[label=below:{$2N-2$}][u](2n-2)[right of=2n-3]{};
        \node[label=below:{$2N-2$}][u](2n-2')[right of=2n-2]{};
        \node[label=below:{$2N-2$}][sp](sp2n-2)[right of=2n-2']{};
        \node[label=left:{1}][u](u1)[above of=2n-2]{};
        \node[label=above:{8}][uf](uf)[above of=u1]{};
        \node[label=above:{4}][sof](sof)[above of=sp2n-2]{};
        \draw[dotted](1)--(2n-3);
        \draw(2n-3)--(2n-2);
        \draw(2n-2)--(2n-2');
        \draw(2n-2')--(sp2n-2);
        \draw(sp2n-2)--(sof);
        \draw(2n-2)--(u1);
        \draw(u1)--(uf);
        \end{tikzpicture}
        \end{scriptsize}\\\hline
    \end{tabular}
        \caption[Magnetic quivers for $X_{N}^{1,1}$ theory.]{Magnetic quivers corresponding to the maximal subdivisions for the $X_{N}^{1,1}$ theory.}
        \label{XN11quiversx}
    \end{table}
		The magnetic quivers for the $X_{N}^{1,1}$ theory are given in Table \ref{XN11quiversx} and the Coulomb branch Hilbert series are tabulated in Table \ref{XN11HSx}.
	\begin{table}
\centering
\begin{tabular}{|c|C{4.9cm}|C{4.9cm}|C{2.8cm}|} \hline
\rowcolor{Grayy}
   & Unitary magnetic quiver & \multicolumn{2}{c|}{Orthosymplectic magnetic quiver} \\ \cline{2-4}
	\rowcolor{Grayy}
  \multirow{-2}{*}{MS}
 & HS($t$)  & HS($t;\vec{m} \in \mathbb{Z}$) & HS($t;\vec{m} \in \mathbb{Z}+\tfrac{1}{2}$)  \\ \hline
	$\text{(I)}_{N=1}$ & \footnotesize{$\begin{array}{l} \dfrac{P_{19}(t)}{(1-t)^{2}\,(1-t^3)\,(1-t^4)^3} \\\\ = 1 + 4 t + 13 t^2 + 33 t^3\\ + 80 t^4 + 165 t^5+\cdots  \end{array}$}  & \footnotesize{$\begin{array}{l} \dfrac{P_{19}(t)}{(1-t)^{2}\,(1-t^3)\,(1-t^4)^3} \\\\ = 1 + 4 t + 13 t^2 + 33 t^3\\ + 80 t^4 + 165 t^5+\cdots  \end{array}$} & \footnotesize{not required} \\ \hline
	$\text{(I)}_{N=2}$ & \footnotesize{$1 + 16 t + 151 t^2 + 1039 t^3 + 5750 t^4 + 26954 t^5+\cdots$}  & \footnotesize{$1 + 16 t + 151 t^2 + 1039 t^3 + 5750 t^4 + 26954 t^5+\cdots$} & \footnotesize{not required} \\ \hline
	$\text{(II)}_{N=2}$ & \footnotesize{$1 + 16 t + 151 t^2 + 1004 t^3 + 5198 t^4 + 22184 t^5+\cdots$}  & \footnotesize{$1 + 16 t + 151 t^2 + 1004 t^3 + 5198 t^4 + 22184 t^5+\cdots$} & \footnotesize{not required} \\ \hline
	$\text{(II)}_{N=3}$ & \footnotesize{$1 + 36 t + 701 t^2 + 9659 t^3+\cdots$}  & \footnotesize{$1 + 36 t + 701 t^2 + 9659 t^3+\cdots$} & \footnotesize{not required} \\ \hline
\end{tabular}
\caption[Coulomb branch HS for magnetic quivers of $X_{N}^{1,1}$ theory.]{Coulomb branch Hilbert series for the unitary and orthosymplectic magnetic quivers presented in Table \ref{XN11quiversx}. The explicit form of $P_{19}(t)$ is provided in Appendix \ref{sec:app2}.}
\label{XN11HSx}
\end{table}
\section{\texorpdfstring{Magnetic quivers from O5$^-$ - O5$^+$}{TEXT}}
\label{sec:O5-O5+}
So far we have focused on the configurations where the two asymptotic orientifold planes are of the same type. It is possible to consider cases with an O5$^-$-plane on one end and an O5$^+$-plane on the other end. We will consider such examples in this section. 
\subsection{The \texorpdfstring{$+_N^{3,1}$}{TEXT} theory}
For obtaining the configuration which has an O5$^-$-plane on one end and an O5$^+$-plane on the other end, we decouple one $\text{USp}(2)$ part from the quiver in \eqref{electric +NM OSp} for $M=3$. 
An IR quiver description of the theory is
\be\label{electric +1N31 O5-O5+ OSp}
\begin{array}{c}\begin{tikzpicture}
\node {[1]-$\text{SO}(6)-\text{USp}(2)-\text{SO}(6)-\cdots-\text{USp}(2)-\text{SO}(6)-\text{USp}(2)-[3].$};
 \draw [thick,decorate,decoration={brace,amplitude=6pt},xshift=0pt,yshift=10pt]
(-4.75,0) -- (4.5,0)node [black,midway,xshift=0pt,yshift=20pt] {
$2N-2$};
\end{tikzpicture}
\end{array}
\ee
The orientifold web for the $+_N^{3,1}$ theory is presented in Figure \ref{fig:+N31 orientifold web}. 
\begin{figure}[!ht]
    \centering
\begin{scriptsize}
    \begin{tikzpicture}[scale=.8]
    \draw[thick](-3,0)--(7,0);
    \draw[thick,dashed](-3,0)--(-5,0);
    \draw[thick,dashed](7,0)--(9,0);
    \node[7brane]at(2,0){};
    \node[7brane]at(3,0){};
    \node[7brane]at(4,0){};
    \node[7brane]at(5,0){};
    \node[7brane]at(6,0){};
    \node[7brane]at(7,0){};
    \node[7brane]at(-2,0){};
    \node[7brane]at(-3,0){};
    \draw[thick](0,0)--(0,3);
    \node[7brane]at(0,2){};
    \node[7brane]at(0,3){};
    \node at (0,3.4){$.$};
    \node at (0,3.5){$.$};
    \node at (0,3.6){$.$};
    \draw[thick](0,4)--(0,6);
    \node[7brane]at(0,4){};
    \node[7brane]at(0,5){};
    \node[7brane]at(0,6){};
    \node at (7.5,-.3){$\text{O5}^-$};
    \node at (-4.5,-.3){$\text{O5}^+$};
    \node at (1.5,.25){3};
    \node at (2.5,.25){$\frac{5}{2}$};
    \node at (3.5,.25){2};
    \node at (4.5,.25){$\frac{3}{2}$};
    \node at (5.5,.25){1};
    \node at (6.5,.25){$\frac{1}{2}$};
    \node at (-1.5,.25){1};
    \node at (-2.5,.25){1};
    \node at (.25,5.5){1};
    \node at (.25,4.5){2};
    \node at (.75,2.5){$2N-2$};
    \node at (.75,1.5){$2N-1$};
    \end{tikzpicture}
    \hspace{.5cm}
    \begin{tikzpicture}[scale=.7]
    \draw[thick,blue](0,0)--(1,0);
    \draw[thick,red](0,0.3)--(2,0.3);
    \draw[thick,red](0,0.5)--(3,0.5);
    \node[7brane]at(1,-.05){};
    \node[7brane]at(2,.2){};
    \node[7brane]at(3,.4){};
    \draw[thick,blue](-1,-.1)--(1,-.1);
    \draw[thick,red](0,-.2)--(1,-.2);
    \draw[thick,dotted,blue](-1,0)--(0,0);
    \draw[thick,blue](-1,.1)--(1,.1);
    \draw[thick,red](0,-1)--(0,1);
    \draw[thick,green](0.2,-1)--(0.2,1);
    \node[7brane]at(.1,1){};
    \node[7brane]at(.1,-1){};
    \node[7brane]at(-1,0){};
    \end{tikzpicture}
    \end{scriptsize}
    \caption{Orientifold web diagram for the $+_N^{3,1}$ theory.}
    \label{fig:+N31 orientifold web}
\end{figure}
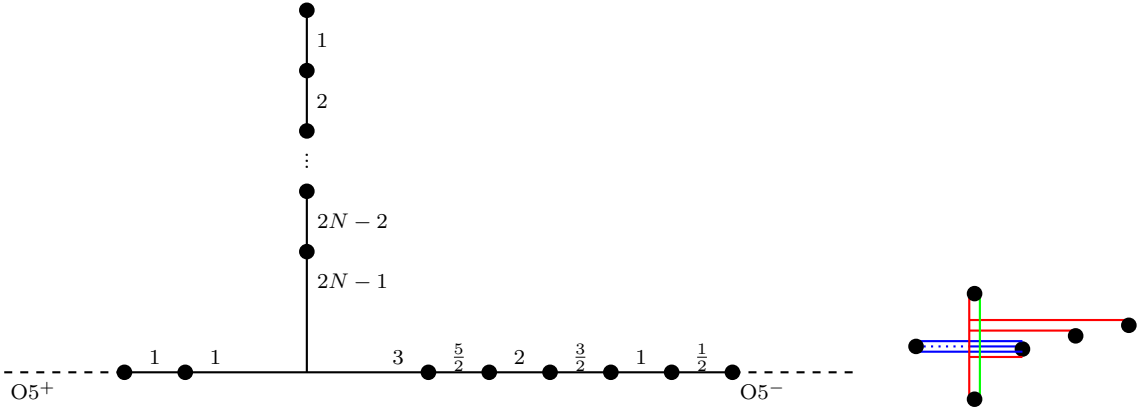
\begin{figure}[ht]
    \centering
    \begin{scriptsize}
        \begin{tikzpicture}[scale=.75]
        \draw[thick](-3,0)--(3,0);
        \draw[thick](0,0)--(0,5);
        \draw[thick](0,0)--(-1.5,-3);
        \node[label=below:{$(1,2)$}][7brane]at(-1.5,-3){};
        \node[7brane]at(1.5,0){};
        \node[7brane]at(3,0){};
        \node[7brane]at(-1.5,0){};
        \node[7brane]at(-3,0){};
        \node[7brane]at(0,2){};
        \node[7brane]at(0,3){};
        \node[7brane]at(0,4){};
        \node[7brane]at(0,5){};
        \node[7brane]at(-1,-2){};
        \node at(3.4,0){$.$};
        \node at(3.5,0){$.$};
        \node at(3.6,0){$.$};
        \node at(-3.4,0){$.$};
        \node at(-3.5,0){$.$};
        \node at(-3.6,0){$.$};
        \draw[thick](4,0)--(6,0);
        \draw[thick](-4,0)--(-6,0);
        \node[7brane]at(4,0){};
        \node[7brane]at(5,0){};
        \node[7brane]at(6,0){};
        \node[7brane]at(-4,0){};
        \node[7brane]at(-5,0){};
        \node[7brane]at(-6,0){};
        \node at (0.8,.25){$2N$};
        \node at (2.25,.25){$2N-2$};
        \node at (4.5,.25){$4$};
        \node at (5.5,.25){$2$};
        \node at (-.75,.25){$2N-2$};
        \node at (-2.25,.25){$2N-4$};
        \node at (-4.6,.25){$4$};
        \node at (-5.5,.25){$2$};
        \node at(.25,1.5){4};
        \node at(.25,2.5){3};
        \node at(.25,3.5){2};
        \node at(.25,4.5){1};
        \node at(-1,-2.75){1};
        \node at(-.5,-1.75){2};
        \end{tikzpicture}
    \end{scriptsize}
    \caption{Unitary web diagram for $+_N^{3,1}$ theory.}
    \label{fig:+N31 unitary web}
\end{figure}
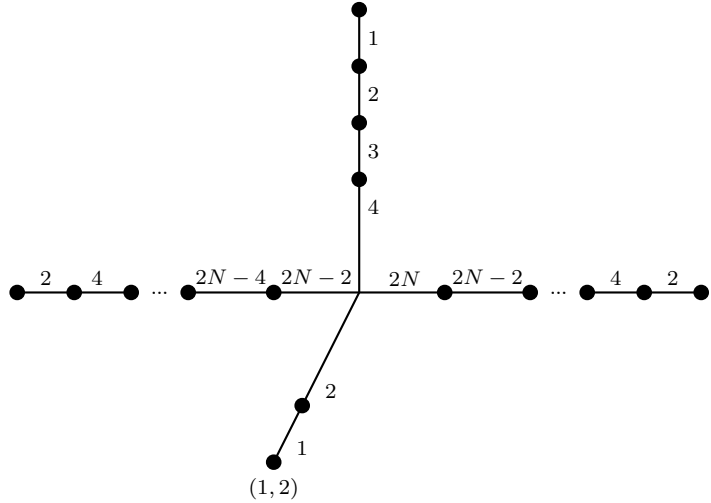


The corresponding unitary web is depicted in Figure \ref{fig:+N31 unitary web}, which is obtained as follows:
We first interpret the $+_N^{3,1}$ theory as a decoupling limit from the $\#_{3,N}$ theory.
The discussion will be clearer when we discuss this decoupling limit in the S-dual description. 
The S-dual description of the $\#_{3,N}$ theory has a low energy description as a 5d $D_3$ quiver gauge theory as given in \eqref{electric +3M unitary}. The corresponding 5-brane web for this 5d $D_3$ quiver gauge theory with $N=2$ is depicted in Figure \ref{fig:D3}, which is related to the web diagram in Figure \ref{fig:+NM} by S-duality as well as ``generalized flop transition'' discussed in \cite{Hayashi:2017btw}. 
The decoupling limit is to move the red D5-branes downward while keeping the other D5-branes' positions intact, as in Figure \ref{fig:D3}.
These D5-branes are the lowest color branes for each gauge node and the lowest flavor brane charged under the central $\text{SU}(2N)$.
By this decoupling limit, we obtain the 5-brane web in Figure \ref{fig:D3decouple}, which is the S-dual description of the $+_N^{3,1}$ theory.
Since $D_3=A_3$, we can consider the corresponding decoupling process also in the ordinary 5-brane web.
The ordinary 5-brane web for the $A_3$ quiver gauge theory is depicted in Figure \ref{fig:A3}. 
The corresponding decoupling limit is again to move the red D5-branes downward while keeping the other D5-branes' positions intact, as in Figure \ref{fig:A3}. Then, we obtain a 5-brane web depicted in Figure \ref{fig:A3decouple}. Since it is obtained from the same decoupling limit from the $\#_{3,N}$ theory, this should correspond to the $+_N^{3,1}$ theory. The strong coupling limit of Figure \ref{fig:A3decouple} gives Figure \ref{fig:+N31 unitary web} after S-duality.  
Although this explanation is for the case with $N=2$, generalization for generic $N$ is straightforward. 

\begin{figure}
\centering
\begin{minipage}{0.45\hsize}
\centering
\begin{scriptsize}
\begin{tikzpicture}[scale=.75]
\draw[thick,dashed](0,0)--(0,6);
\draw[thick](0,5)--(-5,5);
\draw[thick](0,4.9)--(-1,4.9);
\draw[thick](-1,6)--(-1,4.9);
\draw[thick](-2,3.9)--(-1,4.9);
\draw[thick](-2,3.9)--(-5,3.9);

\draw[thick](-2,3.9)--(-2,2.1);
\draw[thick](-3,6)--(-3,0);
\draw[thick](-4,6)--(-4,0);

\draw[thick](-1,0)--(-1,1.1);
\draw[thick](-2,2.1)--(-1,1.1);
\draw[thick](-3,2.1)--(-5,2.1);

\node[label=right:{ON$^-$}]at(0,5.5){};
\node[label=right:{ON$^+$}]at(0,3){};
\node[label=right:{ON$^-$}]at(0,0.5){};
\draw[thick,red](0,1.1)--(-1,1.1);
\draw[thick,red](0,1)--(-5,1);
\draw[thick,red](-2,2.1)--(-3,2.1);
\end{tikzpicture}
\end{scriptsize}
\caption[S-dual of $\#_{3,2}$ theory.]{S-dual of $\#_{3,2}$ theory. We move four red D5-branes downward for decoupling.}
\label{fig:D3}
\end{minipage}
\hspace{5mm}
\begin{minipage}{0.45\hsize}
\centering
\begin{scriptsize}
\begin{tikzpicture}[scale=.5]
\draw[thick,dashed](0,-4)--(0,6);
\draw[thick](0,5)--(-5,5);
\draw[thick](0,4.9)--(-1,4.9);
\draw[thick](-1,6)--(-1,4.9);
\draw[thick](-2,3.9)--(-1,4.9);
\draw[thick](-2,3.9)--(-5,3.9);

\draw[thick](-2,3.9)--(-2,1.1);
\draw[thick](0,-2.9)--(-2,1.1);
\draw[thick](-3,6)--(-3,2.1);
\draw[thick](-2,1.1)--(-3,2.1);
\draw[thick](-4,6)--(-4,-4);

\draw[thick](-3,2.1)--(-5,2.1);

\node[label=right:{ON$^-$}]at(0,5.5){};
\node[label=right:{ON$^+$}]at(0,3){};
\node[label=right:{ON$^-$}]at(0,-3.5){};
\end{tikzpicture}
\end{scriptsize}
\caption[5-brane web for $+_N^{3,1}$ theory obtained from $\#_{3,2}$ theory.]{5-brane web for $+_N^{3,1}$ theory obtained from $\#_{3,2}$ theory by the decoupling limit.}
\label{fig:D3decouple}
\end{minipage}
\end{figure}
\begin{figure}
\centering
\begin{minipage}{0.4\hsize}
\centering
\begin{scriptsize}
    \begin{tikzpicture}[scale=.5]
\draw[thick](0,0)--(0,1);
\draw[thick,red](-1,1)--(0,1);
\draw[thick](1,2)--(0,1);
\draw[thick,red](1,2)--(3,2);
\draw[thick](4,2)--(3,2);
\draw[thick](1,2)--(1,3);
\draw[thick](-1,3)--(1,3);
\draw[thick](-1,2.75)--(.75,2.75);
\draw[thick](4,2.75)--(1.25,2.75);

\draw[thick](2,4)--(1,3);
\draw[thick](2,4)--(6.75,4);
\draw[thick](9,4)--(7.25,4);
\draw[thick](9,3.75)--(7,3.75);
\draw[thick](7,3.75)--(7,4.75);
\draw[thick](7,3.75)--(6,2.75);
\draw[thick](6,1.75)--(6,2.75);
\draw[thick](6,1.75)--(9,1.75);
\draw[thick](6,1.75)--(5,.75);
\draw[thick](5,0)--(5,.75);
\draw[thick](6.25,2)--(9,2);
\draw[thick](5.75,2)--(4,2);
\draw[thick](4,2.75)--(6,2.75);
\draw[thick](2,4)--(2,5);
\draw[thick,red](-1,.75)--(-.25,.75);
\draw[thick,red](0.25,.75)--(5,.75);

\draw[thick](3,0)--(3,5);
\draw[thick](4,0)--(4,5);
    \end{tikzpicture}
\end{scriptsize}
\caption[Usual 5-brane web for $A_3$ quiver for $N=2$]{Usual 5-brane web for $A_3$ quiver for $N=2$.  We move four red D5-branes downward for decoupling.}
\label{fig:A3}
\end{minipage}
\hspace{5mm}
\begin{minipage}{0.45\hsize}
\centering
\begin{scriptsize}
    \begin{tikzpicture}[scale=.5]

\draw[thick](4,2)--(3,2);
\draw[thick](1,2)--(1,3);
\draw[thick](-1,3)--(1,3);
\draw[thick](-1,2.75)--(.75,2.75);
\draw[thick](4,2.75)--(1.25,2.75);

\draw[thick](2,4)--(1,3);
\draw[thick](2,4)--(6.75,4);
\draw[thick](9,4)--(7.25,4);
\draw[thick](9,3.75)--(7,3.75);
\draw[thick](7,3.75)--(7,4.75);
\draw[thick](7,3.75)--(6,2.75);
\draw[thick](6,1.75)--(6,2.75);
\draw[thick](6,1.75)--(9,1.75);
\draw[thick](6,1.75)--(4,-.25);
\draw[thick](3,-2.25)--(4,-.25);
\draw[thick](6.25,2)--(9,2);
\draw[thick](5.75,2)--(4,2);
\draw[thick](4,2.75)--(6,2.75);
\draw[thick](2,4)--(2,5);

\draw[thick](3,2)--(3,5);
\draw[thick](3,2)--(1,0);
\draw[thick](1,2)--(1,0);
\draw[thick](0,-2)--(1,0);
\draw[thick](4,-.25)--(4,5);
    \end{tikzpicture}
\end{scriptsize}
\caption{5-brane web obtained from the $A_3$ quiver by the decoupling.}
\label{fig:A3decouple}
\end{minipage}
\end{figure}

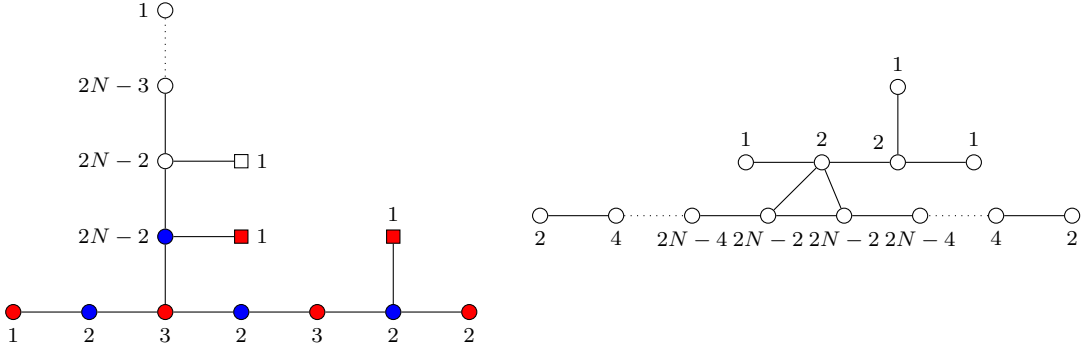
\begin{figure}[!ht]
    \centering
    \begin{minipage}{.45\textwidth}
    \begin{scriptsize}
        \begin{tikzpicture}
        \node[label=left:{1}][u](1){};
        \node[label=left:{$2N-3$}][u](2n-3)[below of=1]{};
        \node[label=left:{$2N-2$}][u](2n-2)[below of=2n-3]{};
        \node[label=right:{$1$}][uf](uf)[right of=2n-2]{};
        \node[label=left:{$2N-2$}][sp](sp2n-2)[below of=2n-2]{};
        \node[label=right:{$1$}][sof](so1)[right of=sp2n-2]{};
        \node[label=below:{$3$}][so](so3)[below of=sp2n-2]{};
        \node[label=below:{$2$}][sp](sp2)[left of=so3]{};
        \node[label=below:{$1$}][so](o1)[left of=sp2]{};
        \node[label=below:{$2$}][sp](sp2')[right of=so3]{};
        \node[label=below:{$3$}][so](so3')[right of=sp2']{};
        \node[label=below:{$2$}][sp](sp2'')[right of=so3']{};
        \node[label=below:{$2$}][so](so2)[right of=sp2'']{};
        \node[label=above:{$1$}][sof](so1')[above of=sp2'']{};
        \draw[dotted](1)--(2n-3);
        \draw(2n-3)--(2n-2);
        \draw(2n-2)--(sp2n-2);
        \draw(2n-2)--(uf);
        \draw(sp2n-2)--(so1);
        \draw(sp2n-2)--(so3);
        \draw(so3)--(sp2);
        \draw(sp2)--(o1);
        \draw(so3)--(sp2');
        \draw(sp2')--(so3');
        \draw(so3')--(sp2'');
        \draw(sp2'')--(so1');
        \draw(sp2'')--(so2);
        \end{tikzpicture}
    \end{scriptsize}
    \end{minipage}
        \begin{minipage}{.45\textwidth}
        \begin{scriptsize}
            \begin{tikzpicture}
            \node[label=below:{2}][u](2){};
            \node[label=below:{4}][u](4)[right of=2]{};
            \node[label=below:{$2N-4$}][u](2n-4)[right of=4]{};
            \node[label=below:{$2N-2$}][u](2n-2)[right of=2n-4]{};
            \node[label=below:{$2N-2$}][u](2n-2')[right of=2n-2]{};
            \node[label=below:{$2N-4$}][u](2n-4')[right of=2n-2']{};
            \node[label=below:{$4$}][u](4')[right of=2n-4']{};
            \node[label=below:{$2$}][u](2')[right of=4']{};
            \node[label=above:{$2$}][u](u2)[above right of=2n-2]{};
            \node[label=above:{$1$}][u](u1)[left of=u2]{};
            \node[label=above left:{$2$}][u](u2')[right of=u2]{};
            \node[label=above:{$1$}][u](u11)[right of=u2']{};
            \node[label=above:{$1$}][u](u11')[above of=u2']{};
            \draw(2)--(4);
            \draw[dotted](4)--(2n-4);
            \draw(2n-4)--(2n-2);
            \draw(2n-2)--(2n-2');
            \draw(2')--(4');
            \draw[dotted](4')--(2n-4');
            \draw(2n-4')--(2n-2');
            \draw(2n-2)--(u2);
            \draw(2n-2')--(u2);
            \draw(u2)--(u1);
            \draw(u2)--(u2');
            \draw(u2')--(u11);
            \draw(u2')--(u11');
            \end{tikzpicture}
        \end{scriptsize}
    \end{minipage}
    \caption{Magnetic quivers for the $+_N^{3,1}$ theory.}
    \label{fig:magpN31}
\end{figure}

The main part of the maximal subdivision of the orientifold web is given in Figure \ref{fig:+N31 orientifold web}.
One of the new features in this example compared to the ones in the previous sections is the dotted blue line on the O5$^+$-plane connected to the ordinary blue line on the O5$^-$-plane. 
The dotted blue line represents one of the four half D5-branes included in the O5$^+$-plane, which is based on the interpretation discussed around \eqref{SI-charge-1}, while the ordinary blue line is the half D5-brane coming from the three full D5-branes on the O5$^-$-plane.
Together with the other two ordinary blue half D5-branes, it contributes as a SO(3) gauge node in the OSp magnetic quiver in Figure \ref{fig:magpN31}. 
Since three half D5-branes are consumed to construct the SO(3) gauge node, three red half D5-branes remain on the O5$^-$-plane. 
Due to the charge conservation and the s-rule, the red part, as well as O5-plane, should be treated as one subweb,
where the number of half D5 branes reduce by one as we go over half D7-branes to the right. 
Due to this red subweb, the SO(3) gauge node appears from the part where there are originally two full D5-branes on the O5$^-$-plane
because a single half D5-brane is used as a part of the red subweb, and only three half D5-branes remain in this region. 
This red subweb cannot be detached from the O5-plane because there is no mirror pair
and thus contributes as fundamental hypermultiplet to various gauge nodes.
The number of half hypermultiplets is obtained by computing the stable intersection number between the subweb corresponding to the considered gauge node
and this red subweb, which includes both the original and its mirror image. 
For example, the stable intersection number between the red subweb and the green NS5-brane is three.
However, they both attach to the same (0,1) 7-branes from the same direction at the two places, including the mirror image
so that the number of the half-hypermultiplets for the $\text{USp}(2N-2)$ gauge node is $3-2=1$. 
Or, equivalently, we could have considered that the green NS5-branes are placed to the left of the red subweb so that it does not intersect with it. 
In this case, we should regard that the green NS5-branes are intersecting with O5$^+$-plane.
Then, the rule discussed below \eqref{SI-charge-1} enables us to reinterpret
the stable intersection number with the red subweb as a stable intersection number with the 3 half D5-branes inside O5$^+$-plane, 
whichever interpretation gives a consistent magnetic quiver. 

Some part of the structure in the magnetic quiver discussed above may be more natural to understand 
if we consider Hanany-Witten transition.
Suppose we concentrate only on the red subweb as well as remaining D5-branes on the O5-planes
 while omitting the remaining part. 
By moving three half D7-branes from the right to the left of the red subweb,
we obtain a simpler web diagram,
from which we can straightforwardly read off most of the SO and the USp gauge nodes in the magnetic quiver.
This discussion is parallel to the 5-brane analysis for the Higgs branch of the 5d Sp(N) gauge theory at finite coupling in \cite{Bourget:2020gzi}. 
This observation would be useful to partially support our magnetic quiver 
but if we need to obtain the full structure of the magnetic quiver, 
the original orientifold web would be more convenient.

Finally, to understand if the $\mathbb{Z}_2$ node in the magnetic quiver of Figure \ref{fig:magpN31} is gauge or flavor, we computed the Hilbert series of the Higgs branch in both cases, and compared with the unitary magnetic quiver. We find that only when we take $\mathbb{Z}_2$ node to be gauge, the match is recovered. The Higgs branch Hilbert series reads
\begin{equation}
\text{HS}_{H}(t^2)=1 + t^2 + 2 t^3 + 5 t^4 + 6 t^5 + O(t^6).
\end{equation}

\FloatBarrier
\subsection{The \texorpdfstring{$\tilde{K}_N^1$}{TEXT} theory}
The  $\tilde{K}_N^1$ theory in the class of O5$^-$-O5$^+$ is obtained from the $+_{N+1}^{3,1}$ theory by decoupling a flavor from \eqref{electric +1N31 O5-O5+ OSp}.  At low energies, there is a gauge theory description
\be\label{electric KN1 O5+O5- OSp}
\begin{array}{c}\begin{tikzpicture}
\node {$\text{SO}(6)-\text{USp}(2)-\text{SO}(6)-\cdots-\text{USp}(2)-\text{SO}(6)-\text{USp}(2)-[3]$};
 \draw [thick,decorate,decoration={brace,amplitude=6pt},xshift=0pt,yshift=10pt]
(-4.75,0) -- (4.0,0)node [black,midway,xshift=0pt,yshift=20pt] {
$2N$};
\end{tikzpicture}
\end{array}
\ee
The orientifold web for the $\tilde{K}_N^1$ theory is presented in Figure \ref{fig:KN1 O5+O5- orientifold web}.

In Figure \ref{fig:D3decouple} and Figure \ref{fig:A3decouple}, 
the $SU(2N+1)$ gauge node has $2N+1$ flavor, 
and is coupled to a non-Lagrangian theory, where 
the figures are depicted for the case $N=1$.
Decoupling one flavor from the $\text{SU}(2N+1)$ gauge node,
we obtain the web diagrams in 
Figure \ref{fig:KN1 O5+O5- unitary web}. 

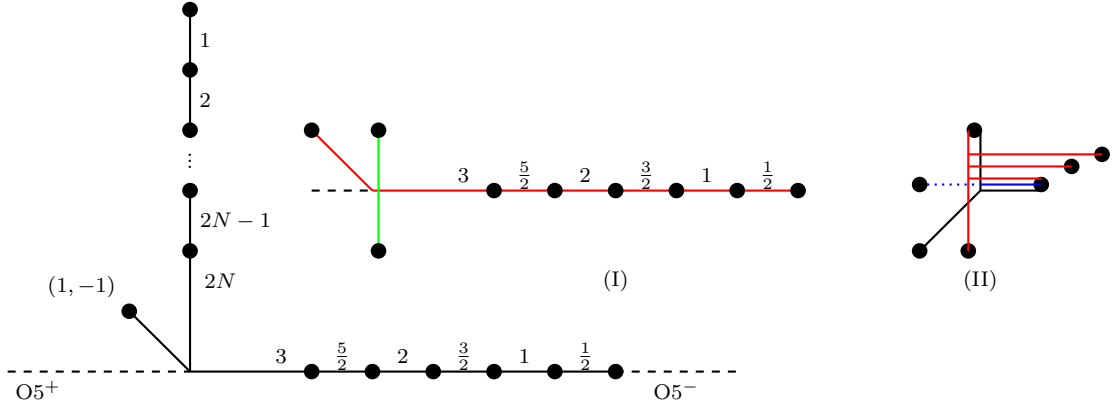
\begin{figure}[!ht]
    \centering
\begin{scriptsize}
    \begin{tikzpicture}[scale=.8]
    \draw[thick](0,0)--(7,0);
    \draw[thick](0,0)--(-1,1);
    \node[label=above left:{$(1,-1)$}][7brane]at(-1,1){};
    \draw[thick,dashed](-3,0)--(0,0);
    \draw[thick,dashed](7,0)--(9,0);
    \node[7brane]at(2,0){};
    \node[7brane]at(3,0){};
    \node[7brane]at(4,0){};
    \node[7brane]at(5,0){};
    \node[7brane]at(6,0){};
    \node[7brane]at(7,0){};
    \draw[thick](0,0)--(0,3);
    \node[7brane]at(0,2){};
    \node[7brane]at(0,3){};
    \node at (0,3.4){$.$};
    \node at (0,3.5){$.$};
    \node at (0,3.6){$.$};
    \draw[thick](0,4)--(0,6);
    \node[7brane]at(0,4){};
    \node[7brane]at(0,5){};
    \node[7brane]at(0,6){};
    \node at (8,-.3){$\text{O5}^-$};
    \node at (-2.5,-.3){$\text{O5}^+$};
    \node at (1.5,.25){3};
    \node at (2.5,.25){$\frac{5}{2}$};
    \node at (3.5,.25){2};
    \node at (4.5,.25){$\frac{3}{2}$};
    \node at (5.5,.25){1};
    \node at (6.5,.25){$\frac{1}{2}$};
    \node at (.25,5.5){1};
    \node at (.25,4.5){2};
    \node at (.75,2.5){$2N-1$};
    \node at (.5,1.5){$2N$};

        \node at (4.5,3.25){3};
    \node at (5.5,3.25){$\frac{5}{2}$};
    \node at (6.5,3.25){2};
    \node at (7.5,3.25){$\frac{3}{2}$};
    \node at (8.5,3.25){1};
    \node at (9.5,3.25){$\frac{1}{2}$};
        \draw[thick,red](3,3)--(10,3);
    \draw[thick,red](3,3)--(2,4);
    \draw[thick,green](3.1,2)--(3.1,4);
    \node[7brane]at(3.1,2){};
    \node[7brane]at(3.1,4){};
    \node[7brane]at(2,4){};
    \draw[thick,dashed](2,3)--(3,3);
    \node[7brane]at(5,3){};
    \node[7brane]at(6,3){};
    \node[7brane]at(7,3){};
    \node[7brane]at(8,3){};
    \node[7brane]at(9,3){};
    \node[7brane]at(10,3){};
        \node at(7,1.5){(I)};

    \draw[thick](13,4)--(13,3);
    \draw[thick](14,3)--(13,3);
    \draw[thick](12,2)--(13,3);
    \node[7brane]at(12.9,4){};
    \node[7brane]at(12.8,2){};
    \node[7brane]at(14,3.1){};
    \node[7brane]at(14.5,3.4){};
    \node[7brane]at(15,3.6){};
    \node[7brane]at(12,2){};
    \draw[thick,red](12.8,4)--(12.8,2);
    \draw[thick,red](14,3.2)--(12.8,3.2);
    \draw[thick,red](14.5,3.4)--(12.8,3.4);
    \draw[thick,red](15,3.6)--(12.8,3.6);
    \draw[thick,blue](14,3.1)--(13,3.1);
    \draw[thick,blue,dotted](12,3.1)--(13,3.1);
		\node[7brane]at(12,3.1){};
    \node at(13,1.5){(II)};
    \end{tikzpicture}
    \end{scriptsize}
    \caption{An orientifold web for the $\tilde{K}_N^1$ theory.}
    \label{fig:KN1 O5+O5- orientifold web}
\end{figure}
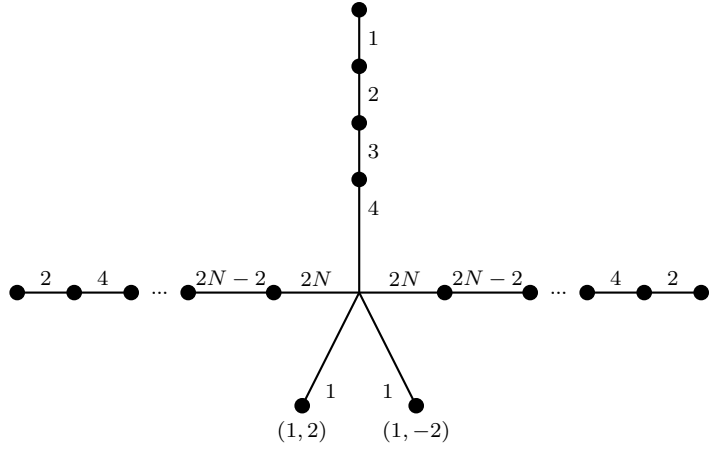
\begin{figure}[ht]
    \centering
    \begin{scriptsize}
        \begin{tikzpicture}[scale=.75]
        \draw[thick](-3,0)--(3,0);
        \draw[thick](0,0)--(0,5);
        \draw[thick](0,0)--(-1,-2);
        \draw[thick](0,0)--(1,-2);
        \node[7brane]at(1.5,0){};
        \node[7brane]at(3,0){};
        \node[7brane]at(-1.5,0){};
        \node[7brane]at(-3,0){};
        \node[7brane]at(0,2){};
        \node[7brane]at(0,3){};
        \node[7brane]at(0,4){};
        \node[7brane]at(0,5){};
        \node[label=below:{$(1,2)$}][7brane]at(-1,-2){};
        \node[label=below:{$(1,-2)$}][7brane]at(1,-2){};
        \node at(3.4,0){$.$};
        \node at(3.5,0){$.$};
        \node at(3.6,0){$.$};
        \node at(-3.4,0){$.$};
        \node at(-3.5,0){$.$};
        \node at(-3.6,0){$.$};
        \draw[thick](4,0)--(6,0);
        \draw[thick](-4,0)--(-6,0);
        \node[7brane]at(4,0){};
        \node[7brane]at(5,0){};
        \node[7brane]at(6,0){};
        \node[7brane]at(-4,0){};
        \node[7brane]at(-5,0){};
        \node[7brane]at(-6,0){};
        \node at (0.8,.25){$2N$};
        \node at (2.25,.25){$2N-2$};
        \node at (4.5,.25){$4$};
        \node at (5.5,.25){$2$};
        \node at (-.75,.25){$2N$};
        \node at (-2.25,.25){$2N-2$};
        \node at (-4.5,.25){$4$};
        \node at (-5.5,.25){$2$};
        \node at(.25,1.5){4};
        \node at(.25,2.5){3};
        \node at(.25,3.5){2};
        \node at(.25,4.5){1};
        \node at(-.5,-1.75){1};
        \node at(.5,-1.75){1};
        \end{tikzpicture}
    \end{scriptsize}
    \caption{A unitary web description of the $\tilde{K}_N^1$ theory.}
    \label{fig:KN1 O5+O5- unitary web}
\end{figure}

The $\tilde{K}^1_N$ theory has two maximal subdivisions, as in Figure \ref{fig:KN1 O5+O5- orientifold web}.
In maximal subdivision (I), the red subweb cannot be detached from the O5-plane and thus, giving 
fundamental hypermultiplets to the $\text{USp}(2N)$ node coming from the $2N$ NS5-branes intersecting with the O5-planes.
In maximal subdivision (II), there is a blue dotted line on the O5$^+$-plane connected to the ordinary blue line on the O5$^-$-plane,
analogous to the case for the $+^{3,1}_N$ theory.
However, this does not become part of a gauge node because there are no other half D5-branes available at the same place.
Therefore, both the blue subweb and the red subweb contribute as fundamental hypermultiplets of various gauge nodes.
The magnetic quivers are given in Table \ref{O5+O5-KN1magneticquivers}. 

To understand if the two $\mathbb{Z}_2$ nodes in the magnetic quiver associated with the maximal subdivision (II) are gauge or flavor, we computed the Hilbert series of the Higgs branch in both cases and compared with the unitary magnetic quiver. We find that only when we take both $\mathbb{Z}_2$ nodes to be flavor, the match is recovered. The Higgs branch Hilbert series reads
\begin{equation}
\text{HS}_{H}(t^2)=1 + 5 t^2 + 8 t^3 + 18 t^4 + 36 t^5 + 71 t^6 + 120 t^7 + \cdots.
\end{equation}
The Coulomb branch Hilbert series is tabulated in Table \ref{O5+O5-KN1CoulombHS}.
%
 \begin{table}[!hbp]
        \centering
      \begin{tabular}{|c|C{6.2cm}|C{7.1cm}|}
			\rowcolor{Grayy}
         \hline MS&Unitary & Orthosymplectic\\\hline
        (I) &\begin{scriptsize}
        \begin{tikzpicture}
        \node[label=below:{2}][u](2){};
        \node[label=below:{4}][u](4)[right of=2]{};
        \node[label=below:{$2N-2$}][u](2n-2)[right of=4]{};
        \node[label=below:{$2N$}][u](2n)[right of=2n-2]{};
        \node[label=below:{$2N-2$}][u](2n-2')[right of=2n]{};
        \node[label=below:{$4$}][u](4')[right of=2n-2']{};
        \node[label=below:{$2$}][u](2')[right of=4']{};
        \node[label=left:{$1$}][u](u1)[above of=2n]{};
        \draw(2)--(4);
        \draw[dotted](4)--(2n-2);
        \draw(2n-2)--(2n);
        \draw(2')--(4');
        \draw[dotted](4')--(2n-2');
        \draw(2n-2')--(2n);
        \draw(u1)--(2n);
        \draw[double distance=4pt](u1)--(2n);
        \draw[double distance=1pt](u1)--(2n);
        \end{tikzpicture}
        \end{scriptsize} &\begin{scriptsize}
        \begin{tikzpicture}
        \node[label=below:{1}](1)[u]{};
        \node[label=below:{$2N-1$}][u](2n-1)[right of=1]{};
        \node[label=below:{$2N$}][sp](sp2n)[right of=2n-1]{};
        \node[label=above:{6}][sof](sof)[above of=sp2n]{};
        \draw(1)[dotted]--(2n-1);
        \draw(2n-1)--(sp2n);
        \draw(sp2n)--(sof);
        \end{tikzpicture}
        \end{scriptsize}\\\hline
        (II)&\begin{scriptsize}
        \begin{tikzpicture}
        \node[label=below:{2}][u](2){};
        \node[label=below:{4}][u](4)[right of=2]{};
        \node[label=below:{$2N-2$}][u](2n-2)[right of=4]{};
        \node[label=below:{$2N-1$}][u](2n)[right of=2n-2]{};
        \node[label=below:{$2N-2$}][u](2n-2')[right of=2n]{};
        \node[label=below:{$4$}][u](4')[right of=2n-2']{};
        \node[label=below:{$2$}][u](2')[right of=4']{};
        \node[label=left:{$1$}][u](u1)[above of=2n-2]{};
        \node[label=right:{$1$}][u](u1')[above of=2n-2']{};
        \node[label=above:{1}][u](u11)[above of=u1]{};
        \node[label=above:{1}][u](u11')[above of=u1']{};
        \node[label=above:{2}][u](u2)[right of=u11]{};
        \draw(2)--(4);
        \draw[dotted](4)--(2n-2);
        \draw(u1)--(u2);
        \draw(u1')--(u2);
        \draw(u11)--(u2);
        \draw(u11')--(u2);
        \draw(2n)--(u1);
        \draw(2n)--(u1);
        \draw(2n)--(u1');
        \draw[double distance=2pt](u1)--(u1');
        \draw(2n-2)--(2n);
        \draw(2')--(4');
        \draw[dotted](4')--(2n-2');
        \draw(2n-2')--(2n);
        \draw(u1)--(2n-2);
        \draw(u1')--(2n-2');
        \end{tikzpicture}
        \end{scriptsize}&\begin{scriptsize}
        \begin{tikzpicture}
        \node[label=below:{1}][u](1){};
        \node[label=below:{$2N-2$}][u](2n-3)[right of=1]{};
        \node[label=below:{$2N-1$}][u](2n-2)[right of=2n-3]{};
        \node[label=below:{$1$}][u](2n-2')[right of=2n-2]{};
        \node[label=above left:{$2N-2$}][sp](sp2n-2)[above of=2n-2]{};
        \node[label=left:{1}][uf](u1)[above left of=2n-2]{};
        \node[label=above:{2}][sof](sof)[above of=sp2n-2]{};
        \node[label=below:{2}][sp](sp2)[right of=2n-2']{};
        \node[label=below:{3}][so](so3)[right of=sp2]{};
        \node[label=below:{2}][sp](sp2')[right of=so3]{};
        \node[label=below:{2}][so](so2)[right of=sp2']{};
        \node[label=above:{2}][uf](uf2)[above of=2n-2']{};
        \node[label=above:{1}][sof](so1)[above of=sp2]{};
        \node[label=above:{1}][sof](so1')[above of=sp2']{};
        \draw(2n-2')--(sp2);
        \draw(sp2)--(so3);
        \draw(so3)--(sp2');
        \draw(sp2')--(so2);
        \draw(so1)--(sp2);
        \draw(so1')--(sp2');
        \draw(uf2)--(2n-2');
        \draw[dotted](1)--(2n-3);
        \draw(2n-3)--(2n-2);
        \draw(2n-2)--(2n-2');
        \draw(2n-2)--(sp2n-2);
        \draw(sp2n-2)--(sof);
        \draw(2n-2)--(u1);
        \end{tikzpicture}
        \end{scriptsize}\\\hline
    \end{tabular}
        \caption[Magnetic quivers for $\tilde{K}_N^1$ theory.]{Magnetic quivers for various maximal subdivisions (MS) of the $\tilde{K}_N^1$ theory.}
        \label{O5+O5-KN1magneticquivers}
    \end{table}
		\begin{table}
\centering
\begin{tabular}{|c|C{4.9cm}|C{4.9cm}|C{2.8cm}|} \hline
\rowcolor{Grayy}
   & Unitary magnetic quiver & \multicolumn{2}{c|}{Orthosymplectic magnetic quiver} \\ \cline{2-4}
	\rowcolor{Grayy}
  \multirow{-2}{*}{MS}
 & HS($t$)  & HS($t;\vec{m} \in \mathbb{Z}$) & HS($t;\vec{m} \in \mathbb{Z}+\tfrac{1}{2}$)  \\ \hline
	$\text{(I)}_{N=1}$ & \footnotesize{$\begin{array}{l} \dfrac{1 + t + 2 t^2 + t^3 + t^4}{(1-t)^{4}\,(1+t)^2} \\\\ = 1 + 3 t + 9 t^2 + 18 t^3\\ + 35 t^4 + 57 t^5+\cdots  \end{array}$}  & \footnotesize{$\begin{array}{l} \dfrac{1 + t + 2 t^2 + t^3 + t^4}{(1-t)^{4}\,(1+t)^2} \\\\ = 1 + 3 t + 9 t^2 + 18 t^3\\ + 35 t^4 + 57 t^5+\cdots  \end{array}$} & \footnotesize{not required} \\ \hline
	$\text{(I)}_{N=2}$ & \footnotesize{$1 + 15 t + 135 t^2 + 888 t^3 + 4709 t^4 + 21144 t^5+\cdots$}  & \footnotesize{$1 + 15 t + 135 t^2 + 888 t^3 + 4709 t^4 + 21144 t^5+\cdots$} & \footnotesize{not required} \\ \hline
	$\text{(II)}_{N=1}$ & \footnotesize{$1 + 19 t + 173 t^2 + 24 t^{5/2} + 1042 t^3 + 328 t^{7/2} + 4760 t^4 +  2312 t^{9/2} + 17908 t^5+\cdots$}  & \footnotesize{$1 + 19 t + 173 t^2 + 24 t^{5/2} + 1042 t^3 + 328 t^{7/2} + 4760 t^4 +  2312 t^{9/2} + 17908 t^5+\cdots$} & \footnotesize{not required} \\ \hline
	$\text{(II)}_{N=2}$ & \footnotesize{$1 + 31 t + 495 t^2 + 5443 t^3 + 48 t^{7/2} + 46260 t^4 + 
 1472 t^{9/2} + 323154 t^5+\cdots$}  & \footnotesize{$1 + 31 t + 495 t^2 + 5443 t^3 + 48 t^{7/2} + 46260 t^4 + 
 1472 t^{9/2} + 323154 t^5+\cdots$} & \footnotesize{not required} \\ \hline
\end{tabular}
\caption[Coulomb branch HS for magnetic quivers of $\tilde{K}_{N}^{1}$ theory.]{Coulomb branch Hilbert series for the unitary and orthosymplectic magnetic quivers presented in Table \ref{O5+O5-KN1magneticquivers} for the $\tilde{K}_{N}^{1}$ theory.}
\label{O5+O5-KN1CoulombHS}
\end{table}
\FloatBarrier
\section{Magnetic quivers from $\widetilde{\text{O5}}^+$}
\label{sec:O5tilde+}
Finally we consider some examples where we have an $\widetilde{\text{O5}}^+$-plane on one or two of the ends in the orientifold web configuration. 
\subsection{\texorpdfstring{$\widetilde{\text{O5}}^+$ - $\widetilde{\text{O5}}^+$}{TEXT}}

\begin{figure}[!ht]
    \centering
    \begin{scriptsize}
    \begin{tikzpicture}[scale=.8]
    \draw[thick](-2,0)--(2,0);
    \draw[thick,dashed](-4,0)--(-2,0);
    \draw[thick,dashed](4,0)--(2,0);
    \node[7brane]at(2,0){};
    \node[7brane]at(-2,0){};
    \draw[thick](0,0)--(0,3);
    \node[7brane]at(0,2){};
    \node[7brane]at(0,3){};
    \node at (0,3.4){$.$};
    \node at (0,3.5){$.$};
    \node at (0,3.6){$.$};
    \draw[thick](0,4)--(0,6);
    \node[7brane]at(0,4){};
    \node[7brane]at(0,5){};
    \node[7brane]at(0,6){};
    \node at (3.5,-.3){$\widetilde{\text{O5}}^+$};
    \node at (-3.5,-.3){$\widetilde{\text{O5}}^+$};
    \node at (1.5,.25){1};
    \node at (-1.5,.25){1};
    \node at (.25,5.5){1};
    \node at (.25,4.5){2};
    \node at (.75,2.5){$2N-1$}; 
    \node at (.5,1.5){$2N$};
    \end{tikzpicture}
    \end{scriptsize}
    \hspace{.5cm}
        \begin{scriptsize}
        \begin{tikzpicture}[scale=.7]
        \draw[thick](-1.5,-1.5)--(1.5,1.5);
        \draw[thick](-2,2)--(2,-2);
        \node[7brane]at(-1,1){};
        \node[7brane]at(-2,2){};
        \node[7brane]at(1,-1){};
        \node[7brane]at(2,-2){};
        \node[7brane]at(-1.5,-1.5){};
        \node[7brane]at(1.5,1.5){};
        \node at (-2.4,2.4){$.$};
        \node at (-2.5,2.5){$.$};
        \node at (-2.6,2.6){$.$};
        \node at (2.4,-2.4){$.$};
        \node at (2.5,-2.5){$.$};
        \node at (2.6,-2.6){$.$};
        \draw[thick](-3,3)--(-5,5);
        \draw[thick](3,-3)--(5,-5);
        \node[7brane] at (-3,3){};
        \node[7brane] at (-4,4){};
        \node[7brane] at (-5,5){};
        \node[7brane] at (3,-3){};
        \node[7brane] at (4,-4){};
        \node[7brane] at (5,-5){};
        \node at(-1.05,.5){$2N$};
        \node at(-2.45,1.5){$2N-2$};
        \node at(-3.85,3.5){$4$};
        \node at(-4.85,4.5){$2$};
        \node at(1.05,-.5){$2N$};
        \node at(2.45,-1.5){$2N-2$};
        \node at(3.85,-3.5){$4$};
        \node at(4.85,-4.5){$2$};
        \node at(-1.25,-.75){2};
        \node at(1.25,.75){2};
        \end{tikzpicture}
        \end{scriptsize}
    \caption[Orientifold and unitary web diagrams for $+_N$ theory.]{Orientifold web(left), and ordinary web(right) for the $+_N$ theory. The maximal subdivision is the trivial one and therefore not shown explicitly.}
    \label{fig:+N O5tilde-O5tilde}
   \end{figure}
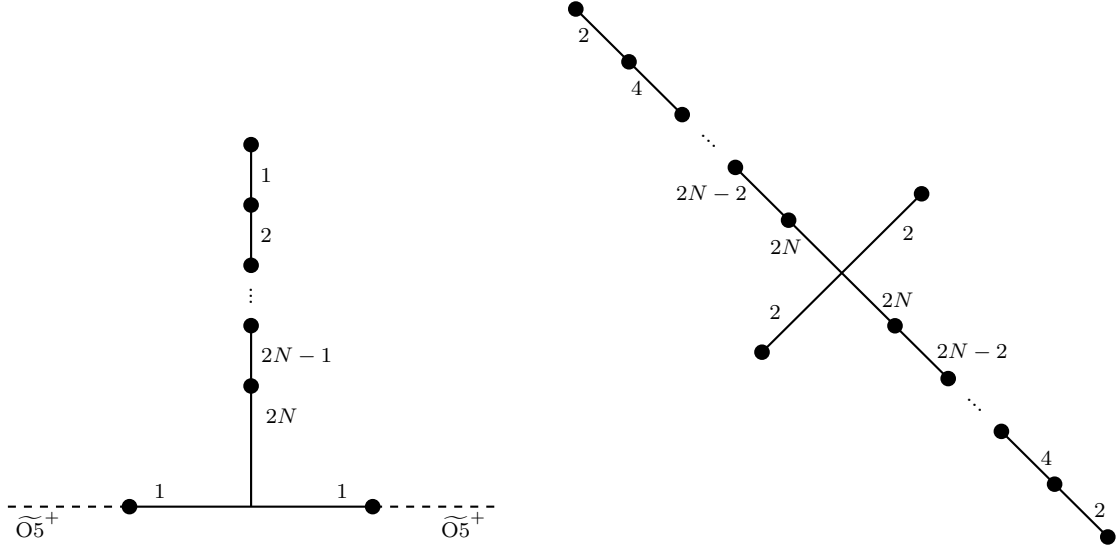

We begin with an example where we have $\widetilde{\text{O5}}^+$-planes on the two ends of the brane configuration. For constructing such a configuration we start from the theory considered in section \ref{sec:+1NO5+O5+}, which yields the orthosymplectic quiver \eqref{electric +1N O5+O5+ OSp} as an IR theory. From the quiver theory \eqref{electric +1N O5+O5+ OSp}, we Higgs the two $SO(6)$ gauge theories on the ends. The resulting IR theory becomes
\be\label{electric +1N O5tilde OSp}
\begin{array}{c}\begin{tikzpicture}
\node {$\text{SO}(5)-{\underset{\underset{\text{\large$\left[\frac{1}{2}\right]$}}{\textstyle\vert}}{\text{USp}(2)}}-\text{SO}(6)-\text{USp}(2)-\text{SO}(6)-\cdots-\text{USp}(2)-\text{SO}(6)-{\underset{\underset{\text{\large$\left[\frac{1}{2}\right]$}}{\textstyle\vert}}{\text{USp}(2)}}-\text{SO}(5)$};
 \draw [thick,decorate,decoration={brace,amplitude=6pt},xshift=0pt,yshift=10pt]
(-7.5,0.5) -- (7.5,0.5)node [black,midway,xshift=0pt,yshift=20pt] {
$2N-1$};
\end{tikzpicture}
\end{array}
\ee
where $\frac{1}{2}$ represents a half-hypermultiplet in the fundamental representation of $\text{USp}(2)$. 
It is possible to construct the corresponding orientifold web diagram by performing the same Higgsing to the web in Figure \ref{fig:+1M O5+-O5+}. The orientifold web diagram is depicted in Figure \ref{fig:+N O5tilde-O5tilde} which has asymptotically $\widetilde{\text{O5}}^+$-planes. From the orientifold web configuration the presence of a half D7-brane accounts for the half-hypermultiplet. 

The theory in \eqref{electric +1N O5tilde OSp} can also be constructed by gluing $2$ copies of $\text{SO}(5)$ with 1 vector and 1 singlet, 
and $N-2$ copies of $\text{SO}(6)$ with two vectors via successive gauging of $\text{USp}(2)$ subgroup of the global symmetry. This allows us to propose a unitary web in Figure \ref{fig:+N O5tilde-O5tilde} by gluing together $2$ copies of $\text{USp}(4)$ with one antisymmetric hypermultiplet and one singlet, 
and $N-2$ copies of $\text{SU}(4)_0$ with two antisymmetric hypermultiplets via gauging $\text{SU}(2)$ subgroups of the global symmetry.

The orthosymplectic and unitary magnetic quiver theories can be read off from the web diagrams in Figure \ref{fig:+N O5tilde-O5tilde}, and they are given in Figure \ref{O5t+t++Nmagneticquivers}. The corresponding Coulomb branch Hilbert series are given in Table \ref{O5t+t++NCoulombHS}.
		\begin{figure}[!htb]
\centering
   \begin{subfigure}{0.49\linewidth} \centering
    \begin{scriptsize}
    \begin{tikzpicture}
    \node[label=below:{1}][u](1){};
    \node[label=below:{2}][u](2)[right of=1]{};
    \node[label=below:{$2N-1$}][u](2n-1)[right of=2]{};
    \node[label=below:{$2N$}][sp](sp2n)[right of=2n-1]{};
    \node[label=above:{3}][sof](sof)[above of=sp2n]{};
    \node[label=below:{3}][so](so3)[right of=sp2n]{};
    \draw(1)--(2);
    \draw[dotted](2)--(2n-1);
    \draw(2n-1)--(sp2n);
    \draw(sp2n)--(sof);
    \draw(sp2n)--(so3);
    \end{tikzpicture}
    \end{scriptsize}
     \caption{Orthosymplectic quiver}\label{O5t+t++NOSpquiver}
   \end{subfigure}
   \begin{subfigure}{0.49\linewidth} \centering
     \begin{scriptsize}
    \begin{tikzpicture}
    \node[label=below:{2}][u](2){};
    \node[label=below:{4}][u](4)[right of=2]{};
    \node[label=below:{$2N-2$}][u](2n-2)[right of=4]{};
    \node[label=below:{$2N$}][u](2n)[right of=2n-2]{};
    \node[label=below:{$2N-2$}][u](2n-2')[right of=2n]{};
    \node[label=below:{$4$}][u](4')[right of=2n-2']{};
    \node[label=below:{$2$}][u](2')[right of=4']{};
    \node[label=above:{$2$}][u](u2)[above of=2n]{};
    \draw(2)--(4);
    \draw[dotted](4)--(2n-2);
    \draw(2n-2)--(2n);
    \draw(2')--(4');
    \draw[dotted](4')--(2n-2');
    \draw(2n-2')--(2n);
    \draw[double distance=2pt](2n)--(u2);
    \end{tikzpicture}
    \end{scriptsize}
     \caption{Unitary quiver}\label{O5t+t++Nunitaryquiver}
   \end{subfigure}
\caption{Magnetic quivers for the $+_N$ theory} \label{O5t+t++Nmagneticquivers}
\end{figure}
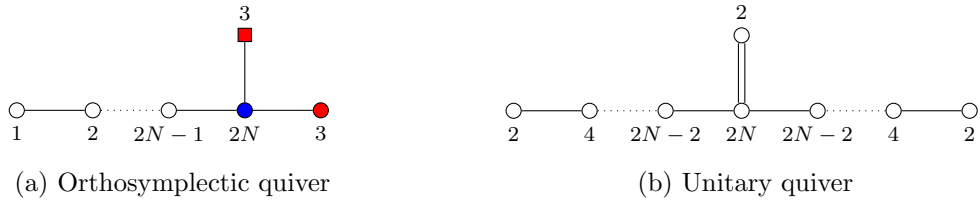
		\begin{table}[!htb]
\centering
\begin{tabular}{|c|C{4.9cm}|C{4.9cm}|C{2.8cm}|} \hline
\rowcolor{Grayy}
   & Unitary magnetic quiver & \multicolumn{2}{c|}{Orthosymplectic magnetic quiver} \\ \cline{2-4}
	\rowcolor{Grayy}
  \multirow{-2}{*}{$+_N$}
 & HS($t$)  & HS($t;\vec{m} \in \mathbb{Z}$) & HS($t;\vec{m} \in \mathbb{Z}+\tfrac{1}{2}$)  \\ \hline
	$+_1$ & \footnotesize{bad theory}  & \footnotesize{bad theory} & \footnotesize{not required} \\ \hline
	$+_2$ & \footnotesize{$1 + 16 t + 168 t^2 + 1315 t^3 + 8329 t^4 + 44491 t^5+\cdots$}  & \footnotesize{$1 + 16 t + 168 t^2 + 1315 t^3 + 8329 t^4 + 44491 t^5+\cdots$} & \footnotesize{not required} \\ \hline
\end{tabular}
\caption[Coulomb branch HS for magnetic quivers of $+_N$ theory.]{Coulomb branch Hilbert series for the unitary and orthosymplectic magnetic quivers presented in Figure \ref{O5t+t++Nmagneticquivers} for the $+_N$ theory.}
\label{O5t+t++NCoulombHS}
\end{table}
\FloatBarrier
\subsection{\texorpdfstring{$\widetilde{\text{O5}}^+$ - O5$^+$}{TEXT}}
It is also possible to consider an example where we have an $\widetilde{\text{O5}}^+$-plane on one end and an O5$^+$-plane on the other end. We again start from the theory considered in section \ref{sec:+1NO5+O5+}, which yields the orthosymplectic quiver \eqref{electric +1N O5+O5+ OSp} as an IR theory. From the quiver theory \eqref{electric +1N O5+O5+ OSp}, we Higgs the $SO(6)$ gauge theory on the right end and decouple flavor from the $SO(6)$ gauge on the left end. Then the resulting IR theory becomes
\be\label{electric KN1 O5tilde OSp}
\begin{array}{c}\begin{tikzpicture}
\node {$[1]-\text{SO}(6)-\text{USp}(2)-\text{SO}(6)-\cdots-\text{SO}(6)-{\underset{\underset{\text{\large$\left[\frac{1}{2}\right]$}}{\textstyle\vert}}{\text{USp}(2)}}-\text{SO}(5)$};
 \draw [thick,decorate,decoration={brace,amplitude=6pt},xshift=0pt,yshift=10pt]
(-4.5,0.5) -- (5.5,0.5)node [black,midway,xshift=0pt,yshift=20pt] {
$2N-1$};
\end{tikzpicture}
\end{array}
\ee
We can construct the corresponding orientifold web diagram by performing the same Higgsing and the decoupling to the web in Figure \ref{fig:+1M O5+-O5+}. The orientifold web diagram is depicted in Figure \ref{fig:KN1 O5tilde+O5+} which has asymptotically an $\widetilde{\text{O5}}^+$-plane and an O5$^+$-plane. 

This theory in \eqref{electric KN1 O5tilde OSp} can also be made by gluing $1$ copy of $\text{SO}(5)$ with 1 vector and 1 singlet, 
and $N-1$ copies of $\text{SO}(6)$ with two vectors via successive gauging of $\text{USp}(2)$ subgroup of the global symmetry. This allows us to propose the corresponding unitary web in Figure \ref{fig:KN1 O5tilde+O5+} by gluing together $1$ copy of $\text{USp}(4)$ with one antisymmetric hypermultiplet and one singlet, 
and $N-1$ copies of $\text{SU}(4)_0$ with two antisymmetric hypermultiplets via gauging $\text{SU}(2)$ subgroups of the global symmetry.
\begin{figure}[!ht]
    \centering
        \begin{scriptsize}
    \begin{tikzpicture}[scale=.7]
        \draw[thick](0,0)--(2,0);
        \draw[thick](0,0)--(-1.5,1.5);
    \draw[thick,dashed](-4,0)--(-0,0);
    \draw[thick,dashed](5,0)--(2,0);
    \node[7brane] at (2,0){};
    \node[label=above left:{$(1,-1)$}][7brane] at (-1.5,1.5){};
    \draw[thick](0,0)--(0,3);
    \node[7brane] at (0,2){};
    \node[7brane] at (0,3){};
    \node[thick] at (0,3.4) {$.$};
\node[thick] at (0,3.5) {$.$};
\node[thick] at (0,3.6) {$.$};
\node[7brane] at (0,4){};
\node[7brane] at (0,5){};
\node[7brane] at (0,6){};
\node at (.5,5.5) {1};
\node at (.5,4.5) {2};
\node at (.75,2.5) {$2N-2$};
\node at (.75,1.5) {$2N-1$};
\node at (-1.2,.8){1};
\draw[thick](0,4)--(0,6);

\node at (4,-.25) {$\widetilde{\text{O5}}^+$};
\node at (1.5,0.35) {1};
\node at (-3,-.25) {$\text{O}5^+$};
    
    \end{tikzpicture}
    \end{scriptsize}
    \hspace{.5cm}
        \begin{scriptsize}
        \begin{tikzpicture}[scale=.7]
        \draw[thick](-2,2)--(2,-2);
        \draw[thick](0,0)--(3,1);
        \node[label=right:{$(3,1)$}][7brane]at(3,1){};
        \node[7brane]at(-2,2){};
        \node[7brane]at(2,-2){};
        \draw[thick](2,2)--(-2,-2);
        \node[7brane]at(-2,-2){};
        \node[7brane]at(2,2){};
        \node[7brane]at(-1,-1){};
        \node[7brane]at(1,1){};
        \node at (2.4,2.4){$.$};
        \node at (2.5,2.5){$.$};
        \node at (2.6,2.6){$.$};
        \node at (-2.4,-2.4){$.$};
        \node at (-2.5,-2.5){$.$};
        \node at (-2.6,-2.6){$.$};
        \draw[thick](3,3)--(5,5);
        \node[7brane]at(3,3){};
        \node[7brane]at(4,4){};
        \node[7brane]at(5,5){};
        \draw[thick](-3,-3)--(-5,-5);
        \node[7brane]at(-3,-3){};
        \node[7brane]at(-4,-4){};
        \node[7brane]at(-5,-5){};
        \node at(1.85,-1.5){1};
        \node at(-1.85,1.5){2};
        \node at(.35,.75){$2N-2$};
        \node at(.8,1.6){$2N-4$};
        \node at(3.5,3.85){$4$};
        \node at(4.5,4.85){$2$};
        \node at(-.25,-.75){$2N$};
        \node at(-.8,-1.6){$2N-2$};
        \node at(-3.5,-3.85){$4$};
        \node at(-4.5,-4.85){$2$};
        \end{tikzpicture}
        \end{scriptsize}
    \caption{Orientifold and unitary web diagram for the $\kappa_N^1$ theory.}
    \label{fig:KN1 O5tilde+O5+}
\end{figure}

The orthosymplectic and unitary magnetic quiver theories can be read off from the web diagrams in Figure \ref{fig:KN1 O5tilde+O5+}, and they are given in Figure \ref{O5t+O5+KN1magneticquivers}. The corresponding Coulomb branch Hilbert series is presented in Table  \ref{O5t+O5+KN1CoulombHS}.
		\begin{figure}[!htb]
\centering
   \begin{subfigure}{0.49\linewidth} \centering
    \begin{scriptsize}
      \begin{tikzpicture}
      \node[label=below:{1}][u](1){};
      \node[label=below:{$2N-2$}][u](2n-2)[right of=1]{};
      \node[label=left:{$2N-2$}][sp](sp2n-2)[above of=2n-2]{};
      \node[label=above:{4}][sof](so4f)[above of=sp2n-2]{};
      \node[label=below:{1}][u](u1)[right of=2n-2]{};

      \node[label=above:{3}][uf](uf)[above of=u1]{};
      \draw[dotted](1)--(2n-2);
      \draw(2n-2)--(sp2n-2);
      \draw(sp2n-2)--(so4f);
      \draw(2n-2)--(u1);
      \draw(u1)--(uf);  
      \end{tikzpicture}
      \end{scriptsize}
     \caption{Orthosymplectic quiver}\label{O5t+O5+KN1OSpquiver}
   \end{subfigure}
   \begin{subfigure}{0.49\linewidth} \centering
     \begin{scriptsize}
       \begin{tikzpicture}
       \node[label=below:{2}][u](2){};
       \node[label=below:{4}][u](4)[right of=2]{};
       \node[label=below:{$2N-4$}][u](2n-4)[right of=4]{};
       \node[label=below:{$2N-2$}][u](2n-2)[right of=2n-4]{};
       \node[label=below:{$2N-2$}][u](2n-2')[right of=2n-2]{};
       \node[label=below:{$2N-4$}][u](2n-4')[right of=2n-2']{};
       \node[label=below:{4}][u](4')[right of=2n-4']{};
       \node[label=below:{2}][u](2')[right of=4']{};
       \node[label=left:{1}][u](u1)[above of=2n-2]{};
       \node[label=right:{1}][u](u1')[above of=2n-2']{};
       \draw(2)--(4);
       \draw[dotted](4)--(2n-4);
       \draw(2n-4)--(2n-2);
       \draw(2n-2)--(2n-2');
       \draw(2')--(4');
       \draw[dotted](4')--(2n-4');
       \draw(2n-4')--(2n-2');
       \draw[double distance=2pt](2n-2)--(u1);
       \draw[double distance=2pt](2n-2')--(u1');
       \draw[double distance =3pt](u1)--(u1');
       \draw(u1)--(u1');
       \end{tikzpicture}
       \end{scriptsize}
     \caption{Unitary quiver}\label{O5t+O5+KN1unitaryquiver}
   \end{subfigure}
\caption{Magnetic quivers for the $\kappa_N^1$ theory} \label{O5t+O5+KN1magneticquivers}
\end{figure}
		\begin{table}[!htb]
\centering
\begin{tabular}{|c|C{4.9cm}|C{4.9cm}|C{2.8cm}|} \hline
\rowcolor{Grayy}
   & Unitary magnetic quiver & \multicolumn{2}{c|}{Orthosymplectic magnetic quiver} \\ \cline{2-4}
	\rowcolor{Grayy}
  \multirow{-2}{*}{$K_N^1$}
 & HS($t$)  & HS($t;\vec{m} \in \mathbb{Z}$) & HS($t;\vec{m} \in \mathbb{Z}+\tfrac{1}{2}$)  \\ \hline
	$K_2^1$ & \footnotesize{$1 + 9 t + 53 t^2 + 6 t^{5/2} + 227 t^3 + 60 t^{7/2} + 792 t^4 + 
 318 t^{9/2} + 2358 t^5$}  & \footnotesize{$1 + 9 t + 53 t^2 + 6 t^{5/2} + 227 t^3 + 60 t^{7/2} + 792 t^4 + 
 318 t^{9/2} + 2358 t^5$} & \footnotesize{not required} \\ \hline
	$K_3^1$ & \footnotesize{$1 + 25 t + 349 t^2 + 3499 t^3+\cdots$}  & \footnotesize{$1 + 25 t + 349 t^2 + 3499 t^3+\cdots$} & \footnotesize{not required} \\ \hline
\end{tabular}
\caption[Coulomb branch HS for magnetic quivers of $\kappa_N^1$ theory.]{Coulomb branch Hilbert series for the unitary and orthosymplectic magnetic quivers presented in Figure \ref{O5t+O5+KN1magneticquivers} for the $\kappa_N^1$ theory.}
\label{O5t+O5+KN1CoulombHS}
\end{table}

%% file: KN1U.tex
\begin{scriptsize}
       \begin{tikzpicture}
       \node[label=below:{2}][u](2){};
       \node[label=above:{4}][u](4) [right of=2]{};
       \node[label=below:{$2N-4$}][u](2k-4) [right of=4]{};
       \node[label=above:{$2N-2$}][u](2k-2) [right of=2k-4]{};
       \node (false) [below of=2k-2]{};
       \node (false') [right of=false]{};
       \node[label=below:{$2N-2$}][u](2k-2') [below of=false]{};
       \node[label=above:{$2N-4$}][u](2k-4') [left of=2k-2']{};
       \node[label=below:{$4$}][u](4') [left of=2k-4']{};
       \node[label=above:{$2$}][u](2') [left of=4']{};
       \node[label=right:{2}][u](22)[right of =2k-2]{};
       \node[label=right:{1}][u](11)[above of=22]{};
       \node[label=right:{2}][u](22')[right of =2k-2']{};
       \node[label=right:{1}][u](11')[below of=22']{};
       \node[label=above:{3}][u](3)[right of=false']{};
       \node[label=above:{2}][u](222)[right of=3]{};
       \node[label=above:{1}][u](111)[right of=222]{};
       \draw (2)--(4);
       \draw[dotted](4)--(2k-4);
       \draw(2k-4)--(2k-2);
       \draw(2k-2)--(2k-2');
       \draw(2k-2')--(2k-4');
       \draw[dotted](2k-4')--(4');
       \draw (4')--(2');
       \draw(2k-2)--(22);
       \draw(22)--(11);
       \draw(2k-2')--(22');
       \draw(22')--(11');
       \draw(22)--(3);
       \draw(22')--(3);
       \draw(3)--(222);
       \draw(222)--(111);
       \end{tikzpicture}
    \end{scriptsize}
		

%% file: KN1OSp.tex
\begin{scriptsize}
\begin{tikzpicture}
\node[label=above:{$2$}][so](so2n-2)[right of=sp2]{};
\node[label=below:{$2$}][sp](sp2nn)[right of=so2n-2]{};
\draw (sp2nn)--(so2n-2);
\node[label=above:{$4$}][so](so2n+2)[right of=sp2nn]{};
\draw (sp2n-2)--(so2n+2);
\node[label=below:{$4$}][sp](sp2n-2)[right of=so2n+2]{};
\draw (sp2nn)--(so2n+2);
\draw (sp2n-2)--(so2n+2);
\node[label=above left:{$1$}][u](so2)[above of=sp2n-2]{};
\draw (sp2n-2)--(so2);
\node[label=above right:{$2N-2$}][u](u2k-2)[above right of =so2]{};
\draw (so2)--(u2k-2);
\node[label=above right:{$2N-2$}][sp](sp2k-2)[below right of=u2k-2]{};
\draw (sp2k-2)--(u2k-2);
\node[label=below:{$4$}][so](so2n+2)[below of=sp2k-2]{};
\draw (sp2k-2)--(so2n+2);
\draw (so2n+2)--(sp2n-2);
\node[label=above:{1}][u](u1)[above of=u2k-2]{};
\draw[dotted] (u1)--(u2k-2);
\node[label=above:{$2$}][sp](sp2n)[right of=so2n+2]{};
\draw(sp2n)--(so2n+2);
\node[label=below:{$2$}][so](so2n)[right of=sp2n]{};
\draw (so2n)--(sp2n);
\end{tikzpicture}
\end{scriptsize}

%% file: KN2U.tex
\begin{scriptsize}
    \begin{tikzpicture}
    \node[label=below:{2}][u](2){};
    \node[label=below:{4}][u](4)[right of=2]{};
    \node[label=below:{$2N-4$}][u](2n-4)[right of=4]{};
    \node[label=below:{$2N-2$}][u](2n-2)[right of=2n-4]{};
    \node[label=below:{$2N-2$}][u](2n-2')[right of=2n-2]{};
    \node[label=below:{$2N-4$}][u](2n-4')[right of=2n-2']{};
    \node[label=below:{$4$}][u](4')[right of=2n-4']{};
    \node[label=below:{$2$}][u](2')[right of=4']{};
    \node[label=above:{1}][u](red)[above of=2n-4]{};
    \node[label=above:{1}][u](green)[above of=2n-2']{};
    \node[label=above:{1}][u](blue)[above of=2n-2]{};
    \node[label=left:{2}][u](u2)[above of=blue]{};
    \node[label=left:{2}][u](u2')[above of=u2]{};
    \node[label=left:{1}][u](u1)[above of=u2']{};
    \draw(2)--(4);
    \draw[dotted](4)--(2n-4);
    \draw(2n-4)--(2n-2);
    \draw(2n-2)--(2n-2');
    \draw(2')--(4');
    \draw[dotted](4')--(2n-4');
    \draw(2n-4')--(2n-2');
    \draw(green)--(2n-2');
    \draw(red)--(2n-2);
    \draw(2n-2)--(blue);
    \draw(2n-2')--(blue);
    \draw(green)--(u2);
    \draw(red)--(u2);
    \draw(u2)--(u2');
    \draw(u1)--(u2');
    \draw (blue)to[out=45,in=-45](u2');
    \end{tikzpicture}
    \end{scriptsize}

%% file: KN2OSp.tex
\begin{scriptsize}
    \begin{tikzpicture}
    \node[label=left:{1}][u](1){};
    \node[label=left:{$2N-2$}][u](2n-2)[below of=1]{};
    \node[label=left:{$2N-2$}][sp](sp2n-2)[below of=2n-2]{};
    \node[label=below:{2}][so](so2)[below of=sp2n-2]{};
    \node[label=below:{2}][sp](sp2)[right of=so2]{};
    \node[label=below:{4}][so](so4)[right of=sp2]{};
    \node[label=below:{2}][sp](sp2')[right of=so4]{};
    \node[label=below:{2}][so](so2')[right of=sp2']{};
     \node[label=right:{1}][u](u1)[above of=so4]{};
    \draw[dotted](1)--(2n-2);
    \draw(2n-2)--(sp2n-2);
    \draw(sp2n-2)--(so2);
    \draw(sp2n-2)--(u1);
    \draw(u1)--(2n-2);
    \draw(u1)--(so4);
    \draw(so2)--(sp2);
    \draw(sp2)--(so4);
    \draw(so4)--(sp2');
    \draw(sp2')--(so2');
    \end{tikzpicture}
    \end{scriptsize}

%% file: KN3U.tex
\begin{scriptsize}
    \begin{tikzpicture}
    \node[label=below:{2}][u](2){};
    \node[label=below:{4}][u](4)[right of=2]{};
    \node[label=below:{$2N-4$}][u](2n-4)[right of=4]{};
    \node[label=below:{$2N-2$}][u](2n-2)[right of=2n-4]{};
    \node[label=below:{$2N-2$}][u](2n-2')[right of=2n-2]{};
    \node[label=below:{$2N-4$}][u](2n-4')[right of=2n-2']{};
    \node[label=below:{$4$}][u](4')[right of=2n-4']{};
    \node[label=below:{$2$}][u](2')[right of=4']{};
    \node[label=left:{1}][u](red)[above of=2n-2]{};
    \node[label=right:{1}][u](blue)[above of=2n-2']{};
    \node[label=above:{1}][u](u1)[above of=red]{};
    \node[label=right:{1}][u](u1')[above right of=blue]{};
    \node[label=above:{1}][u](u11)[right of=u1]{};
    \draw(2)--(4);
    \draw[dotted](4)--(2n-4);
    \draw(2n-4)--(2n-2);
    \draw(2n-2)--(2n-2');
    \draw(2')--(4');
    \draw[dotted](4')--(2n-4');
    \draw(2n-4')--(2n-2');
    \draw(2n-2)--(red);
    \draw[double distance=2pt](2n-2')--(blue);
    \draw(red)--(blue);
    \draw(red)--(u1);
    \draw(blue)--(u1');
    \draw(u1)--(u11);
    \draw(u1')--(u11);
    \draw(blue)--(2n-2);
    \end{tikzpicture}
    \end{scriptsize}

%% file: KN3OSp.tex
\begin{scriptsize}
    \begin{tikzpicture}
    \node[label=left:{1}][u](1){};
    \node[label=left:{$2N-2$}][u](2n-2)[below of=1]{};
    \node[label=left:{$2N-2$}][sp](sp2n-2)[below of=2n-2]{};
    \node[label=below:{1}][u](u1)[right of=sp2n-2]{};
    \node[label=below:{2}][sp](sp2)[right of=u1]{};
    \node[label=below:{2}][so](so2)[right of=sp2]{};
    \node[label=above:{2}][so](so2')[above of=sp2]{};
    \node[label=above:{1}][uf](uf)[above of=u1]{};
    \draw[dotted](1)--(2n-2);
    \draw(2n-2)--(sp2n-2);
    \draw[double distance=2pt](sp2n-2)--(u1);
    \draw(u1)--(2n-2);
    \path [draw,snake it](u1)--(uf);
    \draw(u1)--(sp2);
    \draw(sp2)--(so2);
    \draw(sp2)--(so2');
    \end{tikzpicture}
    \end{scriptsize}

%% file: YN11PhaseIU.tex
\begin{scriptsize}
\begin{tikzpicture}
         \node[label=below:{2}][u](2) at (0,0) {};
         \node[label=below:{4}][u](4) [right of=2] {};
         \node[label=below:{$2N-2$}][u](2k-2) [right of=4] {};
         \node[label=below:{$2N$}][u](2k) [right of=2k-2] {};
         \node[label=below:{$2N-2$}][u](2k-2') [right of=2k] {};
         \node[label=below:{$4$}][u](4') [right of=2k-2'] {};
         \node[label=below:{$2$}][u](2') [right of=4'] {};
         \draw (2')--(4');
         \draw (2k-2')--(2k);
         \draw (2k-2)--(2k);
         \draw[dotted] (4)--(2k-2);
         \draw[dotted] (4')--(2k-2');
         \draw (2)--(4);
         \node[label=above right:{2}][u] (22) [above left of=2k]{};
         \node[label=above left:{2}][u] (22') [above right of=2k]{};
         \draw (22)--(22');
         \draw (22)--(2k);
         \draw (2k)--(22');
         \node[label=left:{1}][u](11)[left of=22]{};
         \node[label=left:{1}][u](111)[above of=22]{};
         \draw (11)--(22);
         \draw (111)--(22);
         \node[label=right:{1}][u](11')[right of=22']{};
         \node[label=right:{1}][u](111')[above of=22']{};
         \draw (11')--(22');
         \draw (111')--(22');
         \end{tikzpicture}
\end{scriptsize}				

%% file: YN11PhaseIOSp.tex
\begin{scriptsize}
\begin{tikzpicture}
         \node[label=below:{2}][so](so2) at (0,0) {};
         \node[label=below:{2}][sp](sp2) [right of=so2] {};
          \node[label=below:{4}][so](so4) [right of=sp2] {};
         \node[label=below:{2}][sp](sp2') [right of=so4] {};
         \node[label=below:{2}][so](so2') [right of=sp2'] {};
         \draw (so2)--(sp2);
         \draw (so4)--(sp2);
         \draw (so2')--(sp2');
         \draw (so4)--(sp2');
         \node[label=right:{$2N$}][sp](sp2k) [above right of=so4] {};
         \node[label=left:{$1$}][u](u1) [above left of=so4] {};
         \node[label=right:{$2N-1$}][u](u2k-1) [above of=sp2k] {};
         \node[label=right:{$1$}][u](u1') [above of=u2k-1] {};
         \draw[dotted] (u1')--(u2k-1);
         \draw (u2k-1)--(sp2k);
         \draw (u1)--(so4);
         \draw (u1)--(sp2k);
         \draw (sp2k)--(so4);
         \end{tikzpicture}
\end{scriptsize}				

%% file: YN11PhaseIIU.tex
\begin{scriptsize}
\begin{tikzpicture}
         \node[label=below:{2}][u](2) at (0,0) {};
         \node[label=below:{4}][u](4) [right of=2] {};
         \node (false) [right of=2k-2 ]{};
         \node (false') [right of=false ]{};
         \node (false'') [right of=false' ]{};
         \node[label=below:{$2N-2$}][u](2k-2) [right of=4] {};
         \node[label=below:{$2N-1$}][u](2k) [below of=false'] {};
         \node[label=below:{$2N-2$}][u](2k-2') [right of=false''] {};
         \node[label=below:{$4$}][u](4') [right of=2k-2'] {};
         \node[label=below:{$2$}][u](2') [right of=4'] {};
         \draw (2')--(4');
         \draw (2k-2')--(2k);
         \draw (2k-2)--(2k);
         \draw[dotted] (4)--(2k-2);
         \draw[dotted] (4')--(2k-2');
         \draw (2)--(4);
         \node[label=right:{1}][u](1111')[above of =2k-2']{};
         \node[label=left:{1}][u](1111)[above of =2k-2]{};
         \draw[double distance=1.25 pt] (1111)to[out=0,in=180](1111');
         \draw (1111)--(2k-2);
         \draw (1111')--(2k-2');
         \node[label=above:{1}][u] (22) [above right of=2k]{};
         \node[label=above:{1}][u] (22') [above left of=2k]{};
         \draw (22)--(22');
         \draw (22)--(2k);
         \draw (2k)--(22');
         \node[label=right:{1}][u](11)[above left of=false']{};
         \draw (11)--(1111);
         \node[label=left:{1}][u](111)[above right of=22]{};
         \draw (111)--(1111');
         \draw (11)--(22);
         \draw (111)--(22);
         \node[label=left:{1}][u](11')[above right of=false']{};
         \node[label=right:{1}][u](111')[above left of=22']{};
         \draw (11')--(1111');
         \draw (111')--(1111);
         \draw (11')--(22');
         \draw (111')--(22');
         \end{tikzpicture}
				\end{scriptsize}				

%% file: YN11PhaseIIOSp.tex
\begin{scriptsize}
\begin{tikzpicture}
         \node[label=below:{2}][so](so2) at (0,0) {};
         \node[label=above:{2}][sp](sp2) [right of=so2] {};
          \node[label=below:{2}][so](so22) [right of=sp2] {};
         \node[label=above:{2}][sp](sp2') [right of=so22] {};
         \node[label=below:{2}][so](so2') [right of=sp2'] {};
         \node[label=right:{$2N-2$}][sp](sp2k-2) [above of=so22] {};
         \node[label=above right:{$2N-1$}][u](u2k-1) [above of=sp2k-2] {};
         \node[label=left:{1}][u](u1') [above of=u2k-1]{};
         \node[label=below:{$1$}][u](u1) [below of=sp2] {};
           \node[label=below:{$1$}][u](u11) [below of=sp2'] {};
           \draw[dashed, double distance=2 pt] (u1)--(u11);
           \draw(u1)to[out =-30, in=-150](u11);
           \draw[dotted](u1')--(u2k-1);
           \draw (u2k-1)--(sp2k-2);
           \draw (sp2)--(u1);
           \draw (sp2')--(u11);
           \draw (sp2k-2)--(so22);
         \draw (u1) to[out=180, in=180,controls=+(180:2.5) and +(180:2)]
         (u2k-1);
         \draw (u11) to[out=0, in=0,controls=+(0:2.5) and +(0:2)]
         (u2k-1);
         \draw (so2)--(sp2);
         \draw (so22)--(sp2);
         \draw (so2')--(sp2');
         \draw (so22)--(sp2');
         \end{tikzpicture}
\end{scriptsize}				

%% file: YN11PhaseIIIU.tex
\begin{scriptsize}
\begin{tikzpicture}
         \node[label=below:{2}][u](2) at (0,0) {};
         \node[label=below:{4}][u](4) [right of=2] {};
         \node[label=below:{$2N-2$}][u](2k-2) [right of=4] {};
         \node[label=above:{$2N-2$}][u](2k) [right of=2k-2] {};
         \node[label=below:{$2N-2$}][u](2k-2') [right of=2k] {};
         \node[label=below:{$4$}][u](4') [right of=2k-2'] {};
         \node[label=below:{$2$}][u](2') [right of=4'] {};
         \draw (2')--(4');
         \draw (2k-2')--(2k);
         \draw (2k-2)--(2k);
         \draw[dotted] (4)--(2k-2);
         \draw[dotted] (4')--(2k-2');
         \draw (2)--(4);
         \node[label=above right:{2}][u] (22) [above of=2k-2]{};
         \node[label=above left:{2}][u] (22') [above of=2k-2']{};
         \draw[double distance=2 pt] (22)--(22');
         \draw (22)--(2k-2);
         \draw (2k-2')--(22');
         \node[label=left:{1}][u](11)[left of=22]{};
         \node[label=left:{1}][u](111)[above of=22]{};
         \draw (11)--(22);
         \draw (111)--(22);
         \node[label=right:{1}][u](11')[right of=22']{};
         \node[label=right:{1}][u](111')[above of=22']{};
         \draw (11')--(22');
         \draw (111')--(22');
         \end{tikzpicture}
						\end{scriptsize}				

%% file: YN11PhaseIIIOSp.tex
\begin{scriptsize}
\begin{tikzpicture}
            \node[label=right:{1}][u](1)at(0,0){};
            \node[label=right:{$2N-2$}][u](2k-2)[below of=1]{};
            \node[label=right:{$2N-2$}][u](2k-2')[below of=2k-2]{};
            \node[label=right:{$2N-2$}][sp](sp2k-2)[below of=2k-2']{};
            \node[label=below:{4}][so](so4)[below of=sp2k-2]{};
            \node[label=below:{2}][sp](sp2)[left of=so4]{};
            \node[label=below:{2}][sp](sp2')[right of=so4]{};
            \node[label=below:{2}][so](so2)[left of=sp2]{};
            \node[label=below:{2}][so](so2')[right of=sp2']{};
            \node[label=left:{1}][u](u11)[above left of=sp2k-2]{};
            \draw[dotted](1)--(2k-2);
            \draw (2k-2)--(2k-2');
            \draw (2k-2')--(sp2k-2);
            \draw (sp2k-2)--(so4);
            \draw (so2)--(sp2);
            \draw (sp2)--(so4);
            \draw (so2')--(sp2');
            \draw (sp2')--(so4);
            \draw (u11)--(2k-2);
            \draw[double distance=2pt] (u11)--(so4);
            \end{tikzpicture}
						\end{scriptsize}				

%% file: YN2121UnitaryWeb.tex
\begin{scriptsize}
    \begin{tikzpicture}[scale=.7]
    \draw[thick,](2.25,10.25)--(-2.25,5.75);
\node[thick][7brane] at (-1.5,6.5) {};
\node[thick][7brane] at (1.5,9.5) {};
\node[thick][7brane] at (-2.25,5.75) {};
\node[thick][7brane] at (2.25,10.25) {};
\node at (2.5,10.5) {.};
\node at (2.6,10.6) {.};
\node at (2.7,10.7) {.};
\node[thick][7brane] at (2.95,10.95) {};
\node[thick][7brane] at (3.7,11.7) {};
\node[thick][7brane] at (4.45,12.45) {};
\draw[thick](2.95,10.95)--(4.45,12.45);
\node at (3.8,12.25) {$2$};
\node at (3,11.5) {$4$};
\node at (1.3,10.2) {$2N-2$};

\node at (-2.5,5.5) {.};
\node at (-2.6,5.4) {.};
\node at (-2.7,5.3) {.};
\node[thick][7brane] at (-2.95,5.05) {};
\node[thick][7brane] at (-3.7,4.3) {};
\node[thick][7brane] at (-4.45,3.55) {};
\draw[thick](-2.95,5.05)--(-4.45,3.55);
\node at (-3.8,3.75) {$2$};
\node at (-3,4.5) {$4$};
\node at (-1.3,5.8) {$2N-2$};
\node at (-.8,6.8) {$2N$};

\draw[thick](-2,8)--(2,8);
\node at (1.5,7.75) {$2$};
\draw[thick] (2,8)--(4,8);
\draw[thick] (-2,8)--(-4,8);
\node[thick][7brane] at (-2,8) {};
\node[thick][7brane] at (2,8) {};
\node[thick][7brane] at (-4,8) {};
\node[thick][7brane] at (4,8) {};

\draw[thick](-2,7)--(2,9);
\node[thick][7brane] at (-2,7) {};
\node[thick][7brane] at (2,9) {};

\draw[thick](0,6)--(0,10);
\node[thick][7brane] at (0,6) {};
\node[thick][7brane] at (0,10) {};

\draw[thick,](-5.5,1.5)--(-8.5,-1.5);
\node[thick][7brane] at (-8.5,-1.5) {};
\node[thick][7brane] at (-5.5,1.5) {};

\draw[thick,purple](-9,0)--(-5,0);

\node[thick][7brane] at (-9,0) {};
\node[thick][7brane] at (-5,0) {};

\draw[thick,green](-9,-1)--(-5,1);
\node[thick][7brane] at (-9,-1) {};
\node[thick][7brane] at (-5,1) {};

\draw[thick,blue](-7,-2)--(-7,2);
\node[thick][7brane] at (-7,-2) {};
\node[thick][7brane] at (-7,2) {};

\node at (-7,-3.5) {(I)};

\draw[thick,red](-1.5,-1.5)--(0,0);
\draw[thick](-1.3,-1.5)--(1.7,1.5);
\draw[thick,red](0,0)--(2,0);
\draw[thick,red](0,0)--(0,2);
\node[7brane] at (-1.4,-1.5) {};
\node[7brane] at (2,0.075) {};
\draw[thick,green] (-2,-1)--(2,1);
\node[7brane] at (-2,-1) {};
\node[7brane] at (2,1) {};
\draw[thick,blue](-.15,-2)--(-.15,0);
\draw[thick,blue](-2.15,0)--(-.15,0);
\draw[thick,blue](1.55,1.7)--(-.15,0);
\draw[thick,pink](-2.15,0.15)--(1.85,0.15);
\node[7brane] at (1.6,1.6){};
\node[7brane] at (-2.15,0.075){};
\node[7brane] at (-.15,-2) {};
\node[7brane] at (0,2) {};
\node at (0,-3.5) {(II)};
\draw[thick,green](5,-1)--(7,0);
\draw[thick,green](7,0)--(10,0);
\draw[thick,green](7,0.1)--(9,0.1);
\draw[thick,green](7,0)--(7,2);
\node[7brane] at (10,0){};
\node[7brane] at (5,-1){};
\node[7brane] at (9,0){};
\node[7brane] at (7,2){};

\draw[thick,orange](9,1)--(7,0);
\draw[thick,orange](7,0)--(4,0);
\draw[thick,orange](7,0.1)--(5,0.1);
\draw[thick,orange](7,0)--(7,-2);
\draw[thick](5.5,-1.5)--(8.5,1.5);
\node[7brane] at (7,-2){};
\node[7brane] at (9,1){};
\node[7brane] at (4,0){};
\node[7brane] at (5,0){};
\node[7brane] at (5.5,-1.5){};
\node[7brane] at (8.5,1.5){};

\node at (7,-3.5) {(III)};

\draw[thick,green](-7,-7)--(-5,-7);
\draw[thick,green](-5.5,-5.5)--(-7,-7);
\draw[thick,green](-9,-8)--(-7,-7);

\draw[thick,blue](-7,-7)--(-9.2,-7);
\draw[thick,blue](-7,-7)--(-5,-6);
\draw[thick,blue](-7,-7)--(-8.5,-8.5);

\draw[thick,red](-9.2,-6.8)--(-7.2,-6.8);
\draw[thick,red](-7.2,-8.8)--(-7.2,-6.8);
\draw[thick,red](-5.7,-5.3)--(-7.2,-6.8);

\draw[thick,cyan](-7,-6.9)--(-5,-6.9);
\draw[thick,cyan](-7,-6.9)--(-7,-4.9);
\draw[thick,cyan](-7,-6.9)--(-8.5,-8.4);

\draw[thick](-8.6,-8.4)--(-5.6,-5.4);
\node[7brane]at(-5,-6.9){};
\node[7brane]at(-9,-8){};
\node[7brane]at(-7,-5){};
\node[7brane]at(-5,-6){};
\node[7brane]at(-8.5,-8.4){};
\node[7brane]at(-7.2,-8.8){};
\node[7brane]at(-5.575,-5.375){};
\node[7brane]at(-9.2,-6.9){};
\node at (-7,-10) {(IV)};

\draw[thick,green](0,-7)--(1.5,-7);
\draw[thick,green](0,-6.8)--(2.5,-6.8);
\draw[thick,green](0,-5)--(0,-7);
\draw[thick,green](-2,-8)--(0,-7);
\node[7brane]at(2.5,-6.8){};
\node[7brane]at(1.5,-7){};
\node[7brane]at(-2,-8){};
\node[7brane]at(0,-5){};

\draw[thick,blue](0,-7)--(-2.2,-7);
\draw[thick,blue](0,-7)--(2,-6);
\draw[thick,blue](0,-7)--(-1.5,-8.5);
\node[7brane]at(2,-6){};
\node[7brane]at(-1.5,-8.4){};
\draw[thick,red](-2.2,-6.8)--(-0.2,-6.8);
\draw[thick,red](-0.2,-8.8)--(-0.2,-6.8);
\node[7brane]at(-.2,-8.8){};
\draw[thick,red](1.3,-5.3)--(-0.2,-6.8);
\node[7brane]at(1.4,-5.3){};
\node[7brane]at(-2.2,-6.9){};
\draw[thick](1.4,-5.4)--(-1.6,-8.4);

\node at (0,-10) {(V)};

\draw[thick,green](7,-7)--(9,-7);
\draw[thick,green](8.5,-5.5)--(7,-7);
\draw[thick,green](5,-8)--(7,-7);

\draw[thick,blue](7,-7)--(4.8,-7);
\draw[thick,blue](7,-7)--(9,-6);
\draw[thick,blue](7,-7)--(5.5,-8.5);

\draw[thick,pink](4.8,-6.8)--(9,-6.8);

\draw[thick,red] (7,-5)--(7,-9);
\draw[thick](5.4,-8.4)--(8.4,-5.4);
\node[7brane]at(9,-6.9){};
\node[7brane]at(5,-8){};
\node[7brane]at(7,-5){};
\node[7brane]at(9,-6){};
\node[7brane]at(5.5,-8.4){};
\node[7brane]at(7,-9){};
\node[7brane]at(8.4,-5.5){};
\node[7brane]at(4.8,-6.9){};
\node at (7,-10) {(VI)};
\draw[thick,](-7,-14)--(-8.5,-15.5);
\draw[thick](-7,-12)--(-7,-14);
\draw[thick](-7,-14)--(-6.6,-14);
\draw[thick](-7.6,-15)--(-6.6,-14);
\draw[thick](-4.6,-13)--(-6.6,-14);

\draw[thick,green](-6.75,-13.75)--(-6.35,-13.75);
\draw[thick,green](-4.85,-12.25)--(-6.35,-13.75);
\draw[thick,green](-6.75,-13.75)--(-5.75,-12.75);
\draw[thick,green](-6.35,-15.75)--(-6.35,-13.75);
\draw[thick,green](-6.75,-13.75)--(-8.75,-14.75);

\draw[thick,blue](-9,-13.9)--(-5,-13.9);
\draw[thick,red](-7.75,-15)--(-5.55,-12.75);
\node[7brane]at(-7.65,-15){};
\node[7brane]at(-9,-13.9){};
\node[7brane]at(-5,-13.9){};
\node[7brane]at(-8.5,-15.5){};
\node[7brane]at(-7,-12){};
\node[7brane]at(-4.6,-13){};
\node at (-7,-17.5) {(VII)};
\node[7brane]at(-4.85,-12.25){};
\node[7brane]at(-5.65,-12.75){};
\node[7brane]at(-6.35,-15.75){};
\node[7brane]at(-8.75,-14.75){};
\draw[thick,](0,-14)--(-1.5,-15.5);
\draw[thick](0,-12)--(0,-14);
\draw[thick](0,-14)--(.4,-14);
\draw[thick](-.6,-15)--(.4,-14);
\draw[thick](2.4,-13)--(.4,-14);

\draw[thick,blue](-2,-13.9)--(2,-13.9);
\draw[thick,green](0.2,-13.7)--(2,-13.7);
\draw[thick,green](0.2,-13.7)--(-1.8,-14.7);
\draw[thick,green](0.2,-13.7)--(1.2,-12.7);
\draw[thick,red](-.75,-15)--(1.5,-12.75);
\draw[thick,cyan](.35,-13.75)--(1.35,-12.75);
\draw[thick,cyan](.35,-13.75)--(.35,-15.75);
\draw[thick,cyan](.35,-13.75)--(-1.8,-13.75);
\node[7brane]at(-.65,-15){};
\node[7brane]at(-1.9,-13.8){};
\node[7brane]at(2,-13.8){};
\node[7brane]at(-1.5,-15.5){};
\node[7brane]at(0,-12){};
\node[7brane]at(2.4,-13){};
\node at (0,-17.5) {(VIII)};
\node[7brane]at(1.35,-12.75){};
\node[7brane]at(.35,-15.75){};
\node[7brane]at(-1.75,-14.75){};
    \end{tikzpicture}\end{scriptsize}

%% file: YN2121PhaseIU.tex
\begin{scriptsize}\begin{tikzpicture}
   \node[label=below:{2}][u](2) at (0,0){};
   \node[label=below:{4}][u](4) [right of=2]{};
   \node[label=below:{$2N-2$}][u](2k-2) [right of=4]{};
   \node[label=below:{$2N$}][u](2k) [right of=2k-2]{};
   \node[label=below:{$2N-2$}][u](2k-2') [right of=2k]{};
   \node[label=below:{$4$}][u](4') [right of=2k-2']{};
   \node[label=below:{$2$}][u](2') [right of=4']{};
   \draw (2)--(4);
   \draw[dotted] (2k-2)--(4);
   \draw (2k-2)--(2k);
   \draw (2k-2')--(2k);
   \draw[dotted] (2k-2')--(4');
   \draw (2')--(4');
\node[label=above:{2}][u](22)[above of=2k] {};
\node[label=below:{1}][u](11)[above left of=22] {};
\node[label=below:{1}][u](11')[above right of=22] {};
\node[label=left:{1}][u](111)[above of=4] {};
\node[label=right:{1}][u](111')[above of=4'] {};
\draw (22)--(11);
\draw (22)--(11');
\draw[double distance=2pt] (111) to[out=60, in=120] (111');
\draw (2k)--(22);
\draw (2k)--(111);
\draw (2k)--(111');
\draw (111)--(22);
\draw (111')--(22);
   \end{tikzpicture}\end{scriptsize}

%% file: YN2121PhaseIOSp.tex
\begin{scriptsize}
      \begin{tikzpicture}
      \node[label=below:{2}][so](so2) at (0,0){};
      \node[label=below:{$2N$}][sp](sp2k) [left of=so2]{};
      \node[label=below:{$2N-1$}][u](u2k-1)[left of=sp2k]{};
      \node[label=below:{$2$}][u](u2)[left of=u2k-1]{};
      \node[label=below:{$1$}][u](u1)[left of=u2]{};
      \node[label=below left:{$2$}][u](u2')[above of=sp2k]{};
      \node[label=right:{$1$}][u](u1')[right of=u2']{};
      \node[label=left:{$1$}][u](u11')[left of=u2']{};
      \draw[double distance=1pt] (u2')to[out=135,in=45,loop,looseness=6](u2');
      \draw (u1)--(u2);
      \draw[dotted](u2)--(u2k-1);
      \draw (u2k-1)--(sp2k);
      \draw (sp2k)--(so2);
      \draw (so2)--(u2');
      \draw (u2')--(sp2k);
      \draw (u1')--(u2');
      \draw (u11')--(u2');
      \end{tikzpicture}
   \end{scriptsize}

%% file: YN2121PhaseIIU.tex
\begin{scriptsize}\begin{tikzpicture}
   \node[label=below:{2}][u](2) at (0,0){};
   \node[label=below:{4}][u](4) [right of=2]{};
   \node[label=below:{$2N-2$}][u](2k-2) [right of=4]{};
   \node[label=below:{$2N-1$}][u](2k) [right of=2k-2]{};
   \node[label=below:{$2N-2$}][u](2k-2') [right of=2k]{};
   \node[label=below:{$4$}][u](4') [right of=2k-2']{};
   \node[label=below:{$2$}][u](2') [right of=4']{};
   \node[label=right:{1}][u](1111) [above of=2k]{};
   \node[label=above:{1}][u](11) [above of=1111]{};
   \node (false) [right of=11]{};
   \node (false') [left of=11]{};
   \node[label=left:{1}][u](1) [left of=false']{};
   \node[label=right:{1}][u](1') [right of=false]{};
   \draw[double distance=2pt](1)--(11);
   \draw[double distance=2pt](1')--(11);
   \draw[double distance=2pt](1)to[out=30, in=150](1');
   \node[label=above:{1}][u](111) [below right of=1]{};
   \node[label=above:{1}][u](111') [below left of=1']{};
   \draw (1111)--(111);
   \draw (1111)--(111');
   \draw (1)--(111);
   \draw (1')--(111');
   \draw (1111)--(2k);
   \draw (1111)--(11);
   \draw (2k)to[out=120,in=-120](11);
   \draw (2k-2)--(1);
   \draw (2k-2')--(1');
   \draw (2)--(4);
   \draw[dotted] (2k-2)--(4);
   \draw (2k-2)--(2k);
   \draw (2k-2')--(2k);
   \draw[dotted] (2k-2')--(4');
   \draw (2')--(4');\end{tikzpicture}\end{scriptsize}

%% file: YN2121PhaseIIOSp.tex
\begin{scriptsize}\begin{tikzpicture}
   \node[label=below:{1}][u](1) at (0,0){};
   \node[label=below:{2}][u](2) [right of=1]{};
   \node[label=below:{$2N-1$}][u](2k-1)[right of=2]{};
   \node[label=below:{$2N-2$}][sp](sp2k-2)[right of=2k-1]{};
   \node[label=right:{1}][u](u1)[right of=sp2k-2]{};
   \node[label=right:{1}][u](111)[above of=u1]{};
   \node[label=right:{1}][u](111')[below of=u1]{};
    \node[label=below:{1}][u](11)[below of=111']{};
   \node[label=above:{1}][u](11')[above of=111]{};
   \draw (1)--(2);
   \draw[dotted](2)--(2k-1);
   \draw (2k-1)--(sp2k-2);
   \draw (sp2k-2)--(u1);
   \draw (2k-1)--(11);
   \draw (2k-1)--(11');
   \draw[dashed,double distance=2pt] (11)to[out=45,in=-45](11');
   \draw (11)to[out=30,in=-30](11');
   \draw[dashed,double distance=2pt](11)to[out=120,in=-120](u1);
   \draw[double distance=2pt](u1)to[out=120,in=-120](11');
   \draw (11)--(111');
   \draw (11')--(111);
   \draw[dashed] (111)--(u1);
   \draw (111')--(u1);
   
  \end{tikzpicture}\end{scriptsize}

%% file: YN2121PhaseIIIU.tex
\begin{scriptsize}\begin{tikzpicture}
   \node[label=below:{2}][u](2) at (0,0){};
   \node[label=below:{4}][u](4) [right of=2]{};
   \node[label=below:{$2N-2$}][u](2k-2) [right of=4]{};
   \node[label=below:{$2N$}][u](2k) [right of=2k-2]{};
   \node[label=below:{$2N-2$}][u](2k-2') [right of=2k]{};
   \node[label=below:{$4$}][u](4') [right of=2k-2']{};
   \node[label=below:{$2$}][u](2') [right of=4']{};
   \node[label=above:{1}][u](1) [above of=2k-2]{};
   \node[label=above:{1}][u](11) [above of=2k-2']{};
   \draw[double distance=2 pt] (1)--(2k);
   \draw[double distance=2 pt] (11)--(2k);
   \draw[double distance=1.25 pt] (1)--(11);
   \draw (1) to[out=15,in=165](11);
   \draw (1) to[out=-15,in=-165](11);
   \draw (2)--(4);
   \draw[dotted] (2k-2)--(4);
   \draw (2')--(4');
   \draw[dotted] (2k-2')--(4');
   \draw (2k-2')--(2k);
   \draw(2k-2)--(2k);
   \end{tikzpicture}\end{scriptsize}

%% file: YN2121PhaseIIIOSp.tex
\begin{scriptsize}
      \begin{tikzpicture}
      \node[label=below:{2}][so](so2) at (0,0){};
      \node[label=below:{$2N$}][sp](sp2k) [left of=so2]{};
      \node[label=below:{$2N-1$}][u](u2k-1)[left of=sp2k]{};
      \node[label=below:{$2$}][u](u2)[left of=u2k-1]{};
      \node[label=below:{$1$}][u](u1)[left of=u2]{};
      \node[label=above:{$4$}][sof](u2')[above of=sp2k]{};
      \node[label=above:{$4$}][spf](u1')[right of=u2']{};
      \draw (u1)--(u2);
      \draw[dotted](u2)--(u2k-1);
      \draw (u2k-1)--(sp2k);
      \draw (sp2k)--(so2);
      \draw (so2)--(u1');
      \draw (u2')--(sp2k);
      \end{tikzpicture}
   \end{scriptsize}

%% file: YN2121PhaseIVU.tex
\begin{scriptsize}\begin{tikzpicture}
   \node[label=below:{2}][u](2) at (0,0){};
   \node[label=below:{4}][u](4) [right of=2]{};
   \node[label=below:{$2N-2$}][u](2k-2) [right of=4]{};
   \node[label=below:{$2N-2$}][u](2k) [right of=2k-2]{};
   \node[label=below:{$2N-2$}][u](2k-2') [right of=2k]{};
   \node[label=below:{$4$}][u](4') [right of=2k-2']{};
   \node[label=below:{$2$}][u](2') [right of=4']{};
   \node[label=below right:{1}][u](cyan) [above of=2k-2]{};
   \node[label=left:{1}][u](blue) [above of=cyan]{};
   \node[label=below left:{1}][u](red) [above of=2k-2']{};
   \node[label=right:{1}][u](green) [above of=red]{};
   \node[label=below:{1}][u](1)[above left of=cyan]{};
   \node[label=below:{1}][u](1')[above right of=red]{};
   \draw(blue)--(cyan);
   \draw(green)--(red);
   \draw (red)--(2k-2');
   \draw (cyan)--(2k-2);
   \draw (green) to[out=-50,in=50] (2k-2');
   \draw (blue) to[out=-130,in=130] (2k-2);
   \draw[double distance=2pt](green)--(blue);
   \draw (blue)--(red);
   \draw (green)--(cyan);
   \draw[double distance=2pt](cyan)--(red);
   \draw (cyan)--(1);
   \draw (1) to[out=90, in=120] (green);
    \draw (red)--(1');
   \draw (1') to[out=90, in=60] (blue);
   \draw (2)--(4);
   \draw[dotted](4)--(2k-2);
   \draw (2k-2)--(2k);
   \draw (2')--(4');
   \draw[dotted](4')--(2k-2');
   \draw (2k-2')--(2k);
   \end{tikzpicture}\end{scriptsize}

%% file: YN2121PhaseIVOSp.tex
\begin{scriptsize}
      \begin{tikzpicture}
      \node[label=below:{1}][u](1) at (0,0){};
      \node[label=below:{2}][u](2)[right of=1]{};
      \node[label=above:{$2N-2$}][u](2k-2)[right of=2]{};
      \node[label=below:{$2N-2$}][u](2k-2')[right of=2k-2]{};
      \node[label=above:{$2N-2$}][sp](sp2k-2)[right of=2k-2']{};
      \node[label=below:{2}][so](so2)[right of=sp2k-2]{};
      \node[label=below:{1}][u](u1)[right of=so2]{};
      \node[label=left:{1}][u](u1')[above of=u1]{};
      \node[label=below:{1}][u](u11')[left of=u1']{};
      \node[label=above:{1}][u](u11)[above left of=u11']{};
      \draw[dashed](u1)to[out=120,in=-30](u11);
      \draw(u1)to[out=30,in=-0,controls=+(0:2.5) and +(0:2)](u11);
      \draw(1)--(2);
      \draw[dotted](2)--(2k-2);
      \draw (2k-2)--(2k-2');
      \draw (2k-2')--(sp2k-2);
      \draw (sp2k-2)--(so2);
      \draw (so2)--(u1);
      \draw[double distance=2pt] (so2)--(u11);
      \draw (u11)--(u1');
      \draw[dashed] (u1')--(u1);
      \draw (u1)--(u11');
      \draw (u11')--(u11);
      \draw (2k-2)--(u11);
      \draw (sp2k-2)to[out=-45,in=-135](u1);
      \end{tikzpicture}
   \end{scriptsize}

%% file: YN2121PhaseVU.tex
\begin{scriptsize}\begin{tikzpicture}
   \node[label=below:{2}][u](2) at (0,0){};
   \node[label=below:{4}][u](4) [right of=2]{};
   \node[label=below:{$2N-2$}][u](2k-2) [right of=4]{};
   \node[label=below:{$2N-1$}][u](2k) [right of=2k-2]{};
   \node[label=below:{$2N-2$}][u](2k-2') [right of=2k]{};
   \node[label=below:{$4$}][u](4') [right of=2k-2']{};
   \node[label=below:{$2$}][u](2') [right of=4']{};
   \node[label=above:{1}][u](red)[above of=2k-2]{};
   \node[label=above:{1}][u](green)[above of=2k]{};
   \node[label=above:{1}][u](blue)[above of=2k-2']{};
   \node[label=above:{1}][u](1)[above of=green]{};
   \draw (2)--(4);
   \draw[dotted](4)--(2k-2);
   \draw (2k-2)--(2k);
   \draw (2')--(4');
   \draw[dotted](4')--(2k-2');
   \draw (2k-2')--(2k);
   \draw[double distance=4pt](red)--(green);
   \draw (red)--(green);
   \draw[double distance=4pt](blue)--(green);
   \draw (green)--(blue);
   \draw (red) to[out=45,in=135] (blue);
   \draw (red)--(1);
   \draw (1)--(blue);
   \draw[double distance=2pt] (green)--(2k);
   \draw (blue)--(2k-2');
   \draw (red)--(2k-2);
   
   \end{tikzpicture}\end{scriptsize}

%% file: YN2121PhaseVOSp.tex
\begin{scriptsize}
      \begin{tikzpicture}
      \node[label=below:{1}][u](1){};
      \node[label=below:{2}][u](2)[right of=1]{};
      \node[label=below:{$2N-1$}][u](2k-1)[right of=2]{};
      \node[label=below:{$2N-2$}][sp](sp2k-2)[right of=2k-1]{};
      \node[label=below:{2}][so](so2)[right of=sp2k-2]{};
      \node[label=left:{1}][u](u1)[above left of=sp2k-2]{};
      \node[label=above:{1}][uf](uf)[above of=u1]{};
      \node[label=above:{1}][u](u11)[right of=u1]{};
      \draw(1)--(2);
      \draw[dotted](2)--(2k-1);
      \draw(u1)--(2k-1);
      \draw(2k-1)--(sp2k-2);
      \draw(sp2k-2)--(so2);
      \draw[double distance=2pt](so2)--(u1);
      \draw[double distance=3pt](so2)--(u1);
      \draw(so2)--(u1);
      \path [draw,snake it](u1)--(uf);
      \draw(u1)--(u11);
      \draw[dashed](u1)to[out=30,in=150](u11);
      \end{tikzpicture}
   \end{scriptsize}

%% file: YN2121PhaseVIIU.tex
\begin{scriptsize}
      \begin{tikzpicture}
      \node[label=below:{2}][u](2){};
      \node[label=below:{4}][u](4)[right of=2]{};
      \node[label=below:{$2N-4$}][u](2k-4)[right of=4]{};
      \node[label=below:{$2N-3$}][u](2k-3)[right of=2k-2]{};
      \node[label=below:{$2N-2$}][u](2k-2)[right of=2k-3]{};
      \node[label=below:{$2N-3$}][u](2k-3')[right of=2k-2]{};
      \node[label=below:{$2N-4$}][u](2k-4')[right of=2k-3']{};
      \node[label=below:{$4$}][u](4')[right of=2k-4']{};
      \node[label=below:{$2$}][u](2')[right of=4']{};
      \node[label=above:{2}][u](22)[above of=2k-2]{};
      \node[label=above:{1}][u](11)[left of=22]{};
      \node[label=above:{1}][u](11')[right of=22]{};
      \node[label=above:{1}][u](111)[above of=11]{};
      \node[label=above:{1}][u](111')[above of=11']{};
      \draw(2)--(4);
      \draw[dotted](4)--(2k-4);
      \draw(2k-4)--(2k-3);
      \draw(2k-3)--(2k-2);
      \draw(2k-2)--(2k-3');
      \draw(2k-3')--(2k-4');
      \draw[dotted](2k-4')--(4');
      \draw(4')--(2');
      \draw(2k-2)--(22);
      \draw(22)--(11);
      \draw(22)--(11');
      \draw[double distance=2pt](22)--(111);
      \draw[double distance=2pt](22)--(111');
			\draw[double distance=4pt](111)--(111');
      \draw[double distance=1pt](111)--(111');		
      \draw(111)--(2k-4);
      \draw(111')--(2k-4');
      \end{tikzpicture}
   \end{scriptsize}

%% file: YN2121PhaseVIIOSp.tex
\begin{scriptsize}
      \begin{tikzpicture}
  \node[label=below:{1}][u](1) at (0,0){};
  \node[label=below:{2}][u](2) [right of=1]{};
  \node[label=below:{$2N-3$}][u](2k-3)[right of=2]{};
  \node[label=below:{$2N-2$}][u](2k-2)[right of=2k-3]{};
  \node[label=below:{$2N-3$}][u](2k-3')[right of=2k-2]{};
  \node[label=below:{$2N-4$}][sp](sp2k-4)[right of=2k-3']{};
  \node[label=below:{2}][so](so2)[right of=sp2k-4]{};
  \node[label=above:{1}][u](u1)[above of=so2]{};
  \node[label=below:{2}][u](u2)[left of=u1]{};
  \node[label=above:{1}][u](u1')[left of=u2]{};
  \draw(1)--(2);
  \draw[dotted](2)--(2k-3);
  \draw(2k-3)--(2k-2);
  \draw(2k-2)--(2k-3');
  \draw(2k-3')--(sp2k-4);
  \draw(sp2k-4)--(so2);
        \draw[double distance=4pt] (u2)to[out=135,in=45,loop,looseness=6](u2);
         \draw[double distance=2pt](so2)--(u2);
        \draw[double distance=1pt] (u2)to[out=135,in=45,loop,looseness=6](u2);
  \draw(u2)--(u1);
  \draw(u2)--(u1');
  \draw(u2)--(2k-2);
        \end{tikzpicture}
   \end{scriptsize}
	

%% file: YN2121PhaseVIIIU.tex
\begin{scriptsize}
      \begin{tikzpicture}
      \node[label=below:{2}][u](2){};
      \node[label=below:{4}][u](4)[right of=2]{};
      \node[label=below:{$2N-4$}][u](2n-4)[right of=4]{};
      \node[label=below:{$2N-3$}][u](2n-3)[right of=2n-4]{};
      \node[label=below:{$2N-2$}][u](2n-2)[right of=2n-3]{};
      \node[label=below:{$2N-2$}][u](2n-2')[right of=2n-2]{};
      \node[label=below:{$2N-4$}][u](2n-4')[right of=2n-2']{};
      \node[label=below:{$4$}][u](4')[right of=2n-4']{};
      \node[label=below:{$2$}][u](2')[right of=4']{};
      \node[label=left:{1}][u](cyan)[above left of=2n-2']{};
      \node[label=right:{1}][u](green)[above right of=2n-2']{};
      \node[label=above right:{1}][u](black)[above of=2n-2']{};
      \node[label=above:{1}][u](u1)[above of=cyan]{};
      \node[label=above:{1}][u](u1')[above of=green]{};
      \node[label=above:{1}][u](blue)[above of=black]{};
      \draw(2)--(4);
      \draw[dotted](4)--(2n-4);
      \draw(2n-4)--(2n-3);
      \draw(2n-3)--(2n-2);
      \draw(2n-2)--(2n-2');
      \draw(2n-2')--(2n-4');
      \draw[dotted](2n-4')--(4');
      \draw(4')--(2');
      \draw[double distance=3pt](black)--(green);
      \draw(black)--(green);
      \draw[double distance=3pt](black)--(cyan);
      \draw[double distance=2pt](black)--(blue);
      \draw(green)--(cyan);
      \draw(u1')--(green);
      \draw(u1)--(cyan);
      \draw(u1')--(blue);
      \draw(u1)--(blue);
      \draw(black)--(cyan);
      \draw(black)to[out=135,in=90](2n-4);
      \draw(2n-2')--(green);
      \draw(2n-2')--(cyan);
      \draw(2n-2)to[out=140, in=150](blue);
      \end{tikzpicture}
   \end{scriptsize}

%% file: YN2121PhaseVIIIOSp.tex
\begin{scriptsize}
      \begin{tikzpicture}
      \node[label=below:{1}][u](1)at(0,0){};
      \node[label=below:{2}][u](2)[right of=1]{};
      \node[label=below:{$2N-2$}][u](2k-2)[right of=2]{};
      \node[label=below:{$2N-2$}][u](2k-2')[right of=2k-2]{};
      \node[label=below:{$2N-4$}][sp](sp2k-4)[right of=2k-2']{};
      \node (false) [right of=sp2k-4]{};
      \node[label=below:{1}][u](u1)[below of=false]{};
      \node[label=above:{1}][u](u1')[above of=false]{};
      \node[label=below:{1}][u](11)[right of=u1]{};
      \node[label=above:{1}][u](11')[right of=u1']{};
      \node (false') [right of=false]{};
      \node[label=below:{1}][u](u11) [right of=false']{};
      \draw (1)--(2);
      \draw[dotted] (2k-2)--(2);
      \draw (2k-2)--(2k-2');
      \draw (2k-2')--(sp2k-4);
      \draw[dashed] (2k-2')--(u1);
      \draw[dashed] (2k-2')--(u1');
      \draw (2k-2) to[out=80,in=80](u11);
      \draw(u1)--(11);
      \draw[dashed](11)--(u11);
      \draw[dashed] (u11)--(11');
      \draw (11')--(u1');
      \draw[dashed,double distance=2pt](u1)to[out=120, in=-120](u1');
      \draw (u1)--(u1');
      \draw[double distance=3pt](u1)--(u11);
      \draw[double distance=3pt](u1')--(u11);
        \draw(u1)--(u11);
      \draw(u1')--(u11);
      \end{tikzpicture}
   \end{scriptsize}

%% file: HNU.tex
\begin{scriptsize}
\begin{tikzpicture}
         \node[label=below:{2}][u](22) at (-1,0){}; 
         \node[label=below:{2}][u] (22') at (1,0){}; 
         \draw[double distance=2 pt] (22)--(22');
         \node[label=left:{1}][u](11)[left of=22]{};
         \node[label=left:{1}][u](111)[above of=22]{};
         \draw (11)--(22);
         \draw (111)--(22);
         \node[label=right:{1}][u](11')[right of=22']{};
         \node[label=right:{1}][u](111')[above of=22']{};
         \draw (11')--(22');
         \draw (111')--(22');

         \node at (0, 0.3) {$N$};
         \end{tikzpicture}\end{scriptsize}

%% file: HNOSp.tex
\begin{scriptsize}
\begin{tikzpicture}
            
            \node[label=right:{$1$}][u](u1){};
            \node[label=below:{4}][so](so4)[below of=u1]{};
            \node[label=below:{2}][sp](sp2)[left of=so4]{};
            \node[label=below:{2}][sp](sp2')[right of=so4]{};
            \node[label=below:{2}][so](so2)[left of=sp2]{};
            \node[label=below:{2}][so](so2')[right of=sp2']{};
           
            \draw[double distance=2pt] (u1)--(so4);
            \draw (so2)--(sp2);
            \draw (sp2)--(so4);
            \draw (so2')--(sp2');
            \draw (sp2')--(so4);
           
           \node at (0.3,-0.5) {$N$};
            \end{tikzpicture}\end{scriptsize}

%% file: sec3conclusions.tex
\section{Conclusion}\label{sec:conclusions}
In this paper, we studied infinite coupling Higgs branches of 5d superconformal theories based on 5-brane webs, by constructing 3d magnetic quivers whose Coulomb branch yields the Higgs branch of the 5d system. Our primary focus was 5d theories, which can be engineered by 5-brane webs with O5-planes and also without O5-planes by either S-duality or some gauging subalgebra of flavor symmetry. As a 5-brane web without an O5-plane gives a unitary magnetic quiver, while that with an O5-plane gives an orthosymplectic magnetic quiver, these 5d theories of two different 5-brane web descriptions should yield the same Higgs branch. In other words, the Coulomb branches from the corresponding unitary and orthosymplectic magnetic quivers should agree. We employed this fact to further develop how to construct orthosymplectic magnetic quivers \cite{Bourget:2020gzi} by comparing the counterpart unitary magnetic quiver, in particular, by explicitly checking the Hilbert series for both magnetic quivers. With various decoupling limits for both 5-brane webs with/without O5-planes, we proposed a generalization of the rules for constructing 3d magnetic quivers from 5-brane webs with O5-planes. The novel features that we found include (i) generalized stable intersection number involving subwebs intersecting with its mirror through an O5-plane as well as the stable intersection with an O5-plane, which in turn determines the number of charge 2 matter appearing in the magnetic quivers, (ii) a new type of hypermultiplet transforming in the fundamental-fundamental representation of two gauge nodes, (iii) appearance of matter in the antisymmetric representation of gauge nodes. (iv) possibility of decomposable and not-decomposable for seeming equivalent subwebs depending the discrete theta angles for 5-brane configurations for 5d USp gauge groups, and (v) $\mathbb{Z}_2$ gauge nodes.

We also checked the matching of both the Coulomb branch and the Higgs branch Hilbert series for all the unitary and orthosymplectic magnetic quivers appearing in this work (though we provided the Higgs branch Hilbert series only when it was required). Having checked that each pair of unitary and orthosymplectic magnetic quivers have isomorphic Coulomb and Higgs branch moduli space of vacua, it is natural to suggest a possible duality between each pair. It would be interesting to check this duality more systematically, for instance, via superconformal indices or partition functions. In fact, we checked the partition functions for some 3d theories and found that their partition functions can be mapped to each other by a simple fugacity map. Understanding the 4d origin of this duality, should it exist, would be another exciting direction to pursue. 

Though our construction applies to generic 5d theories of any rank, some lower rank theories possess special dualities~\cite{Jefferson:2018irk}. In particular, 5-brane webs for most of rank 2 superconformal theories have been constructed~\cite{Hayashi:2018lyv}, a systematic study of their magnetic quivers would also shed some light on a better understanding of orthosymplectic magnetic quivers~\cite{ACDHKY}. Some magnetic quivers of the rank 2 theories were also considered in \cite{Bourget:2020asf,Bourget:2020xdz, Closset:2020scj}.

%% file: appendixA.tex
\section{Hilbert series computations}\label{app1}
\subsection{Coulomb branch Hilbert series}
Here we briefly review the computation of the Hilbert series of the Coulomb branch of the moduli space for the $3d$ $\mathcal{N}=4$ quiver theories. For the computation, we use the \emph{monopole formula} prescribed in \cite{Cremonesi:2013lqa} which essentially counts the number of \emph{dressed} monopole operators according to their conformal dimension. We refer the readers to \cite{Cremonesi:2013lqa} for the technical details of the formula and simply quote the result here:
\begin{equation}
\text{HS}_C(t) = \sum_{\vec{m}_1} \sum_{\vec{m}_2} \ldots \sum_{\vec{m}_x}\, t^{\Delta(\vec{m}_1,\ldots,\vec{m}_x)} \prod_{i=1}^x P_{G_i}(t,\vec{m}_i) ~.
\label{monopole-form}
\end{equation}
Let us briefly explain various terms in this formula. The gauge group of the theory under consideration is $G_1 \times G_2 \times \ldots \times G_x$, where each of the group $G_i$ is indicated as a circular node in the quiver description. For a particular group $G$ in the quiver, the monopole operators are specified by the magnetic fluxes $\vec{m} = (m_1,m_2,\ldots m_r)$ which belong to the weight lattice $\Gamma(\hat{G})$ of $\hat{G}$, the GNO (or Langlands) dual group of $G$ ($r$ being the rank of $G$ or $\hat{G}$). The gauge invariant monopole operators are specified by those $\vec{m}$ which take values in the quotient space:
\begin{equation}
\vec{m} \in  \Gamma(\hat{G}) / W(\hat{G}) ~,
\end{equation}   
where $W(\hat{G})$ is the Weyl group of $\hat{G}$. These are precisely the fluxes which contribute in the summation in the monopole formula (\ref{monopole-form}). The Langland duals and the associated magnetic fluxes for some of the Lie groups are given below:
\begin{equation}
\begin{array}{|c|c|c|} \hline
 G & \hat{G} &  \vec{m} \in  \Gamma(\hat{G}) / W(\hat{G}) \\ \hline
\text{U}(r) &  \text{U}(r)  & m_1 \geq m_2 \geq \ldots \geq m_r \geq -\infty   \\
\text{SO}(2r+1) &  \text{USp}(2r)  & m_1 \geq m_2 \geq \ldots \geq m_r \geq 0   \\
\text{USp}(2r) &  \text{SO}(2r+1)  & m_1 \geq m_2 \geq \ldots \geq m_r \geq 0   \\
\text{SO}(2r) & \text{SO}(2r)  & m_1 \geq m_2 \geq \ldots \geq |m_r| \geq 0   \\ \hline
\end{array} ~.
\end{equation} 
The factor $P_G(t,\vec{m})$ is the classical dressing function which counts the gauge invariants of the residual gauge group which is left unbroken by the magnetic flux $\vec{m}$, according to their dimension, and is given as,
\begin{equation}
P_G(t,\vec{m}) = \prod_{i=1}^r  \frac{1}{1-t^{d_i(\vec{m})}} ~,
\end{equation} 
where $d_i(\vec{m})$ are the degrees of Casimir invariants of the unbroken residual gauge group. These functions can be written in a computationally friendly manner by collecting the fluxes which are equal in $\vec{m}$. To do this, let us define an auxiliary sequence of non-increasing fluxes in $\vec{m}$, which we shall denote as $\vec{n} = (n_1,n_2,\ldots,n_r)$. We collect all the repeating fluxes together and define $\vec{n}_{\text{res}} = (a_1^{r_1},\ldots,a_u^{r_u})$ where the notation $a_i^{r_i}$ means that the integer $a_i$ is repeated $r_i$  times (where $r_1+\ldots+r_u = r$). The dressing functions can now be defined for various groups as:
\begin{equation}
\begin{array}{|c|c|c|c|} \hline
 G & \vec{n} = (n_1,n_2,\ldots,n_r) &  \vec{n}_{\text{res}} & P_G \\ \hline
\text{U}(r) &  (m_1,m_2,\ldots,m_r)  & (a_1^{r_1},\ldots,a_u^{r_u}) & \prod_{i=1}^{u} \prod_{k=1}^{r_i} \frac{1}{1-t^{k}}   \\[0.05cm] \hline
\text{SO}(2r+1) &  (m_1,m_2,\ldots,m_r)  & (a_1^{r_1},\ldots,a_{u-1}^{r_{u-1}},0^{r_u}) &  A(r_u) \left( \prod_{i=1}^{u-1} \prod_{k=1}^{r_i} \frac{1}{1-t^{k}} \right)   \\[0.05cm] \hline
\text{USp}(2r) &  (m_1,m_2,\ldots,m_r)  & (a_1^{r_1},\ldots,a_{u-1}^{r_{u-1}},0^{r_u}) & A(r_u) \left(\prod_{i=1}^{u-1} \prod_{k=1}^{r_i} \frac{1}{1-t^{k}} \right)    \\[0.05cm] \hline
\text{SO}(2r) & (m_1,m_2,\ldots,|m_r|)  & (a_1^{r_1},\ldots,a_{u-1}^{r_{u-1}},0^{r_u}) & B(r_u) \left( \prod_{i=1}^{u-1} \prod_{k=1}^{r_i} \frac{1}{1-t^{k}} \right)    \\ \hline
\end{array}
\end{equation} 
where the factors $A(r_u)$ and $B(r_u)$ are explicitly dependent on the number of vanishing fluxes and are given as,
\begin{equation}
A(j) = \prod_{k=1}^{j} \frac{1}{1-t^{2k}} \quad;\quad B(j) = \delta_{j0} + \left( \frac{1}{1-t^{j}} \prod_{k=1}^{j-1} \frac{1}{1-t^{2k}} \right)(1-\delta_{j0}) ~.
\end{equation}
The last thing we need in the monopole formula (\ref{monopole-form}) is the $R$-charge or the conformal dimension $\Delta(\vec{m}_1,\ldots,\vec{m}_x)$ of the bare monopole operators associated with various gauge groups in the quiver specified by the GNO magnetic fluxes $\vec{m}_1,\ldots,\vec{m}_x$. The conformal dimension gets contribution from the vector multiplets and the hyper multiplets present in the theory:
\begin{equation}
\Delta = \Delta_{\text{vector}} + \Delta_{\text{hyper}} ~.
\end{equation}
These contributions are given as follows. Consider a node with group $G$ in the quiver and denote $\alpha$ to be a positive root of $G$. The vector contribution is computed as,
\begin{equation}
\Delta_{\text{vector}}(\vec{m}) = - \sum_{\alpha \in \Delta_{+}} \left|\alpha(\vec{m})\right| ~,
\end{equation}
where the positive roots of $G$ act on the GNO fluxes $\vec{m}$ associated with the weight lattice of GNO dual group $\hat{G}$ and the sum is taken over all positive roots of $G$. By restricting these fluxes to the fundamental Weyl chamber, the above sum can be explicitly performed and is given as,
\begin{equation}
\begin{array}{|c|c|c|} \hline
 G & \Delta_{+}(G) & \Delta_{\text{vector}}(\vec{m}) \\ \hline
\text{U}(r) & \{e_i - e_j \}_{1 \leq i < j \leq r} &  -\sum_{k=1}^r (r+1-2k)m_k  \\
\text{SO}(2r+1) & \{e_i - e_j, e_i+e_j, e_i \}_{1 \leq i < j \leq r} & -\sum_{k=1}^r (2r+1-2k)m_k \\
\text{USp}(2r)  & \{e_i - e_j, e_i+e_j, 2e_i \}_{1 \leq i < j \leq r} & -\sum_{k=1}^r (2r+2-2k)m_k \\
\text{SO}(2r) & \{e_i - e_j, e_i+e_j \}_{1 \leq i < j \leq r} & -\sum_{k=1}^{r-1} (2r-2k)m_k \\ \hline
\end{array}
\end{equation} 
Once we compute this vector term for individual nodes in the quiver, we can simply add them to get the full vector contribution:
\begin{equation}
\Delta_{\text{vector}}(\vec{m}_1,\ldots,\vec{m}_x) = \sum_{i=1}^x \Delta_{\text{vector}}(\vec{m}_i) ~.
\end{equation} 
The term $\Delta_{\text{hyper}}$ is the contribution of hypermultiplets (the links connecting the nodes in the quiver diagram) present in the theory which is given as the sum over the weights of the matter field representation under the gauge groups. To write an explicit formula in terms of the GNO fluxes, consider a hypermultiplet connecting two groups $G_r$ and $G_s$ of ranks $r$ and $s$:
\begin{equation}
\begin{array}{c}\begin{tikzpicture}
    \node[circle,inner sep=4pt,draw,thick](Gr)at(0,0){$G_r$};
    \node[circle,inner sep=4pt,draw,thick](Gs)at(2,0){$G_s$};
    \draw(Gr)--(Gs); 
\end{tikzpicture}\end{array}
\end{equation} 
Further consider the GNO fluxes associated with the two groups as $\vec{a}=(a_1,a_2,\ldots,a_r)$ and $\vec{b}=(b_1,b_2,\ldots,b_s)$. The weights associated with the fundamental representation of the two groups can be written as a tuple of GNO fluxes which we denote as $w(\vec{a})$ and $w(\vec{b})$. For the classical Lie groups, they are given below:  
\begin{equation}
\begin{array}{|c|c|c|} \hline
 G & \vec{m} & w(\vec{m}) \\ \hline
\text{U}(r) & (m_1, \ldots, m_r) & (m_1,\ldots,m_r) \\
\text{SO}(2r+1) & (m_1, \ldots, m_r) & (m_1,\ldots,m_r,0,-m_r,\ldots,-m_1) \\
\text{USp}(2r) & (m_1, \ldots, m_r) & (m_1,\ldots,m_r,-m_r,\ldots,-m_1) \\
\text{SO}(2r) & (m_1, \ldots, m_r) & (m_1,\ldots,m_r,-m_r,\ldots,-m_1) \\ \hline
\end{array} ~.
\end{equation}
The hypermulitplet contribution is now easy to write. If we denote $w_i(\vec{m})$ to be the $i^{\text{th}}$ component in $w(\vec{m})$, the contribution is:
\begin{equation}
\Delta_{\text{hyper}}\left(\vec{a}, \vec{b}\right) = 
\begin{cases} 
\dfrac{1}{4} \sum_{i}\sum_{j} \left|w_i(\vec{a}) - w_j(\vec{b})\right|\, , & \text{when $G_r/G_s=$ SO/USp or USp/SO} \\[0.5cm]
\dfrac{1}{2} \sum_{i}\sum_{j} \left|w_i(\vec{a}) - w_j(\vec{b})\right|\, , & \text{otherwise}
\end{cases} ~,
\end{equation}  
where a hypermultiplet comes with a $1/2$ factor and a half-hypermultiplet (the link connecting a SO type of node to USp type of node) comes with a 1/4 factor in the conformal dimension. Summing over the contributions of all the hypermultiplets present in the theory will finally give the $\Delta_{\text{hyper}}$ for the full quiver. There can also be flavor symmetry groups associated with a gauge group. Such groups are denoted by a square node in the quiver. The GNO fluxes for such nodes are all 0 ($\vec{m} = 0$) and they do not contribute to $\Delta_{\text{vector}}$ in the conformal dimension. The contribution of the link connecting the gauge node with a flavor node can be obtained by simply computing $\Delta_{\text{hyper}}$ as usual and then setting the fluxes of GNO dual to 0:
\begin{equation}
\begin{array}{c}\begin{tikzpicture}
    \node[circle,inner sep=4pt,draw,thick](Gr)at(0,0){$G_r$};
    \node[rectangle,inner sep=6pt,draw,thick](Gs)at(2,0){$G_s$};
    \draw(Gr)--(Gs);
\end{tikzpicture}\end{array} \quad \Longrightarrow \quad \Delta_{\text{hyper}} = \Delta_{\text{hyper}}\left(\vec{a}, \vec{0}\right) ~.
\end{equation}
We may also encounter the cases where we need to put multiple links between two nodes in the quiver. In this scenario, the contribution of the individual links are simply added up. For example, in case of triple hypers, we have:
\begin{equation}
\begin{array}{c}\begin{tikzpicture}
    \node[circle,inner sep=4pt,draw,thick](Gr)at(0,0){$G_r$};
    \node[circle,inner sep=4pt,draw,thick](Gs)at(2,0){$G_s$};
    \draw[double distance=4pt](Gr)--(Gs);
    \draw(Gr)--(Gs);
\end{tikzpicture}\end{array} \quad \Longrightarrow \quad \Delta_{\text{hyper}} = 3 \times \Delta_{\text{hyper}}\left(\vec{a}, \vec{b}\right) ~.
\end{equation}   
We have also proposed an exotic hypermultiplet which transforms under the fundamental-fundamental representations of the unitary gauge nodes it connects. We denote this exotic hyper by a dashed line in the quiver between the two unitary nodes. The contribution of such a hyper to the conformal dimension is given as:
\begin{equation}
\begin{footnotesize}
\begin{array}{c}\begin{tikzpicture}
    \node[circle,inner sep=4pt,draw,thick](Gr)at(0,0){U($r$)};
    \node[circle,inner sep=4pt,draw,thick](Gs)at(2,0){U($s$)};
    \draw[dashed](Gr)--(Gs);
\end{tikzpicture}\end{array} \quad \Longrightarrow \quad \Delta_{\text{hyper}} = \dfrac{1}{2} \sum_{i=1}^r\sum_{j=1}^s \left|w_i(\vec{a}) + w_j(\vec{b})\right| ~,
\end{footnotesize}
\end{equation} 
where $\vec{a}$ and $\vec{b}$ are the GNO fluxes of the two nodes.

\subsection{Higgs branch Hilbert series}
The Hilbert series of the Higgs branch of a $3$d $\mathcal{N}=4$ theory can be easily computed. Suppose the gauge group is $G$, and the matter content is given by $N_h$ hypermultiplets charged under representations $R_i$ ($i=1,\ldots, N_h$) of the gauge group, and possibly charged under representations $R'_i$ of the flavor group. The Hilbert series is then schematically given by

\begin{equation}
\HS(t)=\int_G d\mu_G \  \text{Pfc}(w,t) \text{PE}\left[\sum_{i=1}^{2N_h} \car_{R_i}(w)\car_{R'_i}(x)t^{\frac{1}{2}} \right]
\label{HBHS}
\end{equation}
Let us now review the various terms entering in equation (\ref{HBHS}). 

The term
\begin{equation}
\text{PE}\left[\sum_{i=1}^{2N_h} \car_{R_i}(w)\car_{R'_i}(x)t^{\frac{1}{2}} \right]
\end{equation}
is generating all the possible symmetrized product of all the scalars inside the hypermultiplets. Here $w=(w_1,\cdots, w_r)$ is a collective variable denoting the fugacities of the rank $r$ gauge group, and $x=(x_1,\cdots, x_r)$ are fugacities of the flavor group. Notice that these fugacities take value in the gauge (resp. flavor) symmetry group. Therefore each of the $x$ and $w$ is a complex phase. The fugacity $t$ is a fugacity counting the conformal dimension. The exponent of $t$ is $\frac{1}{2}$ as this is the conformal dimension of one free scalar in $3$d. The term $\car_{R_i}(w)$ (resp. $\car_{Ri_i}(x)$) is the character of the representation under the gauge group (resp. flavor group) under which the $i$-th hypermultiplet is charged. The function $\text{PE}[\cdot]$ is the plethystic exponential, defined as
\begin{equation}
\text{PE}[f(x)]:=\exp\left(\sum_{k=1}^{\infty} \dfrac{1}{k}f\left(x^k\right)\right)
\end{equation} 
for any function $f(x)$ such that $f(0)=0$. The term $\Pfc(w,t)$ is a prefactor, encoding the fact that the symmetrized product of the scalars are not all independent, but they obey some relationships coming from the F-term constraints. This prefactor term is given by
\begin{equation}
\Pfc(w,t)=\text{PE}\left[\sum_{i=1}^{N_r}\car_{R_i''}(w)t^{d_i}\right]^{-1}
\end{equation}
where $\car_{R_i''}(w)$ is the character of the representation of the $i$-th relation, and $d_i$ its degree in conformal dimension. Given the constrained structure of the superpotential in theories with $8$ supercharges, typically $\car_{R_i''}(w)$ will be the character of the adjoint, and the relation will appear at quadratic order in the scalars: $d_i=1$. Finally, the integral over the gauge group is performed in order to count only gauge invariant operators, and not just all the symmetric products of scalars modulo the F-term constraint. Such integral is called the Molien integral. The integration measure $\mu_G$ is given, for any continuous gauge group as

\begin{equation}
\int_G d\mu_G=\dfrac{1}{(2\pi i)^r}\oint_{|w|_1=1}\cdots\oint_{|w|_r=1}\dfrac{dw_1}{w_1}\cdots \dfrac{dw_r}{w_r}\prod_{\alpha\in \Delta^+}\left(1- \prod_{k=1}^r w_k^{\alpha_k}\right)
\label{molien}
\end{equation}
where $\Delta^+$ is the set of the positive roots of the Lie algebra of $G$. Notice that despite the integral is formally performed over the full gauge group, the use of the Molien measure localizes the integral just on the Cartan torus. Every fugacity $w_i$ is then integrated just on the unit circle. The factor $\dfrac{dw_r}{w_r}\prod_{\alpha\in \Delta^+}\left(1- \prod_{k=1}^r w_k^{\alpha_k}\right)$ can be thought of the Jacobian for the change of variable recasting the integral over the gauge group into the integral over the its Cartan torus.

The procedure outlined above can be slightly modified if discrete gauge groups are present. Let us call the discrete gauge group $H$, and its order $|H|$. We can introduce a fugacity $z$ for the discrete gauge symmetry, valued now in $H$. The term generating all the symmetrized products of scalars will now read 
\begin{equation}
\text{PE}\left[\sum_{i=1}^{2N_h} \car_{R_i}(w)\car_{R'_i}(x)zt^{\frac{1}{2}} \right]
\end{equation}
and there will be no prefactor contribution. This is consistent with the fact that there is no vector multiplet associated to a discrete gauge factor, therefore there will be no F-term relations for it. To only retain the singlets for the discrete factor, the Molien integral over $H$ still has to be performed. However, since now the $z$ fugacity takes values is a discrete group, the integral over all the elements of the gauge group is now replaced with discrete sum:

\begin{equation}
\begin{aligned}
&\int_H d\mu_H  \text{Pfc}(w,t)\text{PE}\left[\sum_{i=1}^{2N_h} \car_{R_i}(w)\car_{R'_i}(x)zt^{\frac{1}{2}} \right]=\\
&=\dfrac{1}{|H|}\text{Pfc}(w,t)\sum_h^{|H|}\text{PE}\left[\sum_{i=1}^{2N_h} \car_{R_i}(w)\car_{R'_i}(x)zt^{\frac{1}{2}} \right]_{|z=h}
\end{aligned}
\end{equation}

After performing all the discrete Molien integrations for all the discrete factors of the gauge group, the integrations on the continuous factors still have to be performed, using eq (\ref{molien}).

\section{Palindromic polynomials}
\label{sec:app2}
We summarize in Table \ref{PalPol} explicit forms of parlindromic polynomials which arise in the computation of the Hilbert series of Coulomb branches of some theories discussed in section \ref{sec:O5-O5-} and section \ref{sec:O5+O5+}.

\begin{table}[!htb]
\centering
\begin{tabular}{|c|C{14cm}|} \hline
\rowcolor{Grayy}
  Label & Polynomial \\ \hline
	  $P_0(t)$ & \footnotesize{$1+98 t+3312 t^2+53305 t^3+468612 t^4+2421286 t^5+7664780 t^6+15203076 t^7+19086400 t^8+15203076 t^9+7664780 t^{10}+2421286 t^{11}+468612 t^{12}+53305 t^{13}+3312 t^{14}+98 t^{15}+t^{16}$} \\ \hline
  $P_1(t)$ & \footnotesize{$1+52t+2669t^2+63963t^3+1011274t^4+11000351t^5+88762624t^6+549522302t^7+2698733098t^8+10730180908t^9+35179028314t^{10}+96291284692t^{11}+222448351508t^{12}+436990904921t^{13}+734668194786t^{14}+1061421775571t^{15}+1322191150030t^{16}+1422261110352t^{17}+\ldots \text{palindrome} \ldots+52t^{33}+t^{34}$} \\ \hline
	$P_2(t)$ & \footnotesize{$64t+2560t^2+64768t^3+1006592t^4+11022656t^5+88672768t^6+549832576t^7+2697805824t^8+10732603264t^9+35173452800t^{10}+96302654592t^{11}+222427719168t^{12}+437024347328t^{13}+734619634688t^{14}+1061485070080t^{15}+1322116993536t^{16}+1422339277952t^{17}+\ldots \text{palindrome} \ldots+2560t^{32}+64t^{33}$} \\ \hline
	$P_3(t)$ & \footnotesize{$1+31t+231t^2+595t^3+595t^4+231t^5+31t^6+t^7$} \\ \hline
	$P_4(t)$ & \footnotesize{$1+22t+245t^2+1442t^3+5355t^4+12978t^5+21919t^6+25900t^7+21919t^8+12978t^9+5355t^{10}+1442t^{11}+245t^{12}+22t^{13}+t^{14}$} \\ \hline
	$P_5(t)$ & \footnotesize{$1+14t+91t^2+336t^3+819t^4+1362t^5+1618t^6+1362t^7+819t^8+336t^9+91t^{10}+14t^{11}+t^{12}$} \\ \hline
	$P_6(t)$ & \footnotesize{$1+12t+58t^2+124t^3+170t^4+124t^5+58t^6+12t^7+t^8$} \\ \hline
	$P_7(t)$ & \footnotesize{$8t+48t^2+136t^3+176t^4+136t^5+48t^6+8t^7$} \\ \hline
	$P_8(t)$ & \footnotesize{$1+6t+44t^2+146t^3+446t^4+826t^5+1343t^6+1436t^7+1343t^8+826t^9+446t^{10}+146t^{11}+44t^{12}+6t^{13}+t^{14}$} \\ \hline
	$P_9(t)$ & \footnotesize{$1+6t+35t^2+108t^3+407t^4+1014t^5+2720t^6+5198t^7+10773t^8+16712t^9+27493t^{10}+35046t^{11}+47571t^{12}+50460t^{13}+56752t^{14}+50640t^{15}+47571t^{16}+35046t^{17}+27493t^{18}+16712t^{19}+10773t^{20}+5198t^{21}+2720t^{22}+1014t^{23}+407t^{24}+108t^{25}+35t^{26}+6t^{27}+t^{28}$} \\ \hline
	$P_{10}(t)$ & \footnotesize{$16t^2+80t^3+368t^4+960t^5+2704t^6+5296t^7+10912t^8+16832t^9+27728t^{10}+35184t^{11}+47344t^{12}+50192t^{13}+56608t^{14}+50192t^{15}+47344t^{16}+35184t^{17}+27728t^{18}+16832t^{19}+10912t^{20}+5296t^{21}+2704t^{22}+960t^{23}+368t^{24}+80t^{25}+16t^{26}$} \\ \hline
	$P_{11}(t)$ & \footnotesize{$1+11t+57t^2+170t^3+324t^4+398t^5+324t^6+170t^7+57t^8+11t^9+t^{10}$} \\ \hline
	$P_{12}(t)$ & \footnotesize{$1+3t+38t^2+69t^3+225t^4+240t^5+372t^6+240t^7+225t^8+69t^9+38t^{10}+3t^{11}+t^{12}$} \\ \hline
	$P_{13}(t)$ & \footnotesize{$8t+20t^2+112t^3+156t^4+328t^5+276t^6+328t^7+156t^8+112t^9+20t^{10}+8t^{11}$} \\ \hline
	$P_{14}(t)$ & \footnotesize{$1+2t+6t^2+10t^3+22t^4+26t^5+39t^6+36t^7+39t^8+26t^9+22t^{10}+10t^{11}+6t^{12}+2t^{13}+t^{14}$} \\ \hline
	$P_{15}(t)$ & \footnotesize{$1-2t^2+8t^{5/2}+t^3-8t^{7/2}+t^4-8t^{9/2}+13t^5+8t^{11/2}-24t^6-5t^7+8t^{15/2}+15t^8-24t^{17/2}-t^9+8t^{19/2}+9t^{10}+8t^{21/2}-16t^{11}+8t^{23/2}+9t^{12}+8t^{25/2}-t^{13}-24t^{27/2}+15t^{14}+8t^{29/2}-5t^{15}-24t^{16}+8t^{33/2}+13t^{17}-8t^{35/2}+t^{18}-8t^{37/2}+t^{19}+8t^{39/2}-2t^{20}+t^{22}$} \\ \hline
	$P_{16}(t)$ & \footnotesize{$1-2t^2+t^3+t^4+16t^5-24t^6-11t^7+18t^8+2t^9+51t^{10}-88t^{11}-12t^{12}+47t^{13}+15t^{14}+62t^{15}-144t^{16}+23t^{17}+44t^{18}+44t^{19}+23t^{20}-144t^{21}+62t^{22}+15t^{23}+47t^{24}-12t^{25}-88t^{26}+51t^{27}+2t^{28}+18t^{29}-11t^{30}-24t^{31}+16t^{32}+t^{33}+t^{34}-2t^{35}+t^{37}$} \\ \hline
	$P_{17}(t)$ & \footnotesize{$1 - t - t^2 + t^3 + 4 t^5 - 6 t^6 - 2 t^7 + 4 t^8 + t^9 + 7 t^{10} - 
 15 t^{11} + t^{12} + 6 t^{13} + 4 t^{14} + 6 t^{15} - 20 t^{16} + 6 t^{17} + 
 4 t^{18} + 6 t^{19} + t^{20} - 15 t^{21} + 7 t^{22} + t^{23} + 4 t^{24} - 2 t^{25} - 
 6 t^{26} + 4 t^{27} + t^{29} - t^{30} - t^{31} + t^{32}$} \\ \hline
$P_{18}(t)$ & \footnotesize{$1+8 t+40 t^2+107 t^3+199 t^4+234 t^5+199 t^6+107 t^7+40 t^8+8 t^9+t^{10}$} \\ \hline
$P_{19}(t)$ & \footnotesize{$1 + 2 t + 6 t^2 + 10 t^3 + 22 t^4 + 26 t^5 + 39 t^6 + 36 t^7 + 
 39 t^8 + 26 t^9 + 22 t^{10} + 10 t^{11} + 6 t^{12} + 2 t^{13} + t^{14}$} \\ \hline
\end{tabular}
\caption{Palindromic polynomials appearing in the main sections.}
\label{PalPol}
\end{table}
\FloatBarrier

%% file: charge2examples.tex
\section{More on the number of charge $2$ hypermultiplets}\label{sec:charge2ex}

In section \ref{sec:KNO5-O5-}, we observed that the number of charge $2$ hypermultiplets may be given by \eqref{SI-charge-2}. So far, we have encountered the cases which involve a $(p, 1)$ 5-brane or a $(1, q)$ 5-brane for the origin of charge $2$ hypermultiplets. We here give more support for \eqref{SI-charge-2} by checking further new examples that involve a $(p, q)$ 5-brane with both $p$ and $q$ larger than $1$. In order to consider such examples, we use 5-brane web diagrams whose corresponding orthosymplectic magnetic quiver theories do not have a unitary counterpart. However, we can still check the validity of the 3d orthosymplectic magnetic quivers by comparing their Higgs branch dimension with the dimension of the Coulomb branch moduli space of the original 5d theories.

\begin{figure}[!ht]
    \centering
      \begin{scriptsize}
     \begin{tikzpicture}[scale=0.4]
     
    \draw[thick,dashed](1,0)--(29,0); 
    \node at (2.5,-1) {O5$^-$};
    \node at (27.5,-1) {O5$^-$};
    \draw[thick](6,1)--(8,1);
    \draw[thick](6,1.5)--(7.5,1.5);
    \draw[thick](7.5,5)--(7.5,1.5);
    \draw[thick](7.5,1.5)--(8,1);
    \draw[thick](8,1)--(10,0);

    \draw[thick](12,0)--(14,1);
    \draw[thick](14,1)--(16,1);
    \draw[thick](14,1)--(14.5,1.5);
    \draw[thick](14.5,1.5)--(15.5,1.5);
    \draw[thick](14.5,1.5)--(14.5,2);
    \draw[thick](15.5,2)--(15.5,1.5);
    \draw[thick](15.5,1.5)--(16,1);
    \draw[thick](16,1)--(18,0);
    \draw[thick](14.5,2)--(15.5,2);
    \draw[thick](14.5,2)--(14,2.5);
    \draw[thick](15.5,2)--(16,2.5);
    \draw[thick](14,2.5)--(16,2.5);
    \draw[thick](16,2.5)--(17,3);
    \draw[thick](14,2.5)--(13,3);
    \draw[thick](13,3)--(17,3);
    \draw[thick](17,3)--(18.5,3.5);
    \draw[thick](13,3)--(11.5,3.5);
    \node at (15,4){$\vdots$};
    \draw[thick](10,5)--(20,5);
    \draw[thick](10.5,4.8)--(10,5);
    \draw[thick](19.5,4.8)--(20,5);
    \draw[thick](8.8,5.4)--(10,5);
    \draw[thick](20.6,5.2)--(20,5);

   \draw[dashed] (7.9,5.9) circle [radius=1];
   \draw[thick](7.9-1/1.4,5.9+1/1.4)--(7.9-1/1.4-3,5.9+1/1.4+2);
   \node at (7, 9){$(2k-1,-2)$};
   \draw[dashed] (22.1,5.9) circle [radius=1];
   \draw[thick](21.2,5.4)--(20,5);
   \draw[thick](22.1+1/1.4,5.9+1/1.4)--(22.1+1/1.4+3,5.9+1/1.4+2);
   \node at (23, 9){$(2k-1,2)$};
   \node at (19,6){$(2k-1,1)$};
   \node at (11,6){$(2k-1,-1)$};

    \draw[thick](20,0)--(22,1);
    \draw[thick](22,1)--(24,1);
    \draw[thick](22,1)--(22.5,1.5);
    \draw[thick](22.5,1.5)--(24,1.5);
    \draw[thick](22.5,1.5)--(22.5,5);

    \end{tikzpicture}
    \end{scriptsize}
    \caption[Web diagram with external $(2k-1,2)$ and $(2k-1,-2)$ 5-branes.]{The 5-brane web diagram which has an external $(2k-1,2)$ 5-brane and an external $(2k-1,-2)$ 5-brane. }
    \label{fig:appendixbrane1}
\end{figure}
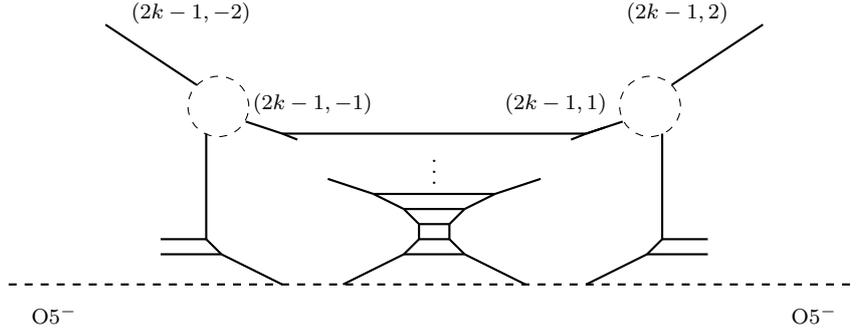

\begin{figure}[!ht]
    \centering\begin{scriptsize}
    \begin{tikzpicture}[scale=.7]
    \draw[thick](-8,0)--(8,0);
    \draw[thick,dashed](-10,0)--(-8,0);
    \draw[thick,dashed](10,0)--(8,0);
    \node[7brane] at (2,0){};
    \node[7brane] at (4,0){};
    \node[7brane] at (6,0){};
    \node[7brane] at (8,0){};
    \node[7brane] at (-2,0){};
    \node[7brane] at (-4,0){};
    \node[7brane] at (-6,0){};
    \node[7brane] at (-8,0){};
\draw[thick](0,0)--(3,2);
\draw[thick](0,0)--(-3,2);
\node[7brane,label=above right:{\vspace{-1cm}$(2k-1,2)$}] at (3,2){};
\node[7brane,label=above left:{$(2k-1,-2)$}] at (-3,2){};

\node at (9,-.25) {$\text{O}5^-$};
\node at (7,0.35) {$\frac{1}{2}$};
\node at (5,0.35) {1};
\node at (3,0.35) {$\frac{3}{2}$};
\node at (1.5,0.35) {2};
\node at (-1.5,0.35) {2};
\node at (-9,-.25) {$\text{O}5^-$};
\node at (-7,0.35) {$\frac{1}{2}$};
\node at (-5,0.35) {1};
\node at (-3,0.35) {$\frac{3}{2}$};
    \end{tikzpicture}\end{scriptsize}
\caption[Orientifold web diagram for theory in Figure \ref{fig:appendixbrane1}.]{The orientifold web diagram for the theory given in Figure \ref{fig:appendixbrane1} at the infinitely strong coupling. }
\label{fig:appendixOSpweb1}
\end{figure}
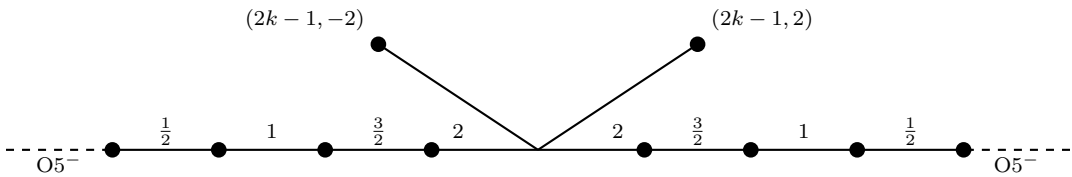

The first example we consider is the 5d theory realized on the 5-brane web diagram in Figure \ref{fig:appendixbrane1}. 
The central gauge theory is an $\text{SO}(4k+2)$ gauge theory and the diagram has an external $(2k-1,2)$ 5-brane and an external $(2k-1, -2)$ 5-brane. The parts surrounded by the dashed circles need to be properly resolved and it introduces $k-1$ faces for each part. Then the dimension of the Coulomb branch moduli space of the $5d$ theory can be counted by the number of faces in the diagram and it is 
\begin{align}\label{dimC1}
\text{dim}\mathcal{M}_C = 2k+1 + 2 + 2(k-1) = 4k + 1.
\end{align}
At the infinitely strong coupling the diagram becomes the one in figure \ref{fig:appendixOSpweb1}.
The brane web has $(2k-1, 2)$ 5-brane and the number of the charge $2$ hypermultiplets coupled to the U(1) gauge node originated from the 5-brane is 
\begin{align}\label{appendixcharge21}
\frac{8k-4}{2} - 2 = 4k-4
\end{align}
according to \eqref{SI-charge-2}. Namely the orientifold web diagram in figure \ref{fig:appendixOSpweb1} yields the orthosymplectic magnetic quiver depicted in \eqref{appendixOSp1}. 
\begin{equation}\label{appendixOSp1}
\begin{array}{c}
\begin{scriptsize}
\begin{tikzpicture}
   
            \node[label=right:{$1$}][u](u1){};
            \node[label=above:{$4k-4$}][uf](uf)[above of=u1]{};
            \node[label=below:{4}][so](so4)[below of=u1]{};
            \node[label=below:{2}][sp](sp2)[left of=so4]{};
            \node[label=below:{2}][sp](sp2')[right of=so4]{};
            \node[label=below:{2}][so](so2)[left of=sp2]{};
            \node[label=below:{2}][so](so2')[right of=sp2']{};
           
            \draw[double distance=2pt] (u1)--(so4);
            \draw (so2)--(sp2);
            \draw (sp2)--(so4);
            \draw (so2')--(sp2');
            \draw (sp2')--(so4);
            \path [draw,snake it](u1)--(uf);

            \end{tikzpicture}\end{scriptsize}
\end{array}
\end{equation}

For counting the dimension of the Higgs branch moduli space of the theory in \eqref{appendixOSp1}, we compute the number of hypermultiplets and the number of vector multiplets and then do subtraction. The number of hypermultipltes is 
\begin{align}
H = 2 + 4 + 4 + 2 + 8 + 4k-4 = 16 + 4k,
\end{align} 
and the number of vector multiplets is 
\begin{align}
V = 1 + 3 + 6 + 3 + 1 + 1 = 15.
\end{align}
Hence the dimension of the Higgs branch is given by
\begin{align}\label{dimH1}
\text{dim}\mathcal{M}_H = 16 + 4k - 15 = 4k + 1.
\end{align}
We find that \eqref{dimH1} is exactly the same as \eqref{dimC1} and this gives support for the number of charge $2$ hypermultiplets counting in \eqref{appendixcharge21}.

\begin{figure}[!ht]
    \centering
      \begin{scriptsize}
     \begin{tikzpicture}[scale=0.4]
     
    \draw[thick,dashed](1,0)--(29,0); 
    \node at (2.5,-1) {O5$^+$};
    \node at (27.5,-1) {O5$^+$};
    
    \draw[thick](6,1)--(8,1);
    \draw[thick](6.5,1.5)--(7.5,1.5);
    \draw[thick](7.5,4.4)--(7.5,1.5);
    \draw[thick](7.5,1.5)--(8,1);
    \draw[thick](8,1)--(10,0);
    \draw[thick](4,0)--(6,1);
    \draw[thick](6.5,1.5)--(6,1);
    \draw[thick](6.5,1.5)--(6.5,4.7);
    
    \draw[thick](12,0)--(14,1);
    \draw[thick](14,1)--(16,1);
    \draw[thick](14,1)--(14.5,1.5);
    \draw[thick](14.5,1.5)--(15.5,1.5);
    \draw[thick](14.5,1.5)--(14.5,2);
    \draw[thick](15.5,2)--(15.5,1.5);
    \draw[thick](15.5,1.5)--(16,1);
    \draw[thick](16,1)--(18,0);
    \draw[thick](14.5,2)--(15.5,2);
    \draw[thick](14.5,2)--(14,2.5);
    \draw[thick](15.5,2)--(16,2.5);
    \draw[thick](14,2.5)--(16,2.5);
    \draw[thick](16,2.5)--(17,3);
    \draw[thick](14,2.5)--(13,3);
    \draw[thick](13,3)--(17,3);
    \draw[thick](17,3)--(18.5,3.5);
    \draw[thick](13,3)--(11.5,3.5);
    \node at (15,4){$\vdots$};
    \draw[thick](10,5)--(20,5);
    \draw[thick](10.5,4.8)--(10,5);
    \draw[thick](19.5,4.8)--(20,5);
    \draw[thick](8.8,5.4)--(10,5);
    \draw[thick](20.6,5.2)--(20,5);

   \draw[dashed] (7.4,5.9) circle [radius=1.5];
   \draw[thick](7.4-2/1.7+0.1,5.9+2/1.7-0.1)--(7.4-2/1.7-5+0.1,5.9+2/1.7+3+0.1);
   \node at (6.5, 9){$(6k-1,-3)$};
   \draw[dashed] (22.6,5.9) circle [radius=1.5];
   \draw[thick](21.2,5.4)--(20,5);
   \draw[thick](22.6+2/1.7-0.1,5.9+2/1.7-0.1)--(22.6+2/1.7-0.1+5,5.9+2/1.7-0.1+3);
   \node at (24, 9){$(6k-1,3)$};
   \node at (19,6){$(6k-1,1)$};
   \node at (11,6){$(6k-1,-1)$};
     
    \draw[thick](23.5,1.5)--(23.5,4.7);
    \draw[thick](24,1)--(23.5,1.5);   
    \draw[thick](24,1)--(26,0);
    \draw[thick](20,0)--(22,1);
    \draw[thick](22,1)--(24,1);
    \draw[thick](22,1)--(22.5,1.5);
    \draw[thick](22.5,1.5)--(23.5,1.5);
    \draw[thick](22.5,1.5)--(22.5,4.4);

    \end{tikzpicture}
    \end{scriptsize}
    \caption[Web diagram with external $(6k-1,3)$ and $(6k-1,-3)$ 5-branes.]{The 5-brane web diagram which has an external $(6k-1,3)$ 5-brane and an external $(6k-1,-3)$ 5-brane.}
    \label{fig:appendixbrane2}
\end{figure}
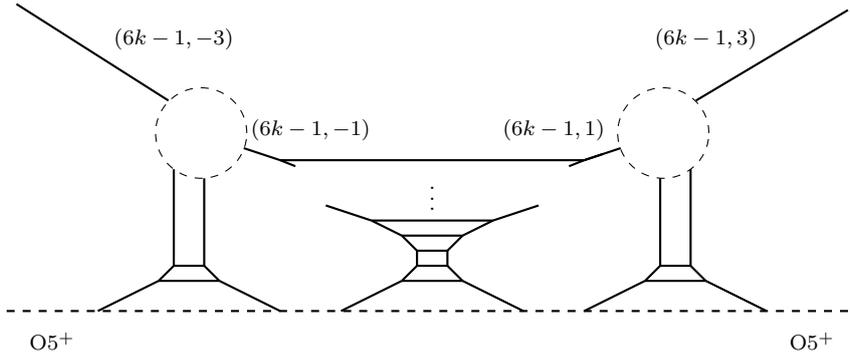

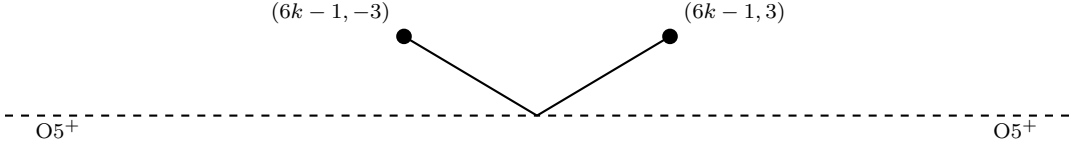
\begin{figure}[!ht]
    \centering\begin{scriptsize}
    \begin{tikzpicture}[scale=.7]
  
    \draw[thick,dashed](-10,0)--(-0,0);
    \draw[thick,dashed](10,0)--(0,0);
\draw[thick](0,0)--(5/2,3/2);
\draw[thick](0,0)--(-5/2,3/2);
\node[7brane,label=above right:{\vspace{-1cm}$(6k-1,3)$}] at (5/2,3/2){};
\node[7brane,label=above left:{$(6k-1,-3)$}] at (-5/2,3/2){};

\node at (9,-.25) {$\text{O}5^+$};
\node at (-9,-.25) {$\text{O}5^+$};

    \end{tikzpicture}\end{scriptsize}
\caption[Orientifold web diagram for the theory in figure \ref{fig:appendixbrane2}.]{The orientifold web diagram for the theory given in Figure \ref{fig:appendixbrane2} at the infinitely strong coupling. }
\label{fig:appendixOSpweb2}
\end{figure}

The first example is in the class of O5$^-$ - O5$^-$ and let us also consider an example in the class of O5$^+$ - O5$^+$. The second example we choose is the 5d theory realized on the 5-brane web diagram in figure \ref{fig:appendixbrane2}. 
The central gauge theory is now an $\text{SO}(12k+2)$ gauge theory and the diagram contains an external $(6k-1,3)$ 5-brane and an external $(6k-1,-3)$ 5-brane. The parts surrounded by the dashed circles again need to be resolved and each part in fact has $6k-2$ faces. In this case the dimension of the Coulomb branch moduli space of the 5d theory is 
\begin{align}\label{dimC2}
\text{dim}\mathcal{M}_C = 6k+1 + 2 + 6 + 2(6k-2) = 18k + 5.
\end{align}
For reading off the magnetic quiver theory we take the infinitely strong coupling limit of the diagram in figure \ref{fig:appendixbrane2} and it becomes the one in figure \ref{fig:appendixOSpweb2}. 
The $(6k-1, 3)$ 5-brane and the $(6k-1, -3)$ 5-brane yields a U(1) gauge theory and the number of charge $2$ hypermultiplets computed by \eqref{charge2O5+} is 
\begin{align}\label{appendixcharge22}
\frac{6(3k-1)}{2} -3 = 18k-6.
\end{align}
Furthermore we also expect the presence of the $3\times 4 = 12$ charge $1$ hypermutiplets which is computed by using \eqref{SI-charge-1}. Then the magentic quiver theory which arises from the diagram in figure \ref{fig:appendixOSpweb2} is given in \eqref{appendixOSp2}. 
\begin{equation}\label{appendixOSp2}
\begin{array}{c}
\begin{scriptsize}
\begin{tikzpicture}
   
            \node[label=below:{$1$}][u](u1){};
            \node[label=above:{$18k-6$}][uf](uf1)[above of=u1]{};
             \node[label=above:{$12$}][uf](uf2)[right of=u1]{};
           
            \draw (u1)--(uf2);
            \path [draw,snake it](u1)--(uf);

            \end{tikzpicture}\end{scriptsize}
\end{array}
\end{equation}

We can compute the dimension of the Higgs branch moduli space of the orthosymplectic magnetic quiver theory in \eqref{appendixOSp2} in a similar way. The number of hypermulitplets in the quiver is 
\begin{align}
H = 18k-6 + 12 = 18k + 6,
\end{align}
and the number of vector multiplets is 
\begin{align}
V = 1, 
\end{align}
which yields
\begin{align}\label{dimH2}
\text{dim}\mathcal{M}_H  = 18k + 5.
\end{align}
We again find the agreement with \eqref{dimC2} which gives another support for \eqref{SI-charge-2}.

%% file: appendixD.tex
\section{The remaining theories in the $Y_N$ family}\label{appYN}
 There are a few more possible configurations in the $Y_N^{p,q}$ family. They do not give rise to any new rules, in addition to those already mentioned so far. They do however serve as working examples, that demonstrate the validity of the rules proposed earlier. Without delving into a technical discussion, we present the web diagrams along with the corresponding magnetic quivers for the $Y_N^{p,q}$ family. We have also checked the matching of the Coulomb branch Hilbert series for the unitary and orthosymplectic quivers for each of the family. These results have been presented in various figures and tables, as summarized in the following. 
\begin{equation*}
\begin{tabular}{|c|c|c|c|c|} \hline
\rowcolor{Grayy}
Theory & Unitray web & Orientifold web & Magnetic quivers & Coulomb branch HS \\
$\text{Y}_{N}^{2,1}$ & Figure \ref{YN21unitaryweb} & Figure \ref{YN21orientifoldweb} & Table \ref{YN21magneticquivers} & Table \ref{YN21HS} \\[0.12cm]
$\text{Y}_{N}^{3,1}$ & Figure \ref{YN31unitaryweb} & Figure \ref{YN31orientifoldweb} & Figure \ref{YN31magneticquivers} & Table \ref{YN31HS} \\[0.12cm]
$\text{Y}_{N}^{2,2}$ & Figure \ref{YN22unitaryweb} & Figure \ref{YN22orientifoldweb} & Table \ref{YN22magneticquivers} & Table \ref{YN22HS} \\[0.12cm]
$\text{Y}_{N}^{2,3}$ & Figure \ref{YN23unitaryweb} & Figure \ref{YN23orientifoldweb} & Figure \ref{YN23magneticquivers} & Table \ref{YN23HS} \\ \hline
\end{tabular}
\end{equation*}
\begin{figure}[!ht]
    \centering\begin{scriptsize}
    \begin{tikzpicture}[scale=.8]
    \draw[thick](-3,2)--(-1,0);
    \draw[thick] (1,2)--(-3,-2);
    \node[7brane]at (1,2){};
    \node[7brane]at(-2.25,1.25){};
    \node[7brane]at(-2.25,-1.25){};
    \node[7brane] at (-3,2){};
    \node[7brane] at (-3,-2){};
    \node[thick] at (-3.25,2.25){$.$};
    \node[thick] at (-3.45,2.45){$.$};
    \node[thick] at (-3.35,2.35){$.$};
    \draw[thick](-3.7,2.7)--(-5.2,4.2);
    \node[7brane]at(-3.7,2.7){};
    \node[7brane]at(-4.45,3.45){};
    \node[7brane]at(-5.2,4.2){};
    \node at (-3.5,1.5) {$2N-2$};
    \node at (-4.25,3){4};
    \node at (-5,3.75){2};
    \node at(-2.75,-1.5){1};
    \node at(-2,-.75){2};
    \node at (-1.65,1.05) {$2N$};
    \node at (.5,1.3) {1};
    \draw[thick](-3.5,0)--(2,0);
    \node[7brane] at (1,0){};
    \node[7brane] at (2,0){};
    \node[7brane] at (-3.5,0){};
    \node at (0.25,0.25) {$2N+2$};
    \node at(1.5,.25){1};
    \node at (-3,0.25) {1};
    \draw[thick](-1,-2.25)--(-1,2.25);
    \node[7brane] at (-1,2.25){};
    \node[7brane] at (-1,-1.5){};
    \node[7brane] at (-1,-2.25){};
    \draw[thick](-1,-3.25)--(-1,-5.25);
    \node[7brane] at (-1,-3.25){};
    \node[7brane] at (-1,-4.25){};
    \node[7brane] at (-1,-5.25){};
    \node[thick] at (-1,-2.65) {$.$};
\node[thick] at (-1,-2.75) {$.$};
\node[thick] at (-1,-2.85) {$.$};
    \node at (-0.75,2) {$1$};
    \node at (-1.25,-4.75) {2};
    \node at (-1.25,-3.75) {4};
    \node at (-1.6,-1.85) {$2N-2$};
    \node at (-1.25,-1.2) {$2N$};    
    \end{tikzpicture}\end{scriptsize}
    \hspace{1cm}
    \begin{scriptsize}
    \begin{tikzpicture}[scale=.8]
    \node at (0,-2.5){(I)};
    
    \draw[thick](0.1,1.5)--(0.1,0.1);
    \draw[thick](-1.25,-1.25)--(0.1,0.1);
    \draw[thick](.1,.1)--(1.5,.1);
    \node[7brane] at(0.1,1.5){};
    \node[7brane] at(-.2,-1.5){};
    
    \draw[thick, red] (-1.25,1.25)--(-0.2,.2);
    \draw[thick,red](-.2,.2)--(1.5,.2);
    \draw[thick,red](-.2,.2)--(-.2,-1.5);
    \node[7brane]at (-1.25,1.25){};
    \node[7brane] at(-1.25,-1.25){};
    \draw[thick,cyan](-1.3,-1.1)--(1.2,1.4);
    \node[7brane] at (1.2,1.4){};
    \draw[thick,green](-1.5,0)--(1.5,0);
    \node[7brane] at(1.5,0.1){};
    \node[7brane] at(-1.5,0){};
    \node at (5,-2.5){(II)};
    
    \draw[thick](3.75,1.45)--(5.1,.1);
    \draw[thick](5.1,.1)--(5.1,-.1);
    \draw[thick](5.1,-.1)--(7.5,-.1);
    \draw[thick](3.85,-1.35)--(5.1,-.1);
    \draw[thick](5.1,.1)--(6.5,.1);
    \node[7brane]at(7.5,-.1){};
    \node[7brane] at(4.90,1.5){};
    
    \draw[thick, red] (3.75,1.25)--(4.8,.2);
    \draw[thick,red](4.8,.2)--(6.5,.2);
    \draw[thick,red](4.8,.2)--(4.8,-1.5);
    \draw[thick,blue](4.9,-1.5)--(4.9,1.5);
    \node[7brane] at(4.85,-1.5){};
    \node[7brane]at (3.75,1.35){};
    \node[7brane] at(3.75,-1.32){};
    \draw[thick,cyan](3.8,-1.2)--(6.3,1.3);
    \node[7brane] at (6.4,1.4){};
    \draw[thick,green](3.5,0)--(6.5,0);
    \node[7brane] at(6.5,.1){};
    \node[7brane] at(3.5,0){};
    
    \node at (0,-8){(III)};
    
    \draw[thick](-1.25,-3.55)--(0.1,-4.9);
    \draw[thick](0.1,-4.9)--(.1,-5.1);
    \draw[thick](.1,-5.1)--(2.5,-5.1);
    \draw[thick](-1.15,-6.35)--(0.1,-5.1);
    \draw[thick](.1,-4.9)--(1.5,-4.9);
    \node[7brane]at(2.5,-5.1){};
    \draw[thick,blue](0,-6.55)--(0,-4.55);
    \draw[thick,blue](0,-4.55)--(0.2,-4.55);
    \draw[thick,blue](0.2,-4.55)--(1.2,-3.55);
    \draw[thick,blue](0.2,-4.55)--(0.2,-7.5);
    \draw[thick,blue](0,-4.55)--(-1,-3.55);
    \node[7brane]at(0.2,-7.5){};
    \draw[thick, red] (-1.25,-3.75)--(-0.2,-4.8);
    \draw[thick,red](-.2,-4.8)--(1.5,-4.8);
    \draw[thick,red](-.2,-4.8)--(-.2,-6.5);
    \node[7brane] at(-.1,-6.5){};
    \node[big7brane]at (-1.15,-3.6){};
    \node[7brane] at(-1.25,-6.32){};
    \draw[thick,cyan](-1.2,-6.2)--(.3,-4.7);
    \draw[thick,cyan](.3,-4.7)--(.3,-3.5);
    \draw[thick,cyan](.3,-4.7)--(1.4,-4.7);
     \node[7brane] at(0.3,-3.5){};
    \node[7brane] at (1.2,-3.6){};
    \draw[thick,green](-1.5,-5)--(1.5,-5);
    \node[big7brane] at(1.5,-4.85){};
    \node[7brane] at(-1.5,-5){};
    \node at (5,-7.5){(IV)};
    
    \draw[thick,cyan](3.75,-3.55)--(5.1,-4.9);
    \draw[thick,cyan](5.1,-4.9)--(5.1,-5.1);
    \draw[thick,cyan](5.1,-5.1)--(6.5,-5.1);
    \draw[thick,cyan](3.85,-6.35)--(5.1,-5.1);
    \draw[thick,cyan](5.1,-4.9)--(6.5,-4.9);
    \draw[thick, red] (3.75,-3.75)--(4.8,-4.8);
    \draw[thick,red](4.8,-4.8)--(6.5,-4.8);
    \draw[thick,red](4.8,-4.8)--(4.8,-6.5);
    \draw[thick,green](4.9,-6.5)--(4.9,-3.5);
    \node[7brane] at(4.9,-3.5){};
    \node[7brane] at(4.85,-6.5){};
    \node[7brane]at (3.75,-3.65){};
    \node[7brane] at(3.8,-6.4){};
    \draw[thick,cyan](3.8,-5.9)--(6.2,-3.7);
    \node[7brane]at(3.8,-5.9){};
    \node[7brane] at (6.2,-3.7){};
    \draw[thick](3.5,-5)--(6.5,-5);
    \node[big7brane] at(6.5,-4.95){};
    \node[7brane] at(3.5,-5){};
    \end{tikzpicture}
    \end{scriptsize}
    \caption[Ordinary web diagram for $\text{Y}_{N}^{2,1}$ theory.]{Ordinary web for the $\text{Y}_{N}^{2,1}$ theory along with its maximal subdivisions.}
    \label{YN21unitaryweb}
\end{figure}
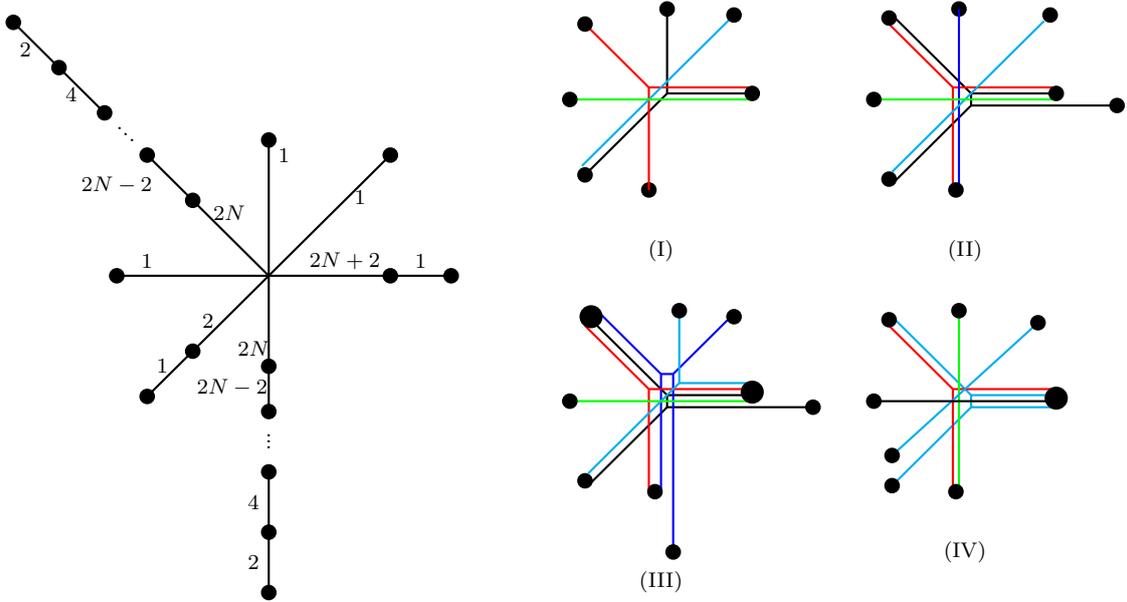
\begin{figure}[!ht]
    \centering
    \begin{scriptsize}
       \begin{tikzpicture}[scale=.75]
       \draw[thick](-5.5,0)--(5.5,0);
       \draw[thick,dashed](-7,0)--(-5.5,0);
       \draw[thick,dashed](7,0)--(5.5,0);
       \node[7brane] at (5.5,0){};
       \node[7brane] at (-5.5,0){};
       \node[7brane] at (4,0){};
       \node[7brane] at (2.5,0){};
       \node[7brane] at (1,0){};
       \node[7brane] at (-3,0){};
       \node[thick] at (-.5,3.4) {$.$};
       \node[thick] at (-.5,3.5) {$.$};
       \node[thick] at (-.5,3.6) {$.$};
       \node[7brane]at (-.5,2){};
       \node[7brane]at (-.5,3){};
       \node[7brane]at (-.5,4){};
       \node[7brane]at (-.5,5){};
       \node[7brane]at (-.5,6){};
       \draw[thick](-.5,0)--(-.5,3);
       \draw[thick](-.5,4)--(-.5,6);
       \draw[thick](-.5,0)--(-2.5,1);
       \draw[thick](-.5,0)--(.75,1.25);
       \node at (-1,5.5){1};
       \node at (-1,4.5){2};
       \node at (-1.25,2.5){$2N-1$};
       \node at (-1,1.5){$2N$};
       
       \node[label=above right:{$(1,1)$}][7brane]at(.75,1.25){};
       \node[label=above left:{$(2,-1)$}][7brane]at(-2.5,1){};
       \node at (-6.25,-.25){O$5^-$};
       \node at (6.25,-.25){O$5^-$};
       \node at (-4,-.3){$\frac{1}{2}$};
       \node at (-1.5,-.25){1};
       \node at (.5,-.25) {2};
       \node at (1.75,-.25){$\frac{3}{2}$};
       \node at (3.25,-.25){$1$};
       \node at (4.75,-.25){$\frac{1}{2}$};
       
       \draw[thick] (3,4.2)--(7,4.2);
       \draw[thick,red](3,3)--(5,4);
       \draw[thick,red](5,4)--(7,4);
       \draw[thick,red](5,4)--(6.25,5.25);
       \draw[thick,blue](5,2.5)--(5,5.5);
       \node[7brane]at(5,5.5){};
       \node[7brane]at(5,2.5){};
       \node[7brane]at(7,4.1){};
       \node[7brane]at(3,4.2){};
       \node[7brane]at(6.25,5.25){};
       \node[7brane]at(3,3){};
       \node at (5,1.5){(I)};
       \end{tikzpicture}
       
       \vspace{.5 cm}
       \begin{tikzpicture}[scale=.7]
       \draw[thick] (-2,0.2)--(2,0.2);
       \draw[thick,red](-2,-1)--(0,0);
       \draw[thick,red](0.2,0)--(2,0);
       \draw[thick,red](0.2,0)--(1.45,-1.25);
       \draw[thick,red](0,0)--(0,1.5);
       \draw[thick,red](0,0)--(0.2,0);
       \draw[thick,red](0.2,0)--(0.2,1.5);
       \draw[thick,blue](0.1,-1.50)--(0.1,1.5);
       \node[7brane]at(0.1,1.5){};
       \node[7brane]at(0.1,-1.5){};
       \node[7brane]at(2,.1){};
       \node[7brane]at(-2,.2){};
       \node[7brane]at(1.45,-1.25){};
       \node[7brane]at(-2,-1){};
       \node at (0,-2.5){(II)};
       \draw[thick,green](4,-1)--(6,0);
       \draw[thick,green](6,0)--(8,0);
       \draw[thick,green](6,0.2)--(8,0.2);
       \draw[thick,red](6.2,0)--(7.45,-1.25);
       \draw[thick,green](6,0)--(6,2);
       \draw[thick,red](4,0)--(6.2,0);
       \draw[thick,red](6.2,0)--(6.2,1.5);
       \draw[thick,green](6.1,-1.50)--(6.1,1.5);
       \node[7brane]at(6.2,1.5){};
       \node[7brane]at(6.1,2){};
       \node[7brane]at(6.1,-1.5){};
       \node[7brane]at(8,.1){};
       \node[7brane]at(7.45,-1.25){};
       \node[7brane]at(4,-1){};
       \node[7brane]at(4,0){};
       \node at (6,-2.5){(III)};
       
       \draw[thick,green](10,-1)--(12,0);
       \draw[thick,green](12,0)--(15,0);
       \draw[thick,green](12,0.2)--(14,0.2);
       \draw[thick,red](12.2,0)--(13.45,-1.25);
       \draw[thick,green](12,0)--(12,1.5);
       \draw[thick,red](10,0)--(12.2,0);
       \draw[thick,red](12.2,0)--(12.2,1.5);
       \draw[thick,blue](12.1,-1.50)--(12.1,1.5);
       \node[7brane]at(12.1,1.5){};
       \node[7brane]at(12.1,-1.5){};
       \node[7brane]at(14,.2){};
       \node[7brane]at(13.45,-1.25){};
       \node[7brane]at(10,-1){};
       \node[7brane]at(15,0){};
       \node[7brane]at(10,0){};
       \node at (12,-2.5){(IV)};

       \end{tikzpicture}
    \end{scriptsize}
    \caption[Orientifold web diagram for $\text{Y}_{N}^{2,1}$ theory.]{Orientifold web for the $\text{Y}_{N}^{2,1}$ theory along with its maximal subdivisions.}
    \label{YN21orientifoldweb}
		\end{figure}
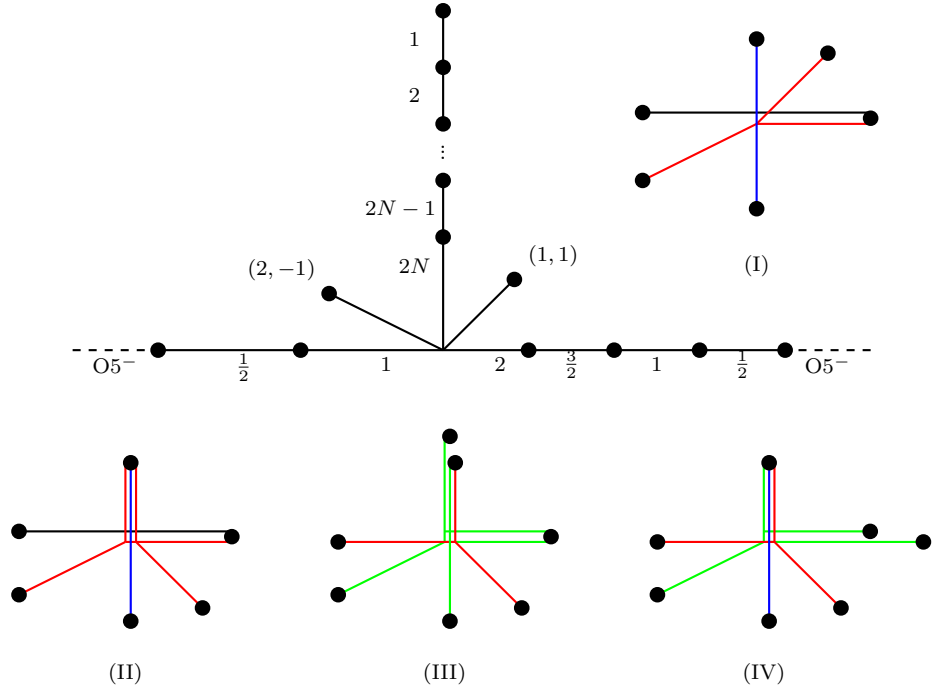
\begin{table}[t]
    \centering
    \begin{tabular}{|c|C{7.5cm}|C{4.8cm}|}\hline
		\rowcolor{Grayy}
    MS&Unitary magnetic&Orthosymplectic magnetic\\\hline
         (I)&\begin{scriptsize}
            \begin{tikzpicture}
            \node[label=below:{2}][u](u2)at (0,0){};
            \node[label=below:{4}][u](u4)[right of=u2]{};
            \node[label=below:{$2N-2$}][u](u2k-2)[right of=u4]{};
            \node[label=below:{$2N$}][u](u2k)[right of=u2k-2]{};
            \node[label=below:{$2N-2$}][u](u2k-2')[right of=u2k]{};
            \node[label=below:{$4$}][u](u4')[right of=u2k-2']{};
            \node[label=below:{$2$}][u](u2')[right of=u4']{};
            \draw(u2)--(u4);
            \draw[dotted](u4)--(u2k-2);
            \draw(u2k-2)--(u2k);
            \draw(u2k)--(u2k-2');
            \draw[dotted](u2k-2')--(u4');
            \draw(u4')--(u2');
            \node[label=left:{1}][u](u11)[above of=u4]{};
            \node[label=above:{1}][u](yellow)[right of=u11]{};
            \node[label=right:{1}][u](brown)[above of=u4']{};
            \node[label=above:{1}][u](u11')[left of=brown]{};
            \node[label=above:{1}][u](orange)[above of=u2k]{};
            \draw(u2k)--(u11);
       \draw(u11)--(yellow);
       \draw(orange)--(u11');
       \draw(u11')--(brown);
       \draw(brown)--(u2k);
       \draw[double distance=2pt](u2k)--(orange);
       \draw(yellow)--(orange);
       \draw(brown)to[out=135, in=45](u11);
            \end{tikzpicture}
         \end{scriptsize}  &\begin{scriptsize}
            \begin{tikzpicture}
            \node[label=below:{1}][u](u1)at(0,0){};
            \node[label=below:{$2N-1$}][u](u2k-1)[right of=u1]{};
            \node[label=below:{$2N$}][sp](sp2k)[right of=u2k-1]{};
            \node[label=below:{2}][so](so2)[right of=sp2k]{};
            \node[label =above:{1}][u](u1')[above of=sp2k]{};
            \node[label=above:{1}][uf](1f)[left of=u1']{};
            \node[label=above:{2}][sp](sp2)[above of=so2]{};
            \node[label=above:{2}][so](so2')[right of=sp2]{};
            \draw[dotted](u1)--(u2k-1);
            \draw(u2k-1)--(sp2k);
            \draw(sp2k)--(so2);
            \draw(so2)--(sp2);
            \draw[double distance=2pt](sp2k)--(u1');
            \path [draw,snake it](u1')--(1f);
            \draw(sp2)--(so2');
            \draw(sp2)--(u1');
            \end{tikzpicture}
         \end{scriptsize}\\\hline
        (II) &\begin{scriptsize}
            \begin{tikzpicture}
            \node[label=below:{2}][u](u2)at (0,0){};
            \node[label=below:{4}][u](u4)[right of=u2]{};
            \node[label=below:{$2N-2$}][u](u2k-2)[right of=u4]{};
            \node[label=below:{$2N-1$}][u](u2k-1)[right of=u2k-2]{};
            \node[label=below:{$2N-2$}][u](u2k-2')[right of=u2k-1]{};
            \node[label=below:{$4$}][u](u4')[right of=u2k-2']{};
            \node[label=below:{$2$}][u](u2')[right of=u4']{};
            \draw(u2)--(u4);
            \draw[dotted](u4)--(u2k-2);
            \draw(u2k-2)--(u2k);
            \draw(u2k)--(u2k-2');
            \draw[dotted](u2k-2')--(u4');
            \draw(u4')--(u2');
            \node[label=above:{1}][u](yellow)[above of=u2k-1]{};
            \node[label=right:{1}][u](pink)[above of=u2k-2']{};
            \node[label=above:{1}][u](u11')[above left of=pink]{};
            \node[label=left:{1}][u](orange)[above of=u2k-2]{};
            \node[label=above:{1}][u](u11)[above right of=orange]{};
       \draw(u11)--(yellow);
       \draw(orange)--(u11);
       \draw(u11')--(pink);
       \draw[double distance=2pt](yellow)--(u2k-1);
       \draw(yellow)--(orange);
       \draw(pink)--(u11');
       \draw(yellow)--(u11');
       \draw[double distance=2pt](orange)to[out=90,in=90,controls=+(90:2) and +(90:2)](pink);
       \draw(yellow)--(pink);
       \draw(u2k-2)--(orange);
       \draw(u2k-2')--(pink);
            \end{tikzpicture}
         \end{scriptsize}&\begin{scriptsize}
            \begin{tikzpicture}
            \node[label=below:{1}][u](u1)at(0,0){};
            \node[label=below:{$2N-1$}][u](u2k-1)[right of=u1]{};
            \node[label=below:{$2N-2$}][sp](sp2k)[right of=u2k-1]{};
            \node[label=below:{2}][so](so2)[right of=sp2k]{};
            \node[label =above:{1}][u](u1')[above of=u2k-1]{};
            \node[label=above:{2}][uf](1f)[left of=u1']{};
            \node[label=above:{2}][sp](sp2)[above of=so2]{};
            \node[label=above:{2}][so](so2')[right of=sp2]{};
            \draw[dotted](u1)--(u2k-1);
            \draw(u2k-1)--(sp2k);
            \draw(sp2k)--(so2);
            \draw(so2)--(sp2);
            \draw[double distance=2pt](u2k-1)--(u1');
            \draw(so2)--(u1');
            \path [draw,snake it](u1')--(1f);
            \draw(sp2)--(so2');
            \draw(sp2)--(u1');
            \end{tikzpicture}
         \end{scriptsize}\\\hline
        (III)&\begin{scriptsize}
            \begin{tikzpicture}
            \node[label=below:{2}][u](u2)at (0,0){};
            \node[label=below:{4}][u](u4)[right of=u2]{};
            \node[label=below:{$2N-2$}][u](u2k-2)[right of=u4]{};
            \node[label=below:{$2N-2$}][u](u2k-2')[right of=u2k-2]{};
            \node[label=below:{$2N-3$}][u](u2k-3)[right of=u2k-2']{};
            \node[label=below:{$2N-4$}][u](u2k-4)[right of=u2k-3]{};
            \node[label=below:{$4$}][u](u4')[right of=u2k-4]{};
            \node[label=below:{$2$}][u](u2')[right of=u4']{};
            \draw(u2)--(u4);
            \draw[dotted](u4)--(u2k-2);
            \draw(u2k-2)--(u2k);
            \draw(u2k)--(u2k-2');
            \draw[dotted](u2k-4)--(u4');
            \draw(u4')--(u2');
            \draw(u2k-3)--(u2k-2');
            \draw(u2k-3)--(u2k-4);
            \node[label=left:{1}][u](black)[above of=u2k-2]{};
            \node[label=above right:{1}][u](blue)[above of=u2k-4]{};
            \node[label=above:{1}][u](cyan)[above of=blue]{};
            \node[label=above:{1}][u](green)[right of=blue]{};
            \node[label=left:{1}][u](u1)[above of=black]{};
            \draw[double distance=3pt](blue)--(black);
            \draw(blue)--(black);
            \draw(blue)--(u2k-2);
            \draw(blue)--(u2k-4);
            \draw(black)--(u1);
            \draw(u1)--(cyan);
            \draw(black)--(u2k-2);
            \draw[double distance=3pt](cyan)--(blue);
            \draw(cyan)--(blue);
            \draw[double distance=2pt](blue)--(green);
            \draw(cyan)--(u2k-2');
            \end{tikzpicture}
         \end{scriptsize}&\begin{scriptsize}
         \begin{tikzpicture}
         \node[label=below:{1}][u](1){};
         \node[label=below:{$2N-3$}][u](2n-3)[right of=1]{};
         \node[label=below:{$2N-2$}][u](2n-2)[right of=2n-3]{};
         \node[label=below:{$2N-2$}][u](2n-2')[right of=2n-2]{};
         \node[label=below:{$2N-4$}][sp](sp2n-4)[right of=2n-2']{};
         \node[label=right:{$1$}][u](u1)[above of=2n-2']{};
         \node[label=above left:{$1$}][u](u1')[above of=2n-2]{};
         \node[label=right:{$2$}][sp](sp2)[above of=u1']{};
         \node[label=right:{$2$}][so](so2)[above of=sp2]{};
         \draw[dotted](1)--(2n-3);
         \draw(2n-3)--(2n-2);
         \draw(2n-2)--(2n-2');
         \draw(2n-2')--(sp2n-4);
         \draw(u1')--(2n-2');
         \draw[dashed](u1')--(2n-2);
         \draw(u1)--(2n-2');
         \draw[double distance=3pt](u1)to[out=150,in=30](u1');
         \draw[dashed,double distance=3pt](u1)to[out=-150,in=-30](u1');
         \draw[dashed](u1)to[out=-150,in=-30](u1');
         \draw(u1)to[out=150,in=30](u1');
         \draw[double distance=2pt](u1')--(sp2);
         \draw(sp2)--(so2);
         \end{tikzpicture}
         \end{scriptsize}\\\hline
              (IV)&\begin{scriptsize}
            \begin{tikzpicture}
            \node[label=above:{2}][u](u2)at (0,0){};
            \node[label=below:{4}][u](u4)[right of=u2]{};
            \node[label=below:{$2N-2$}][u](u2k-2)[right of=u4]{};
            \node[label=below:{$2N-1$}][u](u2k-1)[right of=u2k-2]{};
            \node[label=below:{$2N-2$}][u](u2k-2')[right of=u2k-1]{};
            \node[label=below:{$4$}][u](u4')[right of=u2k-2']{};
            \node[label=below:{$2$}][u](u2')[right of=u4']{};
            \draw(u2)--(u4);
            \draw[dotted](u4)--(u2k-2);
            \draw(u2k-2)--(u2k-1);
            \draw(u2k-1)--(u2k-2');
            \draw[dotted](u2k-2')--(u4');
            \draw(u4')--(u2');
            \node[label=left:{1}][u](cyan)[above of=u2k-2]{};
            \node[label=above right:{1}][u](green)[above of=u2k-2']{};
            \node[label=above:{1}][u](black)[above of=green]{};
            \node[label=left:{1}][u](u1)[above of=cyan]{};
            \draw(u2k-2)--(cyan);
            \draw(cyan)--(u2k-1);
            \draw[double distance=2pt](cyan)--(u1);
            \draw[double distance=3pt](cyan)--(green);
            \draw(cyan)--(green);
            \draw(green)--(black);
            \draw(u1)--(black);
            \draw(green)--(u2k-2');
            \draw(u1)--(u2k-1);
            \end{tikzpicture}
         \end{scriptsize}&\begin{scriptsize}
            \begin{tikzpicture}
            \node[label=left:{1}][u](u1)at(0,0){};
            \node[label=left:{$2N-1$}][u](u2k-1)[below of=u1]{};
            \node[label=left:{$2N-2$}][sp](sp2k)[below of=u2k-1]{};
            \node[label =above:{1}][u](u1')[right of=u1]{};
            \node[label=below:{1}][u](u11)[right of=sp2k]{};
            \node[label=right:{2}][so](so2)[right of=u11]{};
            \draw[dotted](u1)--(u2k-1);
            \draw(u11)--(so2);
            \draw(u2k-1)--(u1');
            \draw(u2k-1)--(u11);
            \draw(u2k-1)--(sp2k);
            \draw(sp2k)--(u11);
            \draw[double distance=2pt](u11)to[out=30,in=-30](u1');
            \draw[dashed, double distance=3pt](u11)to[out=120, in=-120](u1');
            \draw[dashed](u11)to[out=120, in=-120](u1');
            \end{tikzpicture}
         \end{scriptsize}\\\hline 
    \end{tabular}
    \caption[Magnetic quivers for $\text{Y}_{N}^{2,1}$ theory.]{The magnetic quivers derived from various maximal subdivisions (MS) of the unitary and orientifold webs of the $\text{Y}_{N}^{2,1}$ theory.}
    \label{YN21magneticquivers}
\end{table}
\begin{table}[!htb]
\centering
\begin{tabular}{|c|C{3.9cm}|C{3.9cm}|C{3.9cm}|} \hline
\rowcolor{Grayy}
   & Unitary & \multicolumn{2}{c|}{Orthosymplectic} \\ \cline{2-4}
	\rowcolor{Grayy}
  \multirow{-2}{*}{MS} & HS($t$)  & HS($t;\vec{m} \in \mathbb{Z}$) & HS($t;\vec{m} \in \mathbb{Z}+\tfrac{1}{2}$)  \\ \hline
	$(\text{I})_{N=1}$ & \footnotesize{$1+11t+91t^2+552t^3+2654t^4+10598t^5+\cdots$}  & \footnotesize{$1+11t+75t^2+392t^3+1710t^4+6422t^5+\cdots$} & \footnotesize{$16t^2+160t^3+944t^4+4176t^5+\cdots$} \\ \hline
	$(\text{I})_{N=2}$ & \footnotesize{$1+23t+289t^2+2638t^3+19566t^4+124453t^5+\cdots$}  & \footnotesize{$1+23t+289t^2+2590t^3+18510t^4+111893t^5+\cdots$} & \footnotesize{$48t^3+1056t^4+12560t^5+\cdots$} \\ \hline
$(\text{II})_{N=1}$ & \footnotesize{$1+11t+79t^2+405t^3+1644t^4+5572t^5+\cdots$}  & \footnotesize{$1+11t+71t^2+325t^3+1196t^4+3764t^5+\cdots$} & \footnotesize{$8t^2+80t^3+448t^4+1808t^5+\cdots$} \\ \hline
$(\text{II})_{N=2}$ & \footnotesize{$1+23t+289t^2+2622t^3+19178t^4+119256t^5+\cdots$}  & \footnotesize{$1+23t+289t^2+2590t^3+18474t^4+110936t^5+\cdots$} & \footnotesize{$32t^3+704t^4+8320t^5+\cdots$} \\ \hline
$(\text{III})_{N=2}$ & \footnotesize{$1+23t+273t^2+2255t^3+14595t^4+78621t^5+\cdots$}  & \footnotesize{$1+23t+273t^2+2215t^3+13795t^4+70381t^5+\cdots$} & \footnotesize{$40t^3+800t^4+8240t^5+\cdots$} \\ \hline
$(\text{III})_{N=3}$ & \footnotesize{$xx$}  & \footnotesize{$xx$} & \footnotesize{$xx$} \\ \hline
$(\text{IV})_{N=1}$ & \footnotesize{$1+8t+46t^2+184t^3+599t^4 + 1648t^5+\cdots$}  & \footnotesize{$1 + 8 t + 34 t^2 + 108 t^3 + 323 t^4 + 872 t^5+\cdots$} & \footnotesize{$12 t^2 + 76 t^3 + 276 t^4 + 776 t^5+\cdots$} \\ \hline
$(\text{IV})_{N=2}$ & \footnotesize{$1 + 20 t + 224 t^2 + 1843 t^3 + 12276 t^4 + 69526 t^5+\cdots$}  & \footnotesize{$1 + 20 t + 224 t^2 + 1803 t^3 + 11516 t^4 + 61590 t^5+\cdots$} & \footnotesize{$40 t^3 + 760 t^4 + 7936 t^5+\cdots$} \\ \hline
\end{tabular}
\caption[Coulomb branch HS for $\text{Y}_{N}^{2,1}$ theory.]{Coulomb branch Hilbert series for the magnetic quivers of the $Y_{N}^{2,1}$ theory. The corresponding unitary and orthosymplectic quivers are given in table \ref{YN21magneticquivers}.}
\label{YN21HS}
\end{table}
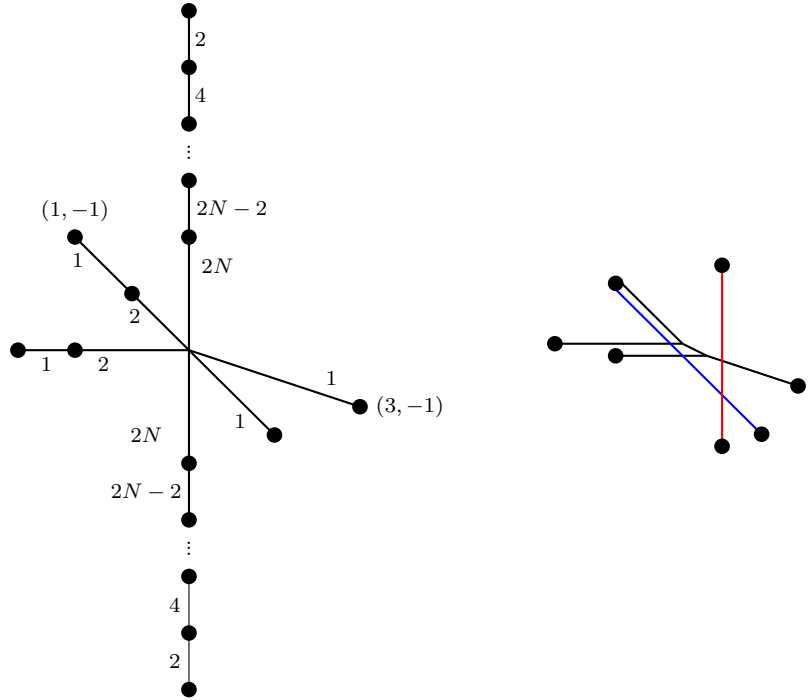
\begin{figure}[!htb]
    \centering
        \begin{scriptsize}
    \begin{tikzpicture}[scale=.75]
\draw[thick,](0,8)--(1.5,6.5);
\node[7brane] at (1.5,6.5) {};
\draw[thick](0,8)--(3,7);
\node[label=right:{$(3,-1)$}][7brane]at(3,7){};
\draw[thick](0,8)--(0,5);
\node[7brane]at(0,6){};
\node[7brane]at(0,5){};
\node at (0,4.6){$.$};
\node at (0,4.5){$.$};
\node at (0,4.4){$.$};
\draw(0,4)--(0,2);
\node[7brane]at(0,4){};
\node[7brane]at(0,3){};
\node[7brane]at(0,2){};
\node at (-0.25,2.5) {$2$};
\node at (-0.25,3.5) {$4$};
\node at (-0.75,5.5) {$2N-2$};
\node at (-0.75,6.5) {$2N$};
\draw[thick](-3,8)--(0,8);
\node at (2.5,7.5) {$1$};
\node at (.9,6.75) {$1$};
\node at (0.2,13.5) {$2$};
\node at (0.2,12.5) {$4$};
\node at (.75,10.5) {$2N-2$};
\node at (0.5,9.5) {$2N$};
\node at (-2.5,7.75) {$1$};
\node at (-1.5,7.75) {$2$};
\node[7brane] at (-2,8) {};
\node[7brane] at (-3,8) {};

\draw[thick](0,8)--(-2,10);
\node[7brane] at (-1,9){};
\node[label=above:{$(1,-1)$}][7brane] at (-2,10){};
\node at (-.95,8.6){2};
\node at (-1.95,9.6){1};

\draw[thick](0,8)--(0,11);
\node[thick][7brane] at (0,10) {};
\node[thick][7brane] at (0,11) {};
\node at (0,11.4){$.$};
\node at (0,11.5){$.$};
\node at (0,11.6){$.$};
\draw[thick](0,12)--(0,14);
\node[7brane]at(0,12){};
\node[7brane]at(0,13){};
\node[7brane]at(0,14){};
        \end{tikzpicture}\end{scriptsize}
        \hspace{1 cm}
        \begin{tikzpicture}[scale=.8]
        \draw[thick](0,0)--(-1.5,.5);
        \draw[thick](-1.9,0.7)--(-1.5,.5);
        \draw[thick](-1.9,0.7)--(-4,.7);
        \draw[thick](-3,0.5)--(-1.5,.5);
        \draw[thick](-1.9,0.7)--(-2.9,1.7);
        \draw[thick,blue](-3.1,1.7)--(-.6,-.8);
        \draw[thick,red](-1.25,-1)--(-1.25,2);
        \node[7brane] at (-4,.7){};
        \node[7brane] at (-3,.5){};
        \node[7brane] at (-1.25,-1){};
        \node[7brane] at (-1.25,2){};
        \node[7brane] at (-.6,-.8){};
        \node[7brane] at (-3,1.7){};
        \node[7brane] at (0,0){};
        \node at(0,-5){};
        \end{tikzpicture}
    \caption[Unitary web diagram for $\text{Y}_{N}^{3,1}$ theory.]{Ordinary web diagram for the $\text{Y}_{N}^{3,1}$ theory along with the maximal subdivision for the centre of the junction.}
    \label{YN31unitaryweb}
\end{figure}
\begin{figure}[!htb]
    \centering
        \begin{scriptsize}
       \begin{tikzpicture}[scale=.8]
       \draw[thick](-.5,0)--(5.5,0);
       \draw[thick,dashed](-.5,0)--(-4,0);
       \draw[thick,dashed](7,0)--(5.5,0);
       \node[7brane] at (5.5,0){};
       \node[7brane] at (4,0){};
       \node[7brane] at (2.5,0){};
       \node[7brane] at (1,0){};
       \node[thick] at (-.5,3.4) {$.$};
       \node[thick] at (-.5,3.5) {$.$};
       \node[thick] at (-.5,3.6) {$.$};
       \node[7brane]at (-.5,2){};
       \node[7brane]at (-.5,3){};
       \node[7brane]at (-.5,4){};
       \node[7brane]at (-.5,5){};
       \node[7brane]at (-.5,6){};
       \draw[thick](-.5,0)--(-.5,3);
       \draw[thick](-.5,4)--(-.5,6);
       \draw[thick](-.5,0)--(-3.5,1);
       \draw[thick](-.5,0)--(.75,1.25);
       \node at (-1,5.5){1};
       \node at (-1,4.5){2};
       \node at (-1.25,2.5){$2N-1$};
       \node at (-1,1.5){$2N$};
       
       \node[label=above right:{$(1,1)$}][7brane]at(.75,1.25){};
       \node[label=above left:{$(3,-1)$}][7brane]at(-3.5,1){};
       \node at (-3.25,-.25){O$5^-$};
       \node at (6.25,-.25){O$5^-$};
       \node at (.5,-.25) {2};
       \node at (1.75,-.25){$\frac{3}{2}$};
       \node at (3.25,-.25){$1$};
       \node at (4.75,-.25){$\frac{1}{2}$};
       \end{tikzpicture}
       \hspace{.5cm}
       \begin{tikzpicture}[scale=.8]
       \draw[thick](0,0)--(1.5,-.5);
       \draw[thick](1.5,-.5)--(1.9,-.7);
       \draw[thick](1.9,-.7)--(2.9,-.7);
       \draw[thick](1.5,-.5)--(2.5,-.5);
       \draw[thick](1.9,-.7)--(3,-1.7);
       \draw[thick,red](1.25,-2)--(1.25,1);
       \node[7brane]at(0,0){};
       \node[7brane]at(1.25,-2){};
       \node[7brane]at(1.25,1){};
       \node[7brane]at(2.9,-1.7){};
       \node[7brane]at(3,-.7){};
       \node[7brane]at(2.5,-.5){};
       \node at (0,-3){};
       \end{tikzpicture}
       \end{scriptsize}
       \caption[Orientifold web diagram for $\text{Y}_{N}^{3,1}$ theory.]{Orientifold web for the $\text{Y}_{N}^{3,1}$ theory along with the maximal subdivision at the centre of the junction.}
    \label{YN31orientifoldweb}
\end{figure}
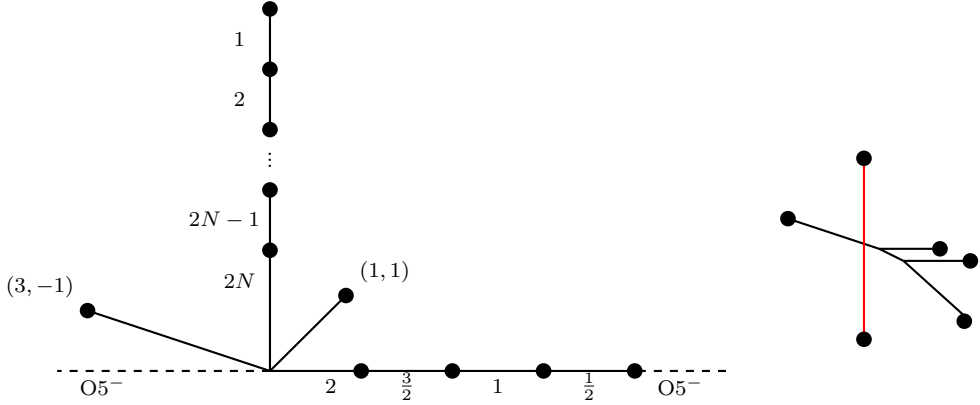
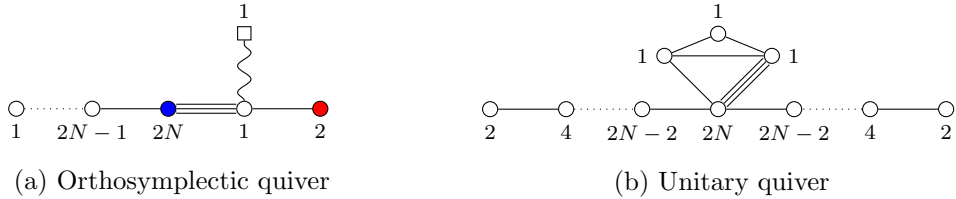
\begin{figure}[!htb]
\centering
   \begin{subfigure}{0.47\linewidth} \centering
     \begin{scriptsize}
    \begin{tikzpicture}
    \node[label=below:{1}][u](1){};
    \node[label=below:{$2N-1$}][u](2n-1)[right of=1]{};
    \node[label=below:{$2N$}][sp](sp2n)[right of=2n-1]{};
    \node[label=below:{1}][u](u1)[right of=sp2n]{};
    \node[label=above:{1}][uf](uf)[above of=u1]{};
    \node[label=below:{2}][so](so2)[right of=u1]{};
    \draw[dotted](1)--(2n-1);
    \draw(2n-1)--(sp2n);
    \draw[double distance=3pt](sp2n)--(u1);
    \draw(sp2n)--(u1);
    \draw(u1)--(so2);
    \path [draw,snake it](u1)--(uf);
    \end{tikzpicture}
    \end{scriptsize}
    \caption{Orthosymplectic quiver}\label{YN31OSp}
    \end{subfigure} 
   \begin{subfigure}{0.47\linewidth} \centering
     \begin{scriptsize}
    \begin{tikzpicture}
    \node[label=below:{2}][u](2){};
    \node[label=below:{4}][u](4)[right of=2]{};
    \node[label=below:{$2N-2$}][u](2n-2)[right of=4]{};
    \node[label=below:{$2N$}][u](2n)[right of=2n-2]{};
    \node[label=below:{$2N-2$}][u](2n-2')[right of=2n]{};
    \node[label=below:{4}][u](4')[right of=2n-2']{};
    \node[label=below:{2}][u](2')[right of=4']{};
    \node[label=left:{1}][u](u1)[above left of=2n]{};
    \node[label=right:{1}][u](u1')[above right of=2n]{};
    \node[label=above:{1}][u](u11)[above of=2n]{};
    \draw(2)--(4);
    \draw[dotted](4)--(2n-2);
    \draw(2n-2)--(2n);
    \draw(2')--(4');
    \draw[dotted](4')--(2n-2');
    \draw(2n-2')--(2n);
    \draw(2n)--(u1);
    \draw[double distance=3pt](u1')--(2n);
    \draw(u1')--(2n);
    \draw(u1)--(u1');
    \draw(u1)--(u11);
    \draw(u1')--(u11);
    \end{tikzpicture}
    \end{scriptsize}
    \caption{Unitary quiver}\label{YN31unitary}
   \end{subfigure} 
\caption{The magnetic quivers for the $\text{Y}_{N}^{3,1}$ theory.} \label{YN31magneticquivers}
\end{figure}
\begin{table}[!htb]
\centering
\begin{tabular}{|c|C{3.9cm}|C{3.9cm}|C{3.9cm}|} \hline
\rowcolor{Grayy}
   & Unitary & \multicolumn{2}{c|}{Orthosymplectic} \\ \cline{2-4}
	\rowcolor{Grayy}
  \multirow{-2}{*}{Theory} & HS($t$)  & HS($t;\vec{m} \in \mathbb{Z}$) & HS($t;\vec{m} \in \mathbb{Z}+\tfrac{1}{2}$)  \\ \hline
	$Y_{1}^{3,1}$ & \footnotesize{$1 + 7 t + 38 t^2 + 145 t^3 + 463 t^4 + 1252 t^5+\cdots$}  & \footnotesize{$1 + 7 t + 30 t^2 + 97 t^3 + 279 t^4 + 716 t^5+\cdots$} & \footnotesize{$8 t^2 + 48 t^3 + 184 t^4 + 536 t^5+\cdots$} \\ \hline
	$Y_{2}^{3,1}$ & \footnotesize{$1 + 19 t + 204 t^2 + 1603 t^3 + 10173 t^4 + 54879 t^5+\cdots$}  & \footnotesize{$1 + 19 t + 204 t^2 + 1579 t^3 + 9741 t^4 + 50543 t^5+\cdots$} & \footnotesize{$24 t^3 + 432 t^4 + 4336 t^5+\cdots$} \\ \hline
\end{tabular}
\caption[Coulomb branch HS for magnetic quivers of $\text{Y}_{N}^{3,1}$ theory.]{Coulomb branch Hilbert series for the magnetic quivers in the figure \ref{YN31magneticquivers} for the $Y_{N}^{3,1}$ theory.}
\label{YN31HS}
\end{table}
\begin{figure}[!htb]
    \centering    \centering\begin{scriptsize}
    \begin{tikzpicture}[scale=.7]
\draw[thick,](-2.25,10.25)--(+2.25,5.75);
\node[thick][7brane] at (1.5,6.5) {};
\node[thick][7brane] at (-1.5,9.5) {};
\node[thick][7brane] at (2.25,5.75) {};
\node[thick][7brane] at (-2.25,10.25) {};
\node at (-2.5,10.5) {.};
\node at (-2.6,10.6) {.};
\node at (-2.7,10.7) {.};
\node[thick][7brane] at (-2.95,10.95) {};
\node[thick][7brane] at (-3.7,11.7) {};
\node[thick][7brane] at (-4.45,12.45) {};
\draw[thick](-2.95,10.95)--(-4.45,12.45);
\node at (-3.8,12.25) {$2$};
\node at (-3,11.5) {$4$};
\node at (-1.3,10.2) {$2N-2$};

\node at (2.5,5.5) {.};
\node at (2.6,5.4) {.};
\node at (2.7,5.3) {.};
\node[thick][7brane] at (2.95,5.05) {};
\node[thick][7brane] at (3.7,4.3) {};
\node[thick][7brane] at (4.45,3.55) {};
\draw[thick](2.95,5.05)--(4.45,3.55);
\node at (3.8,3.75) {$2$};
\node at (3,4.5) {$4$};
\node at (1.3,5.8) {$2N-2$};
\node at (.8,6.8) {$2N$};

\draw[thick](-2,8)--(2,8);
\node at (1.5,7.75) {$1$};
\node[thick][7brane] at (-2,8) {};
\node[thick][7brane] at (2,8) {};

\draw[thick](-1.5,6.5)--(1.5,9.5);
\node[7brane]at (-1.5,6.5){};
\node[7brane]at (1.5,9.5){};

\draw[thick](0,6)--(0,10);
\node[thick][7brane] at (0,6) {};
\node[thick][7brane] at (0,10) {};
        \end{tikzpicture}\end{scriptsize}
        \hspace{1cm}
    \begin{tikzpicture}[scale=.7]
    \draw[thick](0,1.5)--(0,-1.5);
    \node[7brane] at(0,1.5){};
    \node[7brane] at(0,-1.5){};
    
    \draw[thick, red] (-1.25,1.25)--(1.25,-1.25);
    \node[7brane]at (-1.25,1.25){};
    \node[7brane] at(1.25,-1.25){};
    
        \draw[thick, green] (-1.25,-1.25)--(1.25,1.25);
    \node[7brane]at (-1.25,-1.25){};
    \node[7brane] at(1.25,1.25){};
    
    \draw[thick,blue](-1.5,0)--(1.5,0);
    \node[7brane] at(1.5,0){};
    \node[7brane] at(-1.5,0){};
    
    \node at (0,-2.5) {(I)};
      \draw[thick](5,1.5)--(5,0);
     \draw[thick](5,0)--(6.5,0);
     \draw[thick](3.75,-1.25)--(5,0);
    \node[7brane] at(5,1.5){};
    
    \draw[thick, red] (3.75,1.25)--(6.25,-1.25);
    
        \draw[thick, green] (5,0)--(6.25,1.25);
        \draw[thick, green] (5,0)--(3.5,0);
        \draw[thick, green] (5,0)--(5,-1.5);
    \node[7brane]at (3.75,-1.25){};
    \node[7brane] at(6.25,1.25){};
    \node[7brane] at(5,-1.5){};
    \node[7brane] at(6.5,0){};
    \node[7brane] at(3.5,0){};
    \node[7brane] at(6.25,-1.25){};
    \node[7brane]at (3.75,1.25){};
    \node at (5,-2.5) {(II)};
    \draw[thick] (-1.5,-5)--(0,-5);
    \draw[thick](0,-5)--(0,-3.5);
    \draw[thick] (0,-5)--(1.25,-6.25);
    \draw[thick,green](1.35,-6.1)--(-1.15,-3.6);
    \draw[thick,cyan](-1.25,-6.25)--(1.25,-3.75);
    \node[7brane]at(-1.25,-6.25){};
    \node[7brane]at(1.25,-3.75){};
    \node[7brane] at (1.3,-6.2){};
    \node[7brane] at (-1.5,-5){};
    \node[7brane] at (0,-3.5){};
    
    \draw[thick,red] (0,-6.5)--(0,-5);
    \draw[thick,red] (0,-5)--(1.5,-5);
    \draw[thick,red] (0,-5)--(-1.25,-3.75);
    \node[7brane] at (-1.2,-3.7){};
    \node[7brane] at (1.5,-5){};
    \node[7brane] at (0,-6.5){};
    \node at (0,-7.5) {(III)};
    
\end{tikzpicture}
    \caption[Unitary web diagram for $\text{Y}_{N}^{2,2}$ theory.]{Ordinary web for the $\text{Y}_{N}^{2,2}$ theory at the fixed point, along with the possible maximal subdivisions of the centre of the junction.}
    \label{YN22unitaryweb}
\end{figure}
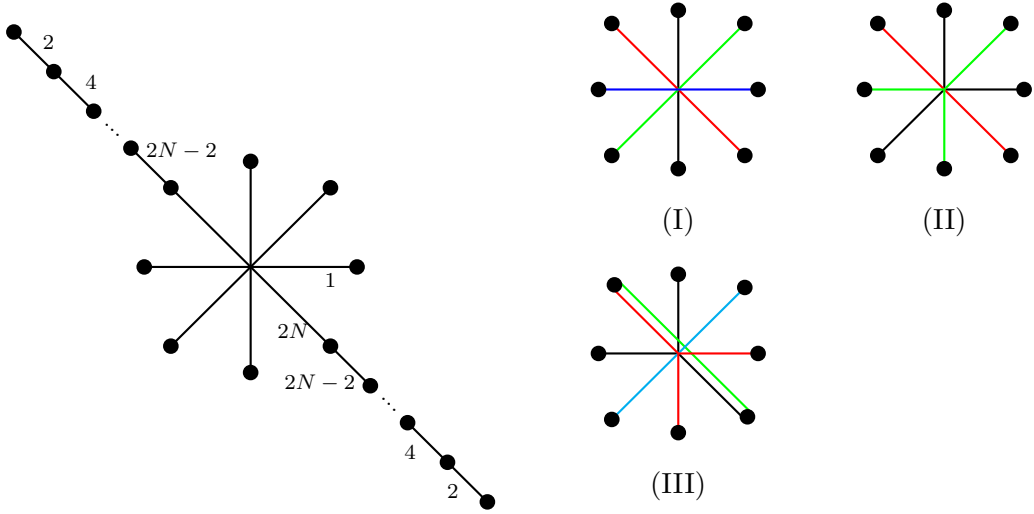
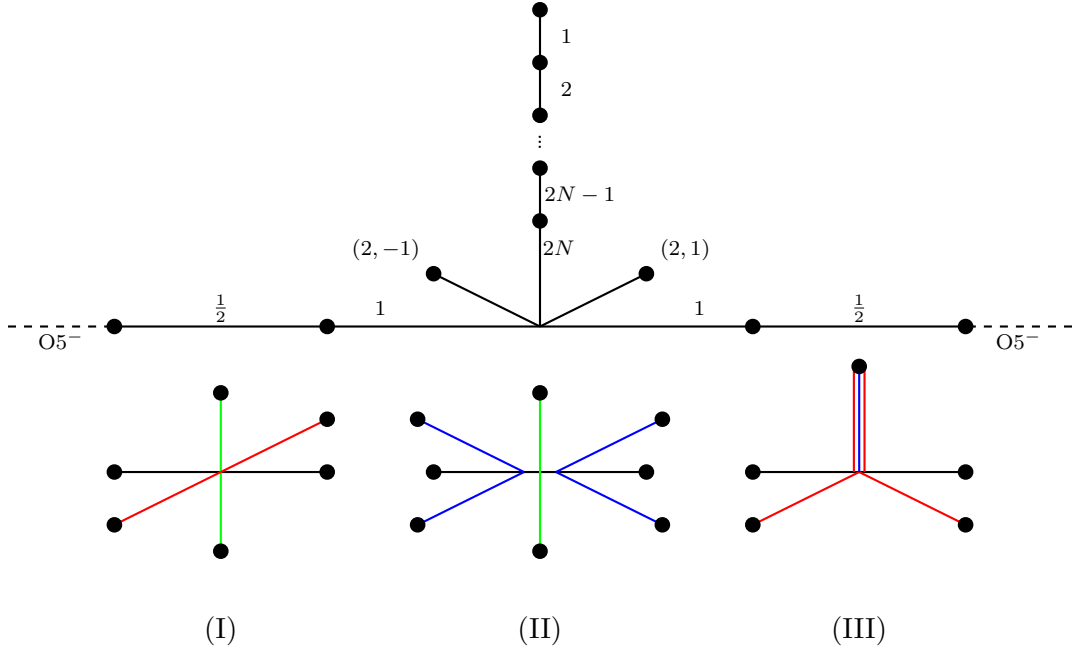
\begin{figure}[!htb]
    \centering\begin{scriptsize}
    \begin{tikzpicture}[scale=.7]
    \draw[thick](-8,0)--(8,0);
    \draw[thick,dashed](-10,0)--(-8,0);
    \draw[thick,dashed](10,0)--(8,0);
    \node[7brane] at (4,0){};
    \node[7brane] at (8,0){};
    \node[7brane] at (-4,0){};
    \node[7brane] at (-8,0){};
    \draw[thick](0,0)--(0,3);
    \node[7brane] at (0,2){};
    \node[7brane] at (0,3){};
    \node[thick] at (0,3.4) {$.$};
\node[thick] at (0,3.5) {$.$};
\node[thick] at (0,3.6) {$.$};
\node[7brane] at (0,4){};
\node[7brane] at (0,5){};
\node[7brane] at (0,6){};
\node at (.5,5.5) {1};
\node at (.5,4.5) {2};
\node at (.75,2.5) {$2N-1$};
\node at (.35,1.5) {$2N$};
\draw[thick](0,4)--(0,6);
\draw[thick](0,0)--(2,1);
\draw[thick](0,0)--(-2,1);
\node[7brane,label=above right:{$(2,1)$}] at (2,1){};
\node[7brane,label=above left:{$(2,-1)$}] at (-2,1){};

\node at (9,-.25) {$\text{O}5^-$};
\node at (6,0.35) {$\frac{1}{2}$};
\node at (3,0.35) {$1$};
\node at (-9,-.25) {$\text{O}5^-$};
\node at (-6,0.35) {$\frac{1}{2}$};
\node at (-3,0.35) {$1$};
\end{tikzpicture}\end{scriptsize}
    \vspace{1cm}
    \begin{tikzpicture}[scale=.7]
\draw[thick](-2,-7)--(2,-7);
\draw[thick,green](0,-8.5)--(0,-5.5);
\draw[thick,red](-2,-8)--(0,-7);
\draw[thick,red](2,-6)--(0,-7);
\node[7brane]at (-2,-8){};
\node[7brane]at (2,-6){};
\node[7brane]at (0,-5.5){};    
\node[7brane]at (0,-8.5){};    
\node[7brane]at (2,-7){};
\node[7brane]at (-2,-7){};
\node at (0,-10){(I)};
\draw[thick](4,-7)--(8,-7);
\draw[thick,green](6,-8.5)--(6,-5.5);
\draw[thick,blue](3.7,-6)--(5.7,-7);
\draw[thick,blue](3.7,-8)--(5.7,-7);
\draw[thick,blue](8.3,-6)--(6.3,-7);
\draw[thick,blue](8.3,-8)--(6.3,-7);
\node[7brane]at (3.7,-6){};
\node[7brane]at (3.7,-8){};
\node[7brane]at (8.3,-6){};
\node[7brane]at (8.3,-8){};
\node[7brane]at (6,-5.5){};
\node[7brane]at (6,-8.5){};    
\node[7brane]at (8,-7){};
\node[7brane]at (4,-7){};
\node at (6,-10){(II)};
\draw[thick](10,-7)--(14,-7);
\draw[thick,red](12.1,-7)--(12.1,-5);
\draw[thick,red](11.9,-7)--(11.9,-5);
\draw[thick,red](10,-8)--(12,-7);
\draw[thick,red](14,-8)--(12,-7);
\draw[thick,blue](12,-7)--(12,-5);
\node[7brane]at (10,-8){};
\node[7brane]at (14,-8){};
\node[7brane]at (12,-5){};    
\node[7brane]at (14,-7){};
\node[7brane]at (10,-7){};
\node at (12,-10){(III)};
\end{tikzpicture}
\caption[Orientifold web diagram for $\text{Y}_{N}^{2,2}$ theory.]{Orientifold web for the $\text{Y}_{N}^{2,2}$ theory along with its maximal subdivisions. In the subdivision marked as (II), the blue and red subweb are related to the discrete theta angle from O5 plane and are thus immobile.}
    \label{YN22orientifoldweb}
\end{figure}
\begin{table}[!htb]
    \centering
		\begin{tabular}{|c|C{7.5cm}|C{4.8cm}|}\hline
			\rowcolor{Grayy}
        MS&Unitary magnetic & Orthosymplectic magnetic \\
        \hline (I)&\begin{scriptsize}
        \begin{tikzpicture}
        \node[label=below:{2}][u](1){};
        \node[label=below:{4}][u](3)[right of=1]{};
        \node[label=below:{$2N-2$}][u](2n-3)[right of=3]{};
        \node[label=below:{$2N$}][u](2n-1)[right of=2n-3]{};
        \node[label=below:{$2N-2$}][u](2n-3')[right of=2n-1]{};
        \node[label=below:{$4$}][u](3')[right of=2n-3']{};
        \node[label=below:{$2$}][u](1')[right of=3']{};
        \node[label=above:{1}][u](green)[above of=2n-1]{};
        \node[label=above:{1}][u](black)[left  of=green]{};
        \node[label=above:{1}][u](blue)[right of=green]{};
        \draw(1)--(3);
        \draw[dotted](3)--(2n-3);
        \draw(2n-3)--(2n-1);
        \draw(2n-3')--(2n-1);
        \draw(1')--(3');
        \draw[dotted](3')--(2n-3');
        \draw[double distance=2pt](2n-1)--(green);
        \draw(2n-1)--(black);
        \draw(green)--(black);
        \draw(green)--(blue);
        \draw(black)to[out=45,in=135](blue);
        \draw(blue)--(2n-1);
        \end{tikzpicture}
        \end{scriptsize} & \begin{scriptsize}
        \begin{tikzpicture}
        \node[label=below:{1}][u](1){};
        \node[label=below:{$2N-1$}][u](2n-1)[right of=1]{};
        \node[label=below:{$2N$}][sp](sp2n)[right of=2n-1]{};
        \node[label=below:{2}][so](so2)[right of=sp2n]{};
        \node[label=above:{1}][u](u1)[above of =sp2n]{};
        \node[label=above:{1}][uf](1f)[left of =u1]{};
        \draw[dotted](1)--(2n-1);
        \draw(2n-1)--(sp2n);
        \draw(sp2n)--(so2);
        \draw(so2)--(u1);
        \draw[double distance=2pt](sp2n)--(u1);
        \path [draw,snake it](u1)--(1f);
        \end{tikzpicture}
        \end{scriptsize}\\\hline
        (II)&
        \begin{scriptsize}
        \begin{tikzpicture}
         \node[label=below:{2}][u](1){};
        \node[label=below:{4}][u](3)[right of=1]{};
        \node[label=below:{$2N-2$}][u](2n-3)[right of=3]{};
        \node[label=below:{$2N$}][u](2n-1)[right of=2n-3]{};
        \node[label=below:{$2N-2$}][u](2n-3')[right of=2n-1]{};
        \node[label=below:{$4$}][u](3')[right of=2n-3']{};
        \node[label=below:{$2$}][u](1')[right of=3']{};
        \node[label=above:{1}][u](green)[above left of =2n-1]{};
        \node[label=above:{1}][u](cyan)[above right of =2n-1]{};
        \draw(1)--(3);
        \draw[dotted](3)--(2n-3);
        \draw(2n-3)--(2n-1);
        \draw(1')--(3');
        \draw[dotted](3')--(2n-3');
        \draw(2n-3')--(2n-1);
        \draw[double distance=2pt](2n-1)--(cyan);
        \draw[double distance=2pt](2n-1)--(green);
        \draw[double distance=2pt](green)--(cyan);
        \end{tikzpicture}
        \end{scriptsize}&\begin{scriptsize}
        \begin{tikzpicture}
        \node[label=below:{1}](1)[u]{};
        \node[label=below:{$2N-1$}][u](2n-1)[right of=1]{};
        \node[label=below:{$2N$}][sp](sp2n)[right of=2n-1]{};
        \node[label=below:{2}][so](so2)[right of=sp2n]{};
        \node[label=above:{4}][sof](sof)[above of=sp2n]{};
        \node[label=above:{2}][spf](spf)[above of=so2]{};
        \draw(1)[dotted]--(2n-1);
        \draw(2n-1)--(sp2n);
        \draw(sp2n)--(so2);
        \draw(sp2n)--(sof);
        \draw(so2)--(spf);
        \end{tikzpicture}
        \end{scriptsize}\\\hline
        (III)&   \begin{scriptsize}
        \begin{tikzpicture}
         \node[label=below:{2}][u](1){};
        \node[label=below:{4}][u](3)[right of=1]{};
        \node[label=below:{$2N-2$}][u](2n-3)[right of=3]{};
        \node[label=below:{$2N-1$}][u](2n-1)[right of=2n-3]{};
        \node[label=below:{$2N-2$}][u](2n-3')[right of=2n-1]{};
        \node[label=below:{$4$}][u](3')[right of=2n-3']{};
        \node[label=below:{$2$}][u](1')[right of=3']{};
        \node[label=above:{1}][u](cyan)[above of =2n-1]{};
        \node[label=left:{1}][u](black)[above of =2n-3]{};
        \node[label=right:{1}][u](red)[above of =2n-3']{};
        \draw(1)--(3);
        \draw[dotted](3)--(2n-3);
        \draw(2n-3)--(2n-1);
        \draw(1')--(3');
        \draw[dotted](3')--(2n-3');
        \draw(2n-3')--(2n-1);
        \draw[double distance=2pt](2n-1)--(cyan);
        \draw[double distance=2pt](red)to[out=120, in=60](black);
        \draw[double distance=2pt](red)--(cyan);
        \draw[double distance=2pt](black)--(cyan);
        \draw(red)--(2n-3');
        \draw(black)--(2n-3);
        \end{tikzpicture}
        \end{scriptsize}&\begin{scriptsize}
        \begin{tikzpicture}
        \node[label=below:{1}][u](1){};
        \node[label=below:{$2N-1$}][u](2n-1)[right of=1]{};
        \node[label=below:{$2N-2$}][sp](sp2n-2)[right of=2n-1]{};
        \node[label=above:{1}][u](u1)[above of=2n-1]{};
        \node[label=left:{2}][uf](uf)[left of=u1]{};
        \node[label=above:{2}][so](so2)[above of=sp2n-2]{};
        \draw[dotted](1)--(2n-1);
        \draw(2n-1)--(sp2n-2);
        \draw[double distance=2pt](2n-1)--(u1);
        \path [draw,snake it](u1)--(uf);
        \draw(so2)--(sp2n-2);
        \draw[double distance=2pt](so2)--(u1);
        \end{tikzpicture}
        \end{scriptsize}\\\hline
    \end{tabular}
    \caption[Magnetic quivers for $\text{Y}_{N}^{2,2}$ theory.]{The magnetic quivers derived from the various maximal subdivisions (MS) of the unitary and orientifold web diagrams of the $\text{Y}_{N}^{2,2}$ theory.}
    \label{YN22magneticquivers}
\end{table}
\begin{table}[!htb]
\centering
\begin{tabular}{|c|C{3.9cm}|C{3.9cm}|C{3.9cm}|} \hline
\rowcolor{Grayy}
   & Unitary & \multicolumn{2}{c|}{Orthosymplectic} \\ \cline{2-4}
	\rowcolor{Grayy}
  \multirow{-2}{*}{MS} & HS($t$)  & HS($t;\vec{m} \in \mathbb{Z}$) & HS($t;\vec{m} \in \mathbb{Z}+\tfrac{1}{2}$)  \\ \hline
	$\text{(I)}_{N=1}$ & \footnotesize{$1 + 5 t + 28 t^2 + 105 t^3 + 339 t^4 + 920 t^5+\cdots$}  & \footnotesize{$1 + 5 t + 20 t^2 + 65 t^3 + 195 t^4 + 512 t^5+\cdots$} & \footnotesize{$8 t^2 + 40 t^3 + 144 t^4 + 408 t^5+\cdots$} \\ \hline
	$\text{(I)}_{N=2}$ & \footnotesize{$1 + 17 t + 168 t^2 + 1233 t^3 + 7427 t^4 + 38575 t^5+\cdots$}  & \footnotesize{$1 + 17 t + 168 t^2 + 1209 t^3 + 7019 t^4 + 34671 t^5+\cdots$} & \footnotesize{$24 t^3 + 408 t^4 + 3904 t^5+\cdots$} \\ \hline
	$\text{(II)}_{N=1}$ & \footnotesize{$1 + 4 t + 15 t^2 + 45 t^3 + 110 t^4 + 239 t^5+\cdots$}  & \footnotesize{$1 + 4 t + 15 t^2 + 45 t^3 + 110 t^4 + 239 t^5+\cdots$} & \footnotesize{not required} \\ \hline
	$\text{(II)}_{N=2}$ & \footnotesize{$1 + 16 t + 151 t^2 + 1041 t^3 + 5810 t^4 + 27652 t^5+\cdots$}  & \footnotesize{$1 + 16 t + 151 t^2 + 1041 t^3 + 5810 t^4 + 27652 t^5+\cdots$} & \footnotesize{not required} \\ \hline
	$\text{(III)}_{N=1}$ & \footnotesize{$1 + 5 t + 20 t^2 + 60 t^3 + 151 t^4 + 331 t^5+\cdots$}  & \footnotesize{$1 + 5 t + 16 t^2 + 40 t^3 + 91 t^4 + 191 t^5+\cdots$} & \footnotesize{$4 t^2 + 20 t^3 + 60 t^4 + 140 t^5+\cdots$} \\ \hline
	$\text{(III)}_{N=2}$ & \footnotesize{$1 + 17 t + 168 t^2 + 1225 t^3 + 7255 t^4 + 36626 t^5+\cdots$}  & \footnotesize{$1 + 17 t + 168 t^2 + 1209 t^3 + 6983 t^4 + 34050 t^5+\cdots$} & \footnotesize{$16 t^3 + 272 t^4 + 2576 t^5+\cdots$} \\ \hline
\end{tabular}
\caption[Coulomb branch HS for magnetic quivers of $\text{Y}_{N}^{2,2}$ theory.]{Coulomb branch Hilbert series for magnetic quivers of the $Y_{N}^{2,2}$ theory. The corresponding unitary and orthosymplectic quivers are given in table \ref{YN22magneticquivers}.}
\label{YN22HS}
\end{table}
\begin{figure}[!htb]
    \centering
    \begin{scriptsize}
        \begin{tikzpicture}[scale=.7]
        \draw[thick,](-2.25,10.25)--(+2.25,5.75);
\node[thick][7brane] at (1.5,6.5) {};
\node[thick][7brane] at (-1.5,9.5) {};
\node[thick][7brane] at (2.25,5.75) {};
\node[thick][7brane] at (-2.25,10.25) {};
\node at (-2.5,10.5) {.};
\node at (-2.6,10.6) {.};
\node at (-2.7,10.7) {.};
\node[thick][7brane] at (-2.95,10.95) {};
\node[thick][7brane] at (-3.7,11.7) {};
\node[thick][7brane] at (-4.45,12.45) {};
\draw[thick](-2.95,10.95)--(-4.45,12.45);
\node at (-3.8,12.25) {$2$};
\node at (-3,11.5) {$4$};
\node at (-1.3,10.2) {$2N-2$};

\node at (2.5,5.5) {.};
\node at (2.6,5.4) {.};
\node at (2.7,5.3) {.};
\node[thick][7brane] at (2.95,5.05) {};
\node[thick][7brane] at (3.7,4.3) {};
\node[thick][7brane] at (4.45,3.55) {};
\draw[thick](2.95,5.05)--(4.45,3.55);
\node at (3.8,3.75) {$2$};
\node at (3,4.5) {$4$};
\node at (1.3,5.8) {$2N-2$};
\node at (.8,6.8) {$2N$};

\draw[thick](-2,8)--(0,8);
\node[thick][7brane] at (-2,8) {};

\draw[thick](-1.5,6.5)--(0,8);
\node[7brane]at (-1.5,6.5){};
\draw[thick](0,8)--(2,9);
\node[7brane][label=above right:{$(2,1)$}]at (2,9){};

\draw[thick](0,6)--(0,10);
\node[thick][7brane] at (0,6) {};
\node[thick][7brane] at (0,10) {};
        \end{tikzpicture}
    \end{scriptsize}
    \caption[Unitary web diagram for $\text{Y}_{N}^{2,3}$ theory.]{Unitary web description of the $Y_N^{2,3}$ theory.}
    \label{YN23unitaryweb}
\end{figure}
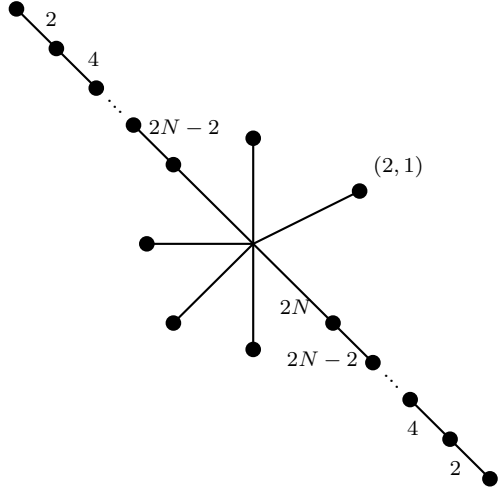
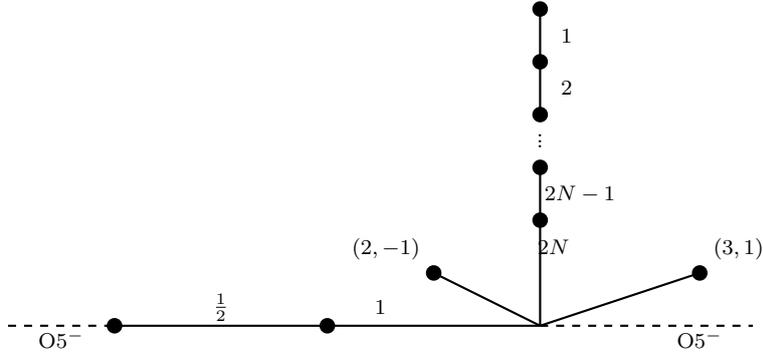
\begin{figure}[!htb]
    \centering
    \begin{scriptsize}
        \begin{tikzpicture}[scale=.7]
         \draw[thick](-8,0)--(0,0);
    \draw[thick,dashed](-10,0)--(-8,0);
    \draw[thick,dashed](0,0)--(4,0);
    \node[7brane] at (-4,0){};
    \node[7brane] at (-8,0){};
    \draw[thick](0,0)--(0,3);
    \node[7brane] at (0,2){};
    \node[7brane] at (0,3){};
    \node[thick] at (0,3.4) {$.$};
\node[thick] at (0,3.5) {$.$};
\node[thick] at (0,3.6) {$.$};
\node[7brane] at (0,4){};
\node[7brane] at (0,5){};
\node[7brane] at (0,6){};
\node at (.5,5.5) {1};
\node at (.5,4.5) {2};
\node at (.75,2.5) {$2N-1$};
\node at (.25,1.5) {$2N$};
\draw[thick](0,4)--(0,6);
\draw[thick](0,0)--(3,1);
\draw[thick](0,0)--(-2,1);
\node[7brane,label=above right:{$(3,1)$}] at (3,1){};
\node[7brane,label=above left:{$(2,-1)$}] at (-2,1){};

\node at (3,-.25) {$\text{O}5^-$};
\node at (-9,-.25) {$\text{O}5^-$};
\node at (-6,0.35) {$\frac{1}{2}$};
\node at (-3,0.35) {$1$};

        \end{tikzpicture}
    \end{scriptsize}
    \caption[Orientifold web diagram for $\text{Y}_{N}^{2,3}$ theory.]{Orientifold web description of the $Y_N^{2,3}$ theory.}
    \label{YN23orientifoldweb}
\end{figure}
\begin{figure}[!htb]
\centering
   \begin{subfigure}{0.49\linewidth} \centering
     \begin{scriptsize}
    \begin{tikzpicture}
    \node[label=below:{1}][u](1){};
    \node[label=below:{$2N-1$}][u](2n-1)[right of=1]{};
    \node[label=below:{$2N$}][sp](sp2n)[right of=2n-1]{};
    \node[label=below:{1}][u](u1)[right of=sp2n]{};
    \node[label=above:{2}][uf](uf)[above of=u1]{};
    \draw[dotted](1)--(2n-1);
    \draw(2n-1)--(sp2n);
    \draw[double distance=3pt](sp2n)--(u1);
    \draw(sp2n)--(u1);
    \path [draw,snake it](u1)--(uf);    
    \end{tikzpicture}
    \end{scriptsize}
     \caption{Orthosymplectic quiver}\label{YN23OSp}
   \end{subfigure}
   \begin{subfigure}{0.49\linewidth} \centering
     \begin{scriptsize}
    \begin{tikzpicture}
    \node[label=below:{2}][u](2){};
    \node[label=below:{4}][u](4)[right of=2]{};
    \node[label=below:{$2N-2$}][u](2n-2)[right of=4]{};
    \node[label=below:{$2N$}][u](2n)[right of=2n-2]{};
    \node[label=below:{$2N-2$}][u](2n-2')[right of=2n]{};
    \node[label=below:{4}][u](4')[right of=2n-2']{};
    \node[label=below:{2}][u](2')[right of=4']{};
    \node[label=left:{1}][u](u1)[above left of=2n]{};
    \node[label=right:{1}][u](u1')[above right of=2n]{};
    \draw(2)--(4);
    \draw[dotted](4)--(2n-2);
    \draw(2n-2)--(2n);
    \draw(2')--(4');
    \draw[dotted](4')--(2n-2');
    \draw(2n-2')--(2n);
    \draw(2n)--(u1);
    \draw[double distance=3pt](u1')--(2n);
    \draw(u1')--(2n);
    \draw[double distance=2pt](u1)--(u1');
    \end{tikzpicture}
    \end{scriptsize}
     \caption{Unitary quiver}\label{YN23unitary}
   \end{subfigure}
\caption{The magnetic quivers for the $\text{Y}_{N}^{2,3}$ theory.} \label{YN23magneticquivers}
\end{figure}
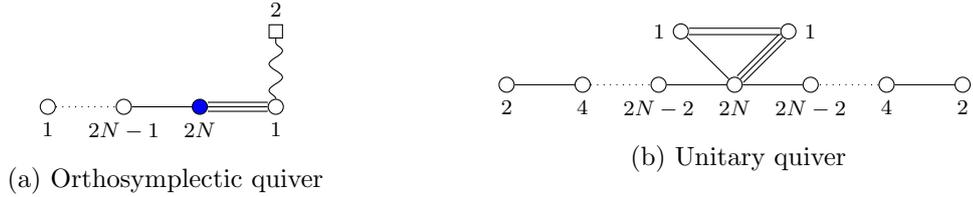
\begin{table}[!htb]
\centering
\begin{tabular}{|c|C{3.9cm}|C{3.9cm}|C{3.9cm}|} \hline
\rowcolor{Grayy}
   & Unitary & \multicolumn{2}{c|}{Orthosymplectic} \\ \cline{2-4}
	\rowcolor{Grayy}
  \multirow{-2}{*}{Theory} & HS($t$)  & HS($t;\vec{m} \in \mathbb{Z}$) & HS($t;\vec{m} \in \mathbb{Z}+\tfrac{1}{2}$)  \\ \hline
	$Y_{1}^{2,3}$ & \footnotesize{$1 + 4 t + 17 t^2 + 47 t^3 + 120 t^4 + 255 t^5+\cdots$}  & \footnotesize{$1 + 4 t + 13 t^2 + 31 t^3 + 72 t^4 + 147 t^5+\cdots$} & \footnotesize{$4 t^2 + 16 t^3 + 48 t^4 + 108 t^5+\cdots$} \\ \hline
	$Y_{2}^{2,3}$ & \footnotesize{$1 + 16 t + 151 t^2 + 1051 t^3 + 5940 t^4 + 28640 t^5+\cdots$}  & \footnotesize{$1 + 16 t + 151 t^2 + 1039 t^3 + 5748 t^4 + 26892 t^5+\cdots$} & \footnotesize{$12 t^3 + 192 t^4 + 1748 t^5+\cdots$} \\ \hline
\end{tabular}
\caption[Coulomb branch HS for magnetic quivers of $\text{Y}_{N}^{2,3}$ theory.]{Coulomb branch Hilbert series for the magnetic quivers in the figure \ref{YN23magneticquivers} for the $Y_{N}^{2,3}$ theory.}
\label{YN23HS}
\end{table}